\documentclass[a4paper,12pt]{article}
\pdfoutput=1
\usepackage{epsfig,amsmath,amsfonts,amsthm}
\usepackage[figuresright]{rotating}
\usepackage[left=28mm,right=20mm,top=30mm,bottom=20mm]{geometry}
\usepackage{subfig}
\usepackage{graphicx}
\graphicspath{{MD_new_figures_with_WT_name_20112018/}}
\usepackage{rotating}
\usepackage{booktabs}
\usepackage{moreverb}

\DeclareGraphicsExtensions{.pdf,.png,.jpg}
\usepackage{natbib}
\numberwithin{equation}{section}

\def\e{\hbox{E}}

\def\var{\hbox{Var}}

\def\max{\hbox{max}}

\providecommand{\keywords}[1]{\textbf{{Keywords}} #1}
\usepackage[nokeyprefix]{refstyle}
\usepackage{varioref}
\usepackage{xr}
\usepackage{hyperref}

\begin{document}

\title{Simulation study of estimating between-study variance and overall effect  in meta-analyses of mean difference}

\author{Ilyas Bakbergenuly, David C. Hoaglin  and Elena Kulinskaya}

\date{\today}

\maketitle



\abstract{Methods for random-effects meta-analysis require an estimate of the between-study variance, $\tau^2$. The performance of estimators of $\tau^2$ (measured by bias and coverage) affects their usefulness in assessing heterogeneity of study-level effects, and also the performance of related estimators of the overall effect. For the effect measure mean difference (MD), we review five point estimators of $\tau^2$ (the popular methods of DerSimonian-Laird, restricted maximum likelihood, and Mandel and Paule (MP); the less-familiar method of Jackson; and a new method (WT) based on the improved approximation to the distribution of the $Q$ statistic by \cite{kulinskaya2004welch}), five interval estimators for $\tau^2$ (profile likelihood, Q-profile, Biggerstaff and Jackson, Jackson, and the new WT method), six point estimators of the overall effect (the five related to the point estimators of $\tau^2$ and an estimator whose weights use only study-level sample sizes), and eight interval estimators for the overall effect (five based on the point estimators for $\tau^2$, the Hartung-Knapp-Sidik-Jonkman (HKSJ) interval, a modification of HKSJ, and an interval based on the sample-size-weighted estimator).  We obtain empirical evidence from extensive simulations and an example.}


\section{Introduction}\label{sec:Intro}
Meta-analysis is a statistical methodology for combining estimated effects from several studies in order to assess their heterogeneity and obtain an overall estimate. In this paper we focus on mean difference as the effect measure.

If the studies can be assumed to have the same true effect, a meta-analysis can use a fixed-effect (FE) model (common-effect model) to combine the estimates. Otherwise, the studies' true effects can depart from homogeneity in a variety of ways. Most commonly, a random-effects (RE) model regards those effects as a sample from a distribution and summarizes their heterogeneity via its variance, usually denoted by $\tau^2$. 
The between-studies variance, $\tau^2$, has a key role in estimates of the mean of the distribution of random effects; but it is also important as a quantitative indication of heterogeneity \citep{higgins2009re}.
 In studying estimation for meta-analysis of MD, we focus first on $\tau^2$ and then proceed to the overall effect.

\cite{veroniki2015methods} provide a comprehensive overview and recommendations  on methods of estimating $\tau^2$ and its uncertainty. Their review, however, has two important limitations. First, the authors study only \lq\lq methods that can be applied for any type of outcome data.'' However, the performance of the methods  varies widely among effect measures. \cite{veroniki2015methods} mention this only in passing (in Section 6.1) as a hypothetical possibility. Second, any review on the topic, such as \cite{veroniki2015methods}, currently can draw on only limited empirical information on the comparative performance of the methods. Table 1 gives the (short) list of previous simulation studies for MD. It shows that only two studies (\cite{viechtbauer2005bias} and \cite{petropoulou2017comparison}) considered and compared several point estimators of $\tau^2$. Unfortunately, these and other studies  (wrongly) used the pooled variance for MD. The pooled variance is equivalent to the unpooled variance, given by equation~(\ref{eq:varMD}), only when the true variances and sample sizes are equal within the study. Also, simulations with one normally distributed effect measure are informative only under this scenario.
\cite{petropoulou2017comparison} use simulation to study 20 estimators of heterogeneity variance and their impact on coverage and length of 95\%
confidence intervals for the overall effect; but to assess bias of the estimators, they use mean absolute error, which is not a measure of bias; it is the linear counterpart of mean squared error.

Several studies have considered  the quality of estimation of MD, but not estimation of heterogeneity variance. \cite{Friedrich2008} report extensive simulations for MD, but they use only the DL method to estimate $\tau^2$ and do not report on its quality. \cite{Lin_2018_PLoSONE_e0204056}  provides similar simulations for MD.   \cite{IntHout2014} study coverage for MD and combinations of different-sized studies, using the standard method and the Hartung-Knapp-Sidik-Jonkman (HKSJ) method based on the DL estimator of $\tau^2$ (\cite{hartung2001refined}, \cite{sidik2002simple}). \cite{Partlett-2017} provide another in-depth study of coverage based on restricted-maximum-likelihood (REML) estimation of $\tau^2$. They also recommend HKSJ confidence intervals.

Only two studies (\cite{Knapp-2006} and \cite{jackson2013confidence}) considered  the quality of interval estimation of $\tau^2$ for  MD.

To address this gap in information on methods of estimating the heterogeneity variance for MD, we use simulation to study four methods recommended by \cite{veroniki2015methods}. These are the well-established methods of \cite{dersimonian1986meta}, restricted maximum likelihood, and \cite{mandel1970interlaboratory}, and the less-familiar method of \cite{jackson2013confidence}. We also include a new method  based on an improved approximation to the distribution of the $Q$ statistic for MD \citep{kulinskaya2004welch} based on the \cite{welch1951comparison} test. We also study coverage of confidence intervals for $\tau^2$ achieved by five methods, including the Q-profile method of \cite{viechtbauer2007confidence}, Q-profile methods based on the improved Welch-type approximation to the distribution of Cochran's $Q$, and profile-likelihood-based intervals.

For each estimator of $\tau^2$,  we also study bias of the corresponding inverse-variance-weighted estimator of the overall effect. 
For comparison, we add an estimator (SSW) whose weights depend only on the sample sizes of the Treatment and Control arms. We study the coverage of the confidence intervals associated with the inverse-variance-weighted estimators, and also the HKSJ interval, the HKSJ interval using the improved estimator of $\tau^2$, and the interval centered at SSW and using the improved $\hat\tau^2$ in estimating its variance.

\begin{sidewaystable}[]
\resizebox{\textwidth}{!}{%
	\begin{tabular}{|l|l|l|l|l|l|l|l|l|l|}\hline
		Study& ($\sigma_T^2$, $\sigma_C^2$)             & $\mu$     & $\tau^2$            & $n$ and/or $\bar{n}$                 & $K$                                                                    & $\hat{\tau}^2$ & $\tau^2$ intervals & $\hat{\mu}$ & $\mu$ intervals                                              \\\hline
		Viechtbauer 2005  & $(10, 10)$        &0,1,2,4   & 0,0.125,0.25,0.5,1           & $\bar{n}=20,40,80,160,320$                                                     & 5, 10,                             & DL                       &                                            &                               &                                                 \\
& pooled variance         &                 &                                 &                            $n_i \sim N(\bar n, (\bar n/3)^2)$                        &       20, 40, 80                                     & ML                                           &                                            &                                               &                                                 \\
		&           &                 &                                 &                                 $n_{iT}=n_{iC}=n_i$                                     &                                            & REML                                         &                                            &                                               &                                                 \\
		&           &                 &                                 &                                                                      &                                            & HE                             &                                            &                                               &                                                 \\
		&           &                 &                                 &                                                                      &                                            & HS                     &                                            &                                               &                                                 \\\hline
Knapp et al. 2006&100,\;10 across studies&0&1, 2.5, 5, 10, 20& 20, 40 &5, 10, 20, 50, 100&&  &&\\
&pooled variance &&&&&&KBH&&\\
\hline

		Friedrich et al 2008          &  $\sigma_T=\sigma_C=\sigma=10,\;40,\;70$           & $0.2\sigma,\;0.5\sigma,\;0.8\sigma$     & $0,\;0.5\sigma$                           & $n=10,\;100$, $n_{T}=n_{C}$;                               & 5,\; 10,\; 30                              & DL                       &                                            & IV                              & IV  \\
&&&&$n_T=n_C/2$,\; $n_T=2n_C$&&&& &\\
&&&&(with $n_T + n_C = 2n$)&&&&& \\
\hline
		Lin 2018    &  $\sigma_{iT}=\sigma_{iC}\sim U(1,5)$       & 0, 0.5, 1, 2, 5                 & 0, 0.25, 1 & U(5,10), U(10,20), & 5,\; 10,\; 20,\;50        & DL                        &                                            & IV                              & IV  \\
		& pooled variance &                 &                                 &            U(20,30),U(30,50)                                                 &                                            &                                              &                                            &                                               &           \\
		&  &                          &                                  &                    U(50,100),U(100,500),                                                 &                                            &                                              &                                            &                                               &           \\
		&  &                          &                                  &                    U(500,1000); $n_{Ti}=n_{Ci}=n_i/2$                                                  &                                            &                                              &                                            &                                               &           \\\hline
		IntHout et al. 2014 &(1,1)&0&depends on $I^2$ and the $n_i$, &$n_{T}=n_{C}=n=25,\; 50, $ &2(1)10(5)20&DL&&IV&IV \\
	     &pooled variance&&$I^2=$0,\;25\%,\; 50\%,\;75\%,\;90\% &100,\; 250,\; 500,\; 1000&&&&& HKSJ\\
         &&&&$\bar{n}$=100,\; 250,\;500,\;1000;&&&&& \\
          &&&&25\%, 50\%, 75\% large (=$10 \times$ small)&&&&& \\\hline	
        Parlett and Riley 2017&one normal mean&1&0.01, 0.05, 0.1, 1&30&3, 5, 7, 10, 100&REML&&IV&IV\\
        &$\sigma^2=0.1n$&&&&&&&&KR\\
        &&&&&&&&&HKSJ,\; HK2\\
        &&&&&&&&&SJ,\; SJ2\\
        \hline
        	Petropoulou and Mavridis 2017             &    (1,1)       &0, 0.5             & 0, 0.01, 0.05, 0.5             &    $n_{T}=n_{C} \sim U(20,\;200)$                                          & 5, 10, 20, 30                              & 17                 &                                            & IV                              & IV  \\
	&    pooled variance       &                 &  $\mu_i \sim N(\mu, \tau^2)$                               &                                                                      &                                            & estimators                      &                                            &                                               &                                                 \\
	&           &                 &   $\mu_i \sim \mathrm{Laplace}(0, 1)$                              &                                                                      &                                            & of $\tau^2$&                                            &                                               &                                                 \\\hline
        Jackson 2013&one normal mean &0&0, 0.029, 0.069, 0.206, 1.302&implicit in $\sigma_i^2$&5; 10, 20, 40 &DL&BJ&&\\
&$(\sigma_1^2, \ldots, \sigma_5^2) =$&&&&&&J&&\\
&(0.009,\;0.046,\;0.122,\;0.265,\;0.600)&&&&&&QP&&\\
\hline
	\end{tabular}
}
\begin{footnotesize}\caption{Simulation studies on meta-analysis of MD. \\
Estimators of $\tau^2$: DL - DerSimonian and Laird, ML - Maximum likelihood, REML - Restricted maximum likelihood, HE - Hedges, HS - Hunter and Schmidt;\\ 
Intervals for $\tau^2$: QP - Q-profile, BJ - Biggerstaff and Jackson, J - Jackson, KBH - Knapp et al. (2006).\\
Estimators of $\theta$:  IV - inverse-variance-weighted;\\
Intervals for $\theta$: all confidence intervals are centered at IV estimator of  $\mu$. IV with z quantiles:  HKSJ, HK2, SJ, SJ2 and KR. Confidence intervals using t quantiles with variance estimators by: HKSJ:  Hartung and Knapp (2001), Sidik and Jonkman(2002);\; HK2: R\"{o}ver et al. (2015);\; SJ and SJ2: Sidik and Jonkman (2006); KR:  Kenward and Roger (1997).
}\end{footnotesize}
\end{sidewaystable}\label{simulations_summary_for_MD}

\section{Study-level estimation of mean difference}
We assume that each of the $K$ studies in the meta-analysis consists of two arms, Treatment and Control, with sample sizes $n_{iT}$ and $n_{iC}$. The total sample size in Study $i$ is $n_i=n_{iT}+n_{iC}$. We denote the ratio of the control sample size to the total by  $q_i=n_{iC}/n_{i}$.  The subject-level data in each arm are assumed to be normally distributed with means $\mu_{iT}$ and $\mu_{iC}$ and variances $\sigma_{iT}^2$ and $\sigma_{iC}^2$. The sample means are $\bar{x}_{ij}$, and the sample variances are $s^2_{ij}$, for $i = 1,\ldots,K$ and $j = C$ or $T$.

The mean difference effect measure  is
$$\mu_{i}=\mu_{iT}-\mu_{iC}, \hbox{    estimated by    } y_i=\bar{x}_{iT}-\bar{x}_{iC},$$
with variance
$\sigma^2_i=\sigma_{iT}^2/n_{iT}+\sigma_{iC}^2/n_{iC}$, estimated by
\begin{equation}\label{eq:varMD}v_i^2 = \hat{\sigma}_i^2 = s_{iT}^2/n_{iT}+s_{iC}^2/n_{iC}.\end{equation}
$s_{iT}^2$ and $s_{iC}^2$ do not depend on $\mu_{iT}$ and $\mu_{iC}$, so $\hat{\sigma}_i^2$ does not involve $\mu_i$. In the best-case scenario for traditional meta-analysis methods, for normal data, the sample means are independent of the sample variances (and therefore of inverse-variance-based weights). However, the relation of the between-study variance $\tau^2$ and the within-study variances $\sigma_i^2$  may affect quality of estimation. Sometimes the  pooled variance is used instead of $v_i^2$ given by Equation~(\ref{eq:varMD}). Unequal variances in the Treatment and Control arms, however, can adversely affect estimation (\cite{kulinskaya2004welch}).

\section{Standard random-effects model} \label{sec:StdREM}
In meta-analysis, the standard random-effects model assumes that within- and between-study variabilities are accounted for by approximately normal distributions of within- and between-study effects.  For a generic measure of effect,
\begin{equation}\label{eq:standardREM}
\hat{\theta}_{i}\sim N(\theta_{i},{\sigma}_{i}^2)\quad\text{and}\quad \theta_{i}\sim N(\theta,\tau^2),
\end{equation}
resulting in the marginal distribution $\hat{\theta}_{i}\sim N(\theta,\sigma_{i}^2+\tau^2)$. $\hat{\theta}_{i}$ is the estimate of the effect in Study $i$, and its within-study variance is $\sigma_{i}^2$, estimated by $\hat{\sigma}_{i}^2$, $i=1,\ldots,K$.  $\tau^{2}$ is the between-study variance, which is estimated by $\hat{\tau}^2$. The overall effect $\theta$ can be estimated by the weighted mean
\begin{equation}\label{eq:WAverChapter6}
\hat{\theta}_{\mathit{RE}}=\frac{\sum\limits_{i=1}^{K}\hat{w}_{i}(\hat{\tau}^2)\hat{\theta}_{i}}{\sum\limits_{i=1}^{K}\hat{w}_{i}(\hat{\tau}^2)},
\end{equation}
where the $\hat{w}_{i}(\hat{\tau}^2)=(\hat{\sigma}_{i}^2+\hat{\tau}^2)^{-1}$ are inverse-variance weights. The FE estimate $\hat{\theta}$  uses weights $\hat{w}_{i}=\hat{w}_{i}(0)$.

If $w_i = 1/\var(\hat{\theta}_i)$, the variance of the weighted mean of the $\hat{\theta}_i$ is $1/\sum w_{i}$. Thus, many authors estimate the variance of $\hat{\theta}_{\mathit{RE}}$ by $\left[\sum_{i=1}^{K}\hat{w}_{i}(\hat{\tau}^2)\right]^{-1}$.  In practice, however, this estimate may not be satisfactory (\cite{sidik2006robust, li1994bias,rukhin2009weighted}).

\section{Point and interval estimation of $\tau^2$ by the  Welch-type and  corrected DerSimonian-Laird methods}

Because the $\hat{w}_i(\hat\tau^2)$ in (\ref{eq:WAverChapter6}) involve the $\hat{\sigma}_{i}^2$, $K - 1$ is an adequate approximation for the expected value of Cochran's $Q$ statistic only for very large sample sizes. However, this approximation is used in all moment methods for estimating $\tau^2$.
As an alternative one can use an improved, effect-measure-specific approximation to the expected value of $Q$. Corrected Mandel-Paule-type moment methods for estimating $\tau^2$ equate the $Q$ statistic, with weights $\hat{w}_i(\tau^2)$, to the first moment of an improved approximate null distribution. The corrected DerSimonian-Laird method estimates $\tau^2$ from the first moment of $Q$ under alternatives.

More-realistic approximations to the null distribution of $Q$  are available for several effect measures. These approximations do not treat the estimates $\hat{\sigma}_{i}^2$ as equal to $\sigma_{i}^2$. For MD, \cite{kulinskaya2004welch} proposed an approximation based on the method of \cite{welch1951comparison}. This method calculates the first two corrected moments of $Q$, $\kappa_1=\e[{Q}]$ and  $\kappa_2=\var[{Q}]$,  under the null hypothesis of homogeneity and then approximates the null distribution of $Q$ by an F distribution: $\hat{c}F_{K-1,\hat{f}_2}$ with matched moments. The estimated degrees of freedom  $\hat{f}_2$ and the scale factor $\hat{c}$ are functions  of $K$, the $n_{iT}$ and $n_{iC}$, and the $\hat{\sigma}_{iT}^2$ and $\hat{\sigma}_{iC}^2$. 

To simplify notation, let $W=\sum{w_i}$, $W_{(k)}=\sum{w_i}^k$, and $p_i= 1-w_i/W$, and let
\begin{equation} \label{eq:gh2}
g_i=\left(\frac{\sigma^4_{iT}}{n_{iT}^2f_{iT}}+ \frac{\sigma^4_{iC}}{n_{iC}^2f_{iC}}\right);
\end{equation}
where $f_{ik}=n_{ik}-1$ is the number of degrees of freedom of for group $k$ of study $i$, $k=T,\; C$.  Then the null moments of $Q$ for MD are \citep{kulinskaya2004welch},
\begin{equation}
\label{eq:nullcontrol}
\kappa_1\approx I-1 +2\sum_iw_i^2g_ip_i^2; \hspace{2em} \kappa_2\approx 2(I-1) +14\sum_iw_i^2g_ip_i^2.
\end{equation}

We propose a new method of estimating $\tau^2$ based on this improved approximation.
Let $E_{WT}({Q})=\kappa_1$ denote the corrected expected value of $Q$.  Then one obtains the WT estimate of $\tau^2$ by iteratively solving
\begin{equation}
Q(\tau^2)=\sum\limits_{i=1}^{K}\frac{(\theta_{i}-\hat{\theta}_{RE})^{2}}{\hat{\sigma}_{i}^2+\tau^2}=E_{WT}({Q}).
\end{equation}
We denote the resulting estimator of $\tau^2$ by $\hat{\tau}_{WT}^2$.

We also propose a new WT confidence interval for the between-study variance. This  interval for $\tau^2$ combines the Q-profile approach and the improved approximation by \cite{kulinskaya2004welch} (i.e.,  the scaled F distribution with $K-1$ and $\hat{f}_2$  degrees of freedom based on the corrected first two  moments of $Q$).

This corrected Q-profile confidence interval can be estimated from the lower and upper quantiles of $F_Q$, the cumulative distribution function for the improved approximation to the distribution of $Q$:
\begin{equation}
Q(\tau_{L}^2)=F_{Q;0.975}\qquad Q(\tau_{U}^2)=F_{Q;0.025}
\end{equation}
The upper and lower confidence limits for $\tau^2$ can be calculated iteratively.

We also propose another new  method of estimating $\tau^2$ based on the improved first moment of $Q$ under alternatives, in the spirit of \cite{dersimonian1986meta}.
Under the alternatives,
$$\e(Q)\approx I-1+2\sum_iw_i^2g_ip_i^2+\tau^2\big(W-W_{(2)}/W\big),$$
and the corrected DerSimonian-Laird (CDL) estimator is given by
$$\hat{\tau}^2_{CDL}=\max\left(\frac{Q-(I-1)-2\sum_iw_i^2g_ip_i^2}{\hat W-\hat W_{(2)}/\hat W}, 0. \right)$$

\section{Sample-size-weighted (SSW)  point and interval  estimators  of $\theta$}

For comparison with the inverse-variance-weights estimators, we include a point estimator whose weights depend only on the studies' sample sizes. For this estimator (SSW),
$w_{i} = \tilde{n}_i = n_{iT}n_{iC}/(n_{iT} + n_{iC})$; $\tilde{n}_i$ is the effective sample size in Study $i$. 
These effective-sample-size-based weights were suggested in \cite [p.110]{hedges1985statistical}.   

The interval estimators corresponding to SSW (SSW WT and SSW CDL) use the SSW point estimator as their center, and their  half-width equals the estimated standard deviation of SSW under the random-effects model times the critical value from the $t$ distribution on $K - 1$ degrees of freedom.  The estimator of the variance of SSW is
\begin{equation}\label{eq:varianceOfSSW}
\widehat{\var}(\hat{\theta}_{\mathit{SSW}})= \frac{\sum \tilde{n}_i^2 (v_i^2 + \hat{\tau}^2)} {(\sum \tilde{n}_i)^2},
\end{equation}
in which $v_i^2$ comes from Equation~(\ref{eq:varMD})  and $\hat{\tau}^2 = \hat{\tau}_{\mathit{WT}}^2$ or $\hat{\tau}_{\mathit{CDL}}^2$, respectively.

\section{Simulation study}\label{sec:simsect}
As mentioned in Section~\ref{sec:Intro}, other studies have used simulation to examine estimators of $\tau^2$ or of the overall effect for MD, but gaps in evidence remain.  For example, the possibility that the variances in the two arms may differ is rarely, if ever, reflected in simulations. Usually,  pooled variances are used \citep{viechtbauer2005bias}, or, equivalently,  only the $y_i$ are simulated, as in  \cite{IntHout2014} and  \cite{Partlett-2017}.


The range of values of $\mu_i$ in previous simulation studies varies from $0$ in  \citep{viechtbauer2005bias} to  $56$ in  \cite{Friedrich2008}, but the value of $\mu_i$ is unimportant because $\hat{\sigma}_{i}^2$ does not involve $\mu_i$. In  \cite{viechtbauer2005bias} $\tau^2$ and the $\sigma_i^2$ are commeasurable and vary from $0.0625$ to $1$.  \cite{IntHout2014} set $\tau^2=C\bar{\sigma}^2$, where $\bar{\sigma}^2$ is the average within-study variance, and vary $C$ from $1/3$ to $9$. In  \cite{Friedrich2008} the $\tau^2$ values, if non-zero, are much larger at $\tau^2 = n\sigma^{2}/4$ for $n=10$ and $100$.  \cite{Partlett-2017} also use comparatively large values of $\tau^2$, varying from $0.01$ to $1$, whereas $\sigma^2_i=0.1/30$ for $n=30$.

\subsection{Design of the simulations}\label{sec:MDsim}
Our simulation study assesses the performance of six methods for point estimation of between-study variance $\tau^2$ (DL, REML, J, MP, WT and CDL) and five methods of interval estimation of $\tau^2$ (Q-profile-based methods corresponding to DerSimonian-Laird and Welch, the generalized Q-profile intervals of \cite{biggerstaff2008exact} and \cite{jackson2013confidence}, and the profile-likelihood confidence interval based on REML).

We also assess the performance of the point and interval estimators of $\mu$  in the random-effects models for MD.

We vary six parameters: the between-study variance ($\tau^2$) and the within-study variances ($\sigma_{T}^2$ and $\sigma_{C}^2$), in addition to the number of studies ($K$), the total sample size ($n$), and the proportion of observations in the Control arm ($q$).  We set the overall true MD $\mu = 0$ because the estimators of $\tau^2$ do not involve $\mu$ and the estimators of $\mu$ are equivariant. Configurations we study are listed  in  Table~\ref{tab:altdataMD}.

To cover both small and large values of the ratio of within-study to between-studies variance, separately from the value of $\tau^2$, we use two series of within-study variances. We generate the within-study sample variances $s_{ij}^2$ ($j=T, \;C$) from chi-squared distributions as $\sigma_{ij}^2\chi_{n_{ij}-1}^2/(n_{ij}-1)$. We generate the estimated mean differences $y_i$ from a normal distribution with mean $\mu$ and variance $\sigma_{iT}^2/n_{iT} + \sigma_{iC}^2/n_{iC} + \tau^2$. We obtain the estimated within-study variances as $v_{i}^2 = s_{iT}^2/n_{iT} + s_{iC}^2/n_{iC}$.

All simulations use the same numbers of studies  $K = 5, \;10, \;30$ and,  for each combination of parameters, the same vector of total sample sizes $n = (n_{1},\ldots, n_{K})$ and the same proportions of observations in the Control arm $q_i = n_{iC}/n_i = .5, \;.75$ for all $i$. The values of $q$ reflect two situations for the two arms of each study: approximately equal (1:1) and quite unbalanced (1:3). The sample sizes in the Treatment and Control arms are $n_{iT}=\lceil{(1 - q_i)n_{i}}\rceil$ and $n_{iC}=n_{i}-n_{iT}$, $i=1,\ldots,K$.

We study equal and unequal study sizes. For equal study sizes $n_i$ is as small as 20, and for unequal study sizes $n_i$ is as small as 12, in order to examine how the methods perform for the extremely small sample sizes that arise in some areas of application.
In choosing unequal study sizes, we follow a suggestion of  \cite{sanches-2000}, who selected study sizes having skewness of 1.464, which they considered typical in behavioral and health sciences. Table~\ref{tab:altdataMD}  gives the details.

The patterns of sample sizes are illustrative; they do not attempt to represent all patterns seen in practice. By using the same patterns of sample sizes for each combination of the other parameters, we avoid the additional variability in the results that would arise from choosing sample sizes at random (e.g., uniformly between 20 and 200).

We use a total of $10,000$ repetitions for each combination of parameters. Thus, the simulation standard error in the estimation of $\mu$ is $0.01$ (for $n=20$) or less for the first series, and $0.02$ or less for the second series of simulations. The simulation standard error for estimated coverage of $\tau^2$ or $\mu$ at the $95\%$ confidence level is roughly $\sqrt{0.95 \times 0.05/10,000}=0.00218$.

The simulations were programmed in R version 3.3.2 using the University of East Anglia 140-computer-node High Performance Computing (HPC) Cluster, providing a total of 2560 CPU cores, including parallel processing and large memory resources. For each configuration, we divided the 10,000 replications into 10 parallel sets of 1000 replications.

\subsection{Analysis of the simulation data}
The structure of the simulations invites an analysis of the results along the lines of a designed experiment, in which the variables are $\tau^2$, $n$, $K$, $q$, $\sigma_C^2$, and $\sigma_T^2$. Most of the variables are crossed, but two have additional structure. Within the two levels of $n$, equal and unequal, the values are nested: $n = 20,\;40,\; 100,\; 250$ and $\bar{n} = 30, \;60, \;100, \;160$. The values of $\sigma_C^2$, and $\sigma_T^2$ consist of a cross of two factors, equal/unequal and small/large ($\sigma_C^2 = 1$ and $\sigma_T^2 = 1$, $\sigma_C^2 = 10$ and $\sigma_T^2 = 10$, $\sigma_C^2 = 1$ and $\sigma_T^2 = 2$, and $\sigma_C^2 = 10$ and $\sigma_T^2 = 20$).  We approach the analysis of the data from the simulations qualitatively, to identify the variables that substantially affect (or do not affect) the performance of the estimators as a whole and the variables that reveal important differences in performance.  We might hope to describe the estimators' performance one variable at a time, but such ``main effects'' often do not provide an adequate summary: important differences are related to certain combinations of two or more variables.

We use this approach to examine bias and coverage in estimation of $\tau^2$ and bias and coverage in estimation of $\mu$. Our summaries of results in Section~\ref{sec:MDresults}  are based on examination of the figures in the corresponding  Appendices.


\begin{table}[ht]
	\caption{\label{tab:altdataMD} \emph{Data patterns in the simulations for MD}}
	\begin{footnotesize}
		\begin{center}
			\begin{tabular}
				{|l|l|l|l|}
				\hline
				Parameter & Equal study sizes& Unequal study sizes& Full results in\\
                &&&Appendix\\
				\hline
				$K$ (number of studies)& 5, 10, 30&5, 10, 30&\\
				$n$ or $\bar{n}$  (average (individual) study size - & 20, 40, 100, 250& 30 (12,16,18,20,84), &\\
                total of the two arms)&&&\\
				For  $K=10$ and $K=30$,  the same set of  	&&60 (24,32,36,40,168), &\\
				unequal study sizes is used twice or six times, 	&&100 (64,72,76,80,208), &\\
					respectively	&&160 (124,132,136,140,268) &\\
				$q$ (proportion of each study in the Control arm) & 1/2, 3/4&1/2, 3/4&\\
				\hline
                $\tau^{2}$ (variance of random effect)&0(0.01)0.1; 0(0.1)1&0(0.01)0.1; 0(0.1)1&  A1, A2;  A3, A4 \\				
				$\sigma_{C}^2,\sigma_{T}^2$(within-study variances)&(1,1), (1,2)&(1,1), (1,2)&\\
			    $\mu^{}$&0&0&  B1, B2;  B3, B4\\
				\hline
				$\tau^{2}$ (variance of random effect)&0(0.1)1&0(0.1)1&  A5, A6\\
				$\sigma_{C}^2,\sigma_{T}^2$ (within-study variances)&(10,10), (10,20)&(10,10), (10,20)&\\
				$\mu^{}$&0&0&   B5, B6\\
				\hline
			\end{tabular}
		\end{center}
	\end{footnotesize}
\end{table}

\section {Methods of estimation of $\tau^2$ and $\mu$ used  in simulations}
\subsection*{Point estimators of $\tau^2$}
\begin{itemize}
\item DL - method of \cite{dersimonian1986meta}
\item J - method of \cite{jackson2013confidence}

\item MP - method of \cite{mandel1970interlaboratory}
\item REML - restricted maximum-likelihood method
\item WT - method based on corrected null moment of $Q$ per \cite{kulinskaya2004welch}
\item CDL - method based on corrected first  moment of $Q$ under alternatives per \cite{kulinskaya2004welch} and \cite{dersimonian1986meta}.
\end{itemize}

\subsection*{Interval estimators of $\tau^2$}
\begin{itemize}
\item BJ - method of \cite{biggerstaff2008exact}
\item J - method of \cite{jackson2013confidence}
\item PL - profile-likelihood confidence interval based on $\hat{\tau}_{REML}^2$
\item QP - Q-profile confidence interval of \cite{viechtbauer2007confidence}
\item WT - Q-profile method based on corrected null distribution of $Q$ per \cite{kulinskaya2004welch}
\end{itemize}

\subsection*{Point estimators of $\mu$}
Inverse-variance-weighted methods with $\tau^2$ estimated by:
\begin{itemize}
\item DL
\item J
\item REML
\item MP
\item WT
\item CDL
\end{itemize}
and
\begin{itemize}
\item SSW - weighted mean with weights that depend only on studies sample sizes
\end{itemize}

\subsection*{Interval estimators of $\mu$}
Inverse-variance-weighted methods using normal quantiles, with  $\tau^2$ estimated by:
\begin{itemize}
\item DL
\item J
\item MP
\item REML
\item WT
\item CDL
\end{itemize}
Inverse-variance-weighted methods with modified variance of $\hat{\mu}$ and t-quantiles as in  \cite{hartung2001refined} and \cite{sidik2002simple}
\begin{itemize}
\item HKSJ (DL) -  $\tau^2$ estimated by DL
\item HKSJ WT -  $\tau^2$ estimated by WT
\end{itemize}
and
\begin{itemize}
\item SSW WT - SSW point estimator of $\mu$ with estimated variance given by (\ref{eq:varianceOfSSW}) with $\hat{\tau}^2 = \hat{\tau}_{\mathit{WT}}^2$ and t-quantiles
\item SSW CDL - SSW point estimator of $\mu$ with estimated variance given by (\ref{eq:varianceOfSSW}) with $\hat{\tau}^2 = \hat{\tau}_{\mathit{CDL}}^2$ and t-quantiles
\end{itemize}

\section{Results}\label{sec:MDresults}

Our full simulation results, comprising $130$ figures, each presenting $12$ combinations of the 4 values of $n$ or $\bar{n}$ and the 3 values of $K$, are provided in Appendices A and B.  A summary  is given below.

\subsection{Bias in estimation of $\tau^2$ (Appendices A1, A3, and A5)}

\noindent All of the estimators (DL, REML, J, MP, CDL and WT) have positive bias when $\tau^2 = 0$.  In favorable situations (e.g., equal $n$'s, $q = .5$, and $\sigma_C^2 = \sigma_T^2 = 1$, as in Figure A1.1), the bias of the estimators other than  WT is slightly greater than 0 when $\tau^2 = 0$, and it decreases to near 0 as $\tau^2 \rightarrow 1$ and becomes closer to 0 as $n$ increases. In some situations, however, the bias is much greater across the range of $\tau^2$; in the most extreme case in our simulations ($n = 20$, $K = 5$, $\sigma_C^2 = 10$, and $\sigma_T^2 = 20$; Figure A5.7), the bias at $\tau^2 = 0$ ranges from 1.6 (MP, J) to 1.8 (DL, REML) and decreases only to 1.4 to 1.7 at $\tau^2 = 1$. Generally, the bias decreases slightly as $K$ increases. Unbalanced arms ($q = .75$) magnify the bias, especially for the smaller values of $n$ and $\bar{n}$ and unequal variances, i.e. when $\sigma_T^2 = 2$ or $\sigma_T^2 = 20$. Among the estimators other than CDL and WT, DL and REML generally have the most bias, followed by J and MP.

The trace in the bias of CDL parallels  that of DL, but it is considerably less biased than all the standard estimators. This favorable difference between CDL and the standard estimators is especially pronounced for small and unequal  sample sizes. CDL is practically unbiased in the case of equal within-arm variances, and its bias is considerably less  than that of DL in the most extreme cases of small and unequal sample sizes combined with unequal variances, as in  Figure A5.7.

The bias of WT follows a strikingly different pattern. For small $\tau^2$ it is positive and smaller than (or equal to) that of the other estimators. As $\tau^2$ increases, the bias of WT becomes negative and takes increasingly more negative values, roughly linearly in $\tau^2$. The crossover point decreases as $K$ increases (e.g., 0.15 at $K = 5$, 0.08 at $K = 10$, and 0.05 at $K = 30$, when $n = 20$, $q = .5$, $\sigma_C^2 = 1$, and $\sigma_T^2 = 1$), but is substantially larger when $\sigma_C^2$ and $\sigma_T^2$ are large (Figures A5.1--A5.8).  The slope against $\tau^2$ flattens substantially as $n$ (or $\bar{n}$) increases.

In summary, except for CDL and WT,  the estimators of $\tau^2$ (DL, REML, J, and MP) have non-negligible positive bias, especially for small sample sizes ($n\leq 40$) and small values of $\tau^2$.
Overall, CDL is the least biased, and WT is the least biased when the values of $\tau^2$ are considerably smaller than the within-study variances $\sigma^2_i$ (say, when $\tau^2\leq 0.2$ for $\sigma^2=1$, and when $\tau^2\leq 1$ for $\sigma^2=10$). All other estimators become acceptable for larger sample sizes $n\geq 100$.\\
\\
\subsection{Coverage in estimation of $\tau^2$ (Appendices A2, A4, and A6)}

\noindent The relation between coverage of the interval estimators (PL, QP, BJ, J, and WT) and $\tau^2$ involves other variables.  When $n$ or $\bar{n}$ is $\geq 100$, most of the interval estimators have coverage close to .95 when $\tau^2 \geq 0.1$, but noticeably above .95 when $\tau^2 = 0$. (When $\sigma_C^2$ and $\sigma_T^2$ are large, PL often has coverage around .97.) When $K = 30$ and $n = 20$ (or $\bar{n} = 30$) or $n = 40$ (or $\bar{n} = 60$), QP, PL, and BJ have quite low coverage at the smallest values of $\tau^2$ (e.g., .67 to .77 when $\tau^2 = 0$ and .83 to .88 when $\tau^2 = 0.1$, and $n = 20$, $K = 30$, $q = .75$, $\sigma_C^2 = 1$, and $\sigma_T^2 = 1$; Figures A4.3 and A2.3); the low coverage sometimes extends to most $\tau^2 \in [0, 1]$, and the departures are substantially greater for $q = .75$ than for $q = .5$. In those situations the coverage of WT is usually close to .95 and is seldom below .90. Otherwise, when $\tau^2 > 0$, the coverage of WT is usually between .95 and .96 and sometimes slightly greater. The impact of $\sigma_C^2$ and $\sigma_T^2$ (large vs. small, unequal vs. equal) is generally small.

In summary, none of the interval estimators of $\tau^2$ (PL, QP, BJ, J, and WT) consistently achieve coverage close to .95 (i.e., between .94 and .96). All have difficulty at $\tau^2 = 0$, usually overcoverage; the departures of PL extend to other small $\tau^2$, and its coverage is often greater than .96 but sometimes less than .94. Meta-analyses in which the studies have small sample sizes are challenging for PL, QP, BJ, and J, which in some situations have coverage well below nominal for all $\tau^2 \in [0, 1]$, especially when the number of studies is larger ($K = 30$ vs. $K = 5$ and $K = 10$). Overall, WT comes closest to providing nominal coverage of $\tau^2$. (The contrast in behavior between the WT interval and point estimators is surprising, but the two are defined in different ways.)\\
\\
\subsection{Bias in estimation of $\mu$ (Appendices B1, B3, and B5)}

\noindent Because the estimated MD and its estimated variance are independent, all the estimators of $\mu$ are practically unbiased in all situations.\\
\\
\subsection{Coverage in estimation of $\mu$ (Appendices B2, B4, and B6)}

\noindent When $\tau^2 \geq 0.1$ and $K \leq 10$, the methods that use critical values from the normal distribution (DL, REML, J, MP, CDL and WT) have coverage substantially below .95 ($< .90$ when $K = 5$ and around .92 when $K = 10$), and those coverages generally change only slightly as $\tau^2$ increases to 1. WT is lowest, in part because it underestimates $\tau^2$ (Figure B1.1). HKSJ and HKSJ WT have coverage close to .95. Coverage of SSW WT exceeds .95 and decreases toward .95 as $\tau^2$ increases. The estimators other than HKSJ- and SSW-type have coverage slightly $> .95$ when $\tau^2 = 0$. HKSJ and HKSJ WT are around .94 when $\tau^2 = 0$, and SSW WT is $> .99$ when $K = 5$, decreasing to around .965 when $K = 30$. For $0<\tau^2<0.1$, all estimators other than  SSW-type have coverage below nominal for small sample sizes, considerably below for unequal sample sizes $\bar n \leq 60$ and $K \leq 10$ (Appendix B4).

When the studies' $n$'s are equal, coverage of most profile-type estimators \textit{decreases} as $n$ increases; this pattern is absent when the $n$'s are unequal. The coverage of WT increases to the level of the other profile-type estimators.

As $K$ increases, the coverage of the profile-type estimators approaches .95 (from below).

When $n = 20$ or $\bar{n} = 30$, the traces for the various estimators show more separation when $q = .75$ than when $q = .5$. This pattern is most noticeable when the $\sigma^2$'s are large and $K= 30$. It is in these circumstances that SSW CDL is considerably better than SSW WT, achieving nominal coverage for larger values of $\tau^2$ or $K$, as in Figure B2.7.  

Undercoverage is somewhat less when $\sigma_C^2 = \sigma_T^2 = 10$ than when $\sigma_C^2 = \sigma_T^2 = 1$.  Coverage differs little between $\sigma_T^2 = 2$ and $\sigma_T^2 = 1$ and between $\sigma_T^2 = 20$ and $\sigma_T^2 = 10$.

In summary, HKSJ and HKSJ WT generally (but not uniformly) have the best coverage.  Their coverage is not always within $\pm .01$ of .95; it may be considerably below nominal for $\tau^2 < 0.1$ when sample sizes are small;  but in situations where clear differences separate the interval estimators, HKSJ and HKSJ WT are much closer to .95.  DL, WT, MP, REML, and J exhibit very serious undercoverage when $K = 5$ and nontrivial undercoverage when $K = 10$.

\section{Discussion: Practical implications for meta-analysis}

The results of our simulations give a rather disappointing picture of the current state of meta-analysis. In brief:\\
Small sample sizes are rather problematic  even for such a well-behaved effect measure as the mean difference, and meta-analyses that involve numerous small studies are especially challenging.

The conventional wisdom is that these deficiencies do not matter, as meta-analysis usually deals with studies that are ``large,'' so all these little problems are automatically resolved.  Unfortunately, this is not true, even in medical meta-analyses; in Issue 4 of the Cochrane Database 2004, the maximum study size was $50$ or less in $25\%$ of meta-analyses that used MD as an effect measure, and less than $110$ in $50\%$ of them \cite{kulinskaya2014combining}.

For MD, the between-study variance, $\tau^2$, is usually overestimated near zero,  but the Welch-type method provides better point estimation for $\tau^2<0.1$, and reliable interval estimation across all values of $\tau^2$, $n$, and $K$.  The estimates of $\mu$ are unbiased, and HKSJ intervals provide good coverage. 

\cite{Arendacka-2012} and  \cite{Liu2017} propose new confidence  intervals  for $\tau^2$ in the one-way heteroscedastic random-effects model. These intervals can be used directly in meta-analysis of means in noncomparative studies. Both publications include extensive simulations and compare their intervals with those of  \cite{Knapp-2006}. Both proposals seem to do very well for normal distributions and very small sample sizes. It should be possible to extend  these methods to MD in comparative two-arm designs; this extension will be pursued elsewhere.


Arguably, the main purpose of a meta-analysis is to provide point and interval estimates of an overall effect.

Usually, after estimating the between-study variance $\tau^2$, inverse-variance weights are used in estimating the overall effect (and, often, its variance). This approach relies on the theoretical result that, for known variances, and given unbiased estimates $\hat{\theta}_i$, it yields a Uniformly Minimum-Variance Unbiased Estimate (UMVUE) of $\theta$.




This results in the unbiased estimation of $\theta$  for the mean differences because the estimated variances are independent of the  estimated effects.  An alternative  approach uses weights that do not involve estimated variances of study-level estimates, for example, weights proportional to the study sizes $n_i$.  \cite{hunter1990methods} and \cite{Shuster-2010}, among others, have proposed such weights. 
We prefer to use weights proportional  to an effective sample size,  $\tilde{n}_i=n_{iT}n_{iC}/n_i$. 
Thus, the overall effect is estimated by $\hat{\theta}_{\mathit{SSW}} = \sum \tilde{n}_i\hat{\theta}_i / \sum \tilde{n}_i$, and its variance is estimated by Equation~(\ref{eq:varianceOfSSW}). 
A good estimator of $\tau^2$, such as MP or CDL  can be used as $\hat{\tau}^2$. Further, confidence intervals for $\theta$ centered at $\hat{\theta}_{\mathit{SSW}}$ with $\hat{\tau}_{\mathit{WT}}^2$ in Equation~(\ref{eq:varianceOfSSW}) can be used.

This approach based on SSW requires further study.  For example, in the confidence intervals we have used critical values from the $t$-distribution on $K - 1$ degrees of freedom, but we have not yet examined the actual sampling distribution of SSW.  The raw material for such an examination is readily available: For each situation in our simulations, each of the $10,000$ replications yields an observation on the sampling distribution of SSW.

\section*{Funding}
The work by E. Kulinskaya was supported by the Economic and Social Research Council [grant number ES/L011859/1].
\section*{Appendices}
\begin{itemize}
	\item  Appendix A. MD: Plots for bias and coverage of $\tau^2$.
	\item Appendix B. MD: Plots for bias, mean squared error, and coverage of estimators of the mean difference $\mu$.
\end{itemize}

\bibliographystyle{plainnat}
\bibliography{MD_SMD_Bib_19Nov18}%

\begin{thebibliography}{28}
\providecommand{\natexlab}[1]{#1}
\providecommand{\url}[1]{\texttt{#1}}
\expandafter\ifx\csname urlstyle\endcsname\relax
  \providecommand{\doi}[1]{doi: #1}\else
  \providecommand{\doi}{doi: \begingroup \urlstyle{rm}\Url}\fi

\bibitem[Arendack{\'a}(2012)]{Arendacka-2012}
Barbora Arendack{\'a}.
\newblock Approximate interval for the between-group variance under
  heteroscedasticity.
\newblock \emph{Journal of Statistical Computation and Simulation}, 82\penalty0
  (2):\penalty0 209--218, 2012.

\bibitem[Biggerstaff and Jackson(2008)]{biggerstaff2008exact}
Brad~J Biggerstaff and Dan Jackson.
\newblock The exact distribution of {C}ochran's heterogeneity statistic in
  one-way random effects meta-analysis.
\newblock \emph{Statistics in Medicine}, 27\penalty0 (29):\penalty0 6093--6110,
  2008.

\bibitem[DerSimonian and Laird(1986)]{dersimonian1986meta}
Rebecca DerSimonian and Nan Laird.
\newblock Meta-analysis in clinical trials.
\newblock \emph{Controlled Cinical Trials}, 7\penalty0 (3):\penalty0 177--188,
  1986.

\bibitem[Friedrich et~al.(2008)Friedrich, Adhikari, and Beyene]{Friedrich2008}
Jan~O Friedrich, Neill~KJ Adhikari, and Joseph Beyene.
\newblock The ratio of means method as an alternative to mean differences for
  analyzing continuous outcome variables in meta-analysis: a simulation study.
\newblock \emph{BMC Medical Research Methodology}, 8:\penalty0 32, 2008.

\bibitem[Hartung and Knapp(2001)]{hartung2001refined}
Joachim Hartung and Guido Knapp.
\newblock A refined method for the meta-analysis of controlled clinical trials
  with binary outcome.
\newblock \emph{Statistics in Medicine}, 20\penalty0 (24):\penalty0 3875--3889,
  2001.

\bibitem[Hedges and Olkin(1985)]{hedges1985statistical}
Larry~V Hedges and Ingram Olkin.
\newblock \emph{Statistical {M}ethods for {M}eta-{A}nalysis}.
\newblock San Diego, California: Academic Press, 1985.

\bibitem[Higgins et~al.(2009)Higgins, Thompson, and
  Spiegelhalter]{higgins2009re}
Julian P~T Higgins, Simon~G Thompson, and David~J Spiegelhalter.
\newblock A re-evaluation of random-effects meta-analysis.
\newblock \emph{Journal of the Royal Statistical Society, Series A},
  172\penalty0 (1):\penalty0 137--159, 2009.

\bibitem[Hunter and Schmidt(1990)]{hunter1990methods}
John~E Hunter and Frank~L Schmidt.
\newblock \emph{Methods of {M}eta-analysis: Correcting {E}rror and {B}ias in
  {R}esearch {F}indings}.
\newblock Sage Publications, Inc, 1990.

\bibitem[IntHout et~al.(2014)IntHout, Ioannidis, and Borm]{IntHout2014}
Joanna IntHout, John P~A Ioannidis, and George~F Borm.
\newblock The {H}artung-{K}napp-{S}idik-{J}onkman method for random effects
  meta-analysis is straightforward and considerably outperforms the standard
  {D}er{S}imonian-{L}aird method.
\newblock \emph{BMC Medical Research Methodology}, 14:\penalty0 25, 2014.

\bibitem[Jackson(2013)]{jackson2013confidence}
Dan Jackson.
\newblock Confidence intervals for the between-study variance in random effects
  meta-analysis using generalised {C}ochran heterogeneity statistics.
\newblock \emph{Research Synthesis Methods}, 4\penalty0 (3):\penalty0 220--229,
  2013.

\bibitem[Knapp et~al.(2006)Knapp, Biggerstaff, and Hartung]{Knapp-2006}
Guido Knapp, Brad~J Biggerstaff, and Joachim Hartung.
\newblock Assessing the amount of heterogeneity in random-effects
  meta-analysis.
\newblock \emph{Biometrical Journal}, 48\penalty0 (2):\penalty0 271--285, 2006.

\bibitem[Kulinskaya et~al.(2004)Kulinskaya, Dollinger, Knight, and
  Gao]{kulinskaya2004welch}
E~Kulinskaya, MB~Dollinger, E~Knight, and H~Gao.
\newblock A {W}elch-type test for homogeneity of contrasts under
  heteroscedasticity with application to meta-analysis.
\newblock \emph{Statistics in Medicine}, 23\penalty0 (23):\penalty0 3655--3670,
  2004.

\bibitem[Kulinskaya et~al.(2014)Kulinskaya, Morgenthaler, and
  Staudte]{kulinskaya2014combining}
Elena Kulinskaya, Stephan Morgenthaler, and Robert~G Staudte.
\newblock Combining statistical evidence.
\newblock \emph{International Statistical Review}, 82\penalty0 (2):\penalty0
  214--242, 2014.

\bibitem[Li et~al.(1994)Li, Shi, and Daniel~Roth]{li1994bias}
Yuanzhang Li, Li~Shi, and H~Daniel~Roth.
\newblock The bias of the commonly-used estimate of variance in meta-analysis.
\newblock \emph{Communications in Statistics--Theory and Methods}, 23\penalty0
  (4):\penalty0 1063--1085, 1994.

\bibitem[Lin(2018)]{Lin_2018_PLoSONE_e0204056}
Lifeng Lin.
\newblock Bias caused by sampling error in meta-analysis with small sample
  sizes.
\newblock \emph{PLoS ONE}, 13\penalty0 (9):\penalty0 e0204056, 2018.

\bibitem[Liu et~al.(2017)Liu, Li, and Hu]{Liu2017}
Xuhua Liu, Na~Li, and Yuqin Hu.
\newblock A new generalized confidence interval for the among-group variance in
  the heteroscedastic one-way random effects model.
\newblock \emph{Communications in Statistics-Simulation and Computation},
  46\penalty0 (3):\penalty0 2299--3110, 2017.

\bibitem[Mandel and Paule(1970)]{mandel1970interlaboratory}
John Mandel and Robert~C Paule.
\newblock Interlaboratory evaluation of a material with unequal numbers of
  replicates.
\newblock \emph{Analytical Chemistry}, 42\penalty0 (11):\penalty0 1194--1197,
  1970.

\bibitem[Partlett and Riley(2017)]{Partlett-2017}
Christopher Partlett and Richard~D Riley.
\newblock Random effects meta-analysis: coverage performance of 95\% confidence
  and prediction intervals following {REML} estimation.
\newblock \emph{Statistics in Medicine}, 36\penalty0 (2):\penalty0 301--317,
  2017.

\bibitem[Petropoulou and Mavridis(2017)]{petropoulou2017comparison}
Maria Petropoulou and Dimitris Mavridis.
\newblock A comparison of 20 heterogeneity variance estimators in statistical
  synthesis of results from studies: a simulation study.
\newblock \emph{Statistics in Medicine}, 36\penalty0 (27):\penalty0 4266--4280,
  2017.

\bibitem[Rukhin(2009)]{rukhin2009weighted}
Andrew~L Rukhin.
\newblock Weighted means statistics in interlaboratory studies.
\newblock \emph{Metrologia}, 46\penalty0 (3):\penalty0 323--331, 2009.

\bibitem[S{\'a}nchez-Meca and Mar{\'\i}n-Mart{\'\i}nez(2000)]{sanches-2000}
Julio S{\'a}nchez-Meca and Fulgencio Mar{\'\i}n-Mart{\'\i}nez.
\newblock Testing the significance of a common risk difference in
  meta-analysis.
\newblock \emph{Computational Statistics \& Data Analysis}, 33\penalty0
  (3):\penalty0 299--313, 2000.

\bibitem[Shuster(2010)]{Shuster-2010}
Jonathan~J Shuster.
\newblock Empirical vs natural weighting in random effects meta-analysis.
\newblock \emph{Statistics in Medicine}, 29\penalty0 (12):\penalty0 1259--1265,
  2010.

\bibitem[Sidik and Jonkman(2002)]{sidik2002simple}
K.~Sidik and J.~N. Jonkman.
\newblock A simple confidence interval for meta-analysis.
\newblock \emph{Statistics in Medicine}, 21\penalty0 (21):\penalty0 3153--3159,
  2002.

\bibitem[Sidik and Jonkman(2006)]{sidik2006robust}
Kurex Sidik and Jeffrey~N Jonkman.
\newblock Robust variance estimation for random effects meta-analysis.
\newblock \emph{Computational Statistics \& Data Analysis}, 50\penalty0
  (12):\penalty0 3681--3701, 2006.

\bibitem[Veroniki et~al.(2016)Veroniki, Jackson, Viechtbauer, Bender, Bowden,
  Knapp, Kuss, Higgins, Langan, and Salanti]{veroniki2015methods}
Areti~Angeliki Veroniki, Dan Jackson, Wolfgang Viechtbauer, Ralf Bender, Jack
  Bowden, Guido Knapp, Oliver Kuss, Julian P~T Higgins, Dean Langan, and
  Georgia Salanti.
\newblock Methods to estimate the between-study variance and its uncertainty in
  meta-analysis.
\newblock \emph{Research Synthesis Methods}, 7\penalty0 (1):\penalty0 55--79,
  2016.

\bibitem[Viechtbauer(2005)]{viechtbauer2005bias}
Wolfgang Viechtbauer.
\newblock Bias and efficiency of meta-analytic variance estimators in the
  random-effects model.
\newblock \emph{Journal of Educational and Behavioral Statistics}, 30\penalty0
  (3):\penalty0 261--293, 2005.

\bibitem[Viechtbauer(2007)]{viechtbauer2007confidence}
Wolfgang Viechtbauer.
\newblock Confidence intervals for the amount of heterogeneity in
  meta-analysis.
\newblock \emph{Statistics in Medicine}, 26\penalty0 (1):\penalty0 37--52,
  2007.

\bibitem[Welch(1951)]{welch1951comparison}
BL~Welch.
\newblock On the comparison of several mean values: an alternative approach.
\newblock \emph{Biometrika}, 38\penalty0 (3/4):\penalty0 330--336, 1951.

\end{thebibliography}

\clearpage
\setcounter{section}{0}
\renewcommand{\thesection}{A.\arabic{section}}
\text{\LARGE{\bf{Appendices}}}
\section*{A1. Bias of $\hat{\tau}^2$ for $\tau^2 = 0.0(0.1)1.0$, $\sigma_{C}^2=1$, $\sigma_{T}^2=1,\;2$.}
For bias of $\hat{\tau}^2$, each figure corresponds to a value of $\mu (= 0, 0.2, 0.5, 1, 2)$, a value of $q (= .5, .75)$, a value of $\tau^2 = 0.0(0.1)1.0$ and a set of values of $n$ (= 20, 40, 100, 250)\\
Each figure contains a panel (with $\tau^2$ on the horizontal axis) for each combination of n (or $\bar{n}$) and $K (=5, 10, 30)$.\\
The point estimators of $\tau^2$ are
\begin{itemize}
	\item DL (DerSimonian-Laird)
	\item REML (restricted maximum likelihood)
	\item MP (Mandel-Paule)
	\item J (Jackson)
	\item CDL (Corrected DerSimonian-Laird)
\end{itemize}

\clearpage
\renewcommand{\thefigure}{A1.\arabic{figure}}

\begin{figure}[t]
	\centering
	\includegraphics[scale=0.33]{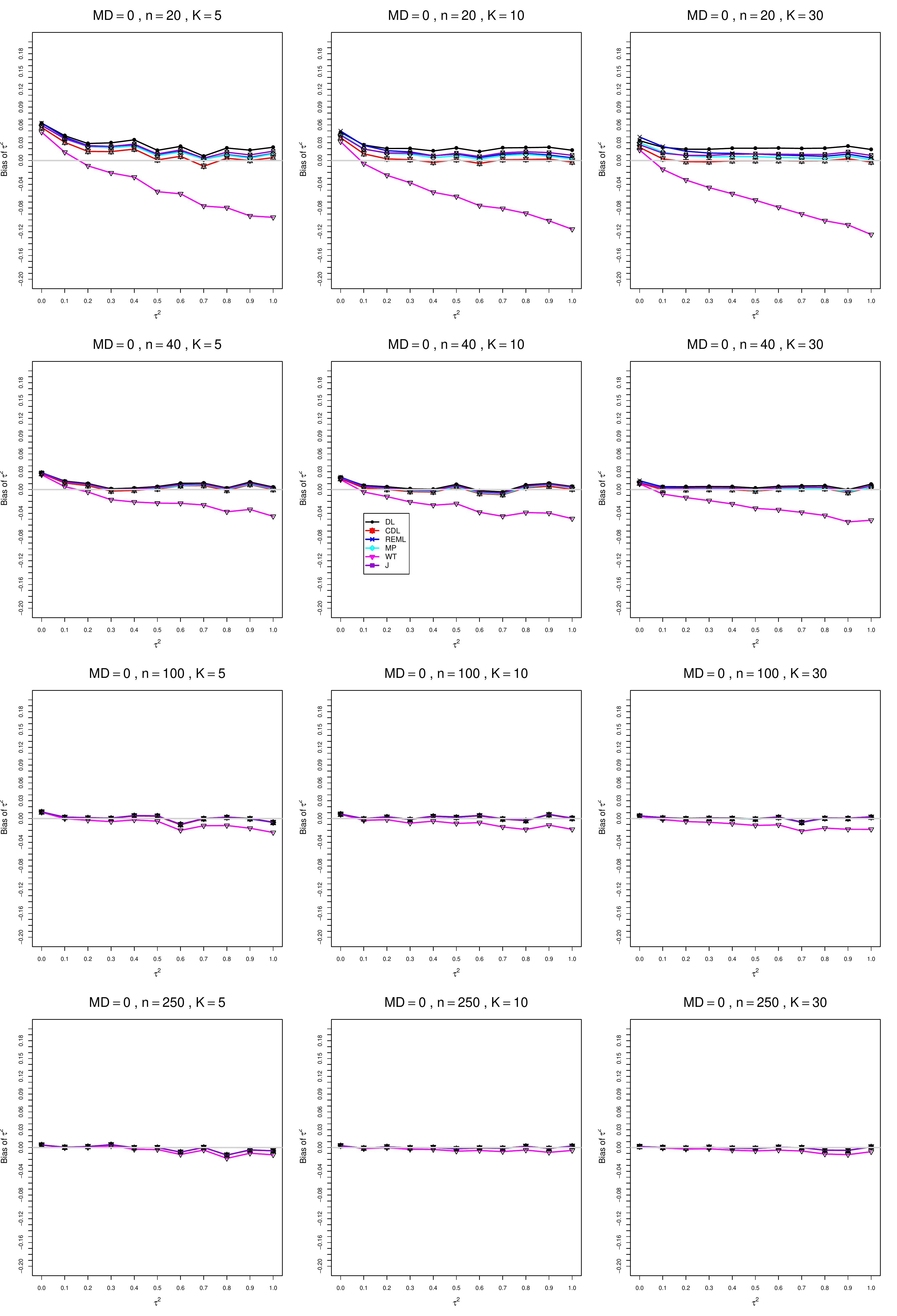}
	\caption{Bias of the estimation of  between-studies variance $\tau^2 = 0.0(0.1)1.0$ for $\mu=0$, $q=0.5$, $\sigma_C^2=1$, $\sigma_T^2=1$,  equal study sizes $n=20,\;40,\;100,\;250$.
		\label{BiasTauMD0_S1_1}}
\end{figure}

\begin{figure}[t]
	\centering
	\includegraphics[scale=0.33]{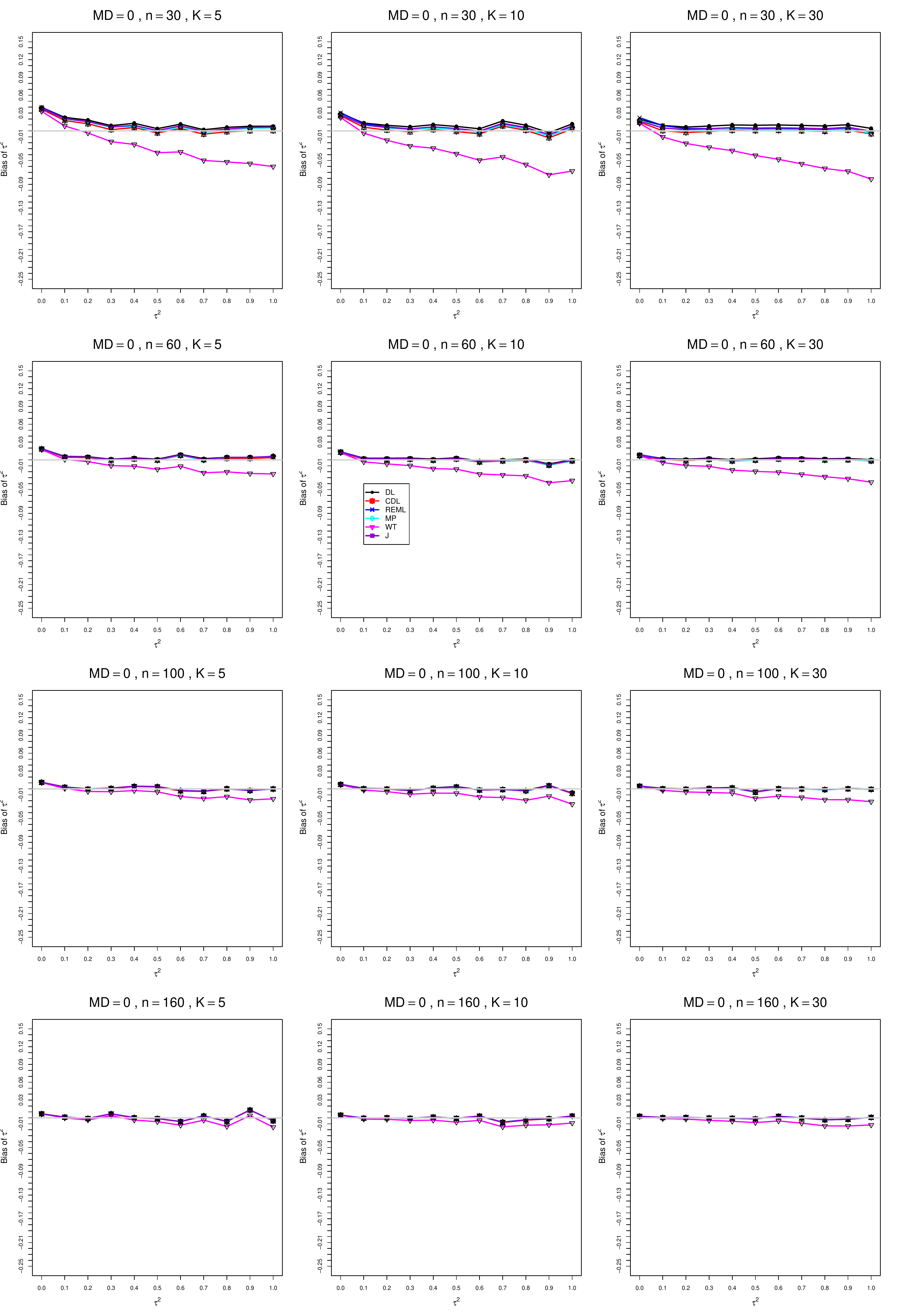}
	\caption{Bias of the estimation of  between-studies variance $\tau^2 = 0.0(0.1)1.0$ for $\mu=0$, $q=0.5$, $\sigma_C^2=1$, $\sigma_T^2=1$, unequal studies of average size $\bar{n}=30,\;60,\;100,\;160$.
		\label{BiasTauMD0_S1_1unequal}}
\end{figure}

\begin{figure}[t]
	\centering
	\includegraphics[scale=0.33]{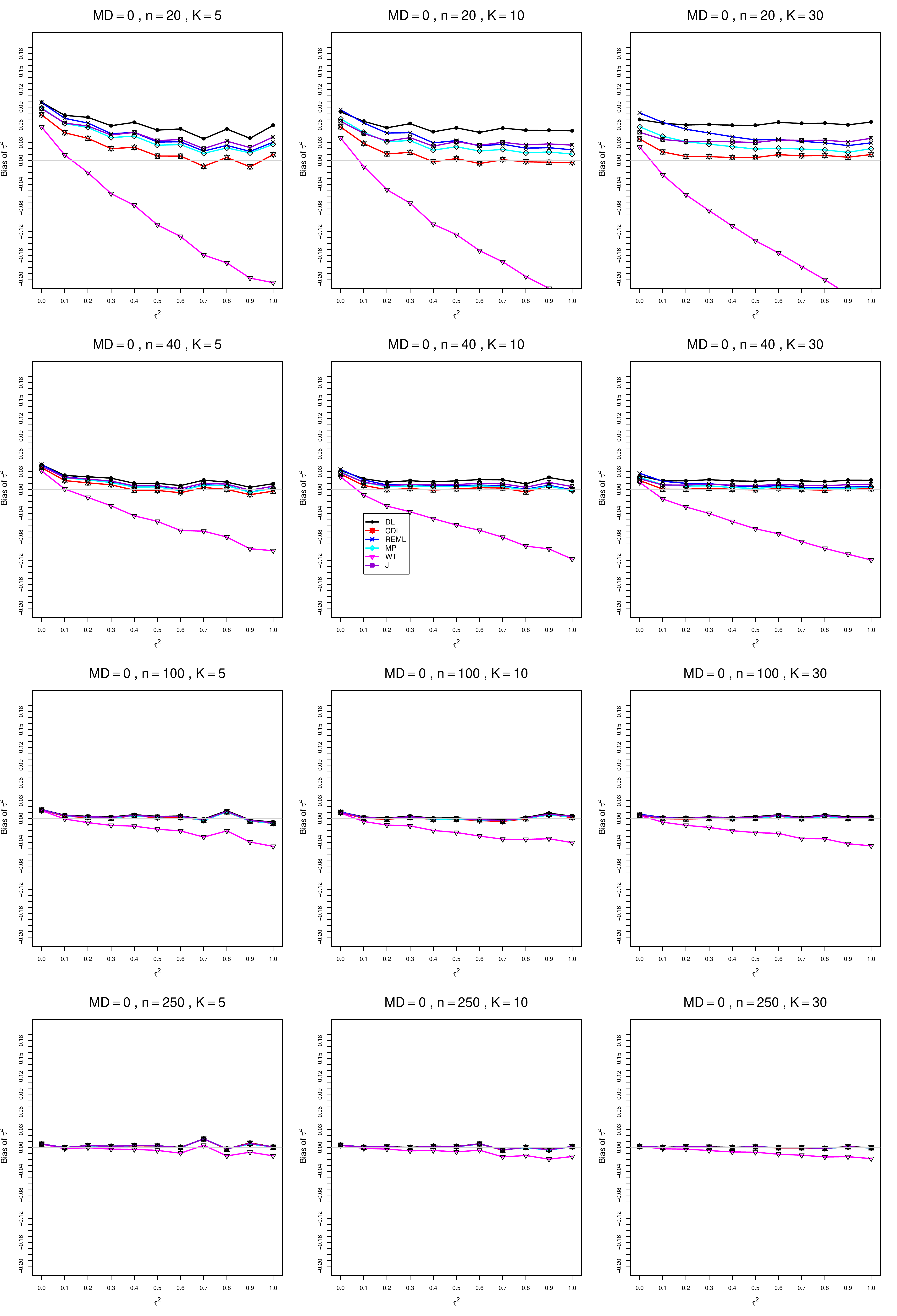}
	\caption{Bias of the estimation of  between-studies variance $\tau^2 = 0.0(0.1)1.0$ for $\mu=0$, $q=0.75$, $\sigma_C^2=1$, $\sigma_T^2=1$,  equal study sizes $n=20,\;40,\;100,\;250$ .
		\label{BiasTauMD0_S1_1q075}}
\end{figure}

\begin{figure}[t]
	\centering
	\includegraphics[scale=0.33]{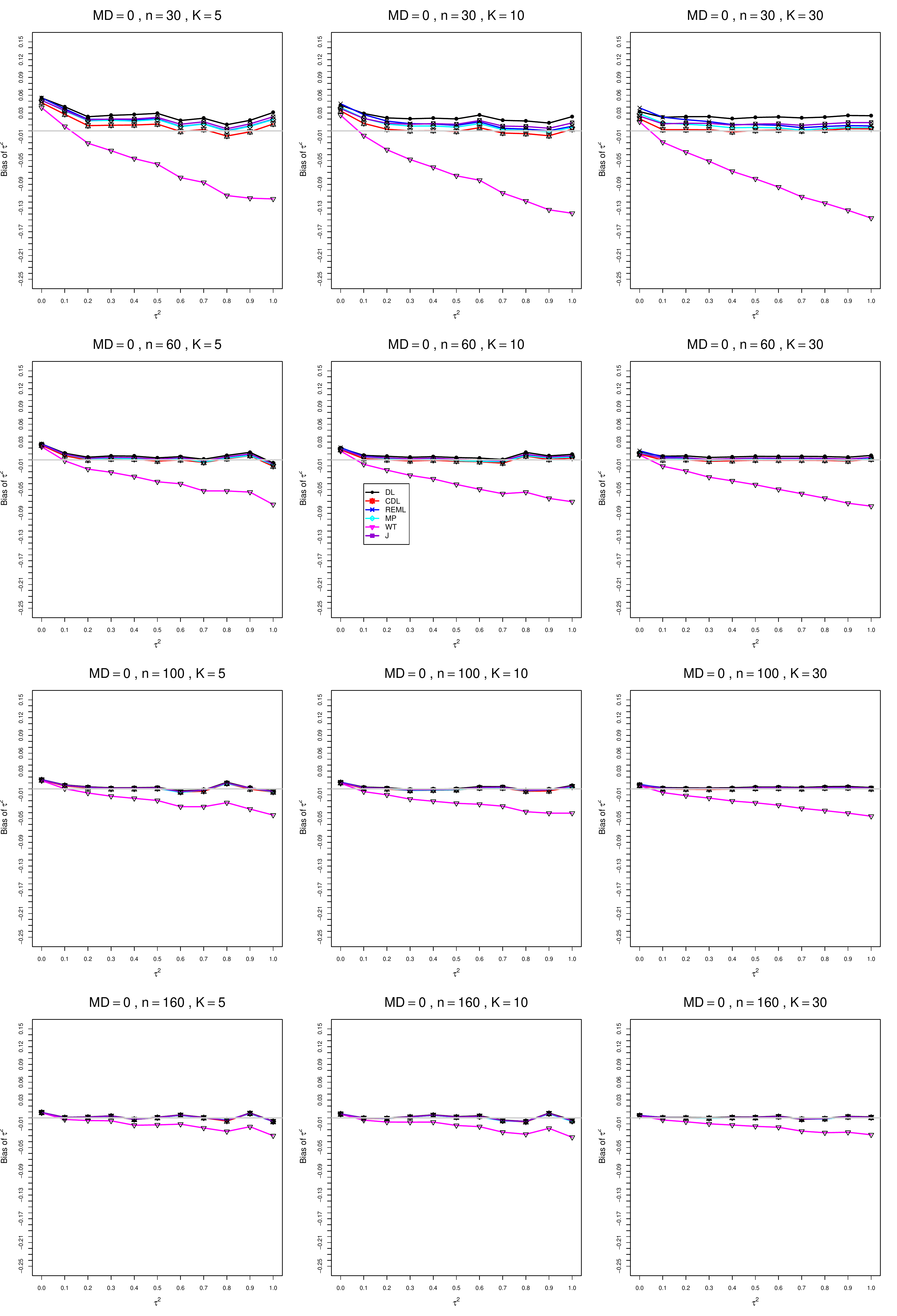}
	\caption{Bias of the estimation of  between-studies variance $\tau^2 = 0.0(0.1)1.0$ for $\mu=0$, $q=0.75$, $\sigma_C^2=1$, $\sigma_T^2=1$, unequal studies of average size  $\bar{n}=30,\;60,\;100,\;160$ .
		\label{BiasTauMD0_S1_1unequalq075}}
\end{figure}

\begin{figure}[t]
	\centering
	\includegraphics[scale=0.33]{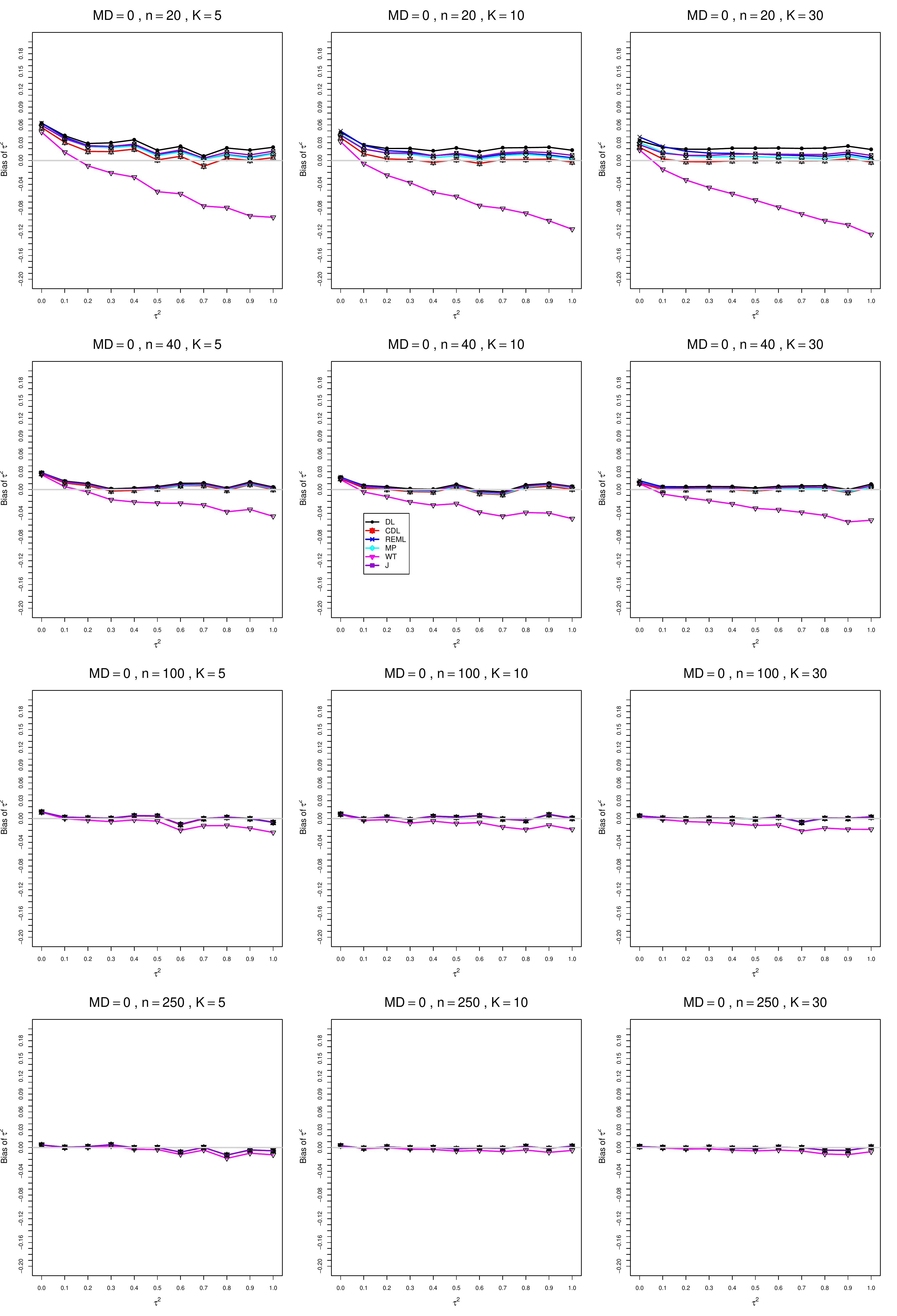}
	\caption{Bias of the estimation of  between-studies variance $\tau^2 = 0.0(0.1)1.0$ for $\mu=0$, $q=0.5$, $\sigma_C^2=1$, $\sigma_T^2=2$,  equal study sizes $n=20,\;40,\;100,\;250$ .
		\label{BiasTauMD0_S1_2}}
\end{figure}

\begin{figure}[t]
	\centering
	\includegraphics[scale=0.33]{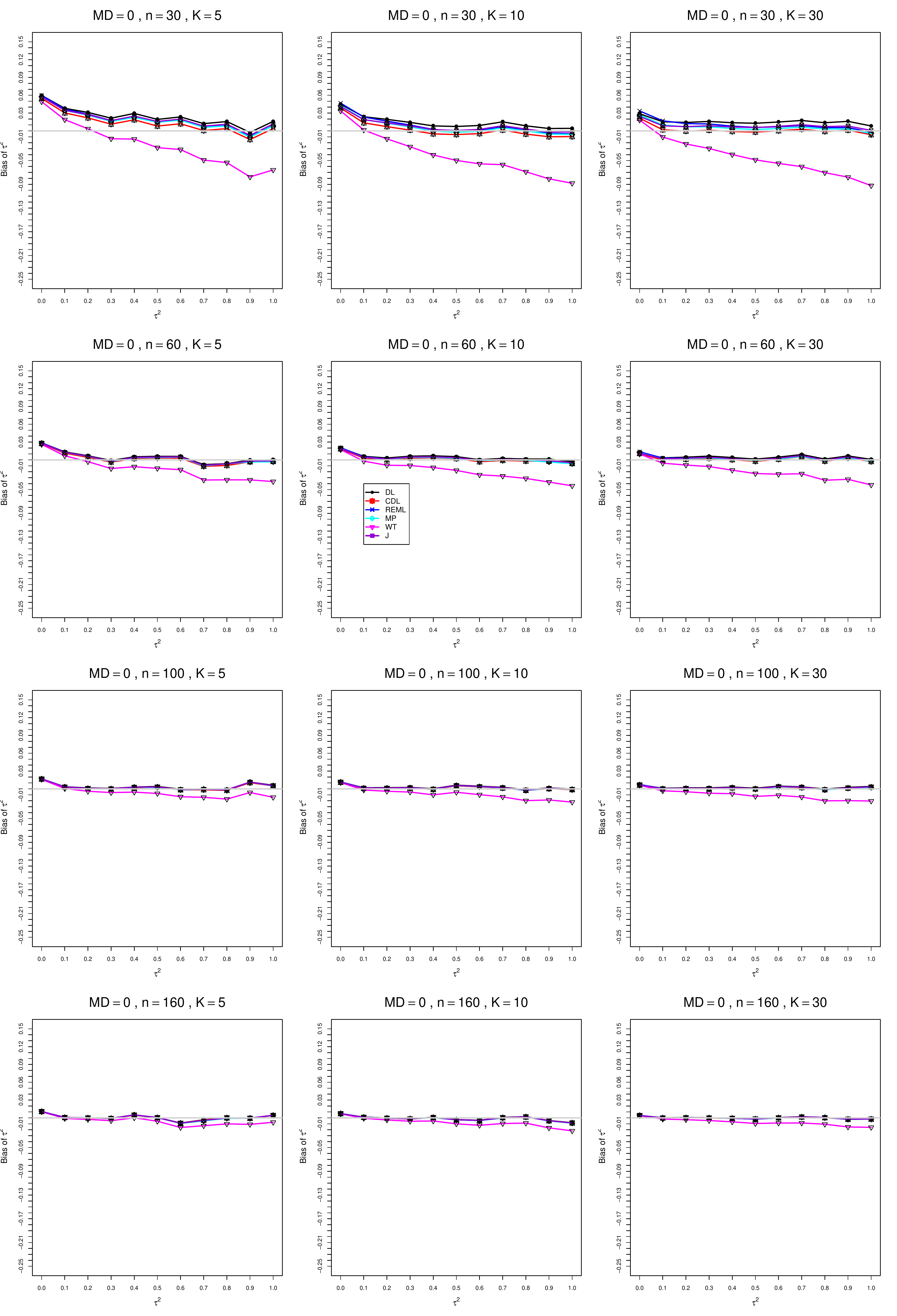}
	\caption{Bias of the estimation of  between-studies variance $\tau^2 = 0.0(0.1)1.0$ for $\mu=0$, $q=0.5$, $\sigma_C^2=1$, $\sigma_T^2=2$, unequal studies of average size  $\bar{n}=30,\;60,\;100,\;160$.
		\label{BiasTauMD0_S1_2unequal}}
\end{figure}

\begin{figure}[t]
	\centering
	\includegraphics[scale=0.33]{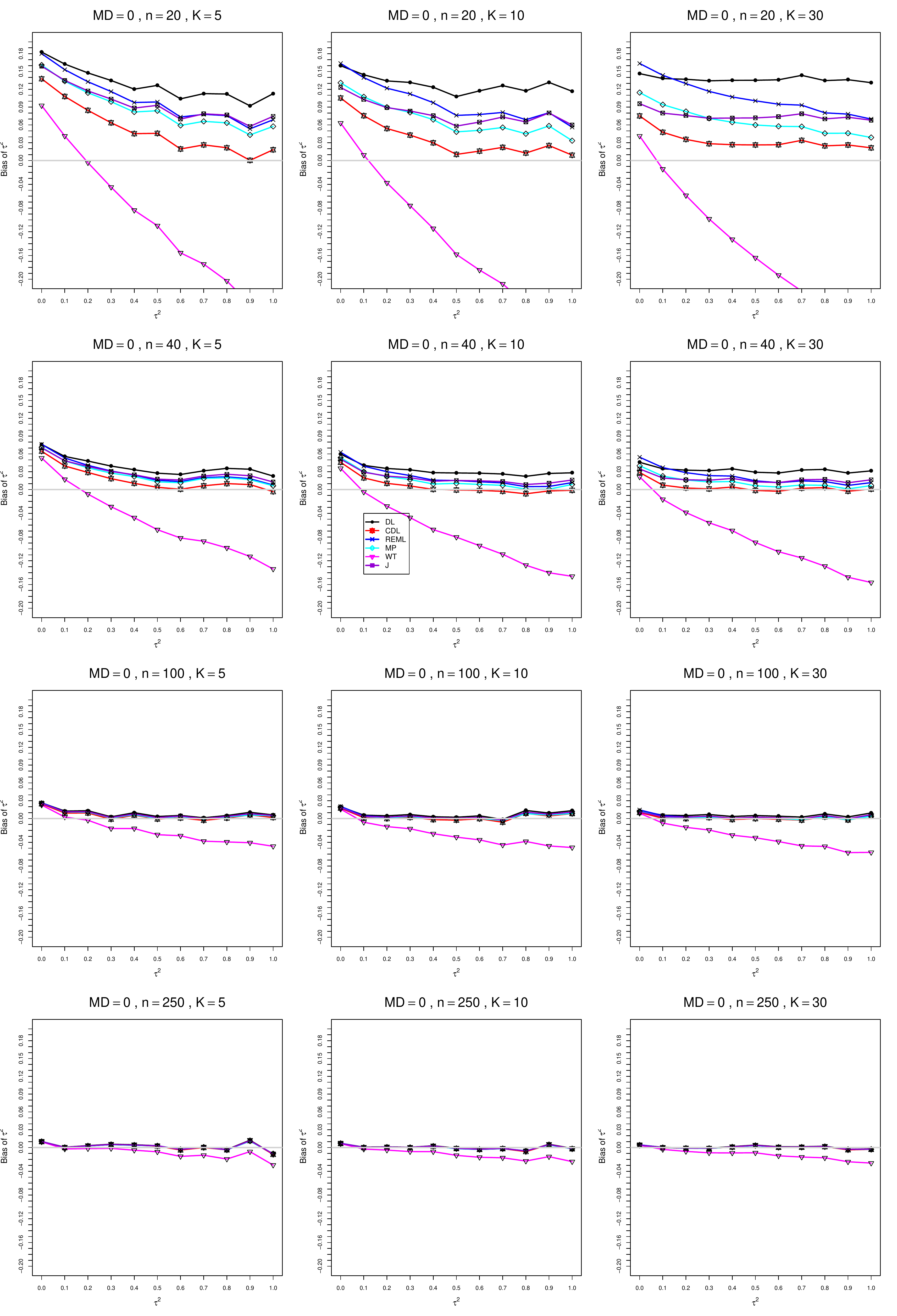}
	\caption{Bias of the estimation of  between-studies variance $\tau^2 = 0.0(0.1)1.0$ for $\mu=0$, $q=0.75$, $\sigma_C^2=1$, $\sigma_T^2=2$,  equal study sizes $n=20,\;40,\;100,\;250$ .
		\label{BiasTauMD0_S1_2q075}}
\end{figure}

\begin{figure}[t]
	\centering
	\includegraphics[scale=0.33]{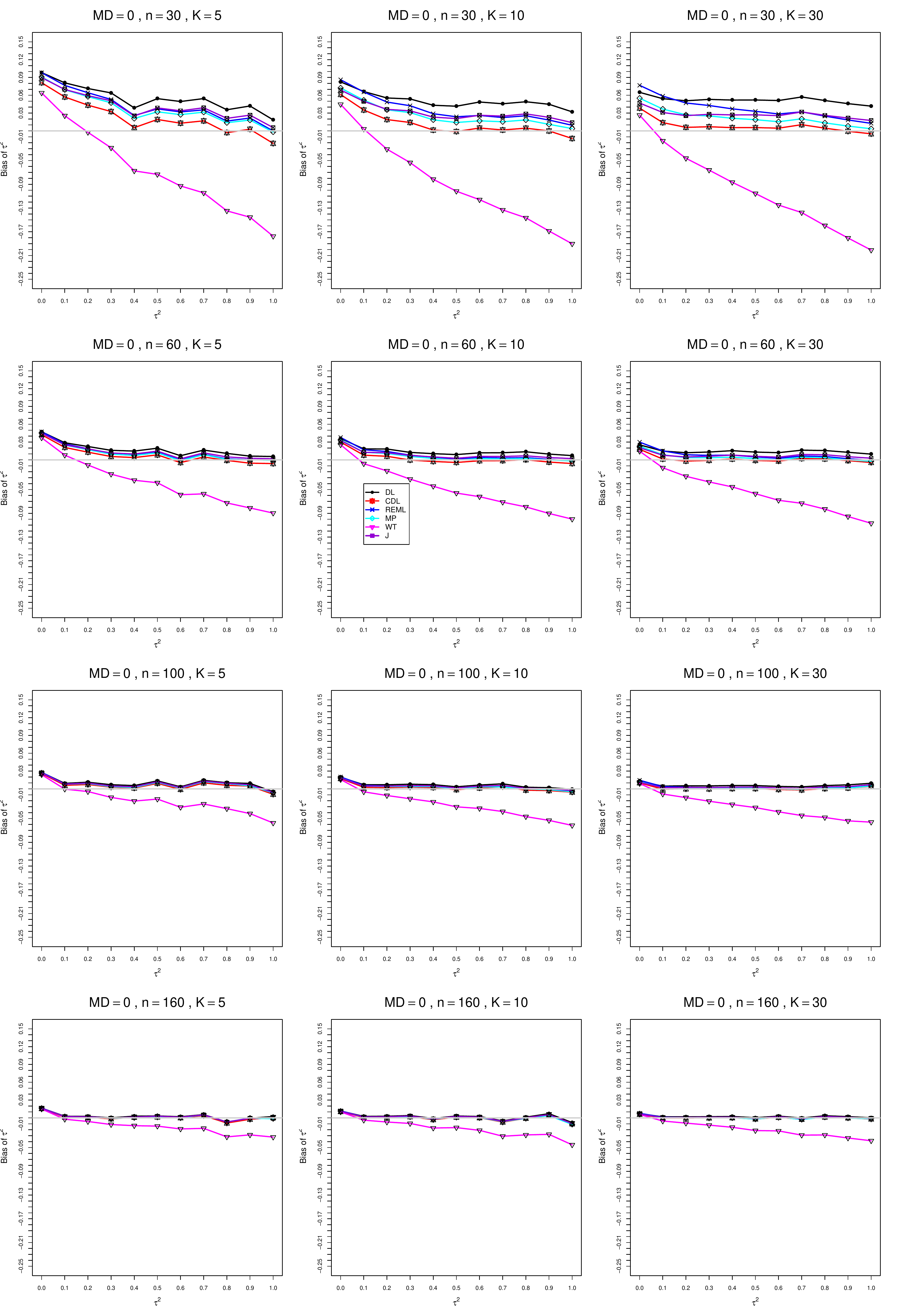}
	\caption{Bias of the estimation of  between-studies variance $\tau^2 = 0.0(0.1)1.0$ for $\mu=0$, $q=0.75$, $\sigma_C^2=1$, $\sigma_T^2=2$, unequal studies of average size  $\bar{n}=30,\;60,\;100,\;160$.
		\label{BiasTauMD0_S1_2unequalq075}}
\end{figure}

\clearpage

\setcounter{section}{0}
\renewcommand{\thefigure}{A2.\arabic{figure}}
\setcounter{figure}{0}
\setcounter{section}{0}
\section*{A2. Coverage of $\hat{\tau}^2$ for $\tau^2 = 0.0(0.1)1.0$, $\sigma_{C}^2=1$, $\sigma_{T}^2=1,\;2$.}
For coverage of $\hat{\tau}^2$, each figure corresponds to a value of $\mu (= 0, 0.2, 0.5, 1, 2)$, a value of $q (= .5, .75)$, a value of $\tau^2 = 0.0(0.1)1.0$ ,  a value of $\sigma_{C}^2=1$, a value of $\sigma_{T}^2=1,\;2$ and a set of values of $n$ (= 20, 40, 100, 250) or $\bar{n} (= 30, 60, 100, 160)$.\\
Each figure contains a panel (with $\tau^2$ on the horizontal axis) for each combination of n (or $\bar{n}$) and $K (=5, 10, 30)$.\\
The interval estimators of $\tau^2$ are
\begin{itemize}
	\item QP (Q-profile confidence interval)
	\item BJ (Biggerstaff and Jackson interval )
	\item PL (Profile likelihood interval)
	\item WT (Corrected Mandel-Paule moment estimator based on Welch-type approximation for Q distribution)
	\item J (Jacksons interval)
\end{itemize}

\begin{figure}[t]
	\centering
	\includegraphics[scale=0.33]{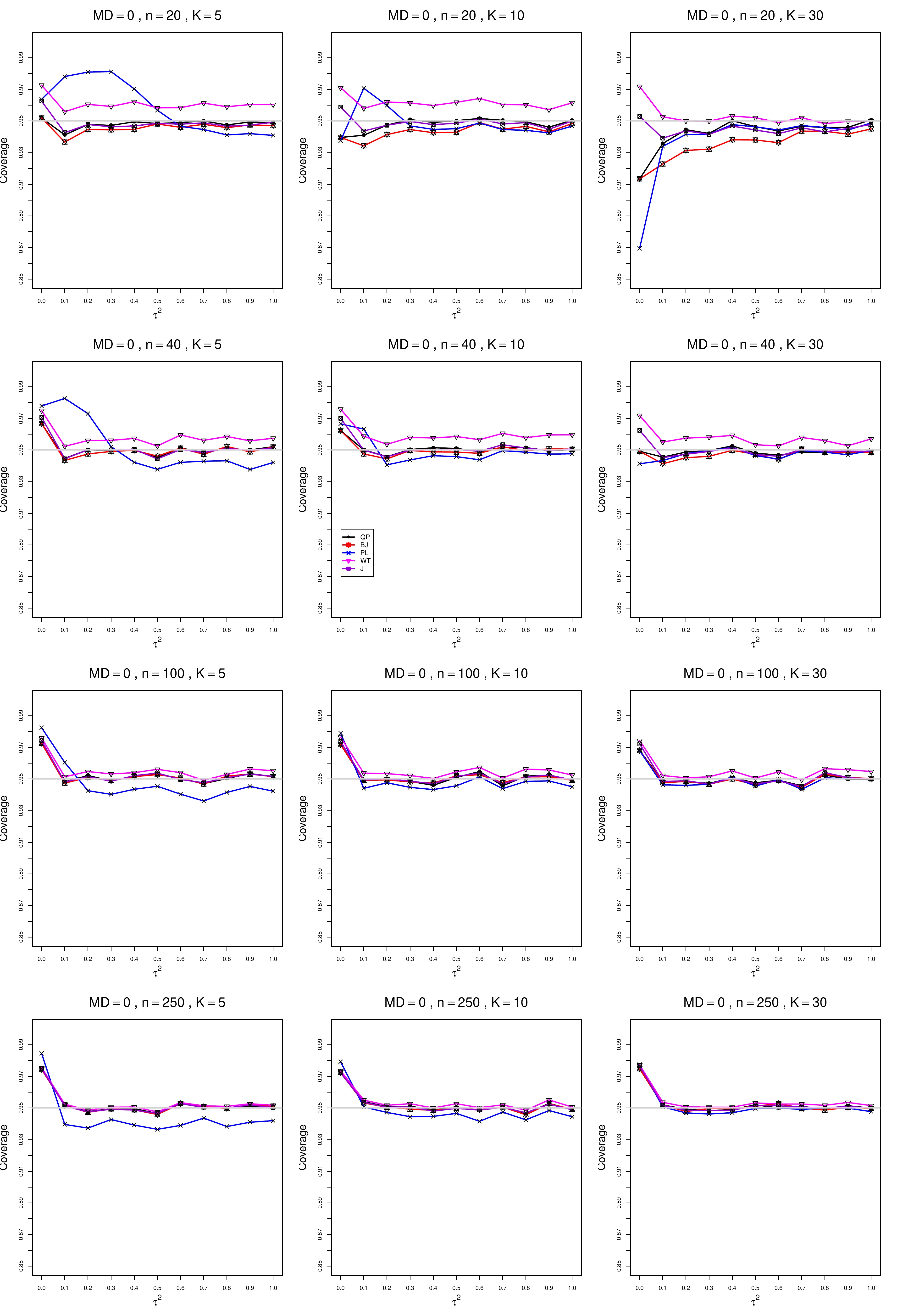}
	\caption{Coverage at  the nominal confidence level of $0.95$ of the  between-studies variance $\tau^2 = 0.0(0.1)1.0$ for $\mu=0$, $q=0.5$, $\sigma_C^2=1$, $\sigma_T^2=1$,  equal study sizes $n=20,\;40,\;100,\;250$.
		\label{CovTauMD0_S1_1}}
\end{figure}

\begin{figure}[t]
	\centering
	\includegraphics[scale=0.33]{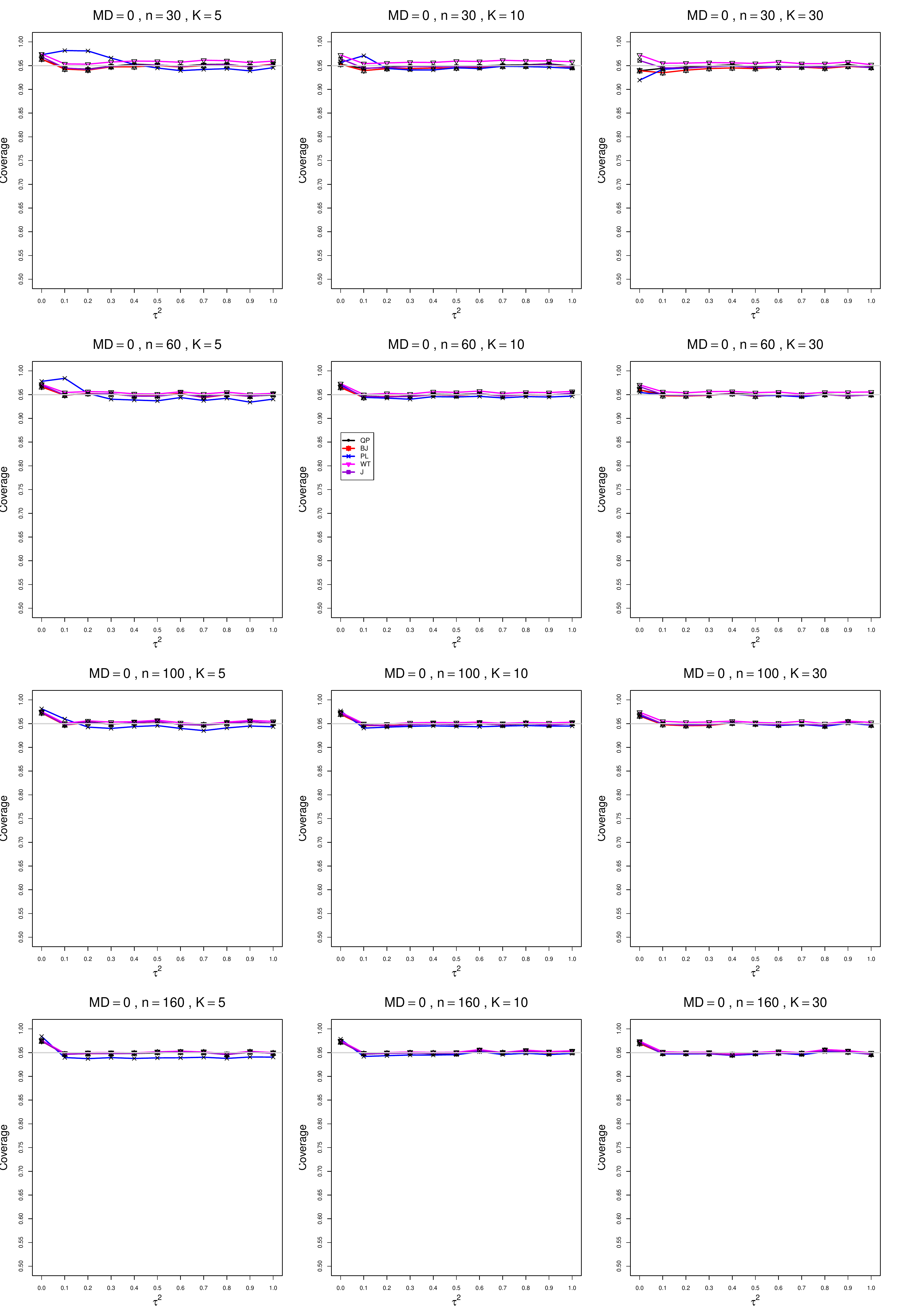}
	\caption{Coverage of 95\% confidence intervals for the  between-studies variance $\tau^2 = 0.0(0.1)1.0$ for $\mu=0$, $q=0.5$, $\sigma_C^2=1$, $\sigma_T^2=1$, unequal studies of average size  $\bar{n}=30,\;60,\;100,\;160$.
		\label{CovTauMD0_S1_1unequal}}
\end{figure}
\clearpage

\begin{figure}[t]
	\centering
	\includegraphics[scale=0.33]{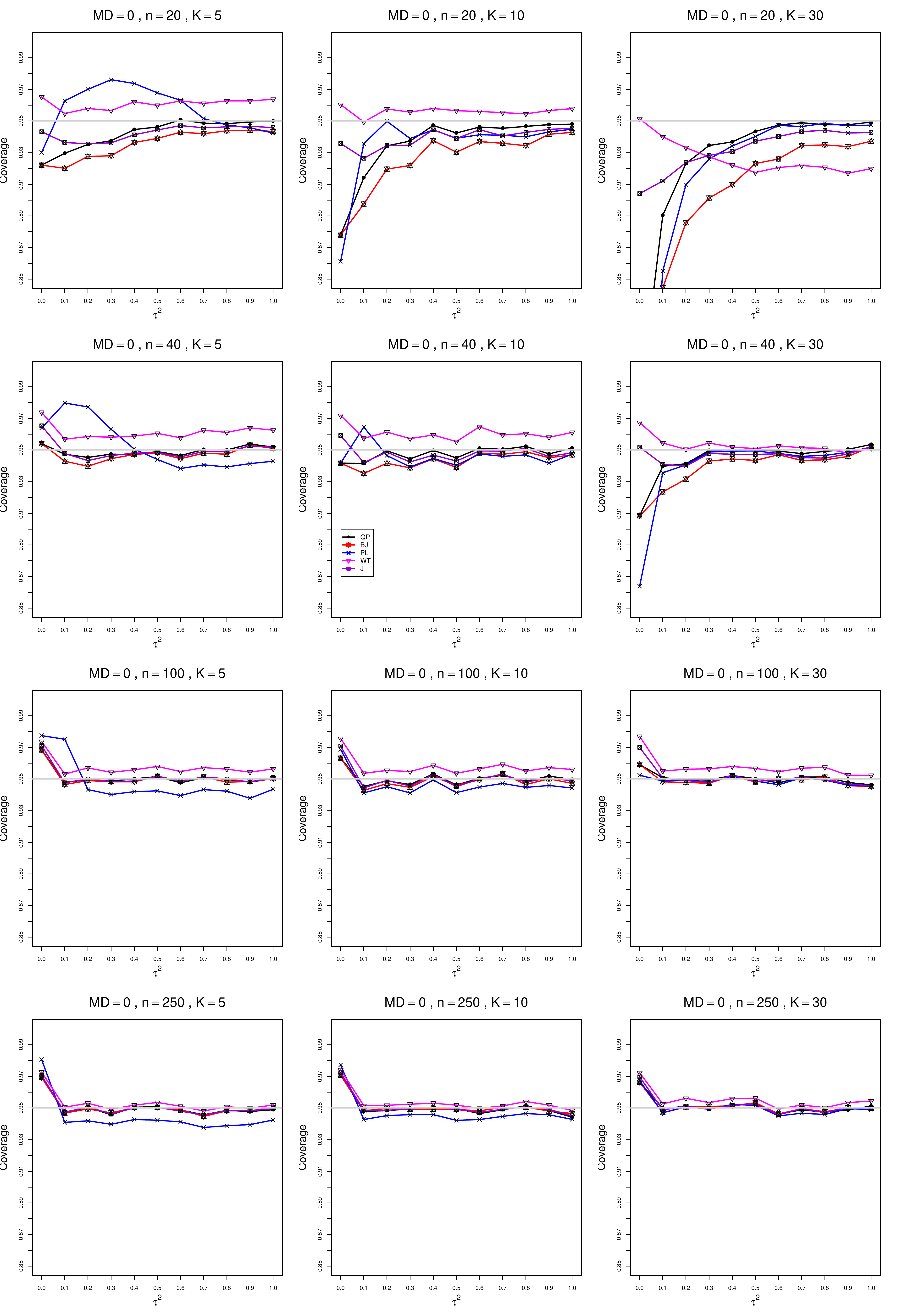}
	\caption{Coverage of 95\% confidence intervals for the  between-studies variance $\tau^2 = 0.0(0.1)1.0$ for $\mu=0$, $q=0.75$, $\sigma_C^2=1$, $\sigma_T^2=1$,  equal study sizes $n=20,\;40,\;100,\;250$.
		\label{CovTauMD0_S1_1q075}}
\end{figure}

\begin{figure}[t]
	\centering
	\includegraphics[scale=0.33]{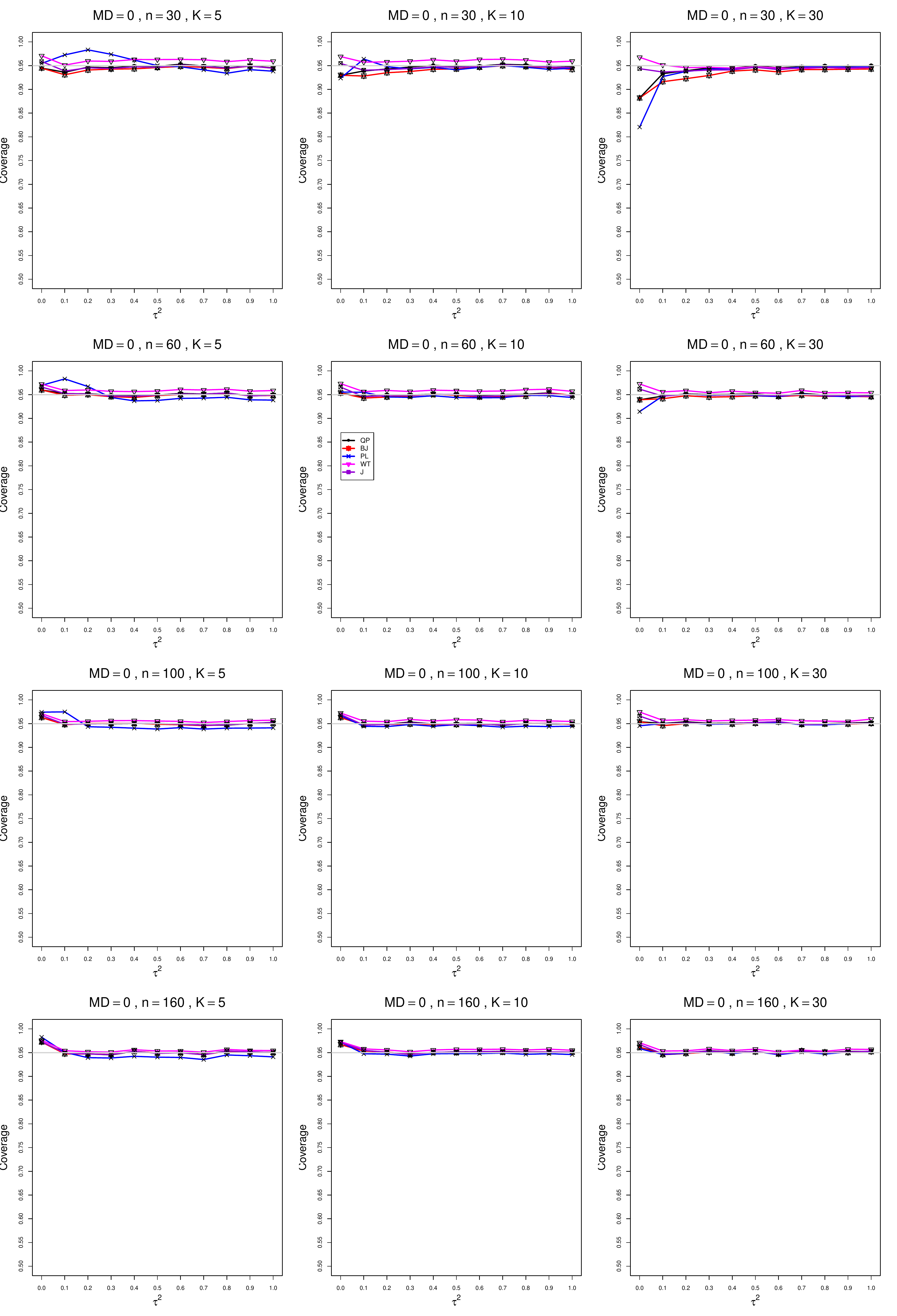}
	\caption{Coverage of 95\% confidence intervals for the  between-studies variance $\tau^2 = 0.0(0.1)1.0$ for $\mu=0$, $q=0.75$, $\sigma_C^2=1$, $\sigma_T^2=1$, unequal studies of average size  $\bar{n}=30,\;60,\;100,\;160$.
		\label{CovTauMD0_S1_1unequalq075}}
\end{figure}
\clearpage

\begin{figure}[t]
	\centering
	\includegraphics[scale=0.33]{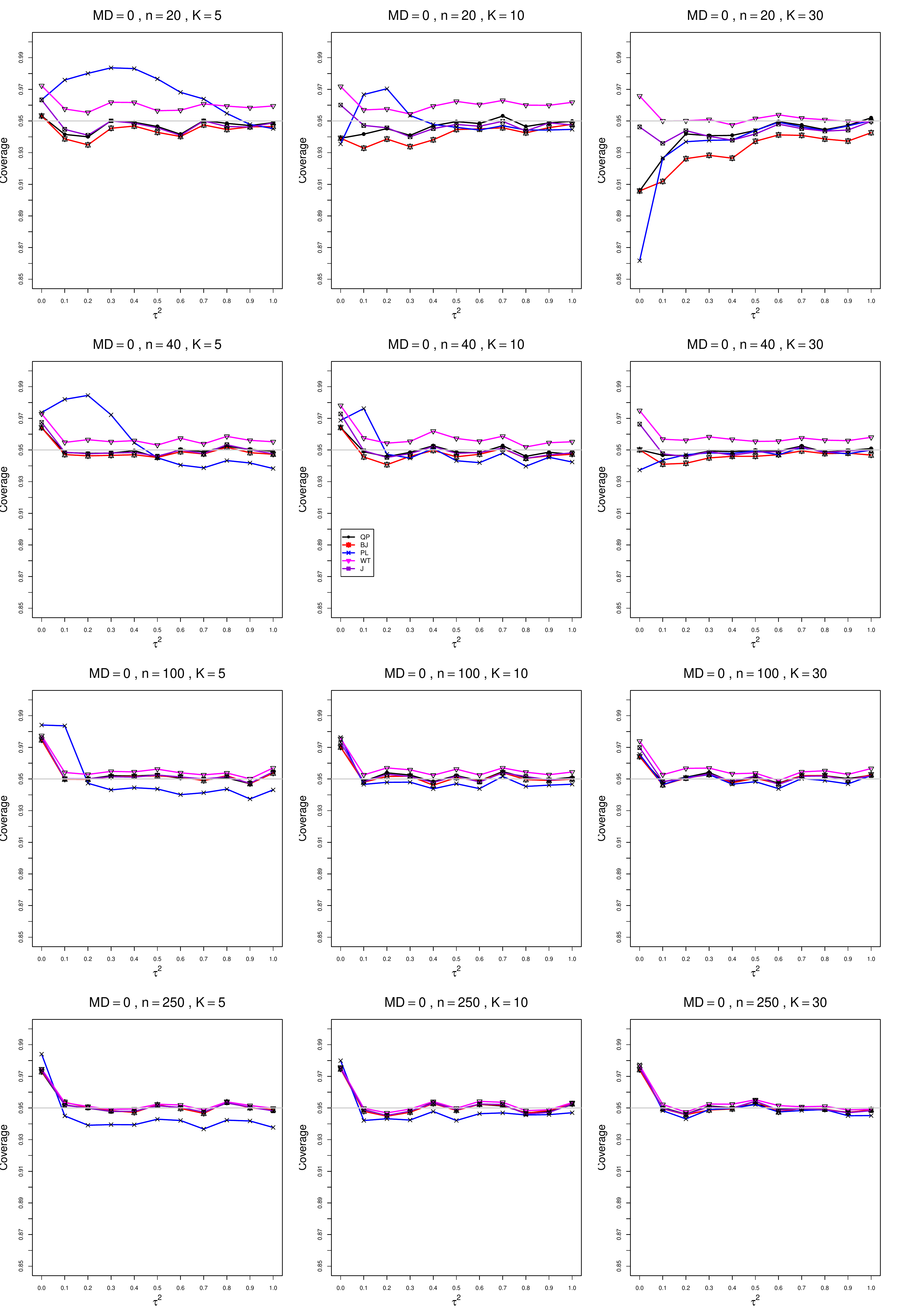}
	\caption{Coverage of 95\% confidence intervals for the  between-studies variance $\tau^2 = 0.0(0.1)1.0$ for $\mu=0$, $q=0.5$, $\sigma_C^2=1$, $\sigma_T^2=2$,  equal study sizes $n=20,\;40,\;100,\;250$.
		\label{CovTauMD0_S2_1}}
\end{figure}

\begin{figure}[t]
	\centering
	\includegraphics[scale=0.33]{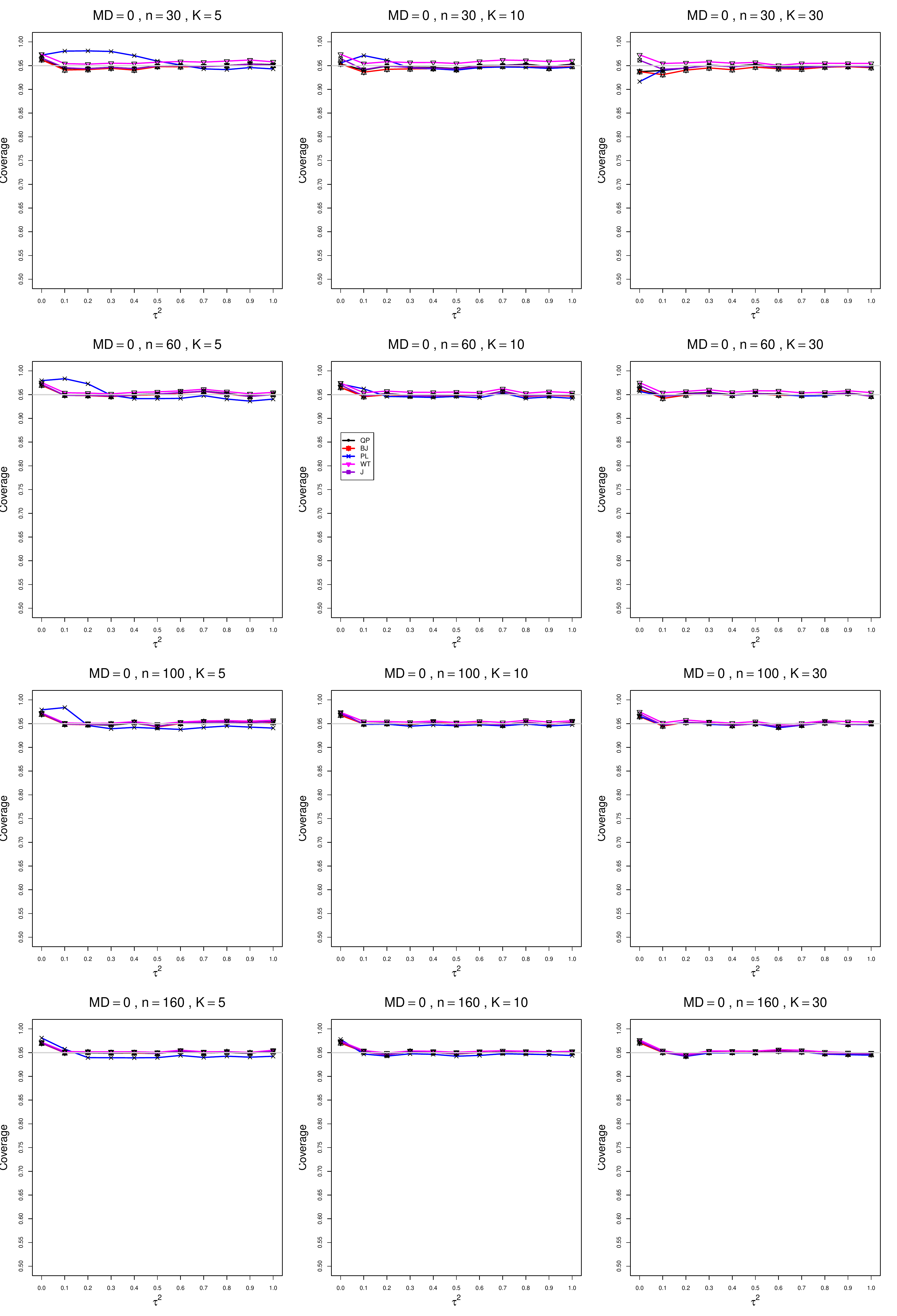}
	\caption{Coverage of 95\% confidence intervals for the  between-studies variance $\tau^2 = 0.0(0.1)1.0$ for $\mu=0$, $q=0.5$, $\sigma_C^2=1$, $\sigma_T^2=2$, unequal studies of average size  $\bar{n}=30,\;60,\;100,\;160$.
		\label{CovTauMD0_S2_1unequal}}
\end{figure}

\clearpage

\begin{figure}[t]
	\centering
	\includegraphics[scale=0.33]{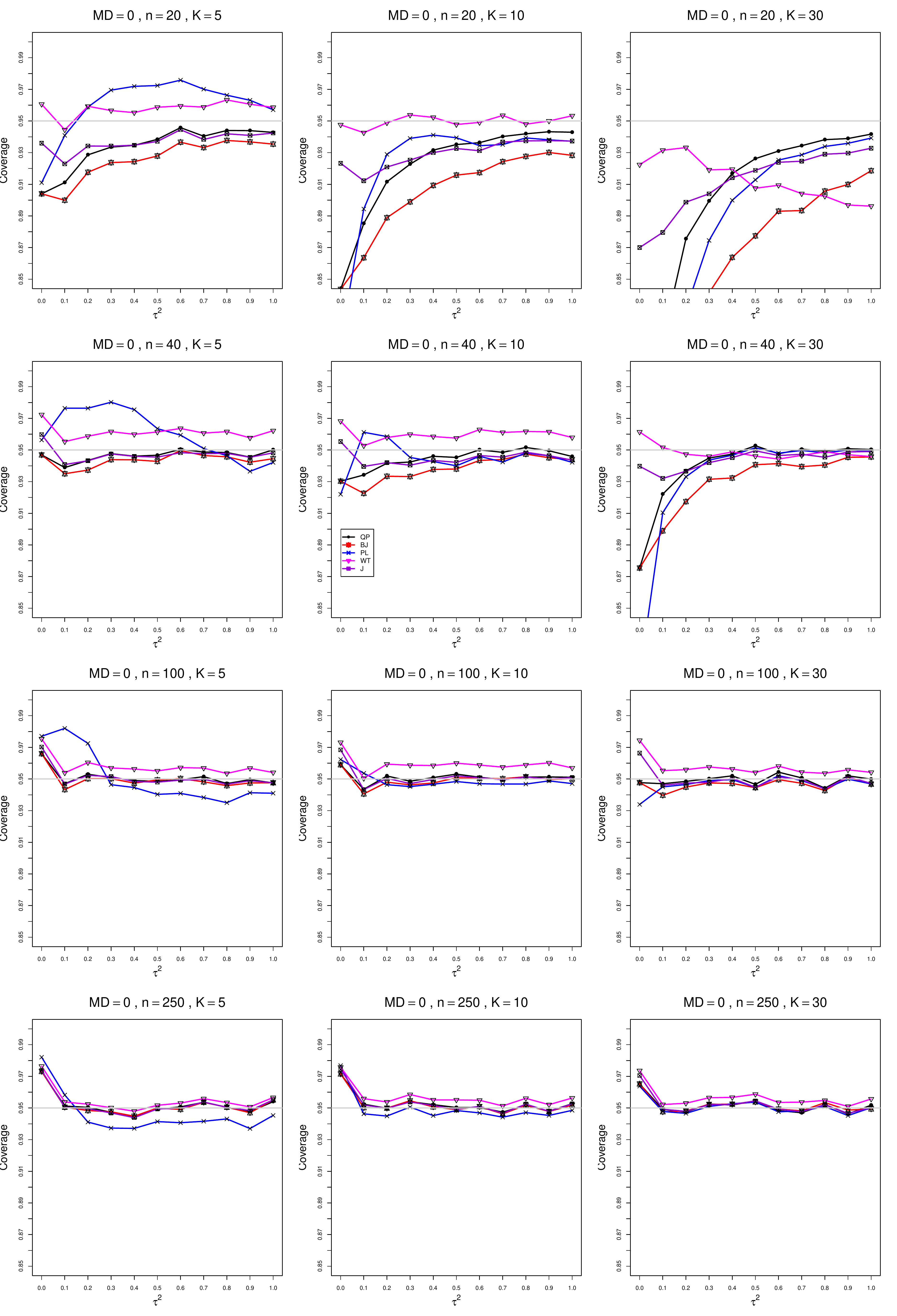}
	\caption{Coverage of 95\% confidence intervals for the  between-studies variance $\tau^2 = 0.0(0.1)1.0$ for $\mu=0$, $q=0.75$, $\sigma_C^2=1$, $\sigma_T^2=2$,  equal study sizes $n=20,\;40,\;100,\;250$.
		\label{CovTauMD0_S2_1q075}}
\end{figure}

\begin{figure}[t]
	\centering
	\includegraphics[scale=0.33]{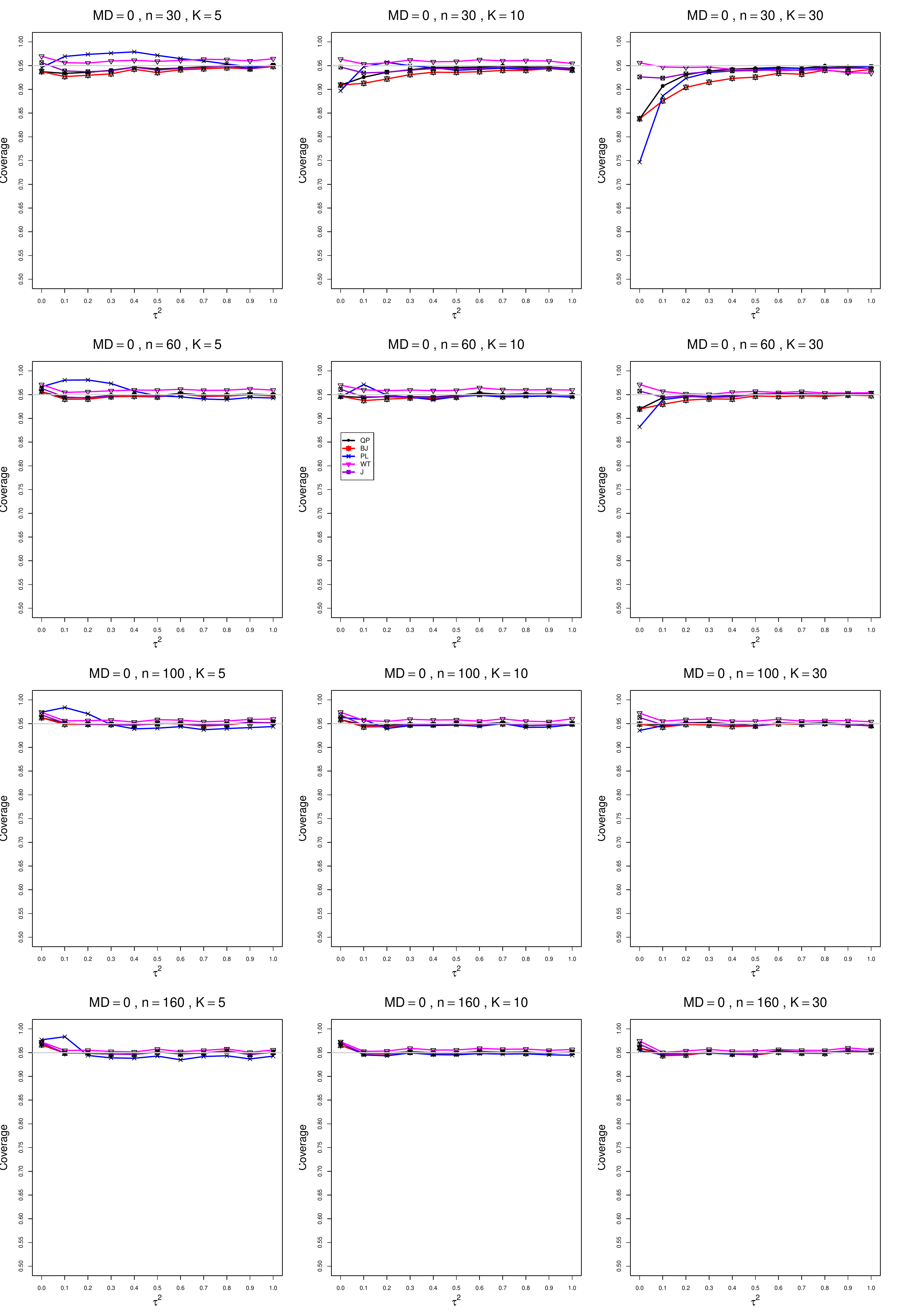}
	\caption{Coverage of 95\% confidence intervals for the  between-studies variance $\tau^2 = 0.0(0.1)1.0$ for $\mu=0$, $q=0.75$, $\sigma_C^2=1$, $\sigma_T^2=2$, unequal studies of average size  $\bar{n}=30,\;60,\;100,\;160$.
		\label{BiasTauMD0_S2_1unequalq075}}
\end{figure}


\clearpage

\renewcommand{\thefigure}{A3.\arabic{figure}}
\setcounter{figure}{0}
\setcounter{section}{0}
\section*{A3. Bias of $\hat{\tau}^2$ for $\tau^2 = 0.0(0.01)0.1$, $\sigma_{C}^2=1$, $\sigma_{T}^2=1,\;2$.}
For bias of $\hat{\tau}^2$, each figure corresponds to a value of $\mu (= 0, 0.2, 0.5, 1, 2)$, a value of $q (= .5, .75)$, a value of $\tau^2 = 0.0(0.01)0.1$, a value of $\sigma_{C}^2=1$, a value of $\sigma_{T}^2=1,\;2$ , and a set of values of $n$ (= 20, 40, 100, 250) or $\bar{n}$ (= 30, 60, 100, 160).\\
Each figure contains a panel (with $\tau^2$ on the horizontal axis) for each combination of n (or $\bar{n}$) and $K (=5, 10, 30)$.\\
The point estimators of $\tau^2$ are
\begin{itemize}
	\item DL (DerSimonian-Laird)
	\item REML (restricted maximum likelihood)
	\item MP (Mandel-Paule)
	\item WT (Corrected Mandel-Paule moment estimator based on Welch-type approximation for Q distribution)
	\item J (Jackson)
	\item CDL (Corrected DerSimonian-Laird)
\end{itemize}

\begin{figure}[t]
	\centering
	\includegraphics[scale=0.33]{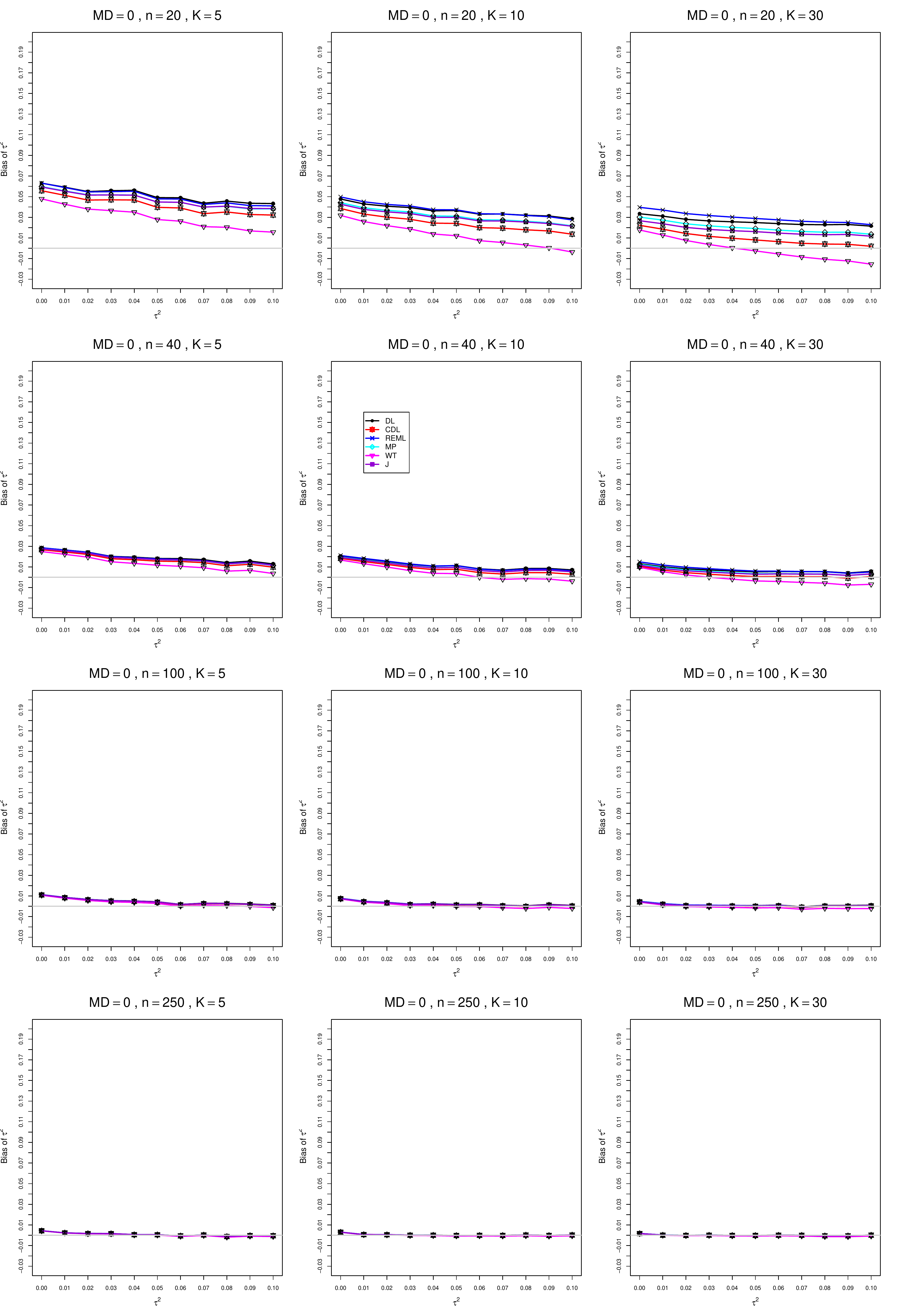}
	\caption{Bias of the estimation of  between-studies variance $\tau^2 = 0.0(0.01)0.1$ for $\mu=0$, $q=0.5$, $\sigma_C^2=1$, $\sigma_T^2=1$,  equal study sizes $n=20,\;40,\;100,\;250$.
		\label{BiasTauMD0_S1_1_small_tau2}}
\end{figure}

\begin{figure}[t]
	\centering
	\includegraphics[scale=0.33]{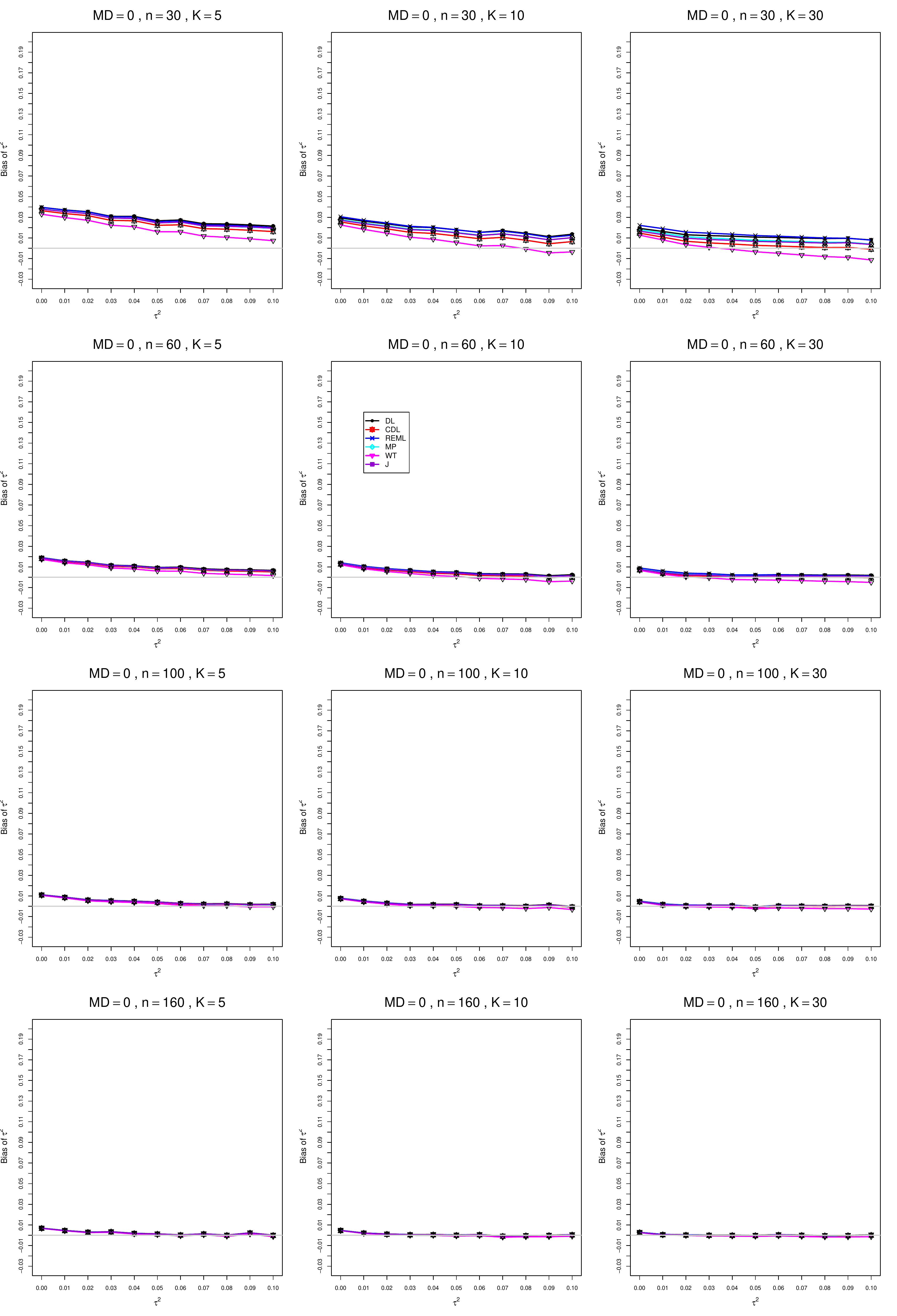}
	\caption{Bias of the estimation of  between-studies variance $\tau^2 = 0.0(0.01)0.1$ for $\mu=0$, $q=0.5$, $\sigma_C^2=1$, $\sigma_T^2=1$, unequal studies of average size  $\bar{n}=30,\;60,\;100,\;160$.
		\label{BiasTauMD0_S1_1unequal_small_tau2}}
\end{figure}

\begin{figure}[t]
	\centering
	\includegraphics[scale=0.33]{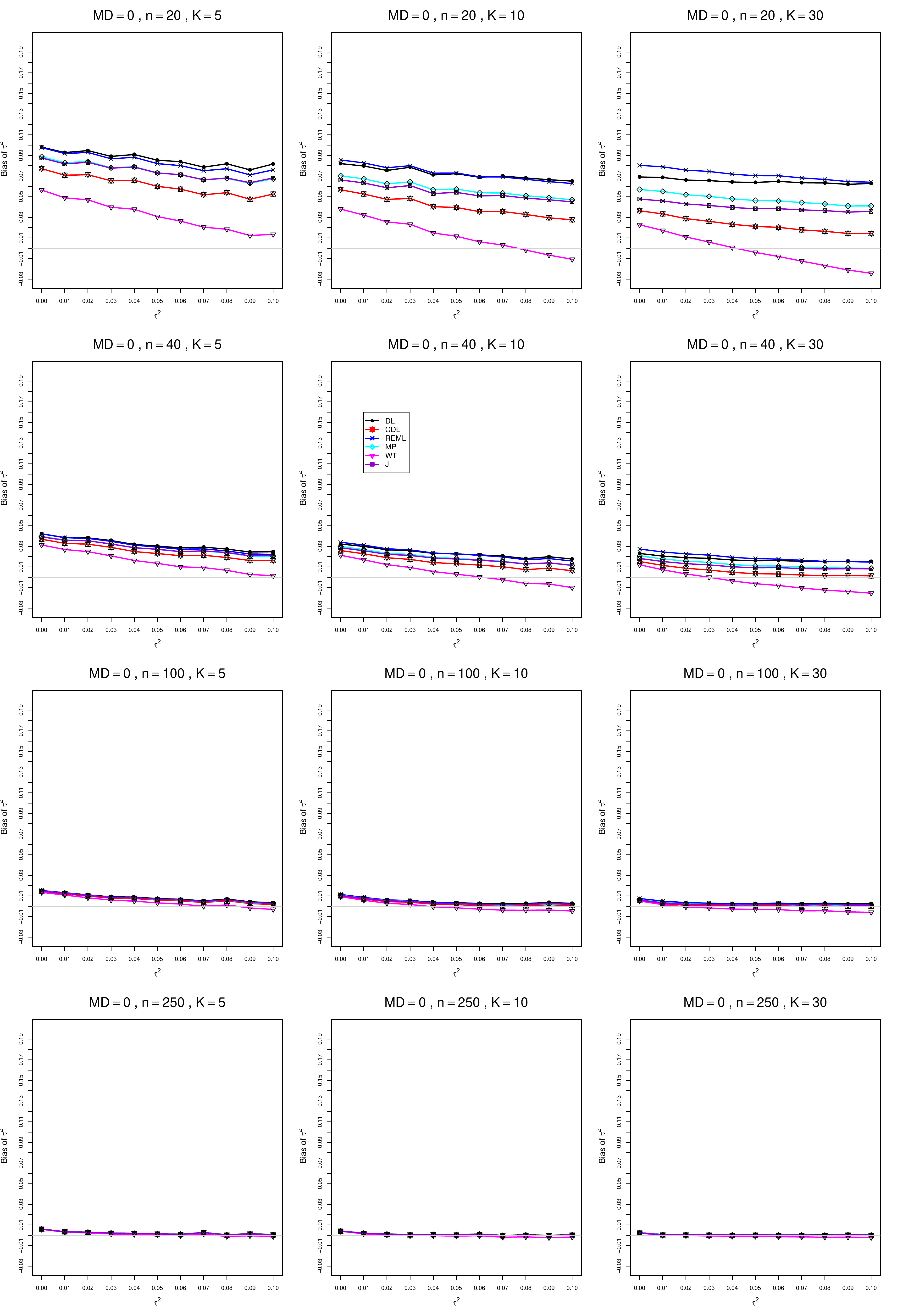}
	\caption{Bias of the estimation of  between-studies variance $\tau^2 = 0.0(0.01)0.1$ for $\mu=0$, $q=0.75$, $\sigma_C^2=1$, $\sigma_T^2=1$,  equal study sizes $n=20,\;40,\;100,\;250$.
		\label{BiasTauMD0_S1_1q075_small_tau2}}
\end{figure}

\begin{figure}[t]
	\centering
	\includegraphics[scale=0.33]{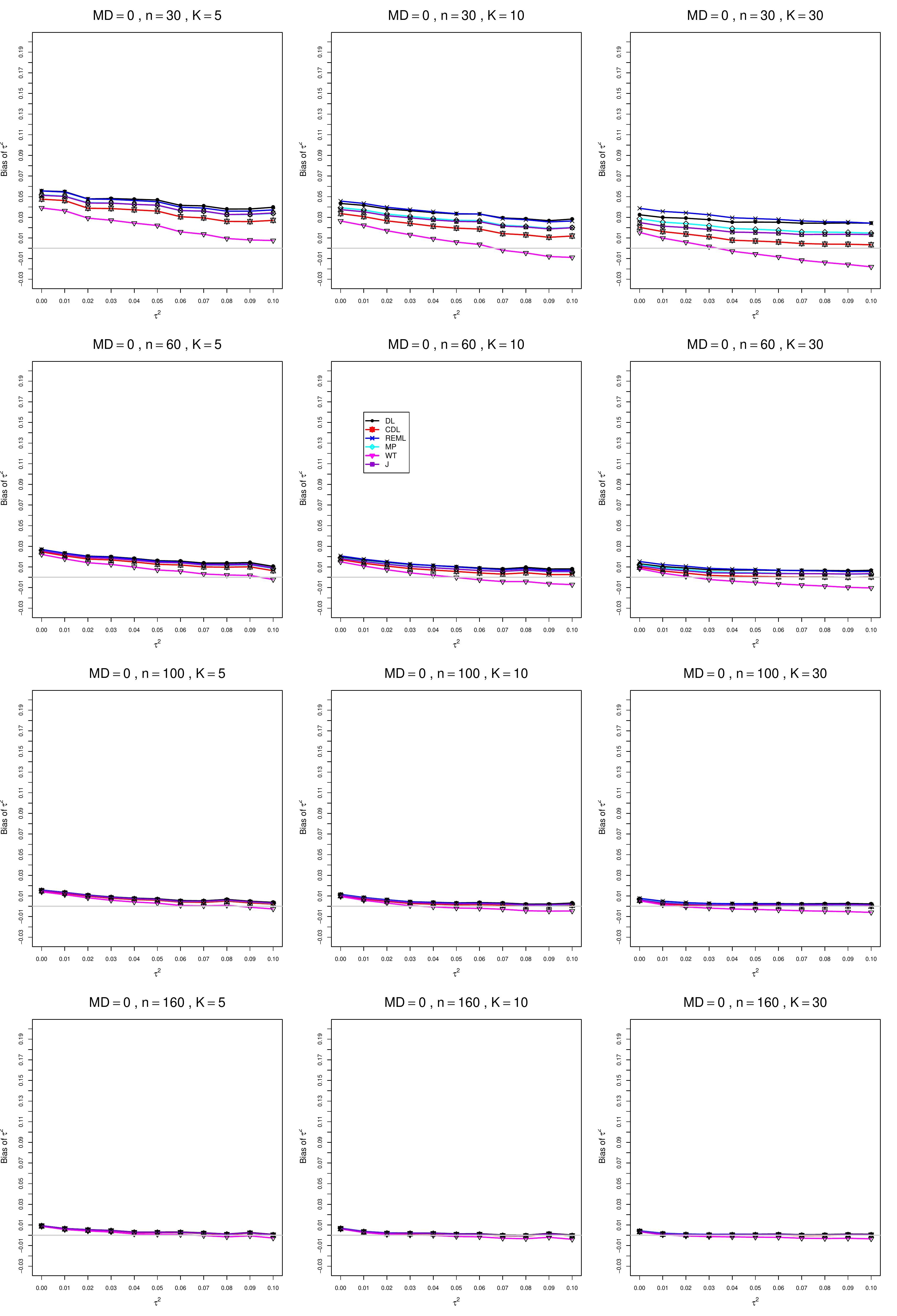}
	\caption{Bias of the estimation of  between-studies variance $\tau^2 = 0.0(0.01)0.1$ for $\mu=0$, $q=0.75$, $\sigma_C^2=1$, $\sigma_T^2=1$, unequal studies of average size  $\bar{n}=30,\;60,\;100,\;160$.
		\label{BiasTauMD0_S1_1unequalq075_small_tau2}}
\end{figure}

\clearpage
\begin{figure}[t]
	\centering
	\includegraphics[scale=0.33]{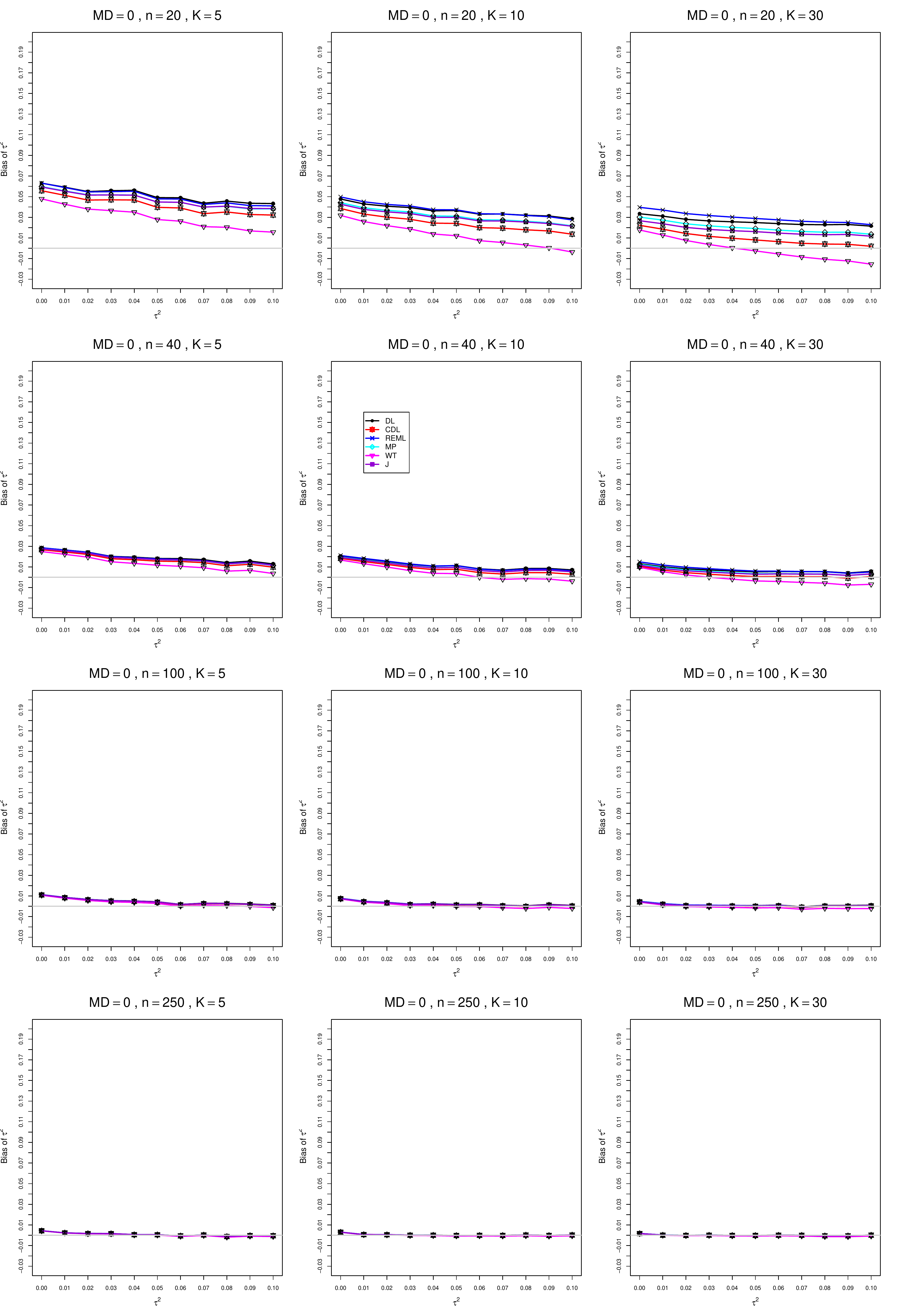}
	\caption{Bias of the estimation of  between-studies variance $\tau^2 = 0.0(0.01)0.1$ for $\mu=0$, $q=0.5$, $\sigma_C^2=1$, $\sigma_T^2=2$,  equal study sizes $n=20,\;40,\;100,\;250$.
		\label{BiasTauMD0_S1_2_small_tau2}}
\end{figure}

\begin{figure}[t]
	\centering
	\includegraphics[scale=0.33]{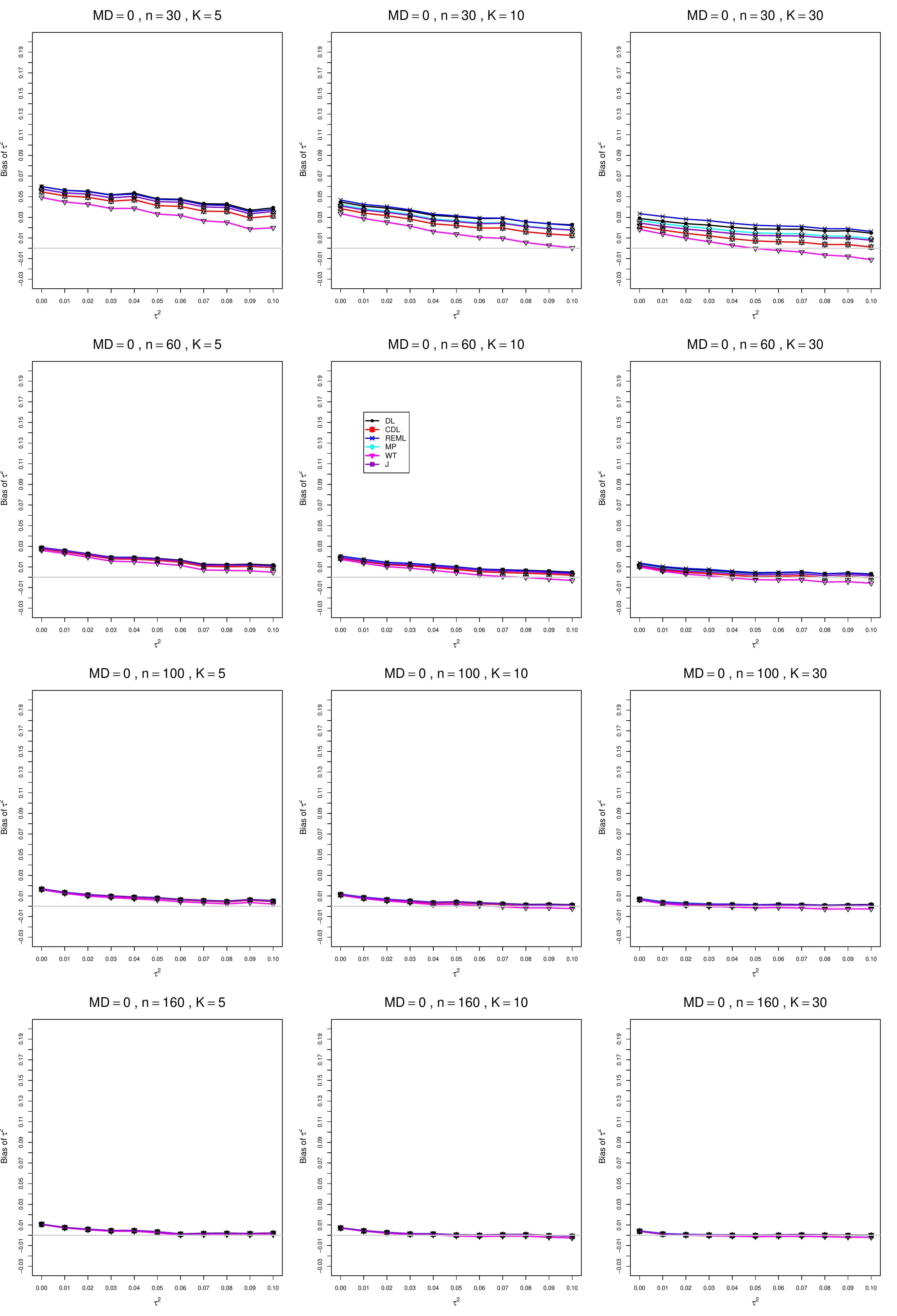}
	\caption{Bias of the estimation of  between-studies variance $\tau^2 = 0.0(0.01)0.1$ for $\mu=0$, $q=0.5$, $\sigma_C^2=1$, $\sigma_T^2=2$, unequal studies of average size  $\bar{n}=30,\;60,\;100,\;160$.
		\label{BiasTauMD0_S1_2unequal_small_tau2}}
\end{figure}


\begin{figure}[t]
	\centering
	\includegraphics[scale=0.33]{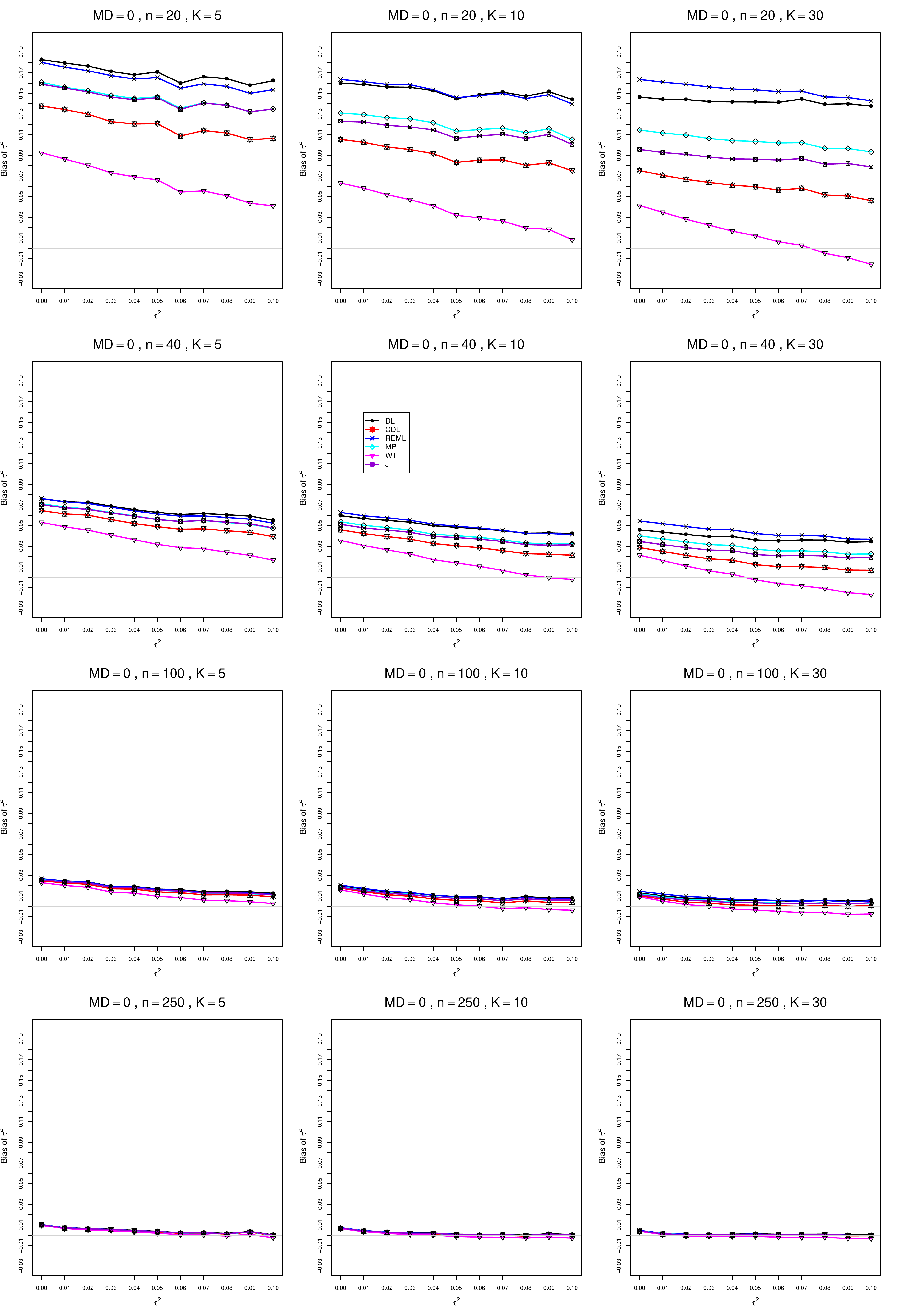}
	\caption{Bias of the estimation of  between-studies variance $\tau^2 = 0.0(0.01)0.1$ for $\mu=0$, $q=0.75$, $\sigma_C^2=1$, $\sigma_T^2=2$,  equal study sizes $n=20,\;40,\;100,\;250$.
		\label{BiasTauMD0_S1_2q075_small_tau2}}
\end{figure}

\begin{figure}[t]
	\centering
	\includegraphics[scale=0.33]{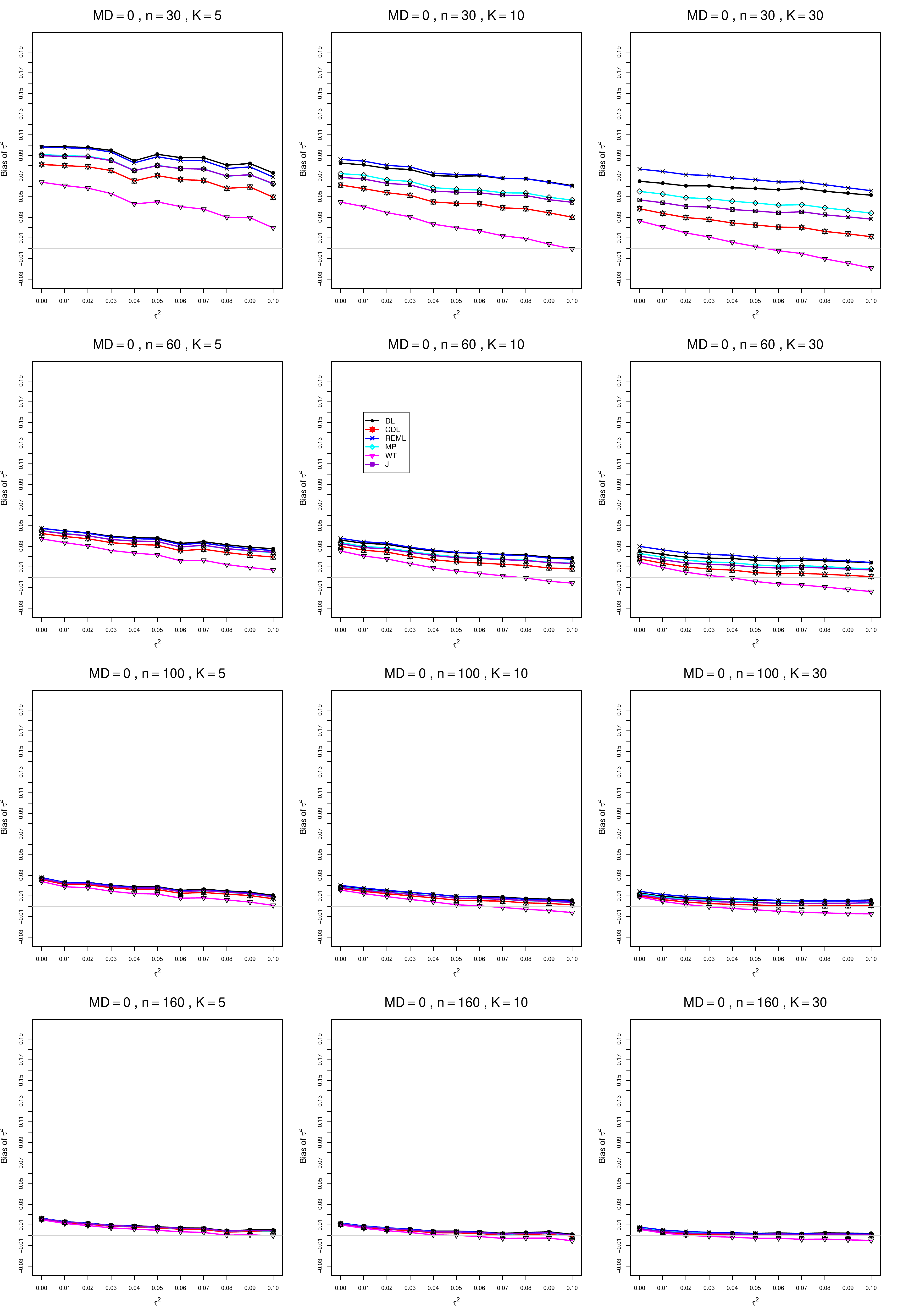}
	\caption{Bias of the estimation of  between-studies variance $\tau^2 = 0.0(0.01)0.1$ for $\mu=0$, $q=0.75$, $\sigma_C^2=1$, $\sigma_T^2=2$, unequal studies of average size  $\bar{n}=30,\;60,\;100,\;160$.
		\label{BiasTauMD0_S1_2unequalq075_small_tau2}}
\end{figure}

\clearpage
\renewcommand{\thefigure}{A4.\arabic{figure}}
\setcounter{figure}{0}
\setcounter{section}{0}
\section*{A4. Coverage of $\hat{\tau}^2$ for $\tau^2 = 0.0(0.01)0.1$, $\sigma_{C}^2=1$, $\sigma_{T}^2=1,\;2$.}
For coverage of $\hat{\tau}^2$, each figure corresponds to a value of $\mu (= 0, 0.2, 0.5, 1, 2)$, a value of $q (= .5, .75)$, a value of $\tau^2 = 0.0(0.01)0.1$ ,  a value of $\sigma_{C}^2=1$, a value of $\sigma_{T}^2=1,\;2$ and a set of values of $n$ (= 20, 40, 100, 250) or $\bar{n} (= 30, 60, 100, 160)$.\\
Each figure contains a panel (with $\tau^2$ on the horizontal axis) for each combination of n (or $\bar{n}$) and $K (=5, 10, 30)$.\\
The interval estimators of $\tau^2$ are
\begin{itemize}
	\item QP (Q-profile confidence interval)
	\item BJ (Biggerstaff and Jackson interval )
	\item PL (Profile likelihood interval)
	\item WT (Corrected Mandel-Paule moment estimator based on Welch-type approximation for Q distribution)
	\item J (Jacksons interval)
\end{itemize}

\begin{figure}[t]
	\centering
	\includegraphics[scale=0.33]{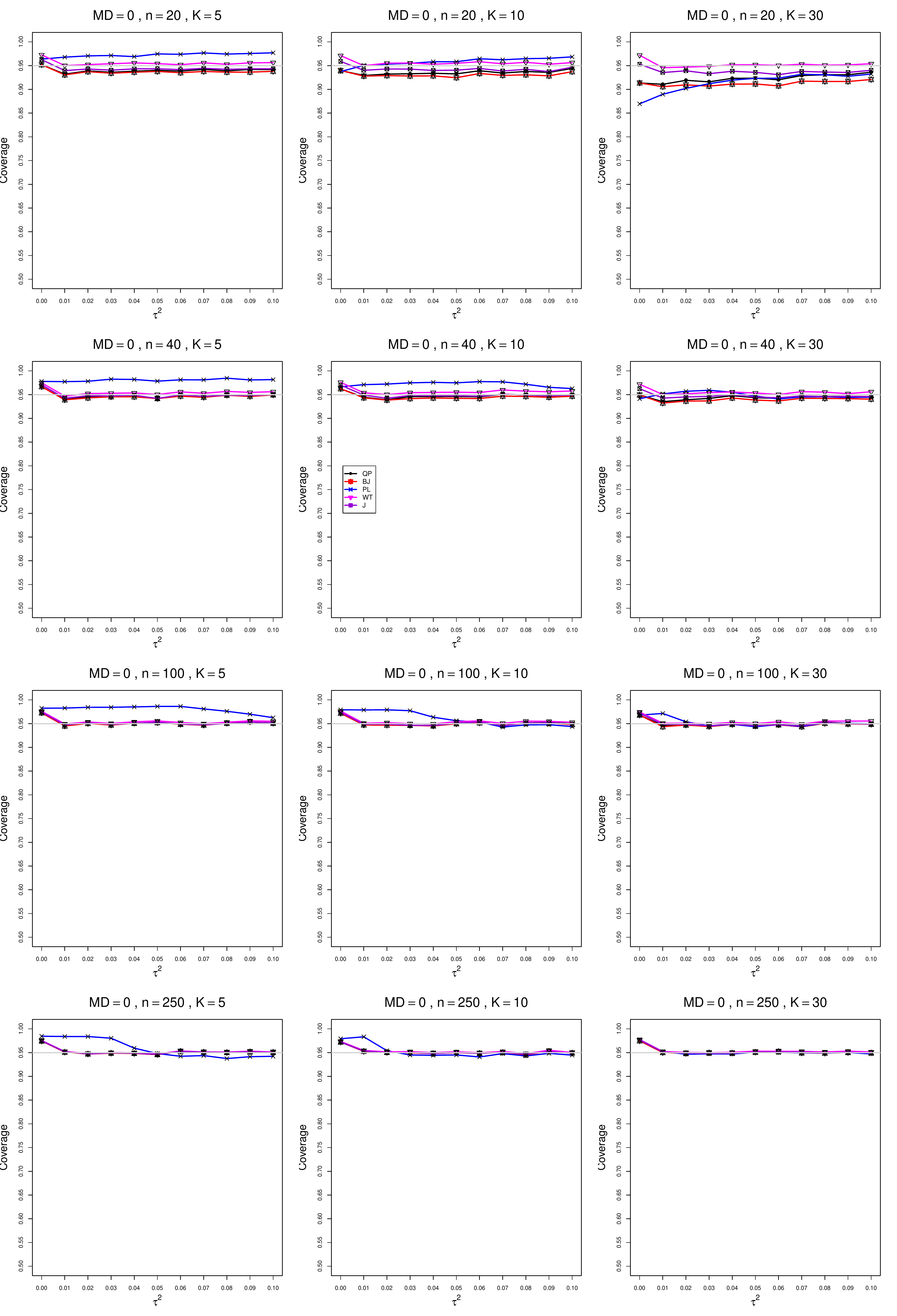}
	\caption{Coverage of 95\% confidence intervals for the  between-studies variance $\tau^2 = 0.0(0.01)0.1$ for $\mu=0$, $q=0.5$, $\sigma_C^2=1$, $\sigma_T^2=1$,  equal study sizes $n=20,\;40,\;100,\;250$.
		\label{CovTauMD0_S1_1_small_tau2}}
\end{figure}

\begin{figure}[t]
	\centering
	\includegraphics[scale=0.33]{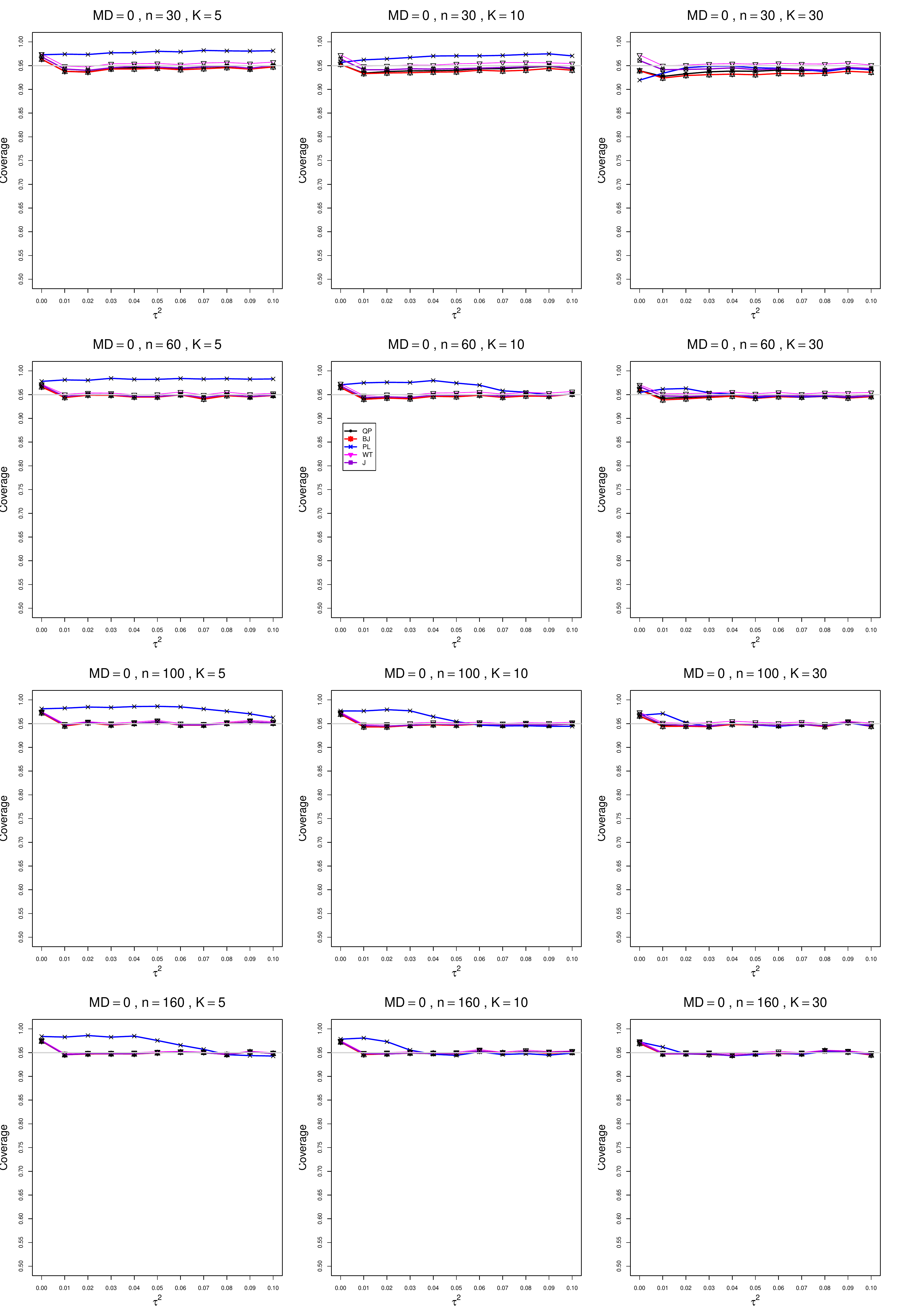}
	\caption{Coverage of 95\% confidence intervals for the  between-studies variance $\tau^2 = 0.0(0.01)0.1$ for $\mu=0$, $q=0.5$, $\sigma_C^2=1$, $\sigma_T^2=1$, unequal studies of average size  $\bar{n}=30,\;60,\;100,\;160$.
		\label{CovTauMD0_S1_1unequal_small_tau2}}
\end{figure}

\clearpage

\begin{figure}[t]
	\centering
	\includegraphics[scale=0.33]{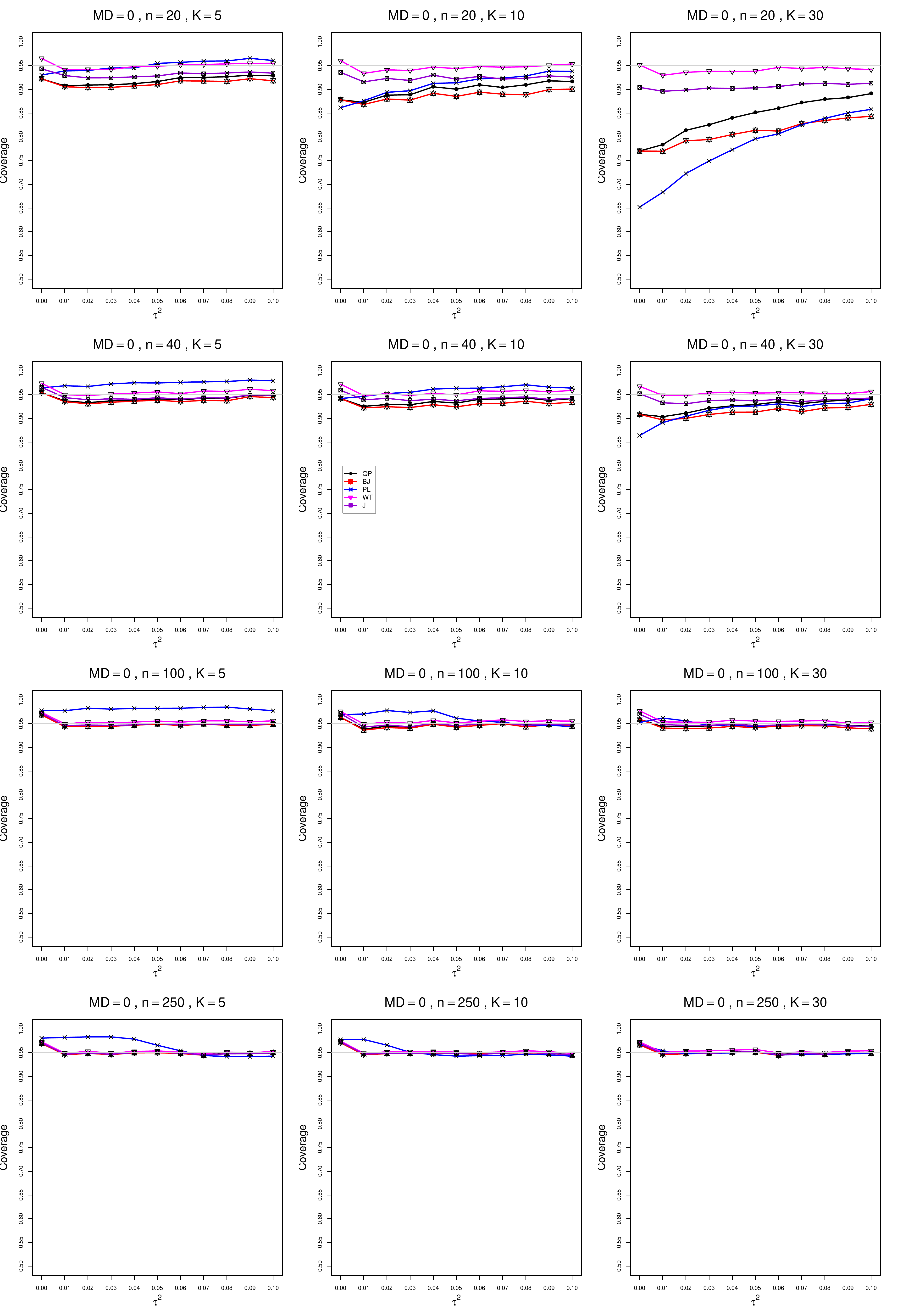}
	\caption{Coverage of 95\% confidence intervals for the  between-studies variance $\tau^2 = 0.0(0.01)0.1$ for $\mu=0$, $q=0.75$, $\sigma_C^2=1$, $\sigma_T^2=1$,  equal study sizes $n=20,\;40,\;100,\;250$.
		\label{CovTauMD0_S1_1q075_small_tau2}}
\end{figure}

\begin{figure}[t]
	\centering
	\includegraphics[scale=0.33]{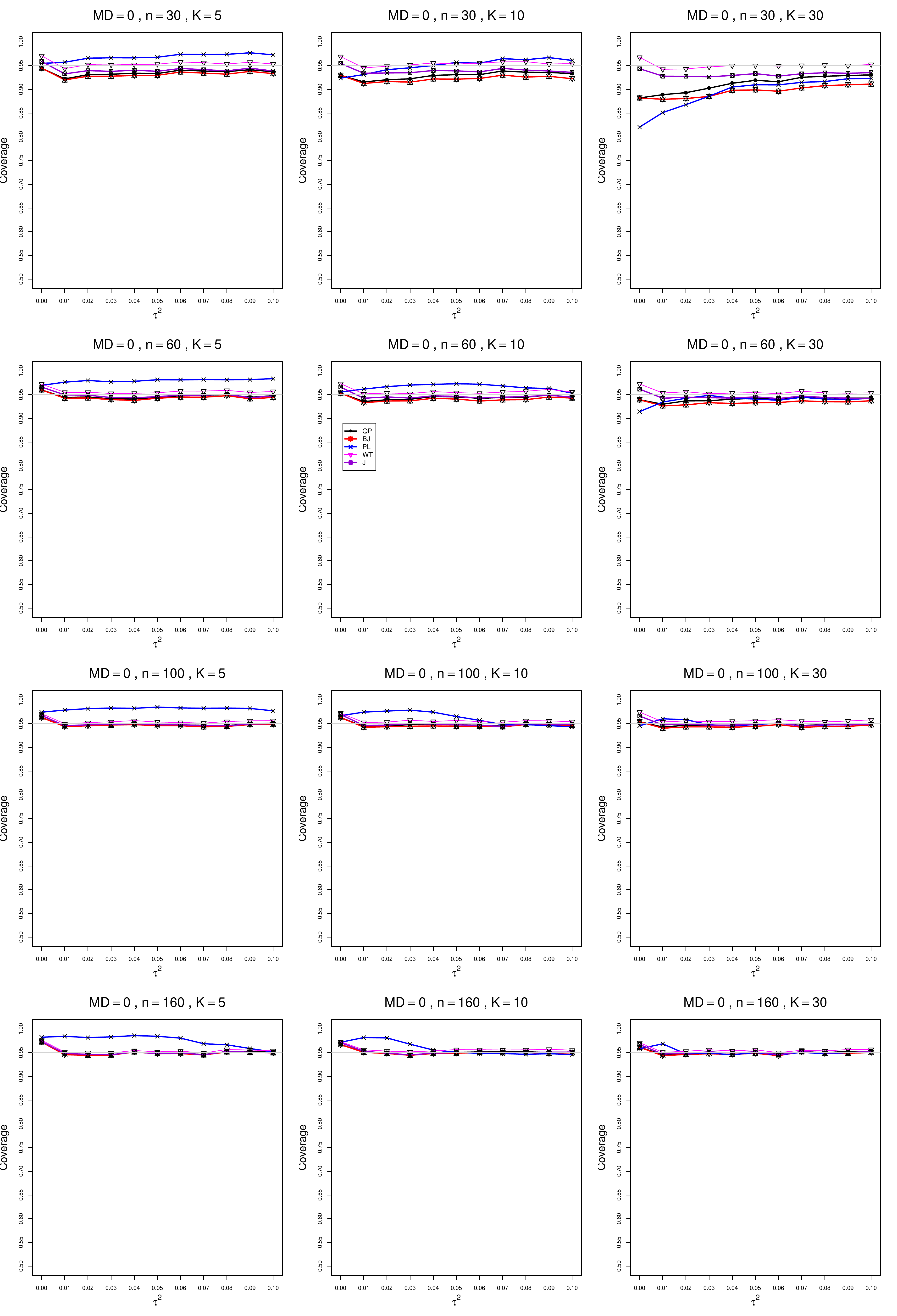}
	\caption{Coverage of 95\% confidence intervals for the  between-studies variance $\tau^2 = 0.0(0.01)0.1$ for $\mu=0$, $q=0.75$, $\sigma_C^2=1$, $\sigma_T^2=1$, unequal studies of average size  $\bar{n}=30,\;60,\;100,\;160$.
		\label{CovTauMD0_S1_1unequalq075_small_tau2}}
\end{figure}
\clearpage

\begin{figure}[t]
	\centering
	\includegraphics[scale=0.33]{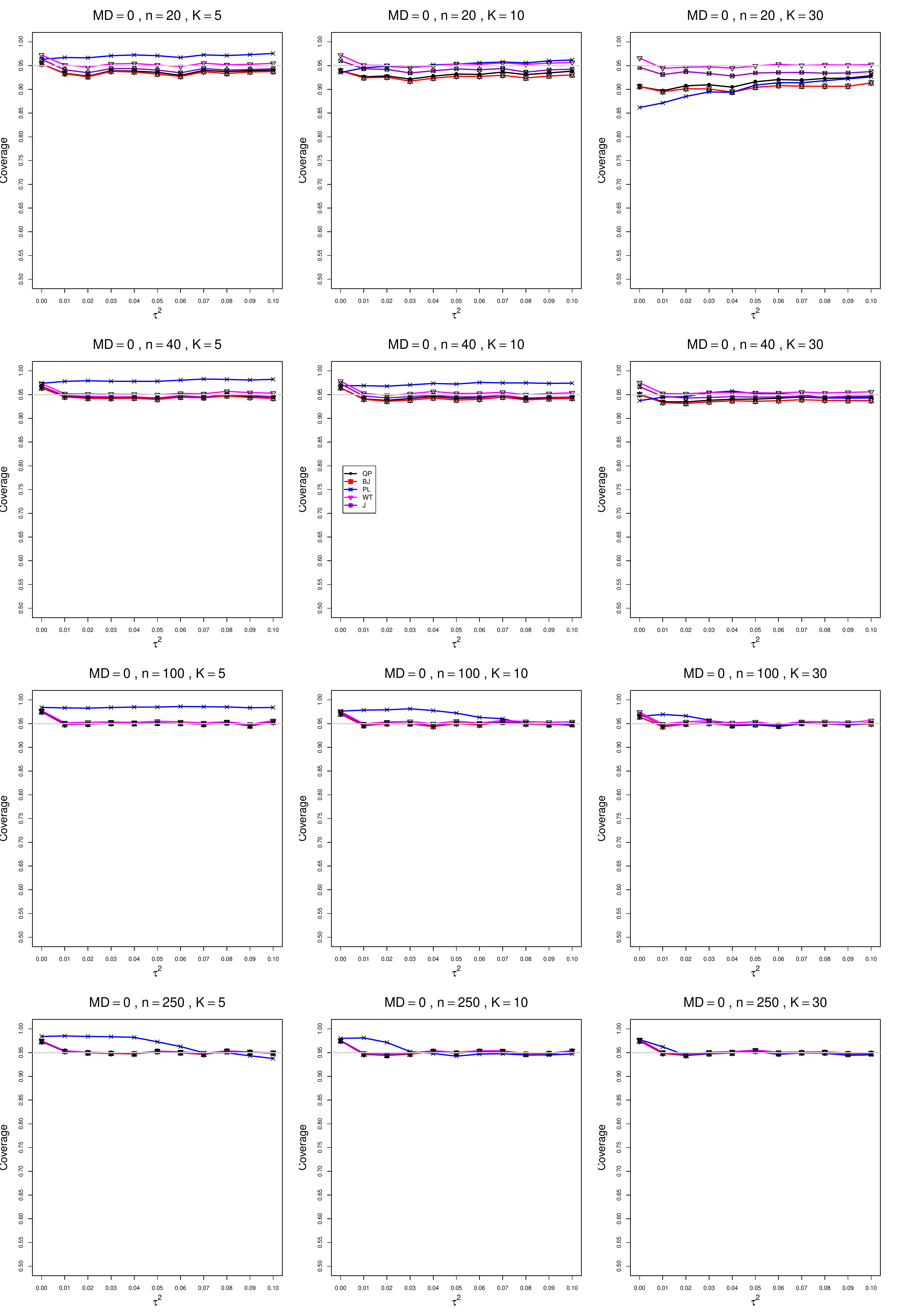}
	\caption{Coverage of 95\% confidence intervals for the  between-studies variance $\tau^2 = 0.0(0.01)0.1$ for $\mu=0$, $q=0.5$, $\sigma_C^2=1$, $\sigma_T^2=2$,  equal study sizes $n=20,\;40,\;100,\;250$.
		\label{CovTauMD0_S2_1_small_tau2}}
\end{figure}

\begin{figure}[t]
	\centering
	\includegraphics[scale=0.33]{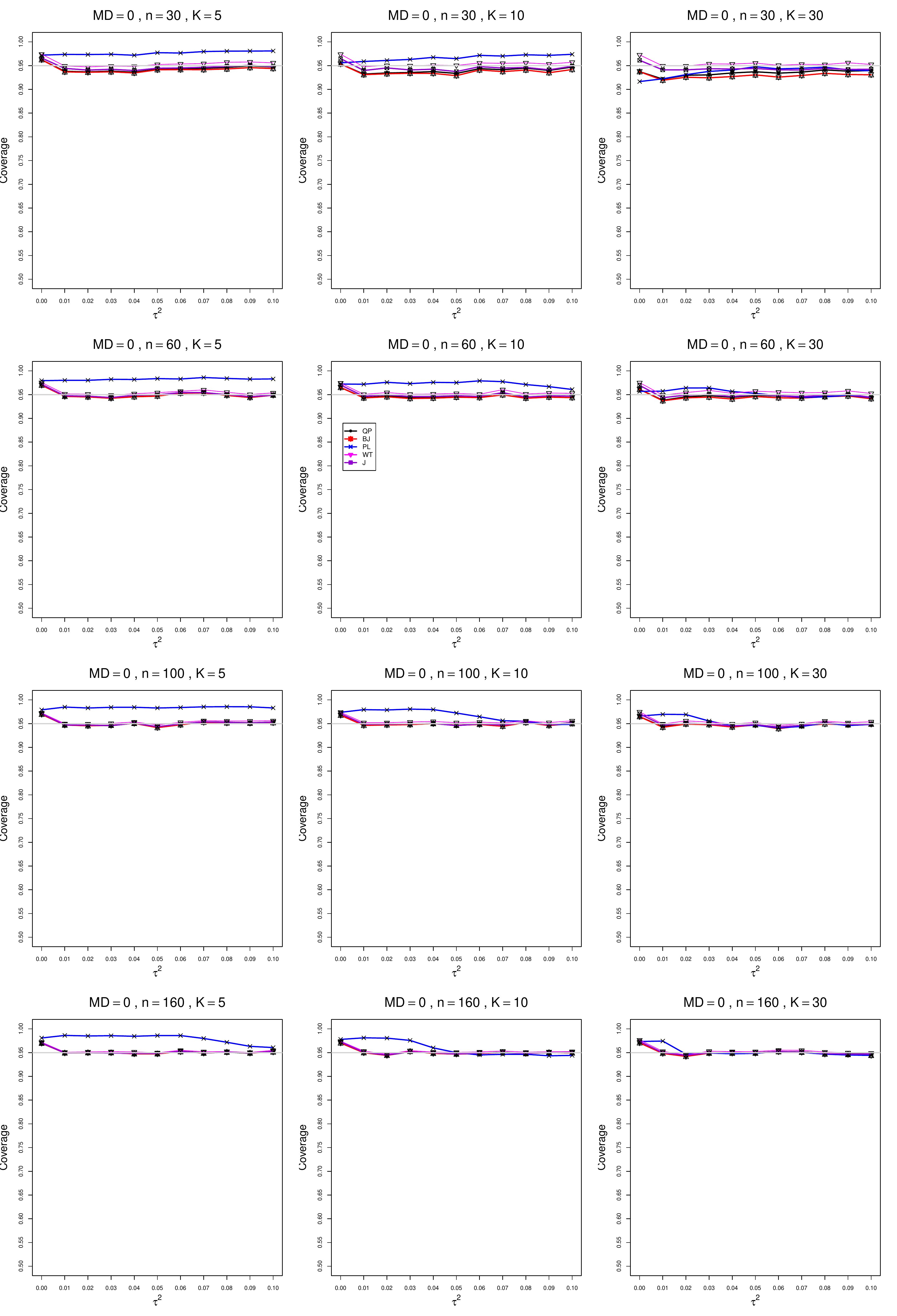}
	\caption{Coverage of 95\% confidence intervals for the  between-studies variance $\tau^2 = 0.0(0.01)0.1$ for $\mu=0$, $q=0.5$, $\sigma_C^2=1$, $\sigma_T^2=2$, unequal studies of average size  $\bar{n}=30,\;60,\;100,\;160$.
		\label{CovTauMD0_S2_1unequal_small_tau2}}
\end{figure}
\clearpage

\begin{figure}[t]
	\centering
	\includegraphics[scale=0.33]{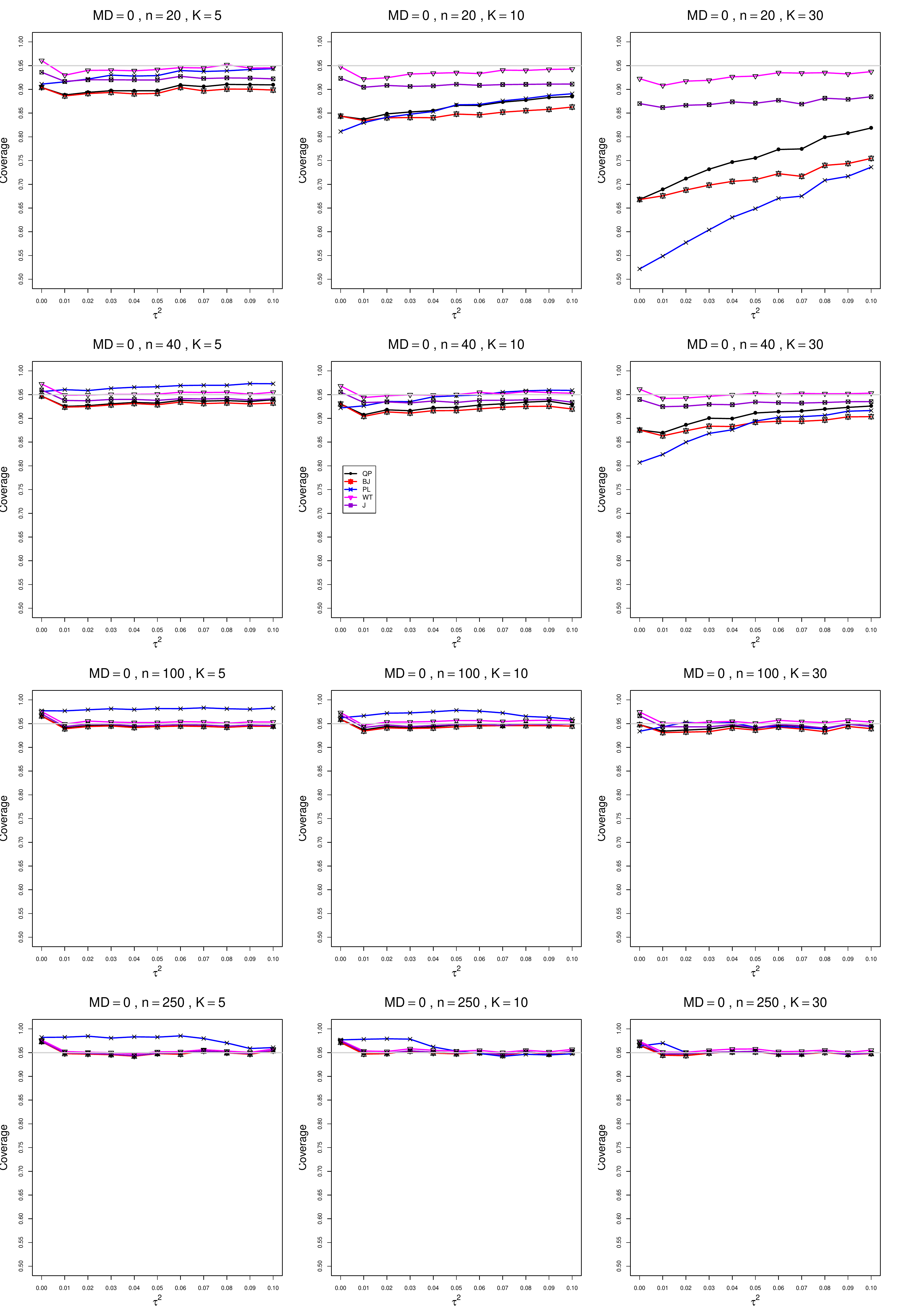}
	\caption{Coverage of 95\% confidence intervals for the  between-studies variance $\tau^2 = 0.0(0.01)0.1$ for $\mu=0$, $q=0.75$, $\sigma_C^2=1$, $\sigma_T^2=2$,  equal study sizes $n=20,\;40,\;100,\;250$.
		\label{CovTauMD0_S2_1q075_small_tau2}}
\end{figure}

\begin{figure}[t]
	\centering
	\includegraphics[scale=0.33]{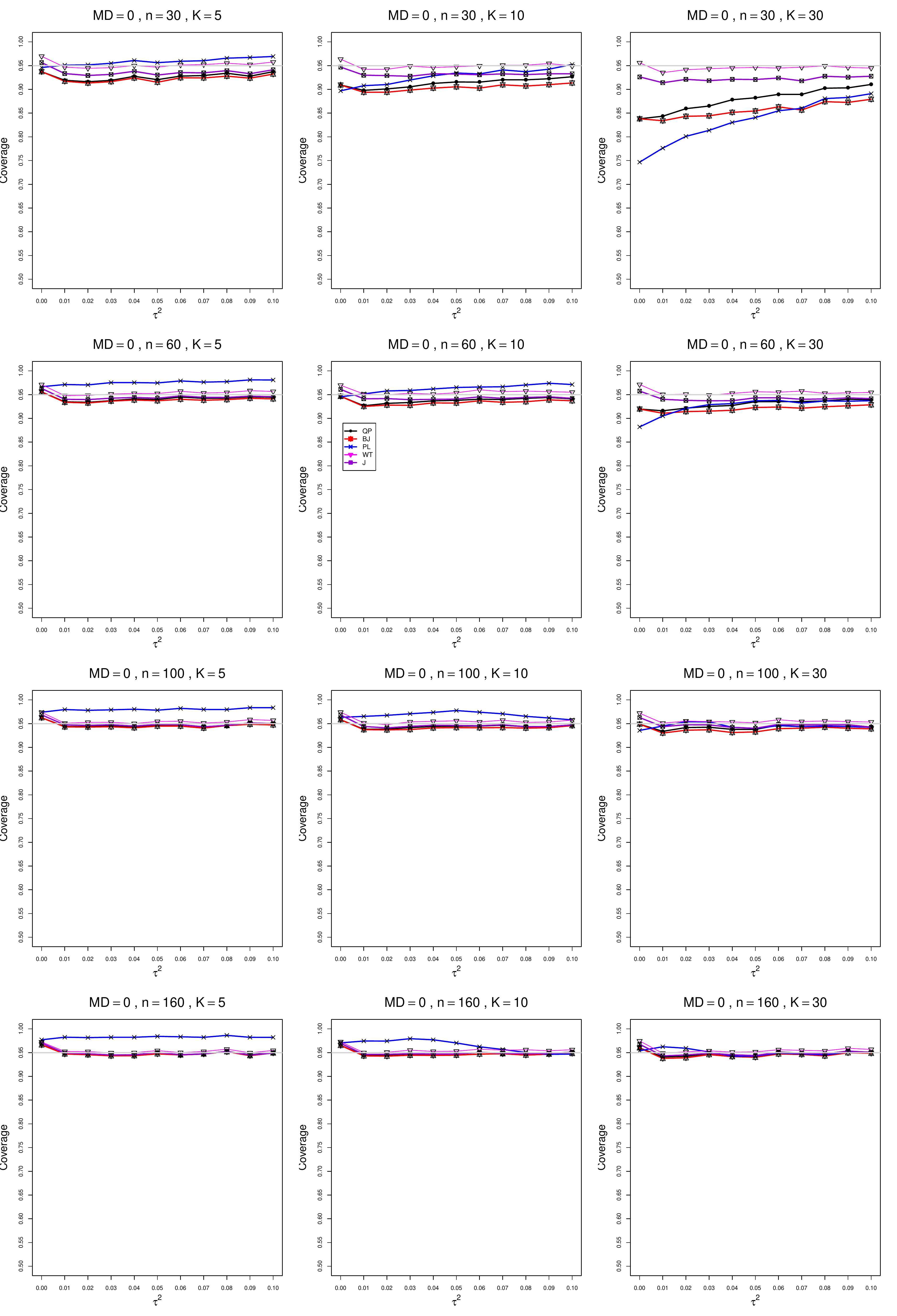}
	\caption{Coverage of 95\% confidence intervals for the  between-studies variance $\tau^2 = 0.0(0.01)0.1$ for $\mu=0$, $q=0.75$, $\sigma_C^2=1$, $\sigma_T^2=2$, unequal studies of average size  $\bar{n}=30,\;60,\;100,\;160$.
		\label{BiasTauMD0_S2_1unequalq075_small_tau2}}
\end{figure}

\clearpage
\renewcommand{\thefigure}{A5.\arabic{figure}}
\setcounter{figure}{0}
\setcounter{section}{0}

\section*{A5. Bias of $\hat{\tau}^2$ for $\tau^2 = 0.0(0.1)1.0$, $\sigma_{C}^2=10$, $\sigma_{T}^2=10,\;20$.}
For bias of $\hat{\tau}^2$, each figure corresponds to a value of $\mu (= 0, 0.2, 0.5, 1, 2)$, a value of $q (= .5, .75)$, a value of $\tau^2 = 0.0(0.1)1.0$, a value of $\sigma_{C}^2=10$, a value of $\sigma_{T}^2=10,\;20$ , and a set of values of $n$ (= 20, 40, 100, 250) or $\bar{n}$ (= 30, 60, 100, 160).\\
Each figure contains a panel (with $\tau^2$ on the horizontal axis) for each combination of n (or $\bar{n}$) and $K (=5, 10, 30)$.\\
The point estimators of $\tau^2$ are
\begin{itemize}
	\item DL (DerSimonian-Laird)
	\item REML (restricted maximum likelihood)
	\item MP (Mandel-Paule)
	\item WT (Corrected Mandel-Paule moment estimator based on Welch-type approximation for Q distribution)
	\item J (Jackson)
	\item CDL (Corrected DerSimonian-Laird)
\end{itemize}

\begin{figure}[t]
	\centering
	\includegraphics[scale=0.33]{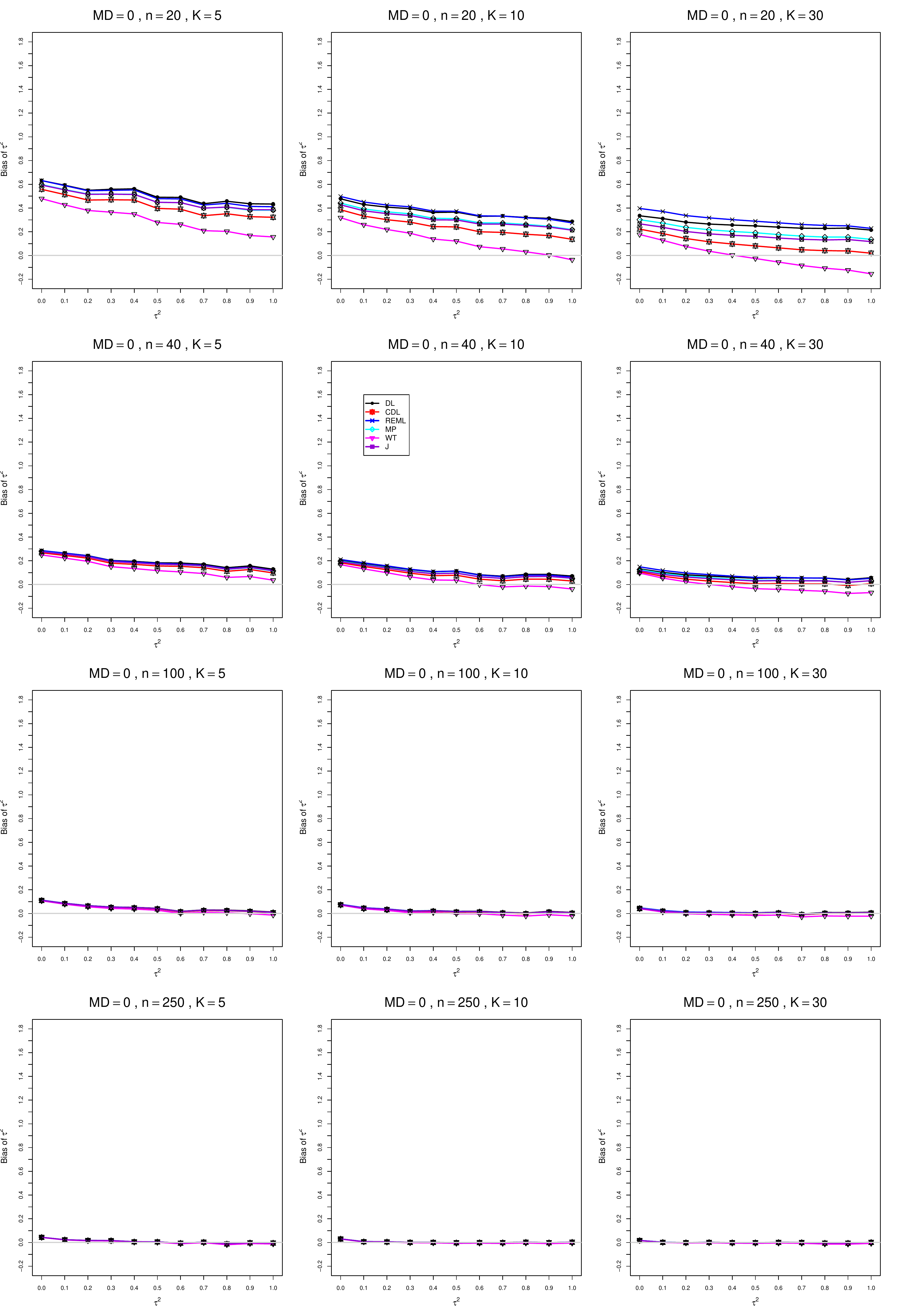}
	\caption{Bias of the estimation of  between-studies variance $\tau^2 = 0.0(0.1)1.0$ for $\mu=0$, $q=0.5$, $\sigma_C^2=10$, $\sigma_T^2=10$,  equal study sizes $n=20,\;40,\;100,\;250$.
		\label{BiasTauMD0_S10_10}}
\end{figure}

\begin{figure}[t]
	\centering
	\includegraphics[scale=0.33]{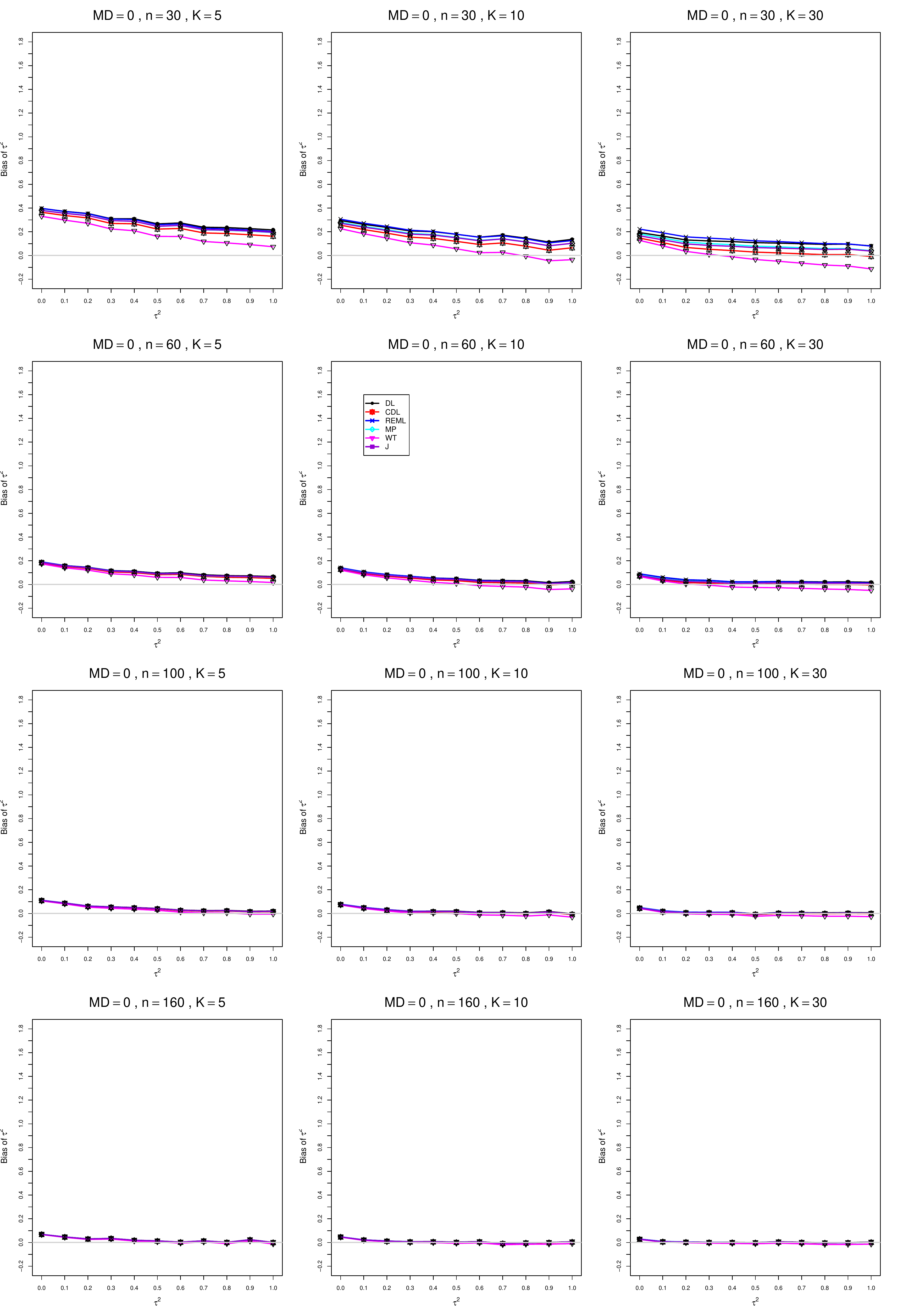}
	\caption{Bias of the estimation of  between-studies variance $\tau^2 = 0.0(0.1)1.0$ for $\mu=0$, $q=0.5$, $\sigma_C^2=10$, $\sigma_T^2=10$, unequal studies of average size  $\bar{n}=30,\;60,\;100,\;160$.
		\label{BiasTauMD0_S10_10unequal}}
\end{figure}


\begin{figure}[t]
	\centering
	\includegraphics[scale=0.33]{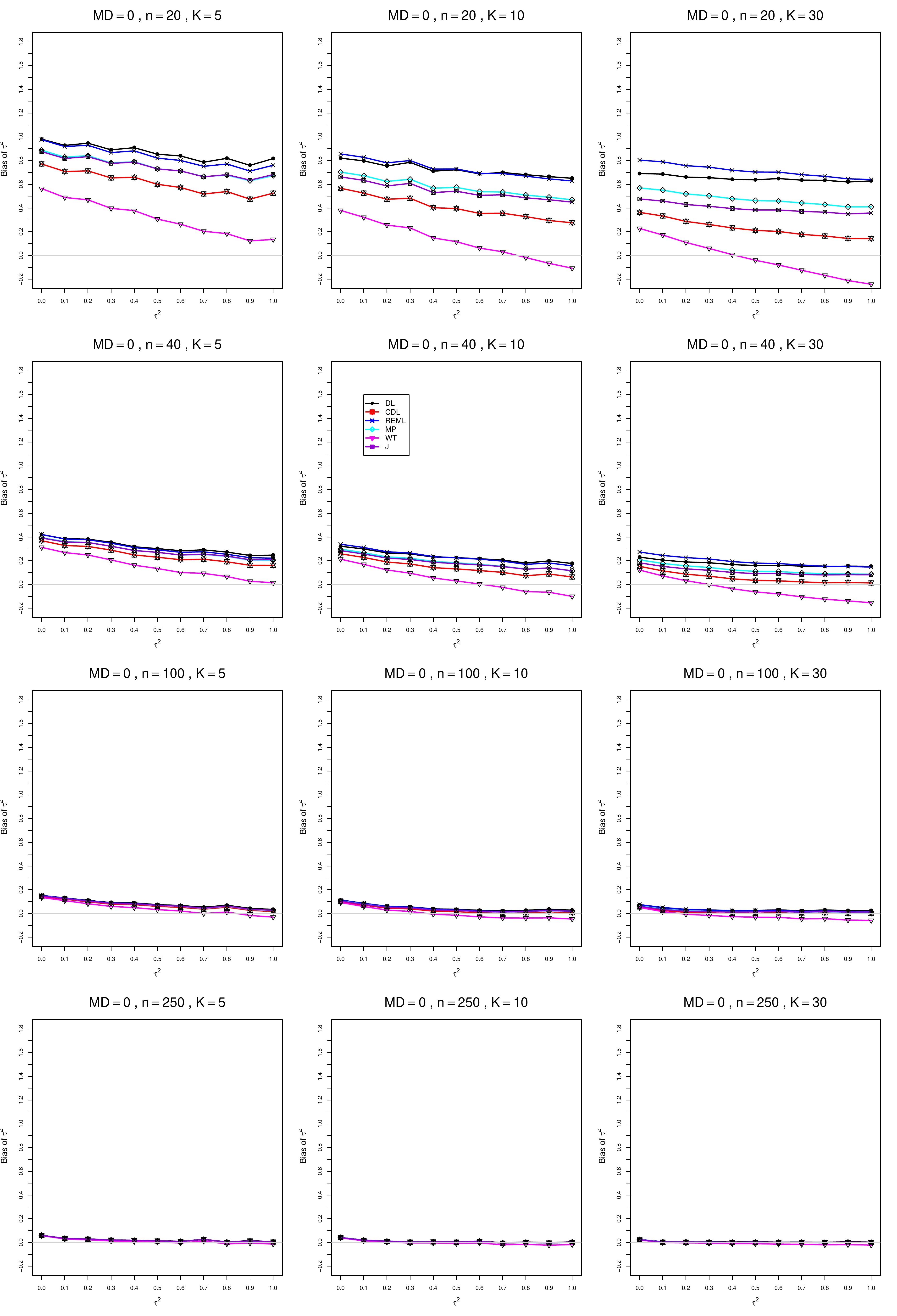}
	\caption{Bias of the estimation of  between-studies variance $\tau^2 = 0.0(0.1)1.0$ for $\mu=0$, $q=0.75$, $\sigma_C^2=10$, $\sigma_T^2=10$,  equal study sizes $n=20,\;40,\;100,\;250$.
		\label{BiasTauMD0_S10_10q075}}
\end{figure}

\begin{figure}[t]
	\centering
	\includegraphics[scale=0.33]{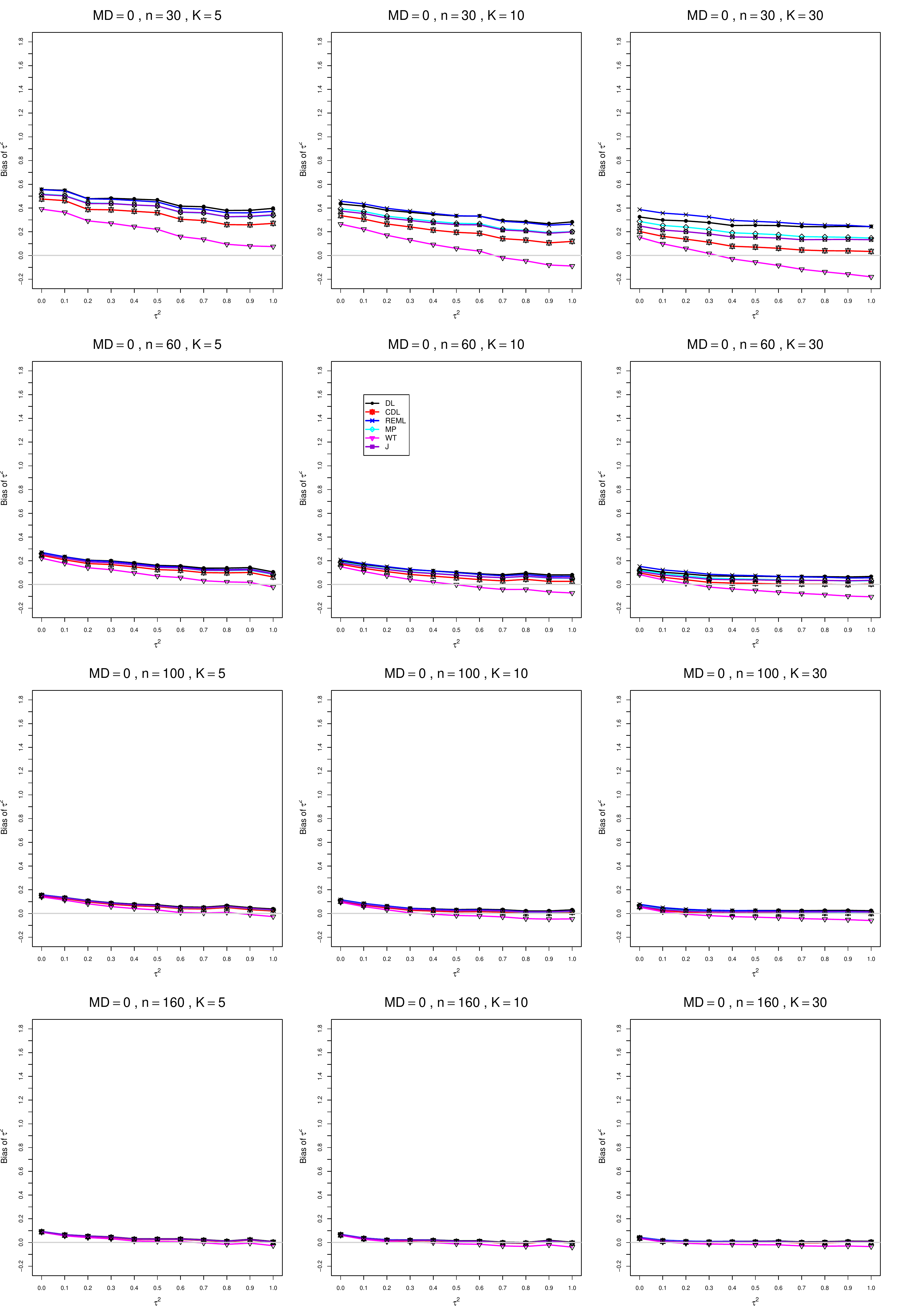}
	\caption{Bias of the estimation of  between-studies variance $\tau^2 = 0.0(0.1)1.0$ for $\mu=0$, $q=0.75$, $\sigma_C^2=10$, $\sigma_T^2=10$, unequal studies of average size  $\bar{n}=30,\;60,\;100,\;160$.
		\label{BiasTauMD0_S10_10unequalq075}}
\end{figure}


\begin{figure}[t]
	\centering
	\includegraphics[scale=0.33]{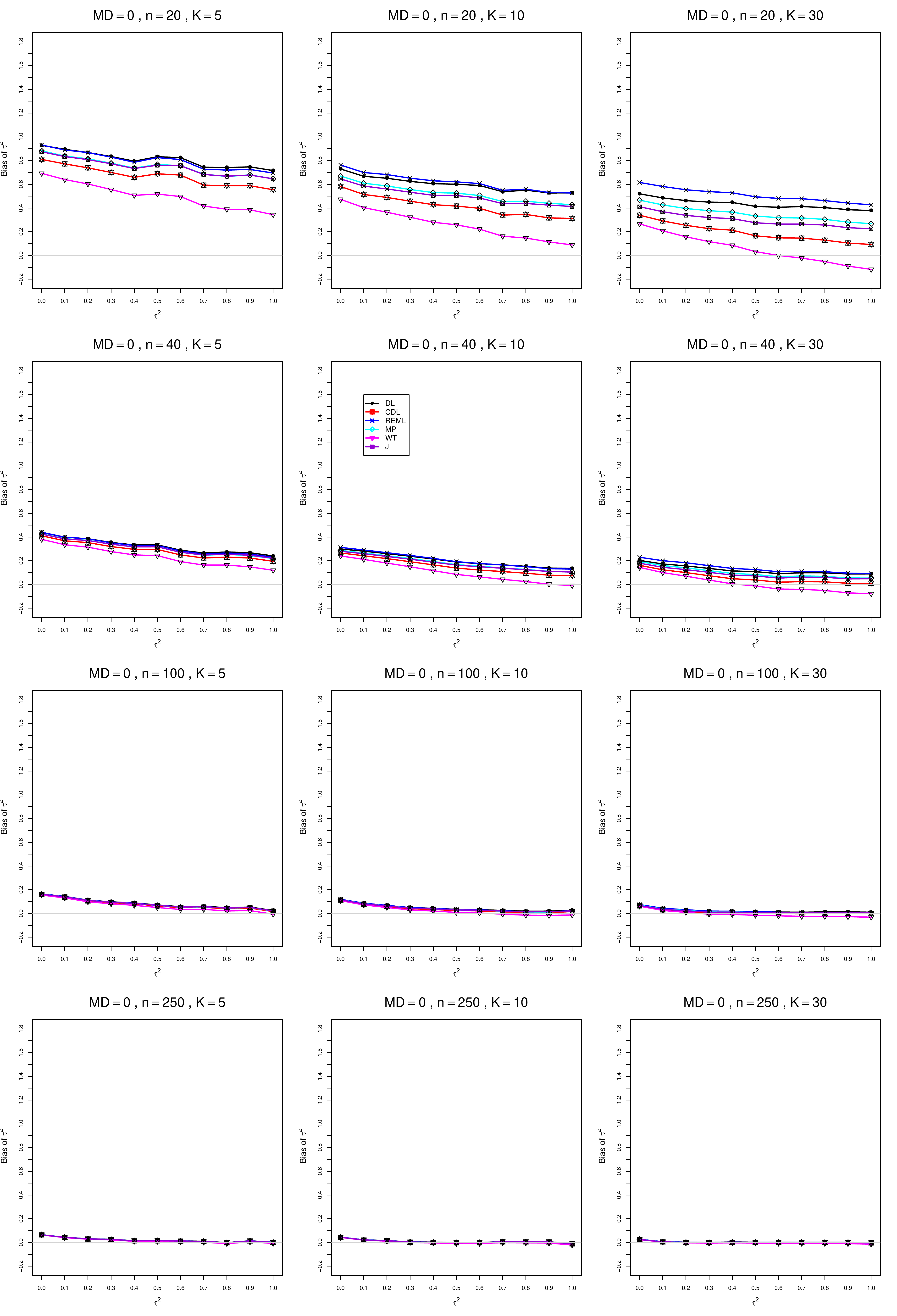}
	\caption{Bias of the estimation of  between-studies variance $\tau^2 = 0.0(0.1)1.0$ for $\mu=0$, $q=0.5$, $\sigma_C^2=10$, $\sigma_T^2=20$,  equal study sizes $n=20,\;40,\;100,\;250$.
		\label{BiasTauMD0_S10_20}}
\end{figure}

\begin{figure}[t]
	\centering
	\includegraphics[scale=0.33]{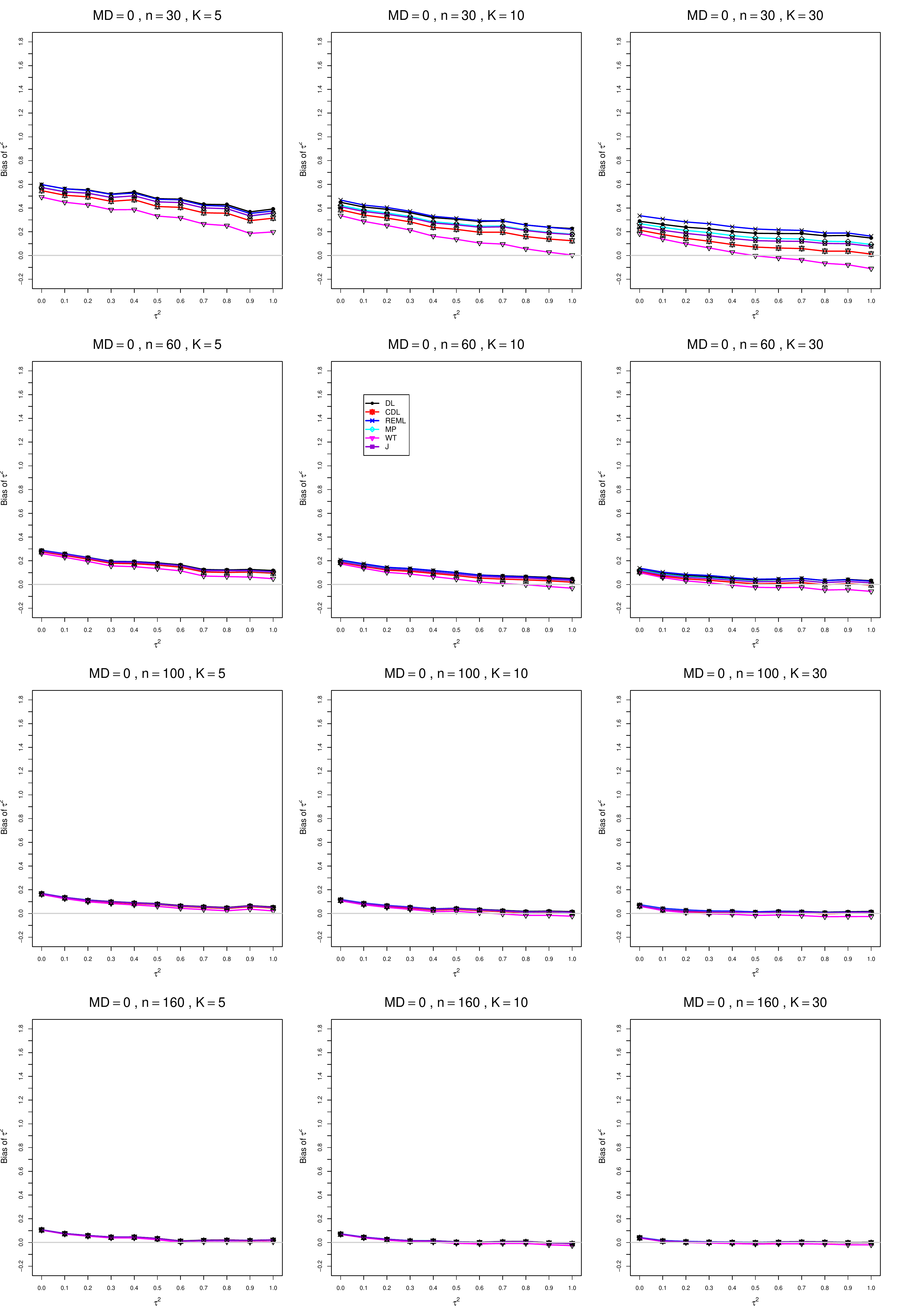}
	\caption{Bias of the estimation of  between-studies variance $\tau^2 = 0.0(0.1)1.0$ for $\mu=0$, $q=0.5$, $\sigma_C^2=10$, $\sigma_T^2=20$, unequal studies of average size  $\bar{n}=30,\;60,\;100,\;160$.
		\label{BiasTauMD0_S10_20unequal}}
\end{figure}


\begin{figure}[t]
	\centering
	\includegraphics[scale=0.33]{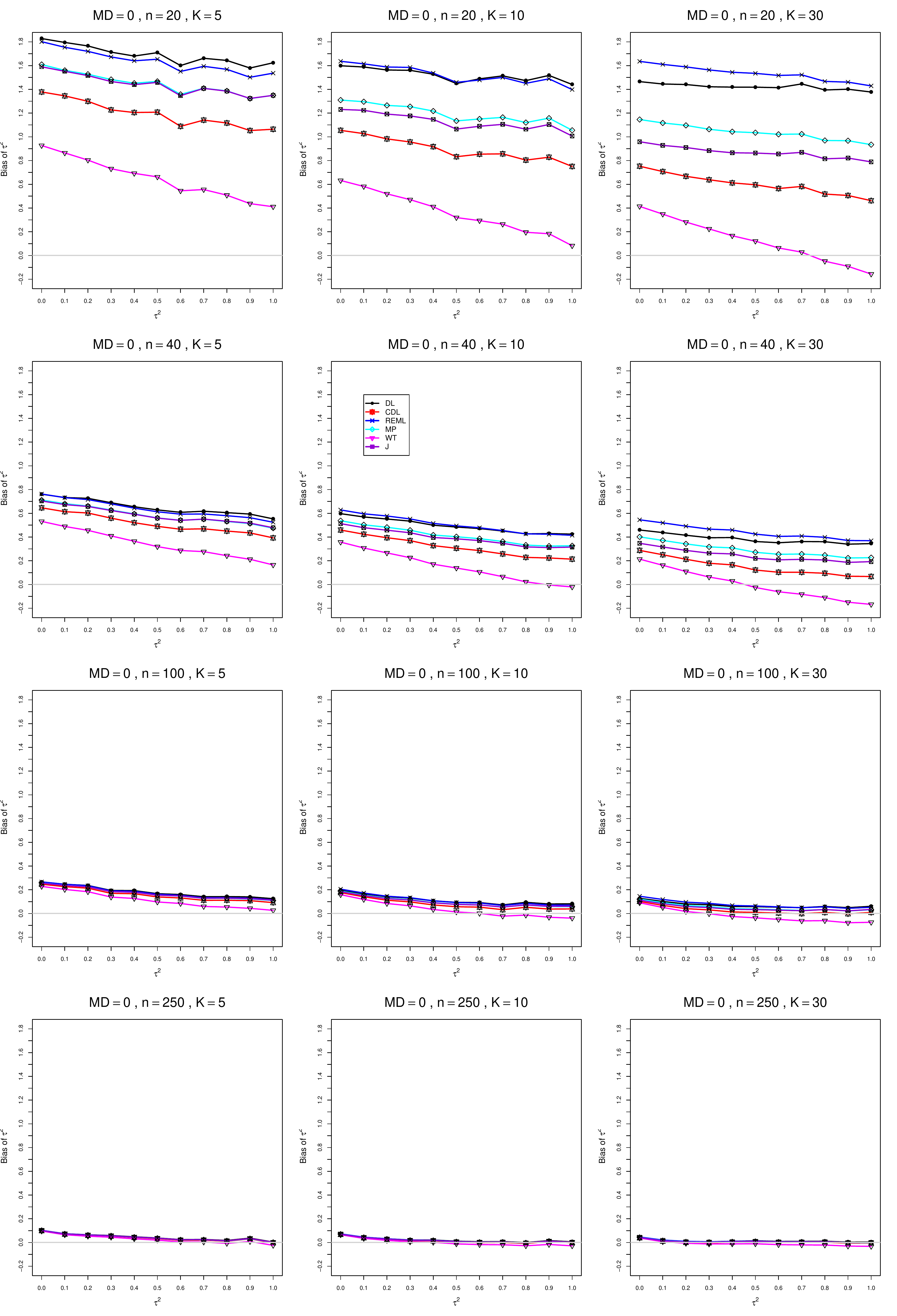}
	\caption{Bias of the estimation of  between-studies variance $\tau^2 = 0.0(0.1)1.0$ for $\mu=0$, $q=0.75$, $\sigma_C^2=10$, $\sigma_T^2=20$,  equal study sizes $n=20,\;40,\;100,\;250$.
		\label{BiasTauMD0_S10_20q075}}
\end{figure}

\begin{figure}[t]
	\centering
	\includegraphics[scale=0.33]{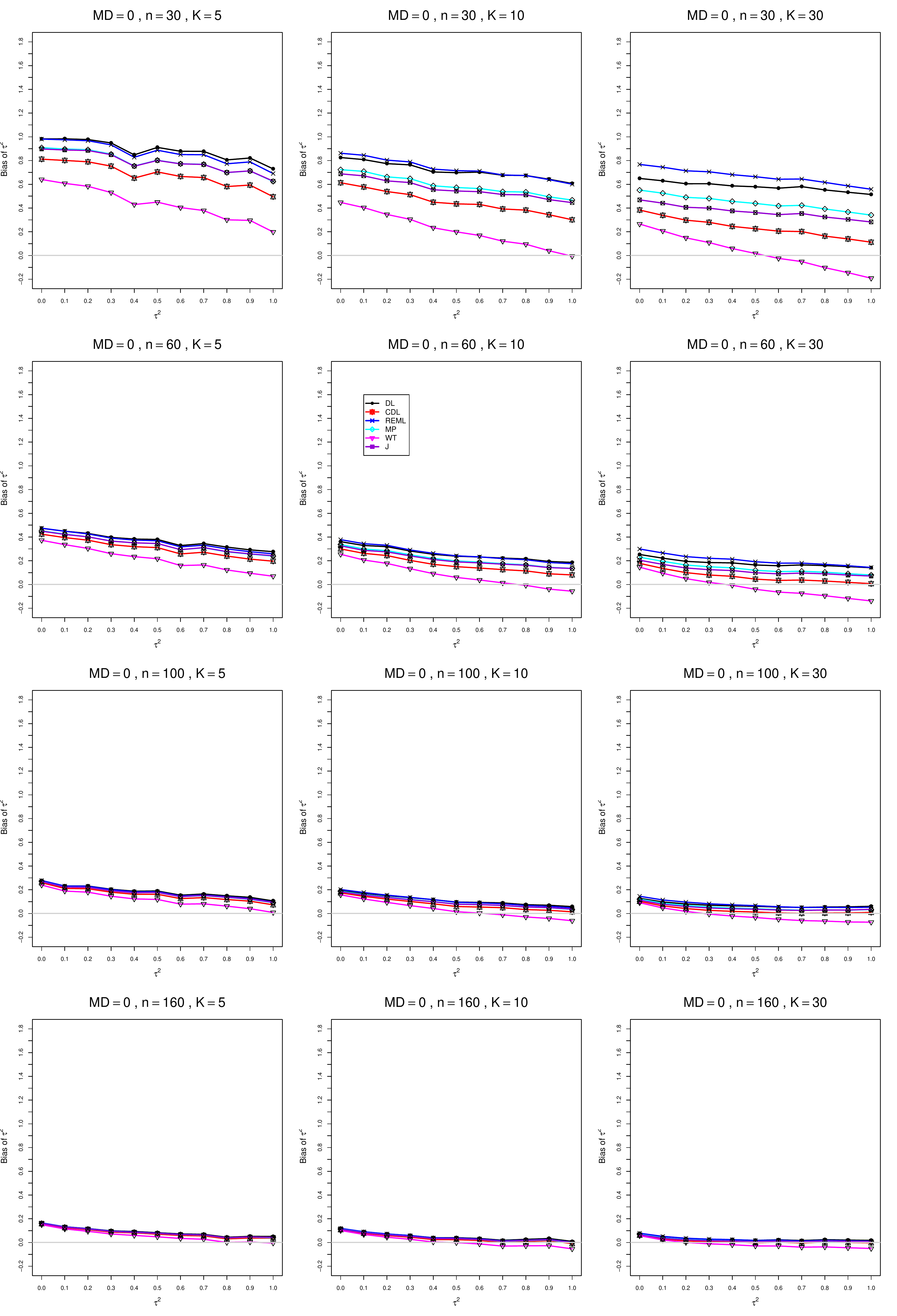}
	\caption{Bias of the estimation of  between-studies variance $\tau^2 = 0.0(0.1)1.0$ for $\mu=0$, $q=0.75$, $\sigma_C^2=10$, $\sigma_T^2=20$, unequal studies of average size  $\bar{n}=30,\;60,\;100,\;160$.
		\label{BiasTauMD0_S10_20unequalq075}}
\end{figure}

\clearpage
\setcounter{section}{0}
\renewcommand{\thefigure}{A6.\arabic{figure}}
\setcounter{figure}{0}
\setcounter{section}{0}
\section*{A6. Coverage of $\hat{\tau}^2$ for $\tau^2 = 0.0(0.1)1.0$, $\sigma_{C}^2=10$, $\sigma_{T}^2=10,\;20$.}
For coverage of $\hat{\tau}^2$, each figure corresponds to a value of $\mu (= 0, 0.2, 0.5, 1, 2)$, a value of $q (= .5, .75)$, a value of $\tau^2 = 0.0(0.1)1.0$ ,  a value of $\sigma_{C}^2=10$, a value of $\sigma_{T}^2=10,\;20$ and a set of values of $n$ (= 20, 40, 100, 250) or $\bar{n} (= 30, 60, 100, 160)$.\\
Each figure contains a panel (with $\tau^2$ on the horizontal axis) for each combination of n (or $\bar{n}$) and $K (=5, 10, 30)$.\\
The interval estimators of $\tau^2$ are
\begin{itemize}
	\item QP (Q-profile confidence interval)
	\item BJ (Biggerstaff and Jackson interval )
	\item PL (Profile likelihood interval)
	\item WT (Corrected Mandel-Paule moment estimator based on Welch-type approximation for Q distribution)
	\item J (Jacksons interval)
\end{itemize}

\begin{figure}[t]
	\centering
	\includegraphics[scale=0.33]{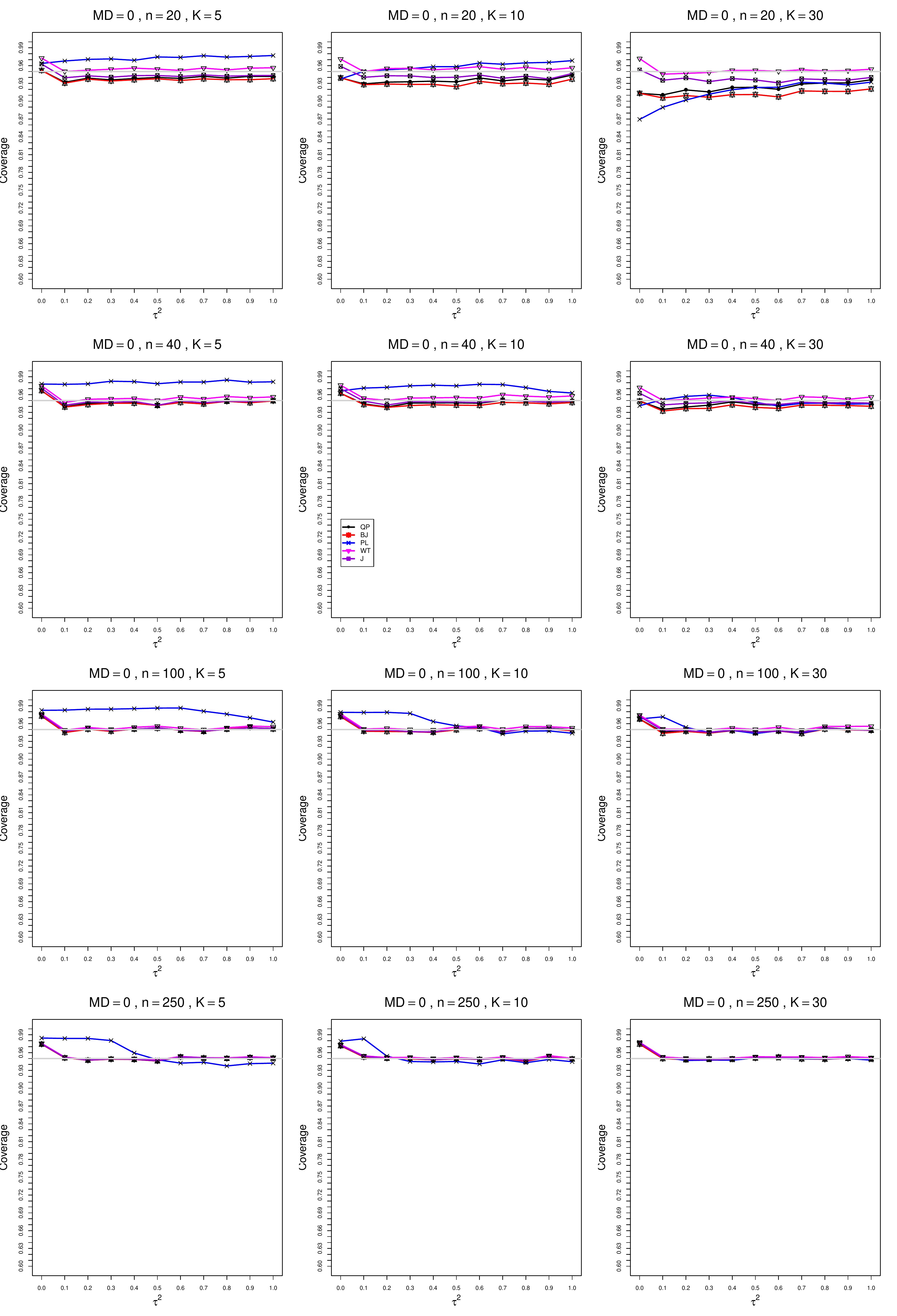}
	\caption{Coverage of 95\% confidence intervals for the  between-studies variance $\tau^2 = 0.0(0.1)1.0$ for $\mu=0$, $q=0.5$, $\sigma_C^2=10$, $\sigma_T^2=10$,  equal study sizes $n=20,\;40,\;100,\;250$.
		\label{CovTauMD0_S10_10}}
\end{figure}

\begin{figure}[t]
	\centering
	\includegraphics[scale=0.33]{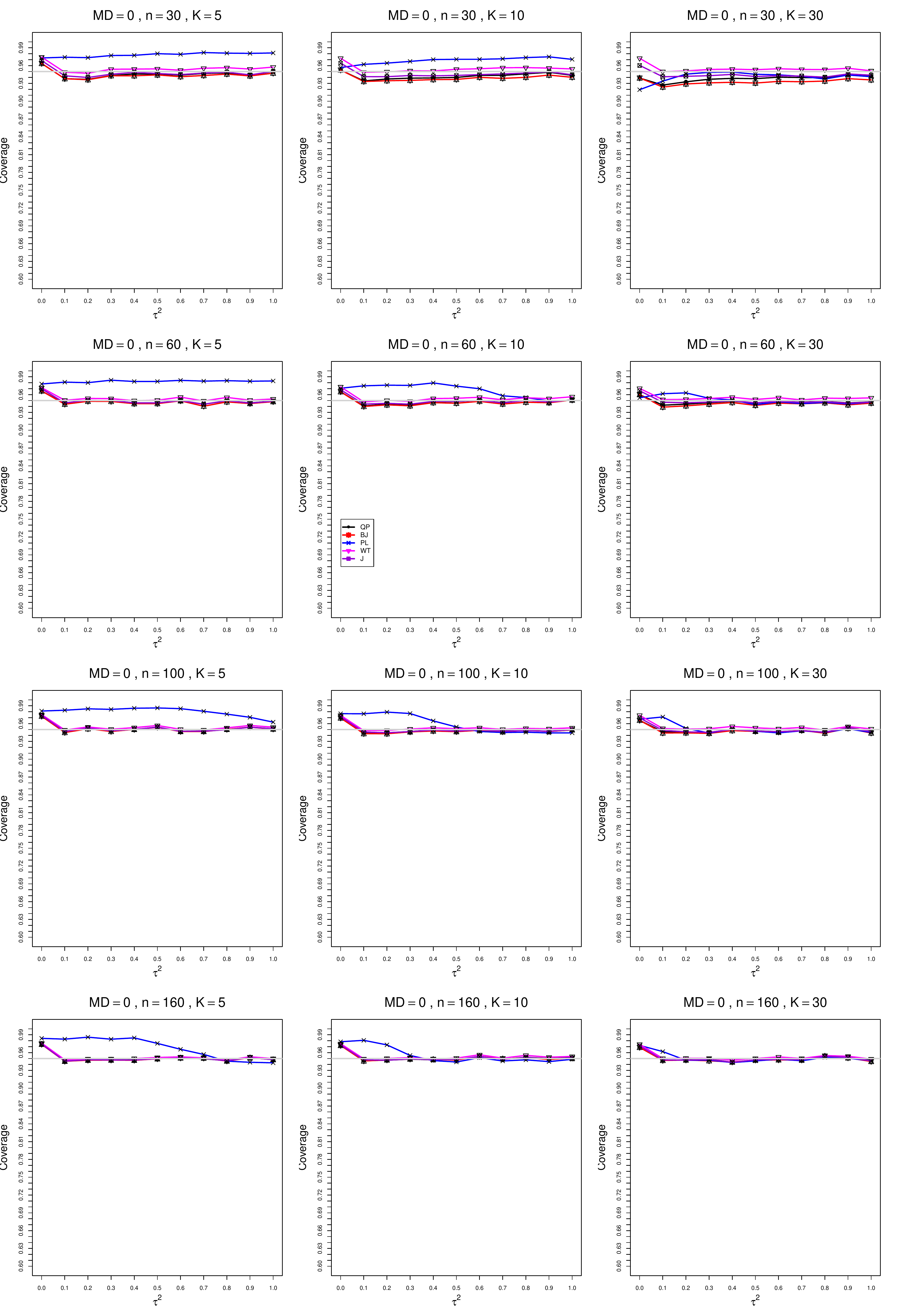}
	\caption{Coverage of 95\% confidence intervals for the  between-studies variance $\tau^2 = 0.0(0.1)1.0$ for $\mu=0$, $q=0.5$, $\sigma_C^2=10$, $\sigma_T^2=10$, unequal studies of average size  $\bar{n}=30,\;60,\;100,\;160$.
		\label{CovTauMD0_S10_10unequal}}
\end{figure}


\begin{figure}[t]
	\centering
	\includegraphics[scale=0.33]{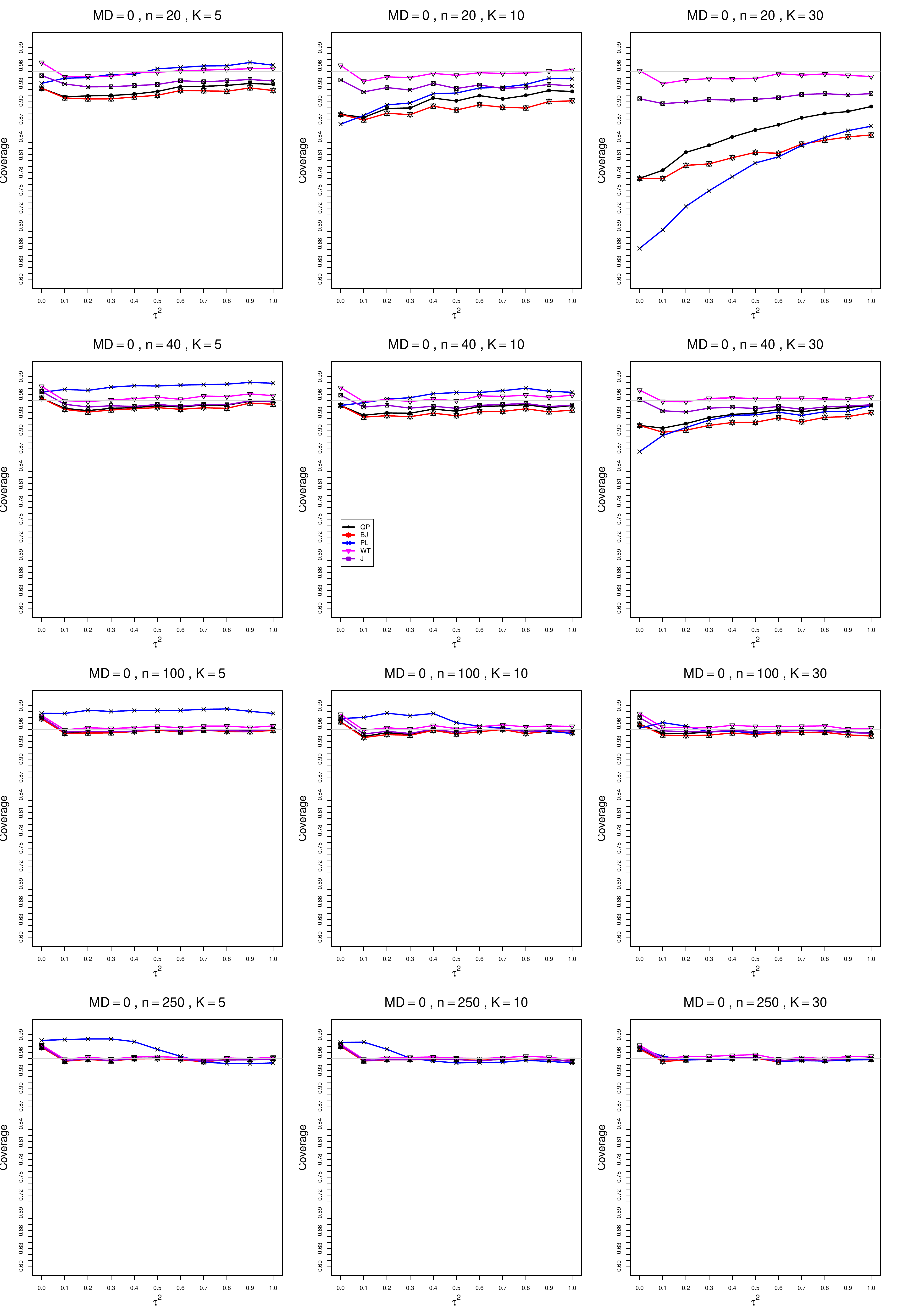}
	\caption{Coverage of 95\% confidence intervals for the  between-studies variance $\tau^2 = 0.0(0.1)1.0$ for $\mu=0$, $q=0.75$, $\sigma_C^2=10$, $\sigma_T^2=10$,  equal study sizes $n=20,\;40,\;100,\;250$.
		\label{CovTauMD0_S10_10q075}}
\end{figure}

\begin{figure}[t]
	\centering
	\includegraphics[scale=0.33]{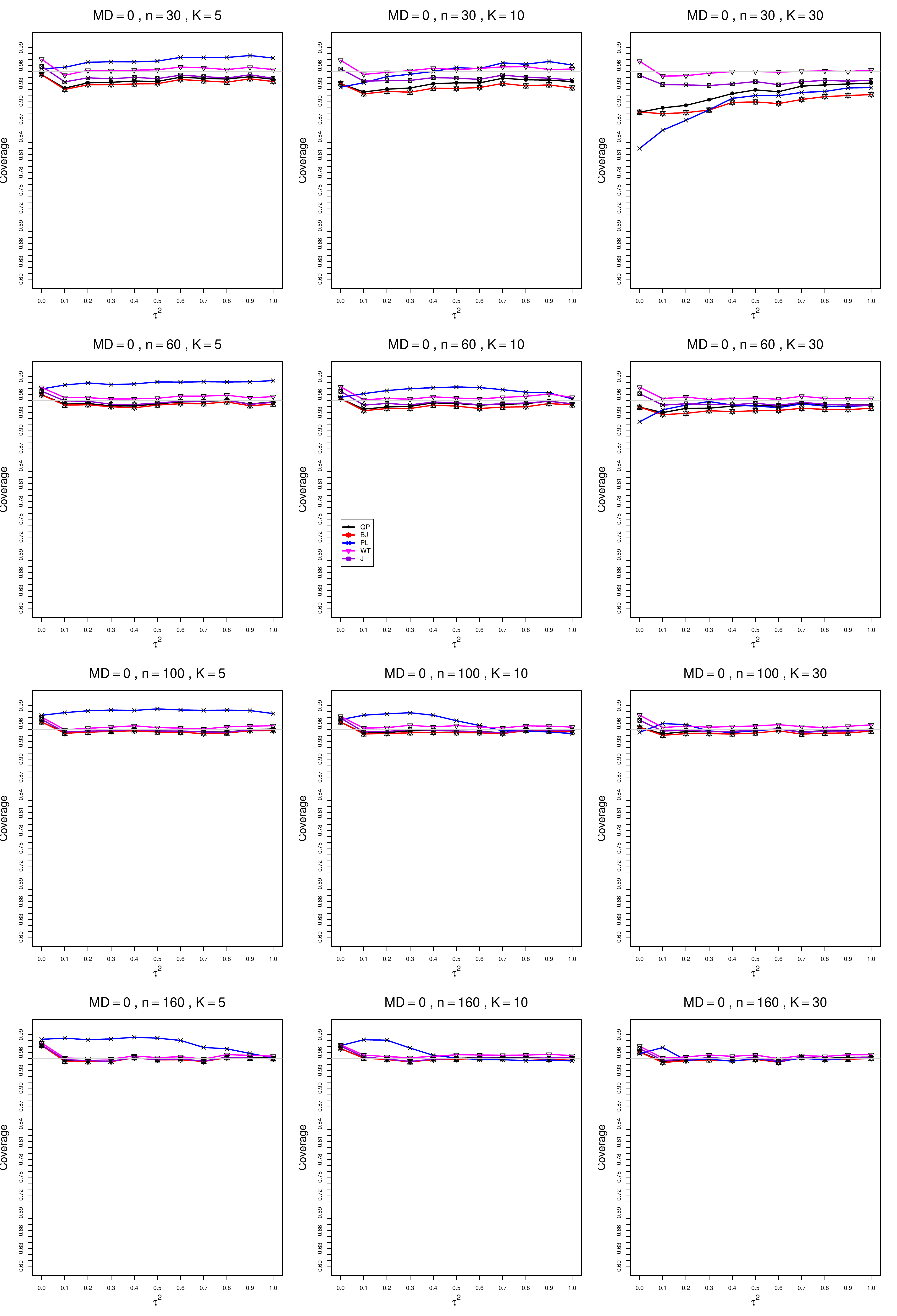}
	\caption{Coverage of 95\% confidence intervals for the  between-studies variance $\tau^2 = 0.0(0.1)1.0$ for $\mu=0$, $q=0.75$, $\sigma_C^2=10$, $\sigma_T^2=10$, unequal studies of average size  $\bar{n}=30,\;60,\;100,\;160$.
		\label{CovTauMD0_S10_10unequalq075}}
\end{figure}

\clearpage

\begin{figure}[t]
	\centering
	\includegraphics[scale=0.33]{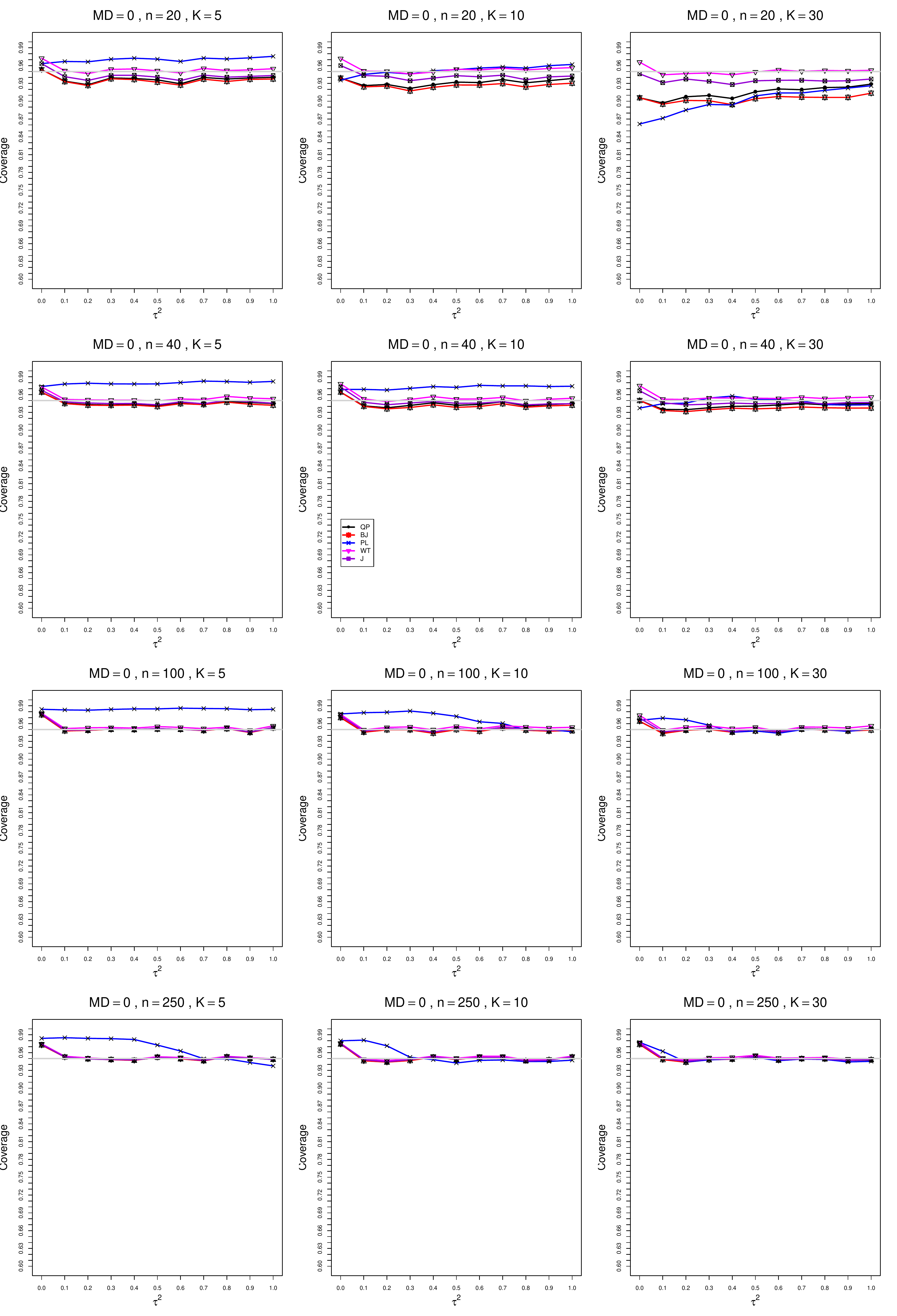}
	\caption{Coverage of 95\% confidence intervals for the  between-studies variance $\tau^2 = 0.0(0.1)1.0$ for $\mu=0$, $q=0.5$, $\sigma_C^2=10$, $\sigma_T^2=20$,  equal study sizes $n=20,\;40,\;100,\;250$.
		\label{CovTauMD0_S20_10}}
\end{figure}

\begin{figure}[t]
	\centering
	\includegraphics[scale=0.33]{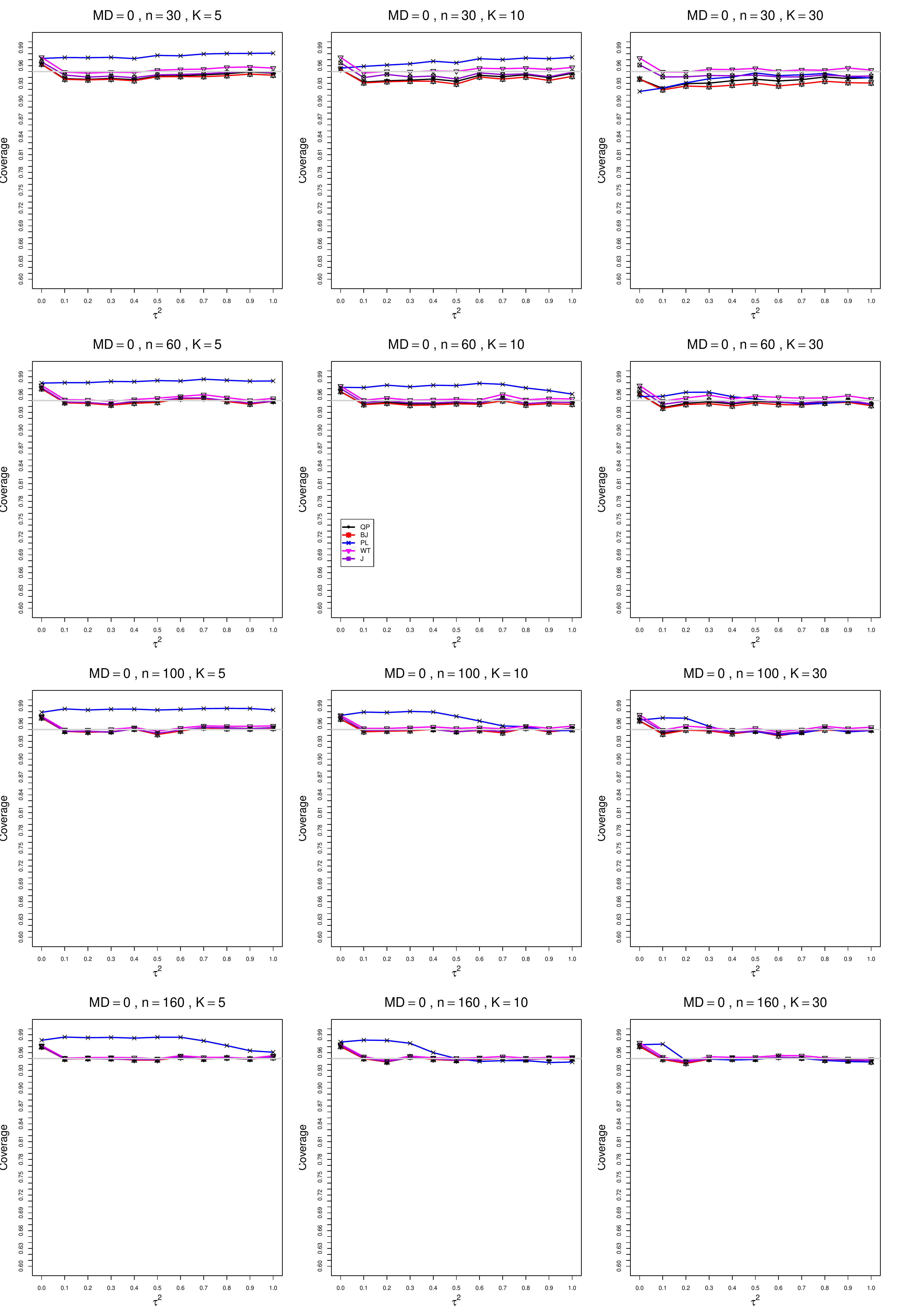}
	\caption{Coverage of 95\% confidence intervals for the  between-studies variance $\tau^2 = 0.0(0.1)1.0$ for $\mu=0$, $q=0.5$, $\sigma_C^2=10$, $\sigma_T^2=20$, unequal studies of average size  $\bar{n}=30,\;60,\;100,\;160$.
		\label{CovTauMD0_S20_10unequal}}
\end{figure}


\begin{figure}[t]
	\centering
	\includegraphics[scale=0.33]{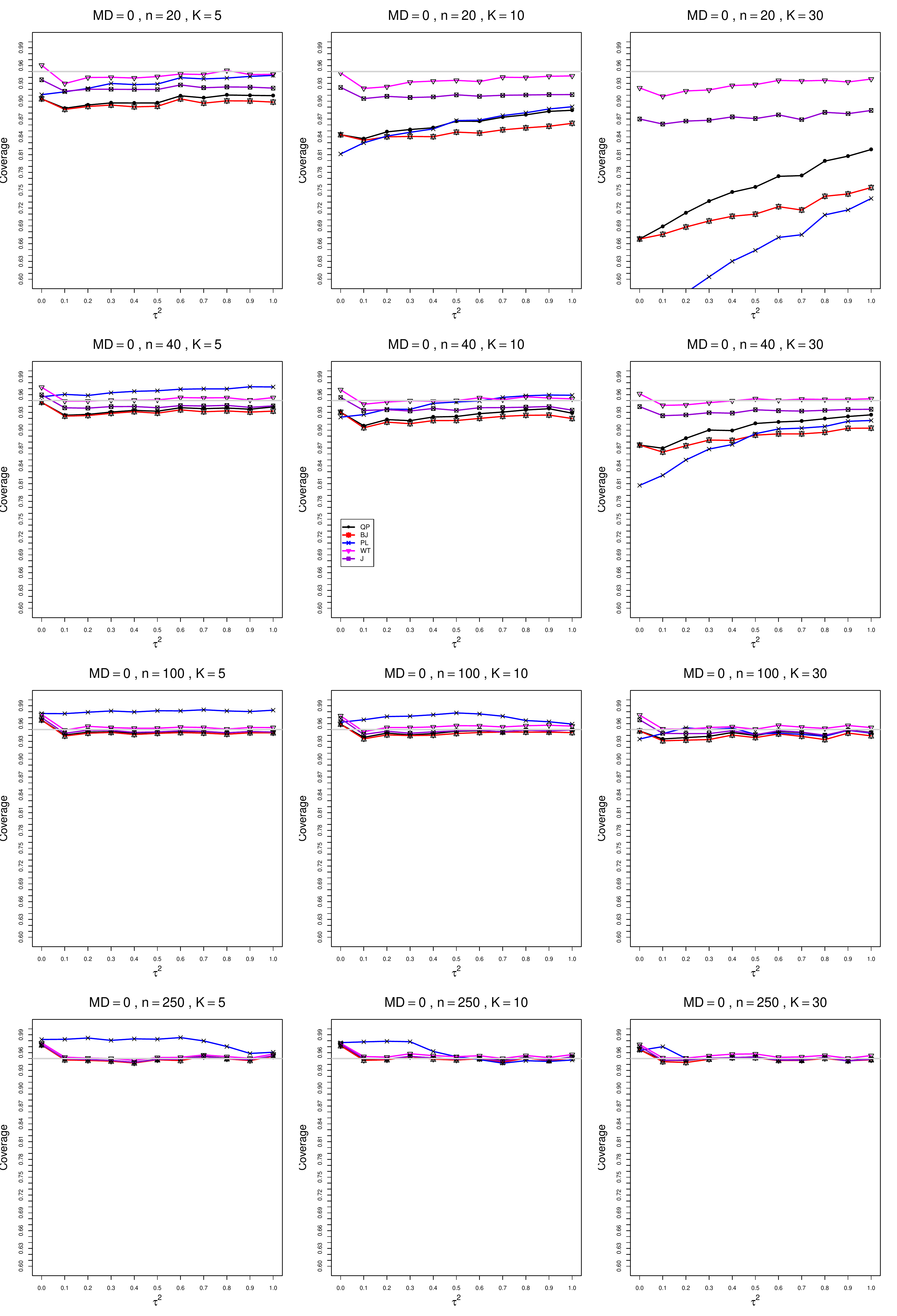}
	\caption{Coverage of 95\% confidence intervals for the  between-studies variance $\tau^2 = 0.0(0.1)1.0$ for $\mu=0$, $q=0.75$, $\sigma_C^2=10$, $\sigma_T^2=20$,  equal study sizes $n=20,\;40,\;100,\;250$.
		\label{CovTauMD0_S20_10q075}}
\end{figure}

\begin{figure}[t]
	\centering
	\includegraphics[scale=0.33]{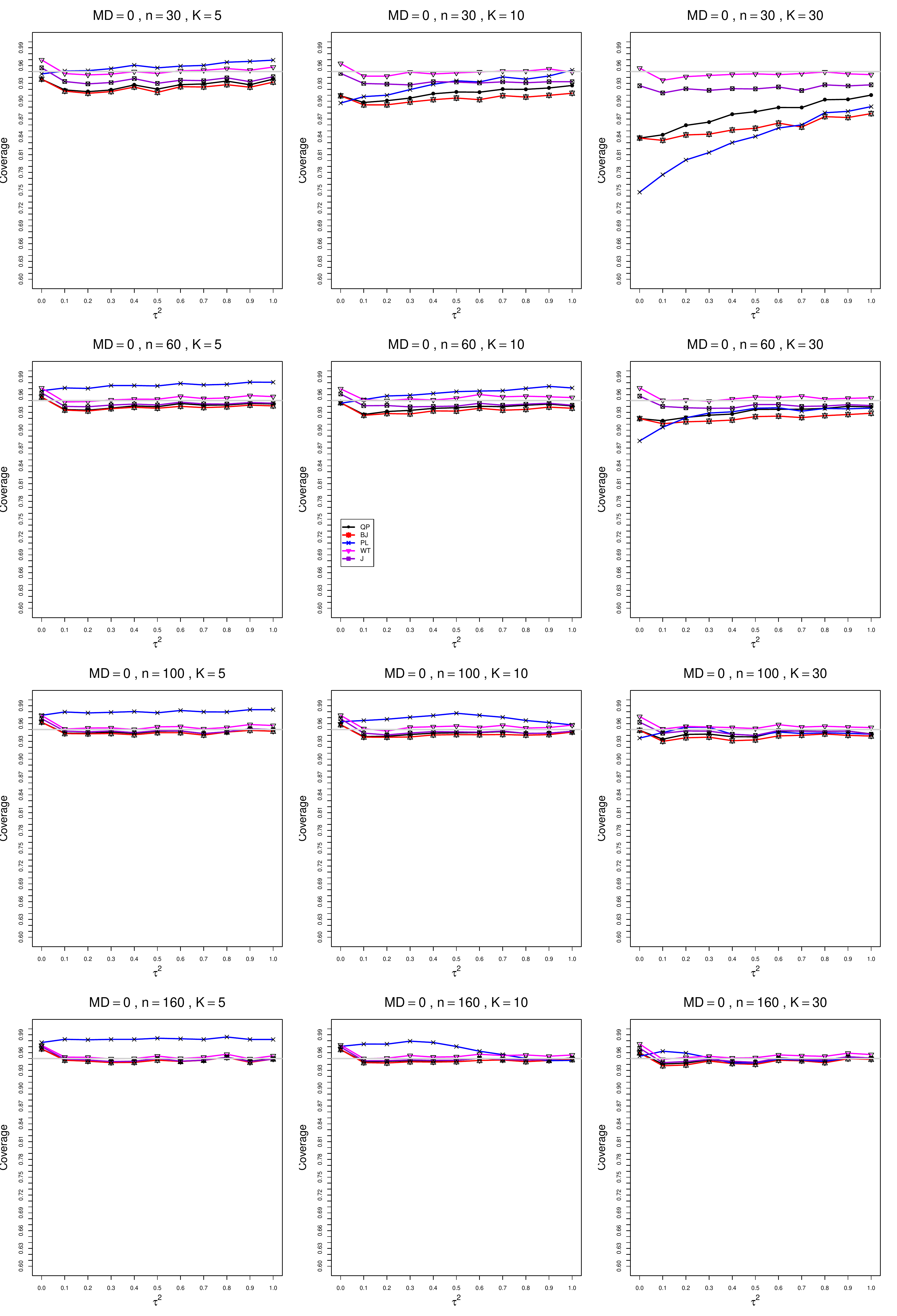}
	\caption{Coverage of 95\% confidence intervals for the  between-studies variance $\tau^2 = 0.0(0.1)1.0$ for $\mu=0$, $q=0.75$, $\sigma_C^2=10$, $\sigma_T^2=20$, unequal studies of average size  $\bar{n}=30,\;60,\;100,\;160$.
		\label{BiasTauMD0_S20_10unequalq075}}
\end{figure}

\clearpage
\section*{B1. Bias and mean squared error of point estimators $\hat{\mu}$ for $\tau^2 = 0.0(0.1)1.0$, $\sigma_{C}^2=1$, $\sigma_{T}^2=1,\;2$.}
For bias of $\mu$, each figure corresponds to a value of $\mu (= 0, 0.2, 0.5, 1, 2)$, a value of $q (= .5, .75)$, a value of $\tau^2 = 0.0(0.1)1.0$, a value of $\sigma_{C}^2=1$, a value of $\sigma_{T}^2=1,\;2$ , and a set of values of $n$ (= 20, 40, 100, 250) or $\bar{n}$ (= 30, 60, 100, 160).\\
Figures for mean squared error (expressed as the ratio of the MSE of SSW to the MSEs of the inverse-variance-weighted estimators that use the MP or WT estimator of $\tau^2$) use the above values of $\mu$ and q but only n = 20, 40, 100, 250.\\
Each figure contains a panel (with $\tau^2$ on the horizontal axis) for each combination of n (or $\bar{n}$) and $K (=5, 10, 30)$.\\
The point estimators of $\mu$ are
\begin{itemize}
	\item DL (DerSimonian-Laird)
	\item REML (restricted maximum likelihood)
	\item MP (Mandel-Paule)
	\item WT (Corrected Mandel-Paule moment estimator based on Welch-type approximation for Q distribution)
	\item J (Jackson)
	\item CDL (Corrected DerSimonian-Laird)
	\item SSW (sample-size weighted)
\end{itemize}

\clearpage
\setcounter{figure}{0}
\renewcommand{\thefigure}{B1.\arabic{figure}}
\begin{figure}[t]
	\centering
	\includegraphics[scale=0.33]{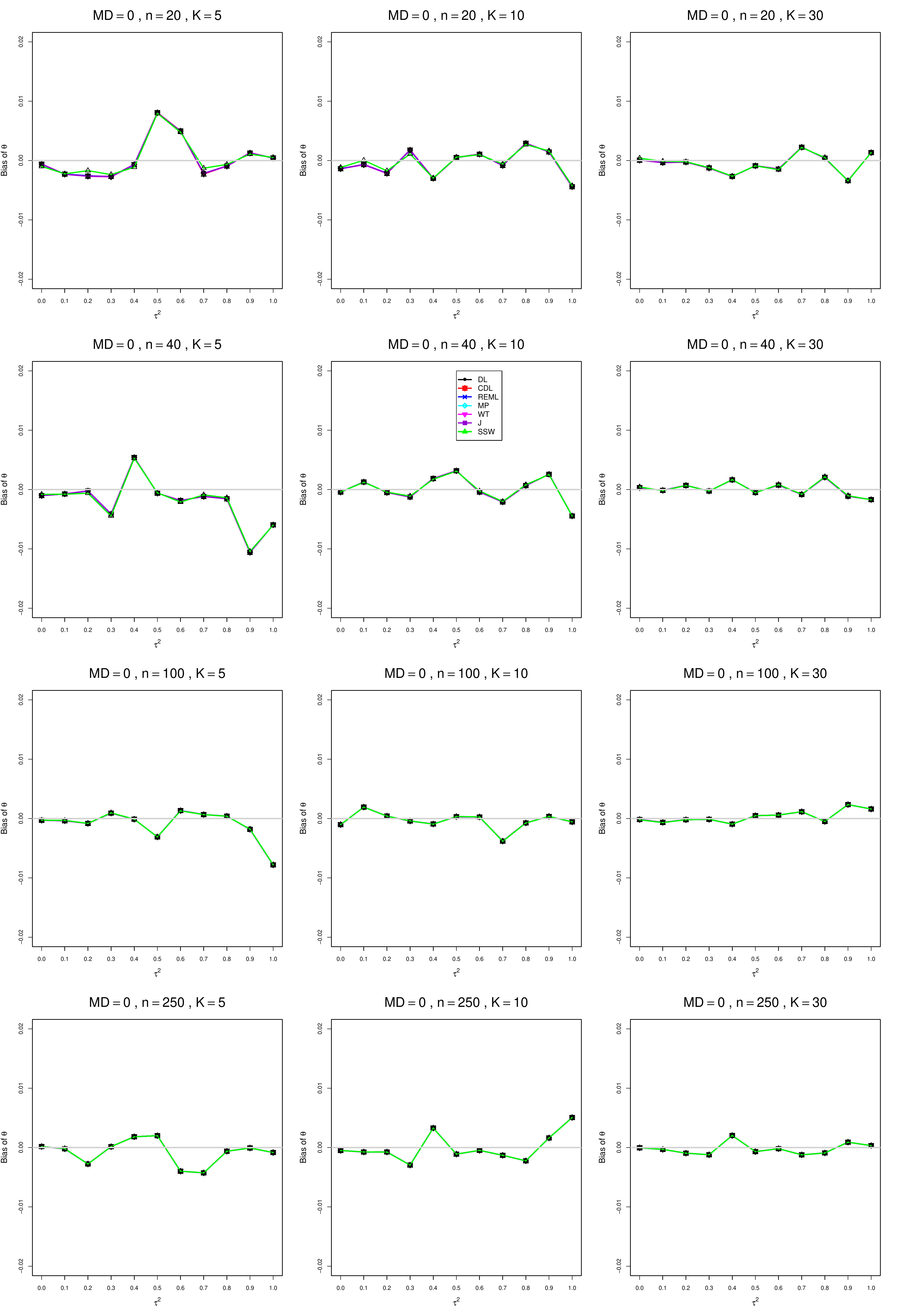}
	\caption{Bias of the estimation of  $\mu = 0$ for between-studies variance $\tau^2 = 0.0(0.1)1.0$, $q=0.5$, $\sigma_C^2=1$, $\sigma_T^2=1$,  equal study sizes $n=20,\;40,\;100,\;250$.
		\label{BiasThetaMD0_S1_1}}
\end{figure}

\begin{figure}[t]
	\centering
	\includegraphics[scale=0.33]{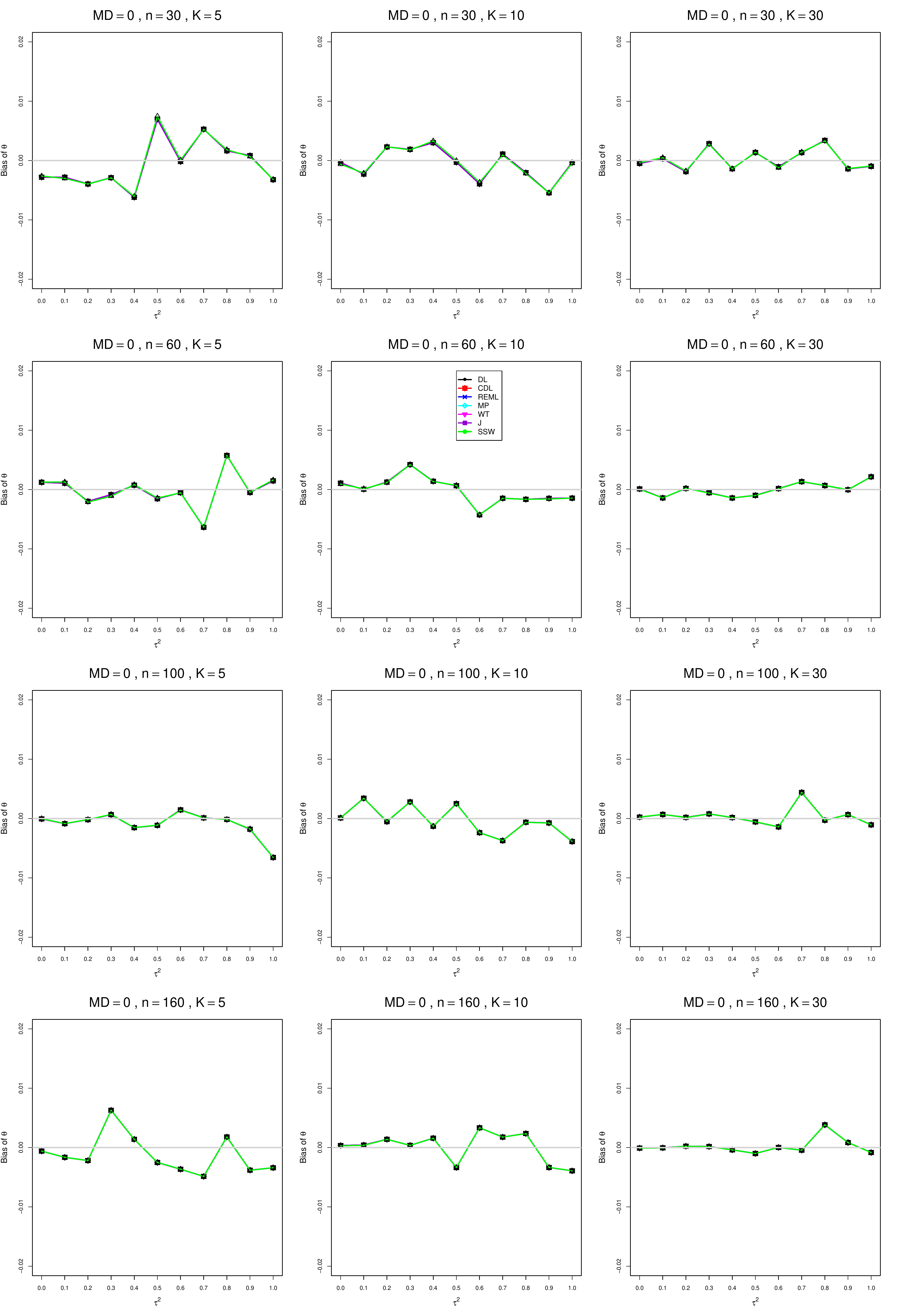}
	\caption{Bias of the estimation of  $\mu = 0$ for between-studies variance $\tau^2 = 0.0(0.1)1.0$, $q=0.5$, $\sigma_C^2=1$, $\sigma_T^2=1$, unequal studies of average size $\bar{n}=30,\;60,\;100,\;160$.
		`
		\label{BiasThetaMD0_S1_1unequal}}
\end{figure}

\begin{figure}[t]\centering
	\includegraphics[scale=0.35]{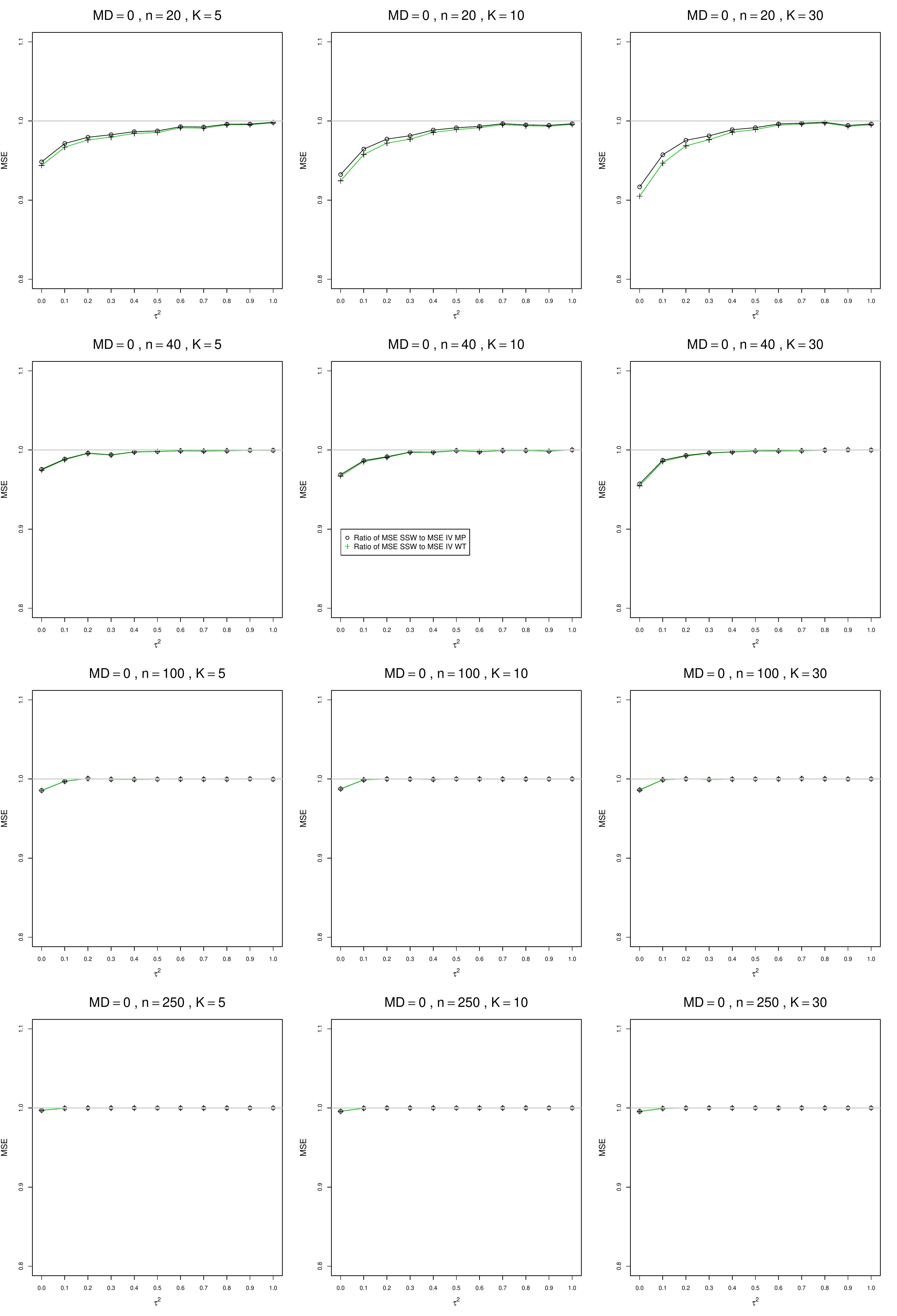}
	\caption{Ratio of mean squared errors of the fixed-weights to mean squared errors of inverse-variance estimator for $\mu=0$, $q=0.5$, $\sigma_C^2=1$, $\sigma_T^2=1$, $n=20,\;40,\;100,\;250$.
		\label{RatioOfMSEwithMD0fromMPandCMPSigma2T1andSigma2C1}}
\end{figure}


\begin{figure}[t]
	\centering
	\includegraphics[scale=0.33]{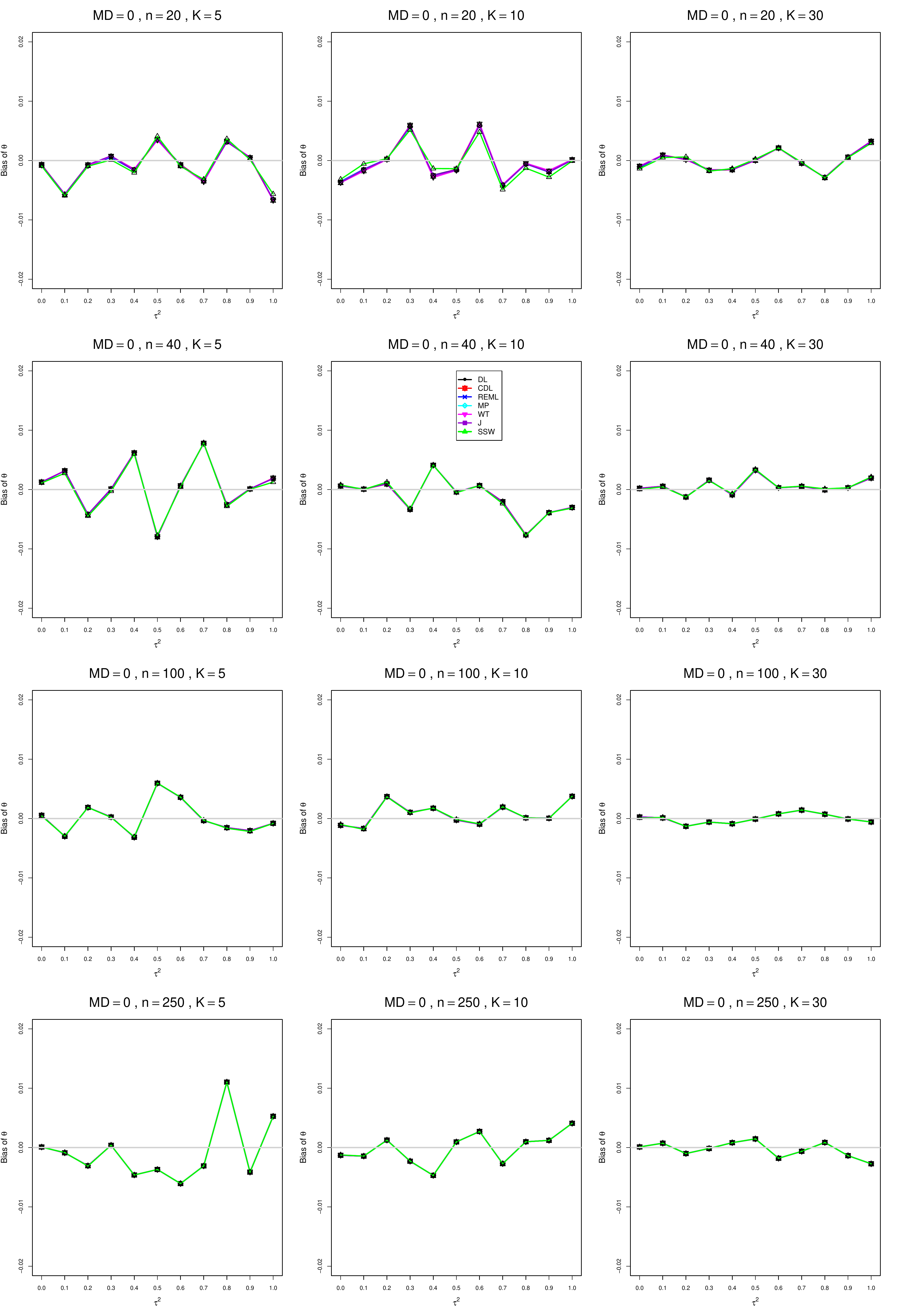}
	\caption{Bias of the estimation of  $\mu = 0$ for between-studies variance $\tau^2 = 0.0(0.1)1.0$, $q=0.75$, $\sigma_C^2=1$, $\sigma_T^2=1$,  equal study sizes $n=20,\;40,\;100,\;250$.
		\label{BiasThetaMD0_S1_1q075}}
\end{figure}

\begin{figure}[t]
	\centering
	\includegraphics[scale=0.33]{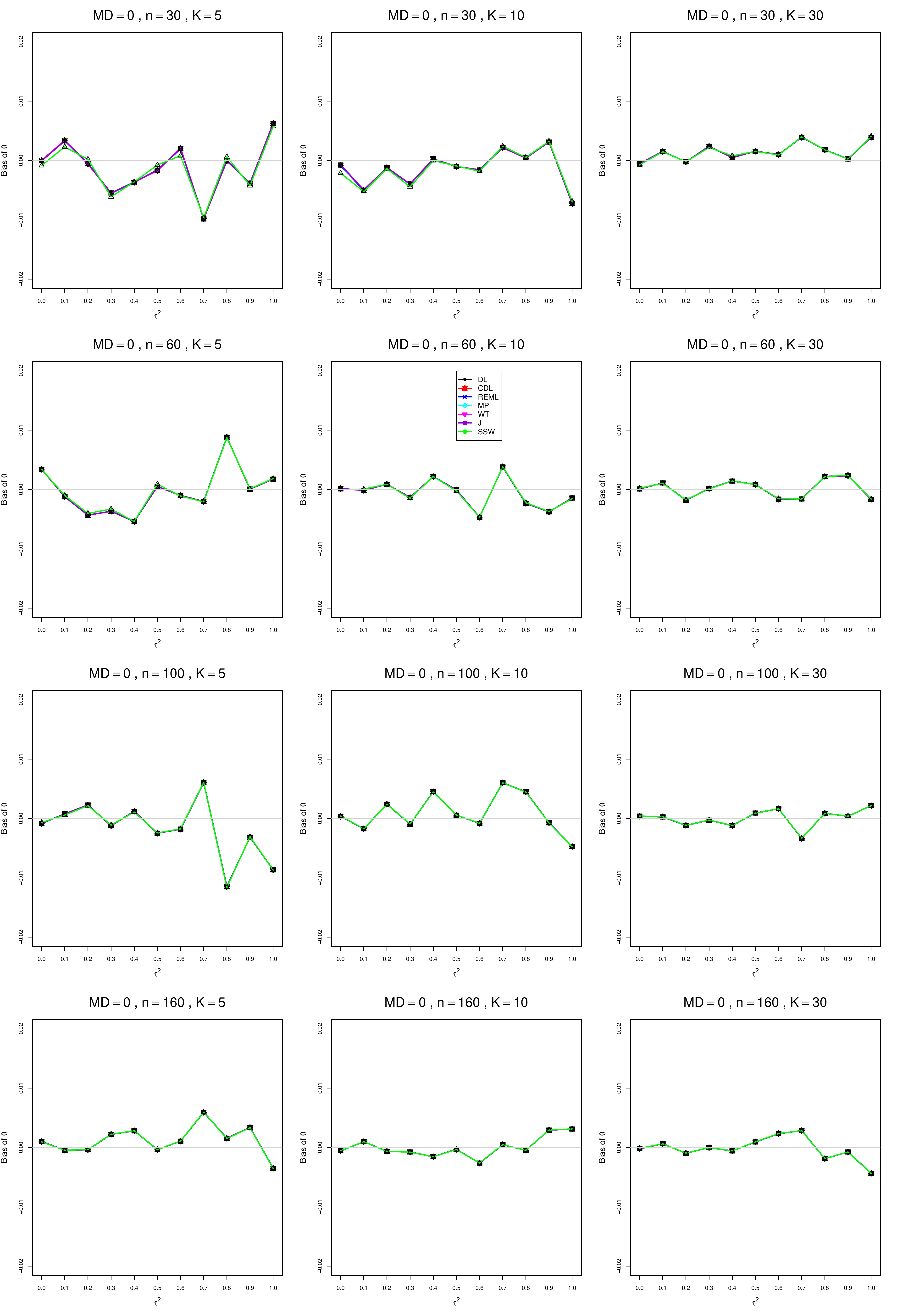}
	\caption{Bias of the estimation of  $\mu = 0$ for between-studies variance $\tau^2 = 0.0(0.1)1.0$, $q=0.75$, $\sigma_C^2=1$, $\sigma_T^2=1$, unequal studies of average size $\bar{n}=30,\;60,\;100,\;160$.
		\label{BiasThetaMD0_S1_1unequalq075}}
\end{figure}

\begin{figure}[t]\centering
	\includegraphics[scale=0.35]{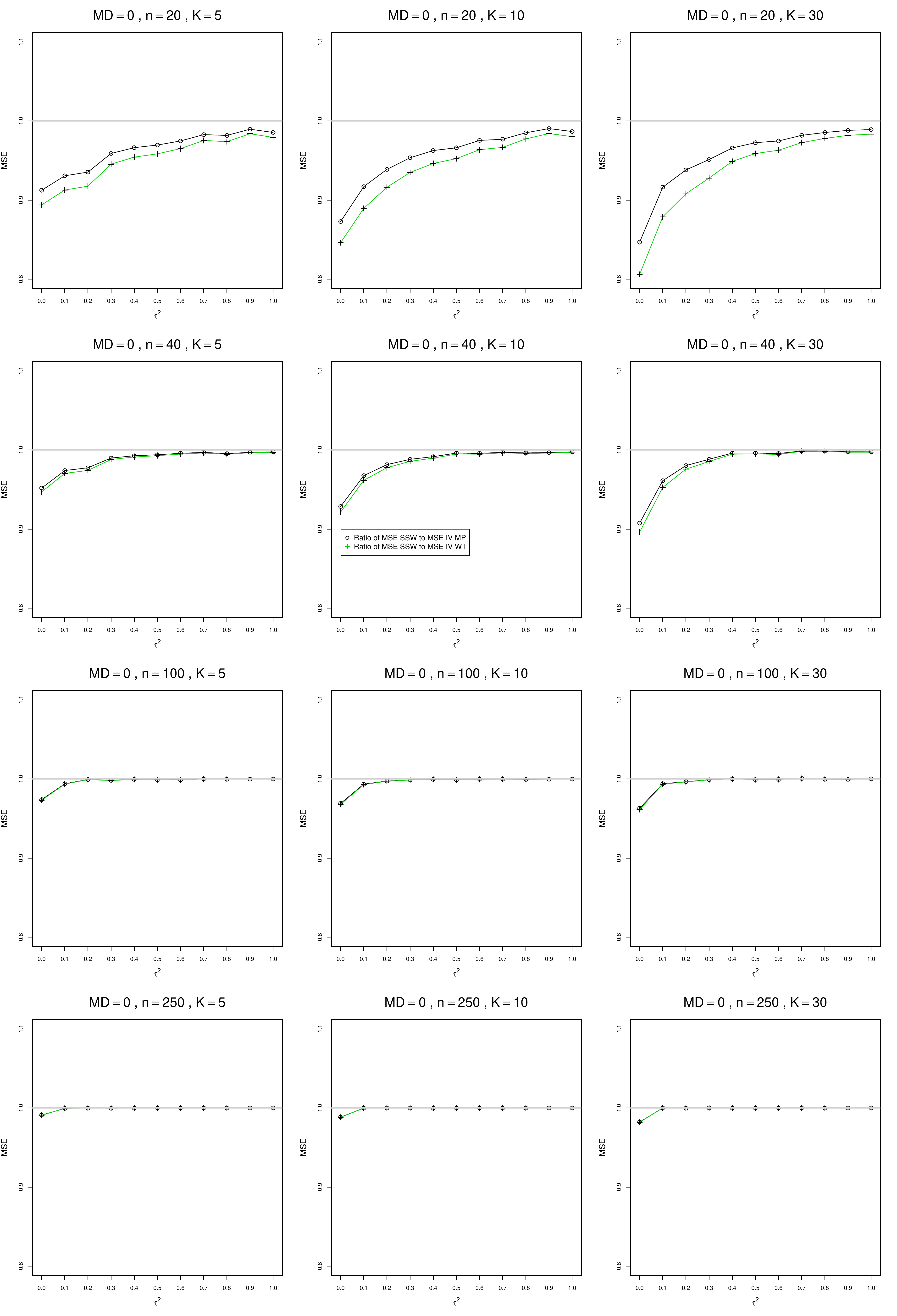}
	\caption{Ratio of mean squared errors of the fixed-weights to mean squared errors of inverse-variance estimator for $\mu=0$, for $q=0.75$,  $\sigma_C^2=1$, $\sigma_T^2=1$, $n=20,\;40,\;100,\;250$. 
		\label{RatioOfMSEwithMD0q075fromMPandCMPSigma2T1andSigma2C1}}
\end{figure}


\begin{figure}[t]
	\centering
	\includegraphics[scale=0.33]{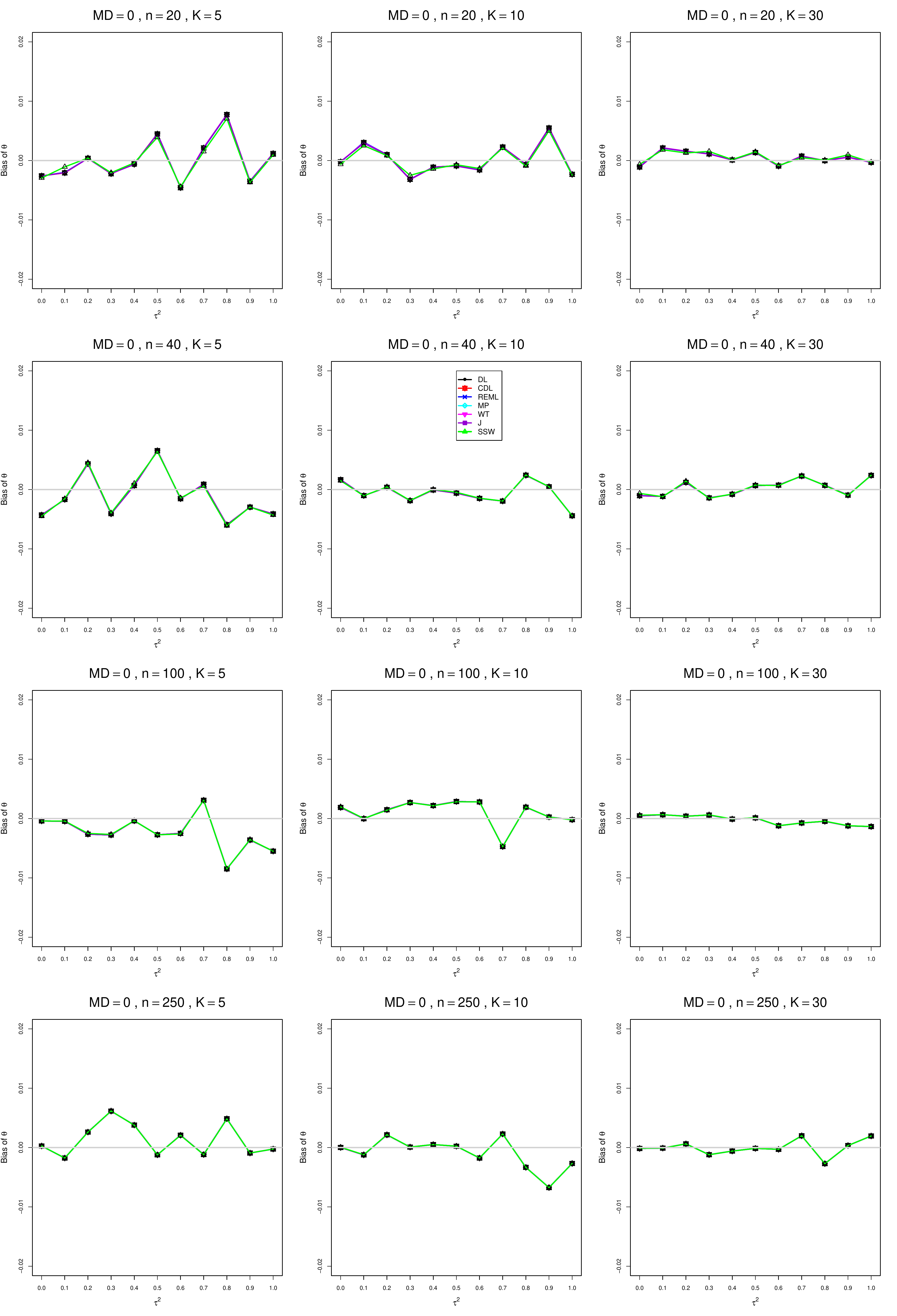}
	\caption{Bias of the estimation of  $\mu = 0$ for between-studies variance $\tau^2 = 0.0(0.1)1.0$, $q=0.5$, $\sigma_C^2=1$, $\sigma_T^2=2$,  equal study sizes $n=20,\;40,\;100,\;250$.
		\label{BiasThetaMD0_S1_2}}
\end{figure}

\begin{figure}[t]
	\centering
	\includegraphics[scale=0.33]{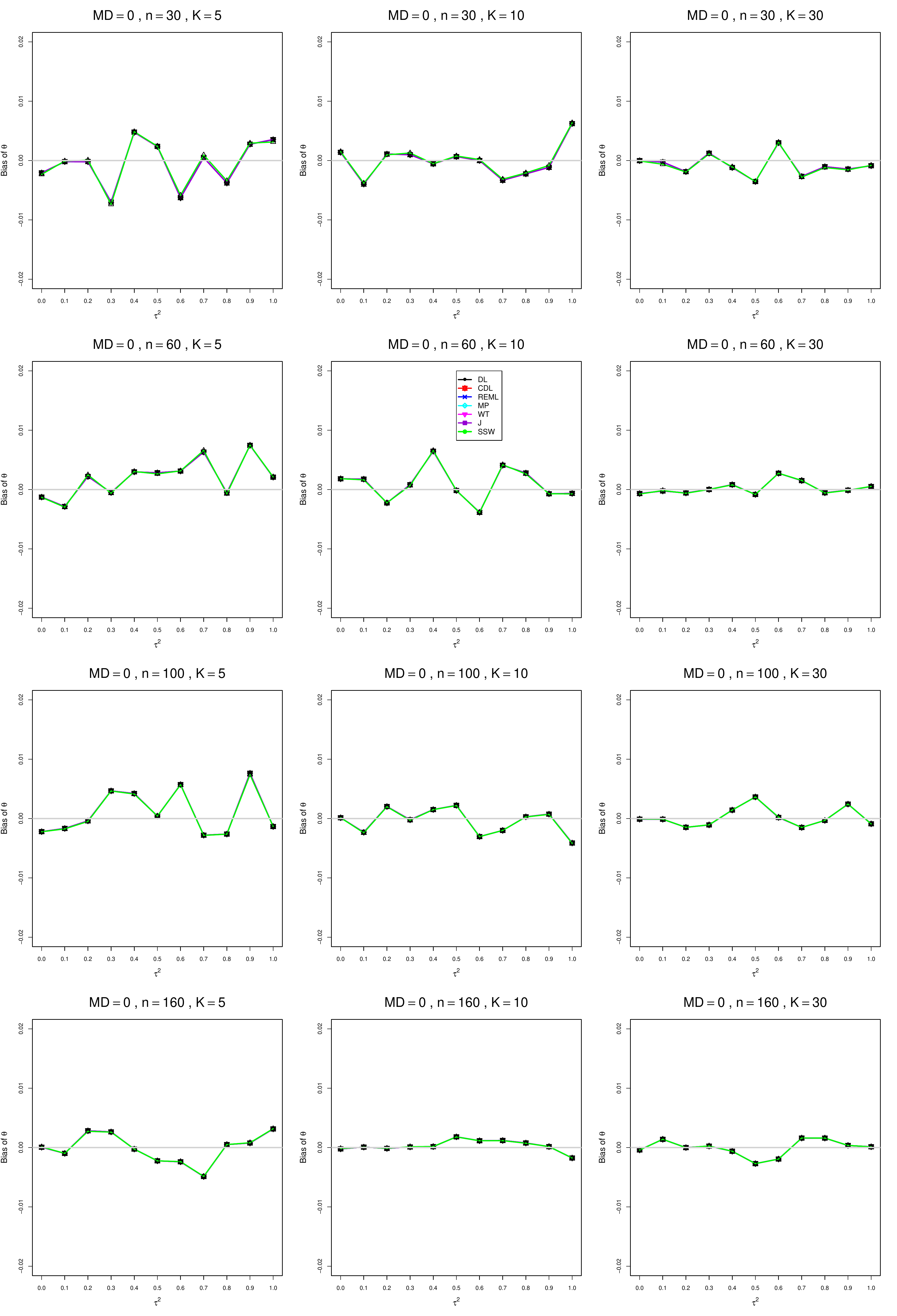}
	\caption{Bias of the estimation of  $\mu = 0$ for between-studies variance $\tau^2 = 0.0(0.1)1.0$, $q=0.5$, $\sigma_C^2=1$, $\sigma_T^2=2$, unequal studies of average size $\bar{n}=30,\;60,\;100,\;160$.
		\label{BiasThetaMD0_S1_2unequal}}
\end{figure}

\begin{figure}[t]\centering
	\includegraphics[scale=0.35]{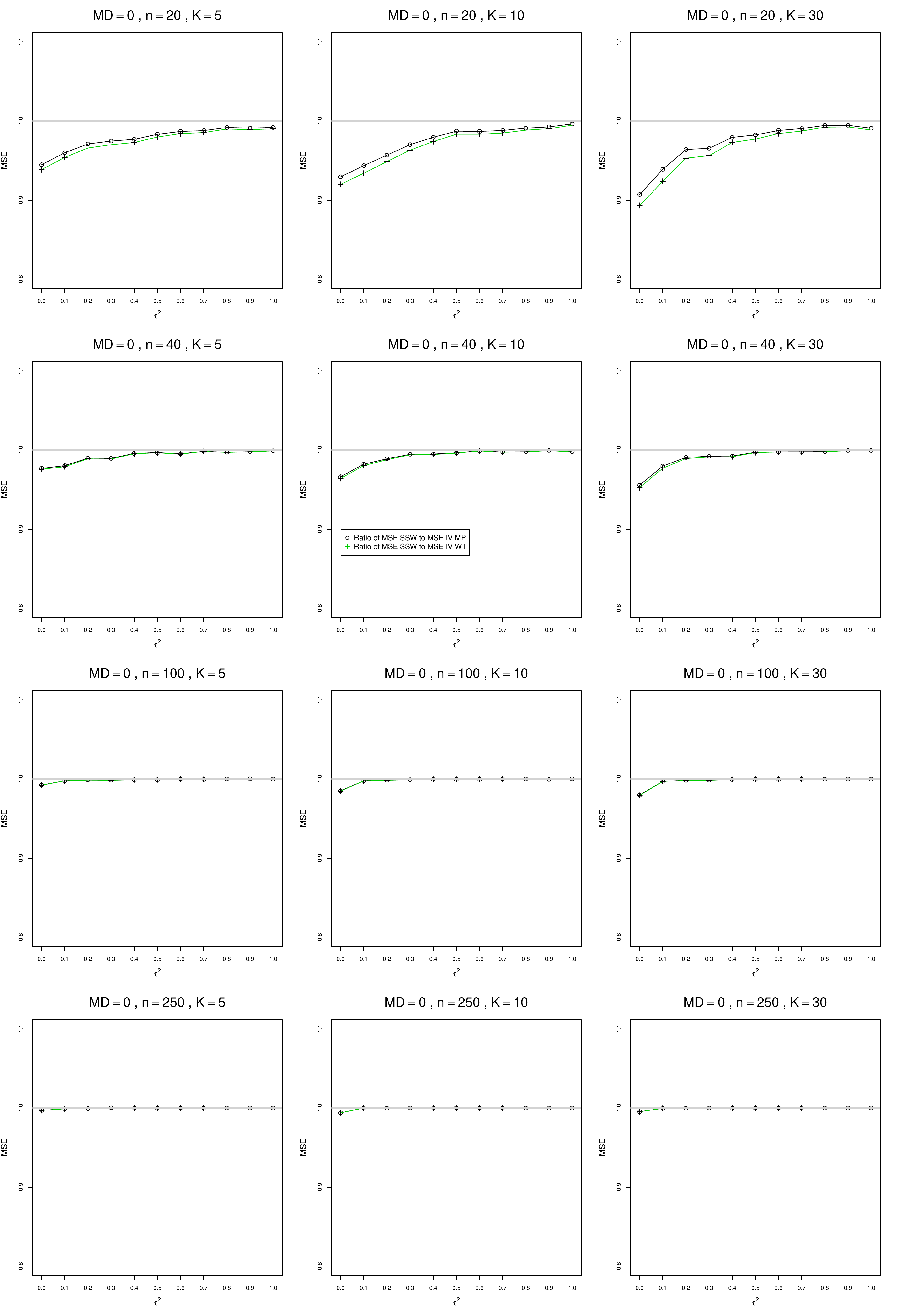}
	\caption{Ratio of mean squared errors of the fixed-weights to mean squared errors of inverse-variance estimator for $\mu=0$, $q=0.5$, $\sigma_C^2=1$, $\sigma_T^2=2$, $n=20,\;40,\;100,\;250$. 
		\label{RatioOfMSEwithMD0fromMPandCMPSigma2T2andSigma2C1}}
\end{figure}

\clearpage

\begin{figure}[t]
	\centering
	\includegraphics[scale=0.33]{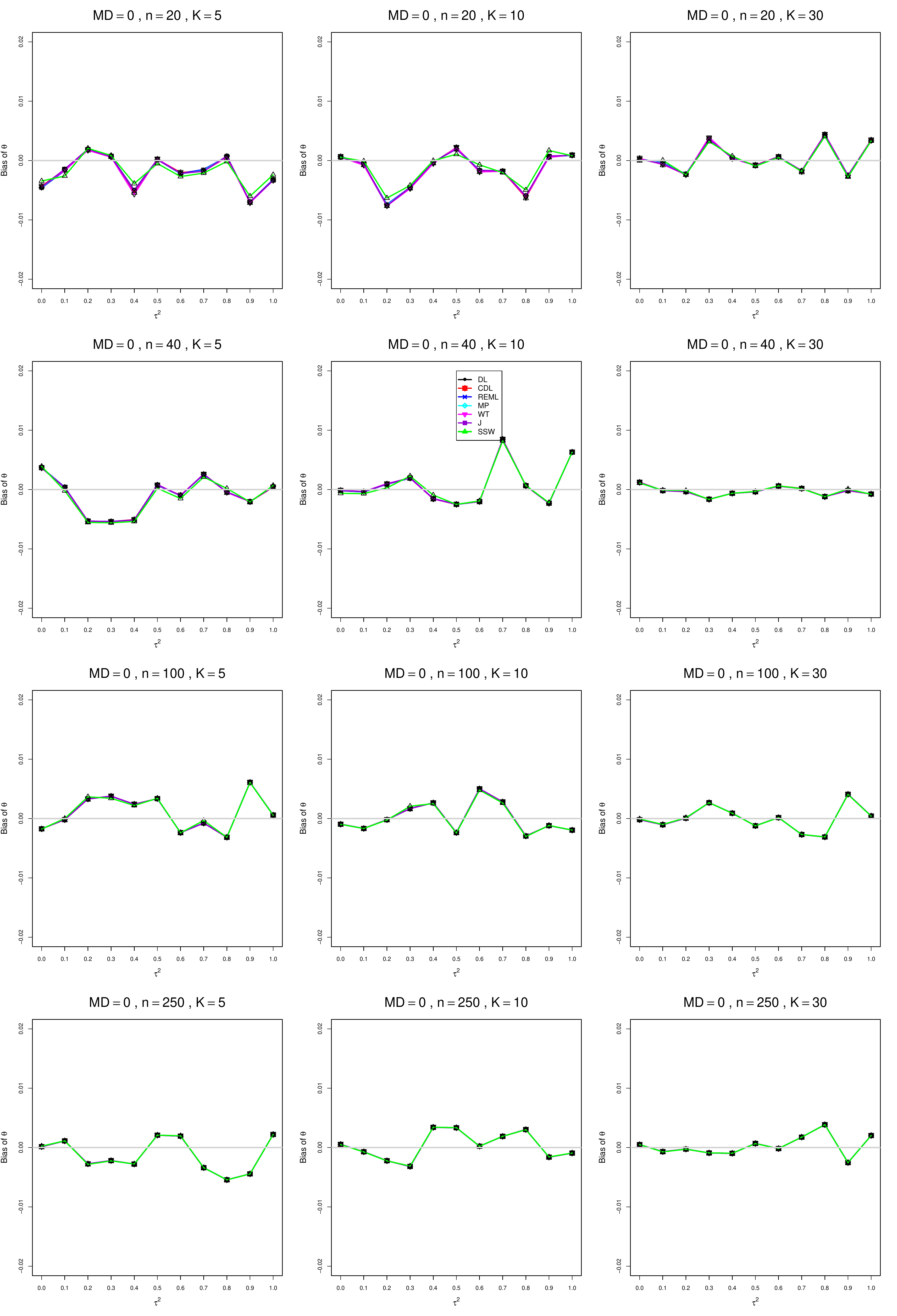}
	\caption{Bias of the estimation of  $\mu = 0$ for between-studies variance $\tau^2 = 0.0(0.1)1.0$, $q=0.75$, $\sigma_C^2=1$, $\sigma_T^2=2$,  equal study sizes $n=20,\;40,\;100,\;250$.
		\label{BiasThetaMD0_S1_2q075}}
\end{figure}

\begin{figure}[t]
	\centering
	\includegraphics[scale=0.33]{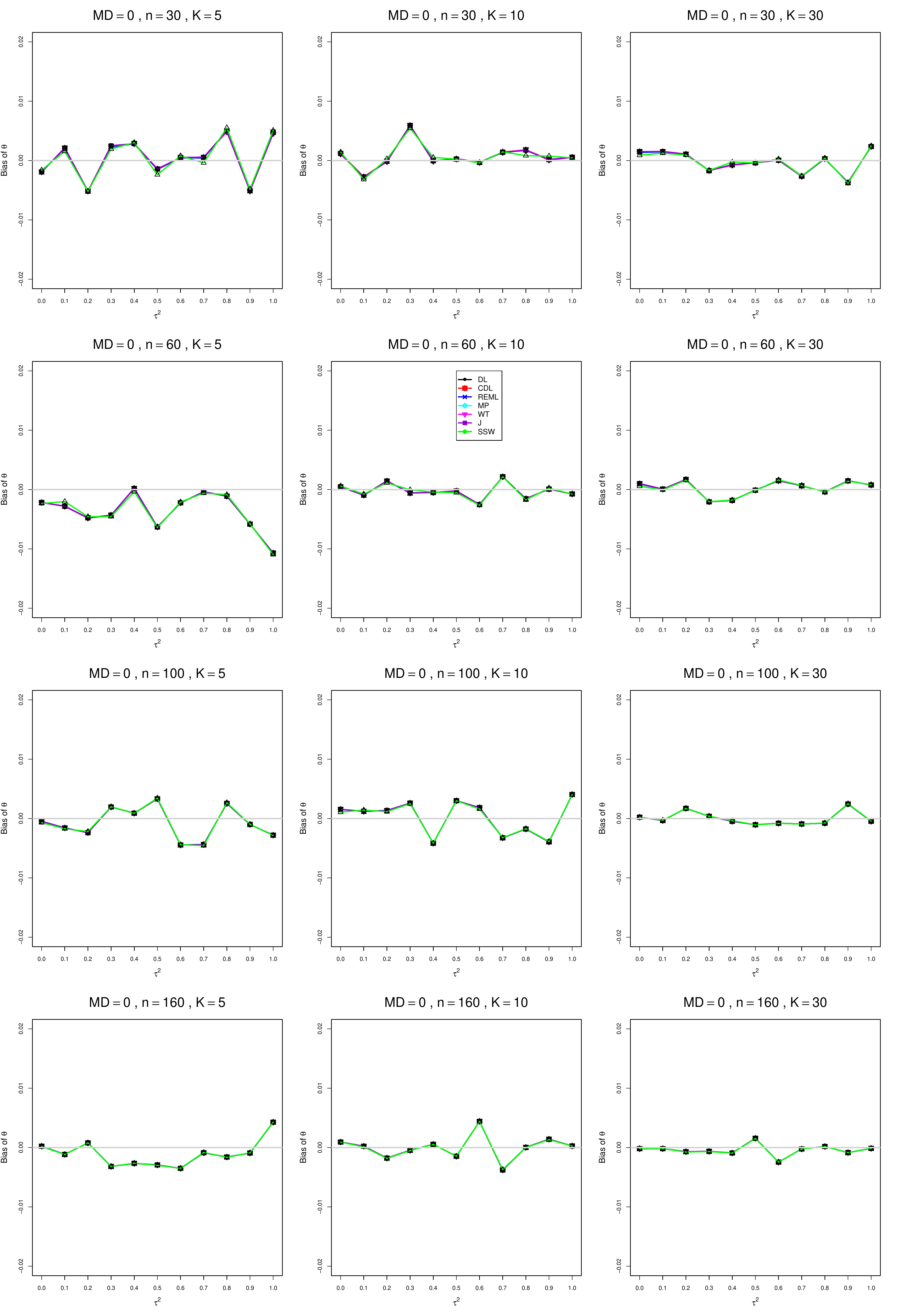}
	\caption{Bias of the estimation of  $\mu = 0$ for between-studies variance $\tau^2 = 0.0(0.1)1.0$, $q=0.75$, $\sigma_C^2=1$, $\sigma_T^2=2$, unequal studies of average size $\bar{n}=30,\;60,\;100,\;160$.
		\label{BiasThetaMD0_S1_2unequalq075}}
\end{figure}

\begin{figure}[t]\centering
	\includegraphics[scale=0.35]{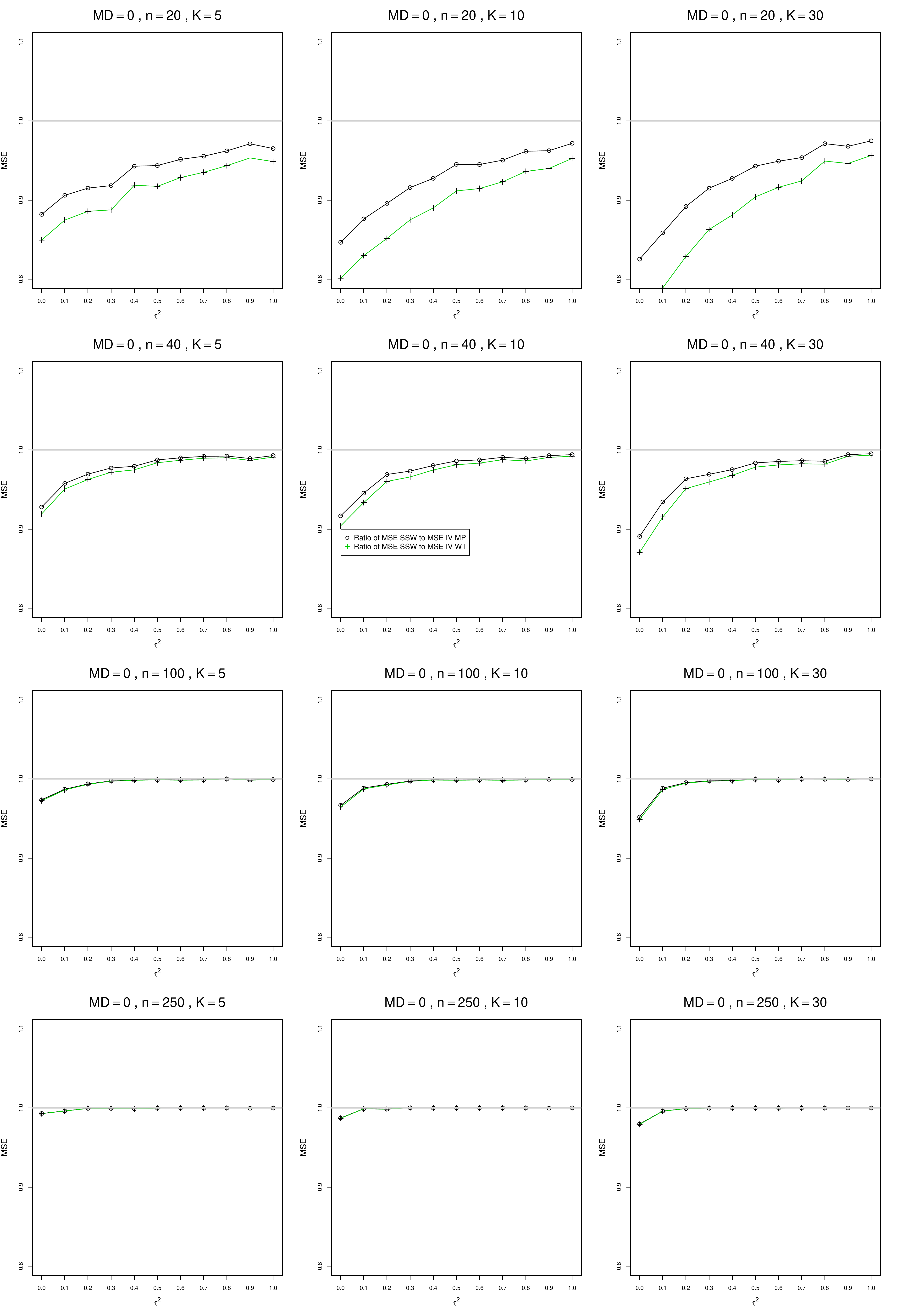}
	\caption{Ratio of mean squared errors of the fixed-weights to mean squared errors of inverse-variance estimator for $\mu=0$, for $q=0.75$,$\sigma_C^2=1$, $\sigma_T^2=2$, $n=20,\;40,\;100,\;250$.
		\label{RatioOfMSEwithMD0q075fromMPandCMPSigma2T2andSigma2C1}}
\end{figure}

\clearpage


\setcounter{section}{0}
\renewcommand{\thefigure}{B2.\arabic{figure}}
\setcounter{figure}{0}
\setcounter{section}{0}

\section*{B2. Coverage of $\hat{\mu}$ for $\tau^2 = 0.0(0.1)1.0$, $\sigma_{C}^2=1$, $\sigma_{T}^2=1,\;2$.}
For coverage of $\mu$, each figure corresponds to a value of $\mu (= 0, 0.2, 0.5, 1, 2)$, a value of $q (= .5, .75)$, a value of $\tau^2 = 0.0(0.1)1.0$, a value of $\sigma_{C}^2=1$, a value of $\sigma_{T}^2=1,\;2$ , and a set of values of $n$ (= 20, 40, 100, 250) or $\bar{n}$ (= 30, 60, 100, 160).\\
Each figure contains a panel (with $\tau^2$ on the horizontal axis) for each combination of n (or $\bar{n}$) and $K (=5, 10, 30)$.\\
The interval estimators of $mu$ are the companions to the inverse-variance-weighted point estimators
\begin{itemize}
	\item DL (DerSimonian-Laird)
	\item REML (restricted maximum likelihood)
	\item MP (Mandel-Paule)
	\item WT (Corrected Mandel-Paule moment estimator based on Welch-type approximation for Q distribution)
	\item J (Jackson)
	\item CDL (Corrected DerSimonian-Laird)
\end{itemize}
and
\begin{itemize}
	\item HKSJ (Hartung-Knapp-Sidik-Jonkman)
	\item HKSJ WT (HKSJ with WT estimator of $\tau^2$)
	\item SSW (SSW as center and half-width equal to critical value from $t_{K-1}$
\end{itemize}
times estimated standard deviation of SSW with $\hat{\tau}^2$ = $\hat{\tau}^2_{WT}$

\begin{figure}[t]
	\centering
	\includegraphics[scale=0.33]{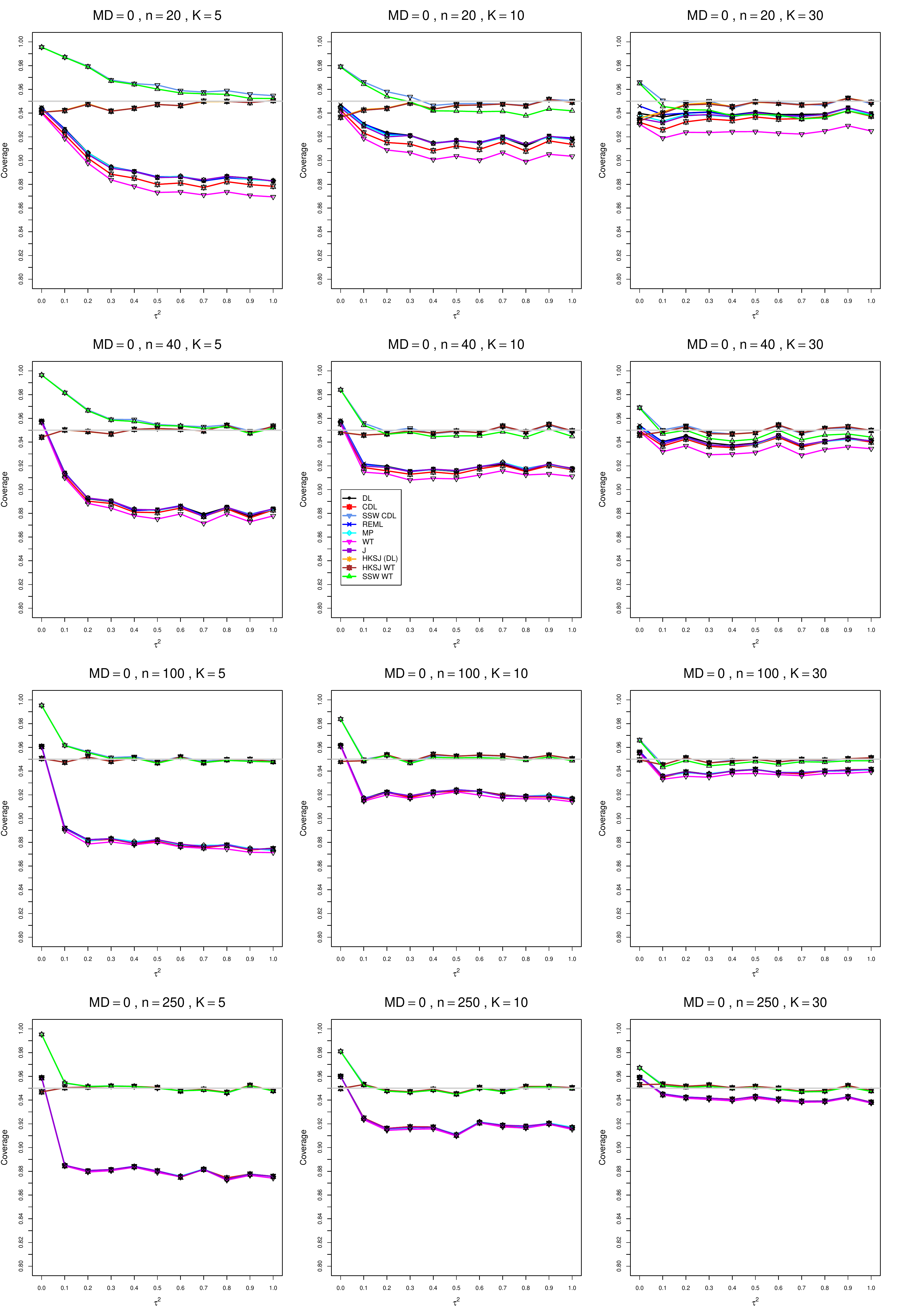}
	\caption{Coverage of 95\% confidence intervals for the $\mu = 0$ for the between-studies variance $\tau^2 = 0.0(0.1)1.0$ for, $q=0.5$, $\sigma_C^2=1$, $\sigma_T^2=1$,  equal study sizes $n=20,\;40,\;100,\;250$.
		\label{CovThetaMD0_S1_1}}
\end{figure}

\begin{figure}[t]
	\centering
	\includegraphics[scale=0.33]{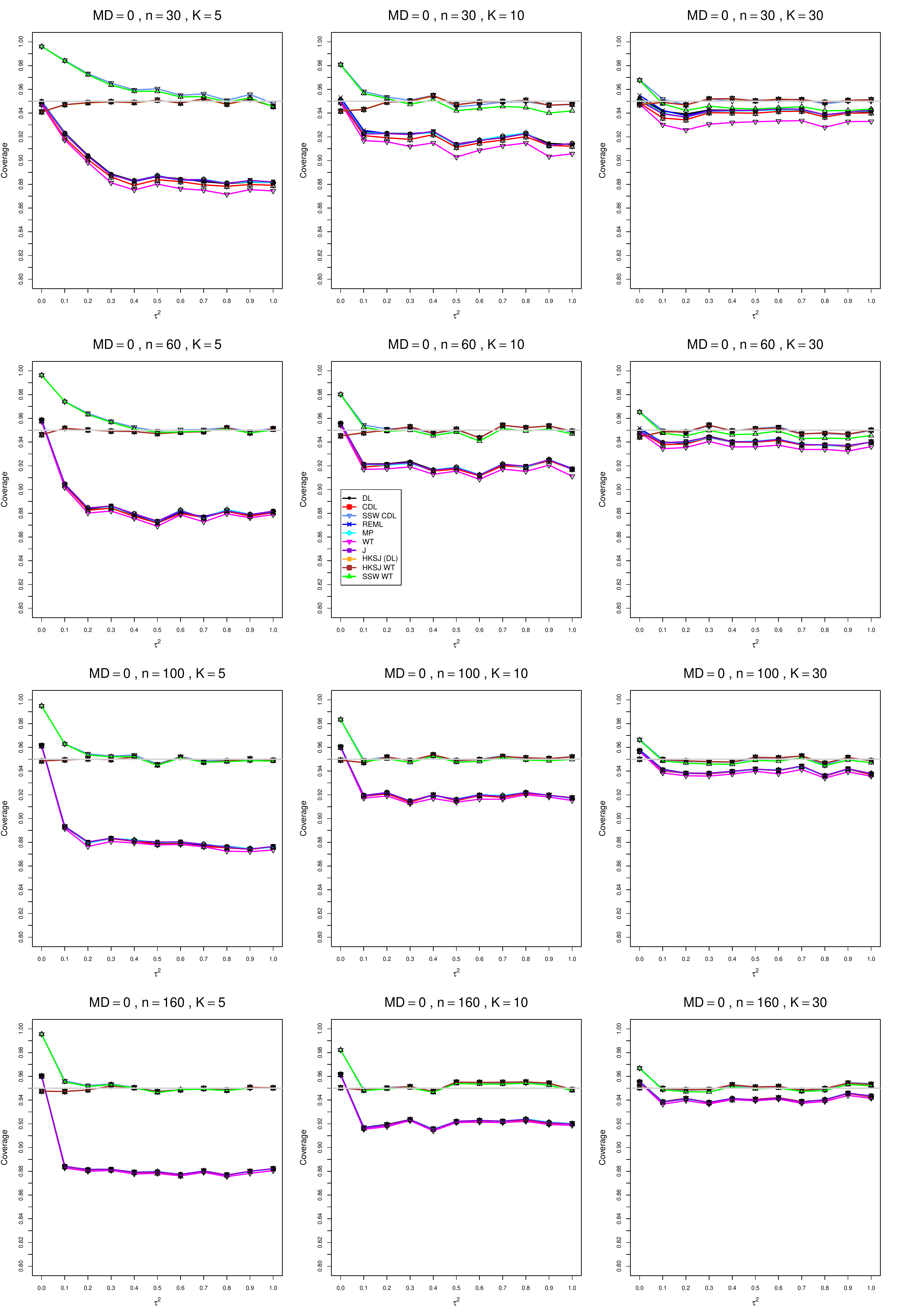}
	\caption{Coverage of 95\% confidence intervals for the $\mu = 0$ for the between-studies variance $\tau^2 = 0.0(0.1)1.0$, $q=0.5$, $\sigma_C^2=1$, $\sigma_T^2=1$, unequal studies of average size $\bar{n}=30,\;60,\;100,\;160$.
		\label{CovThetaMD0_S1_1unequal}}
\end{figure}


\begin{figure}[t]
	\centering
	\includegraphics[scale=0.33]{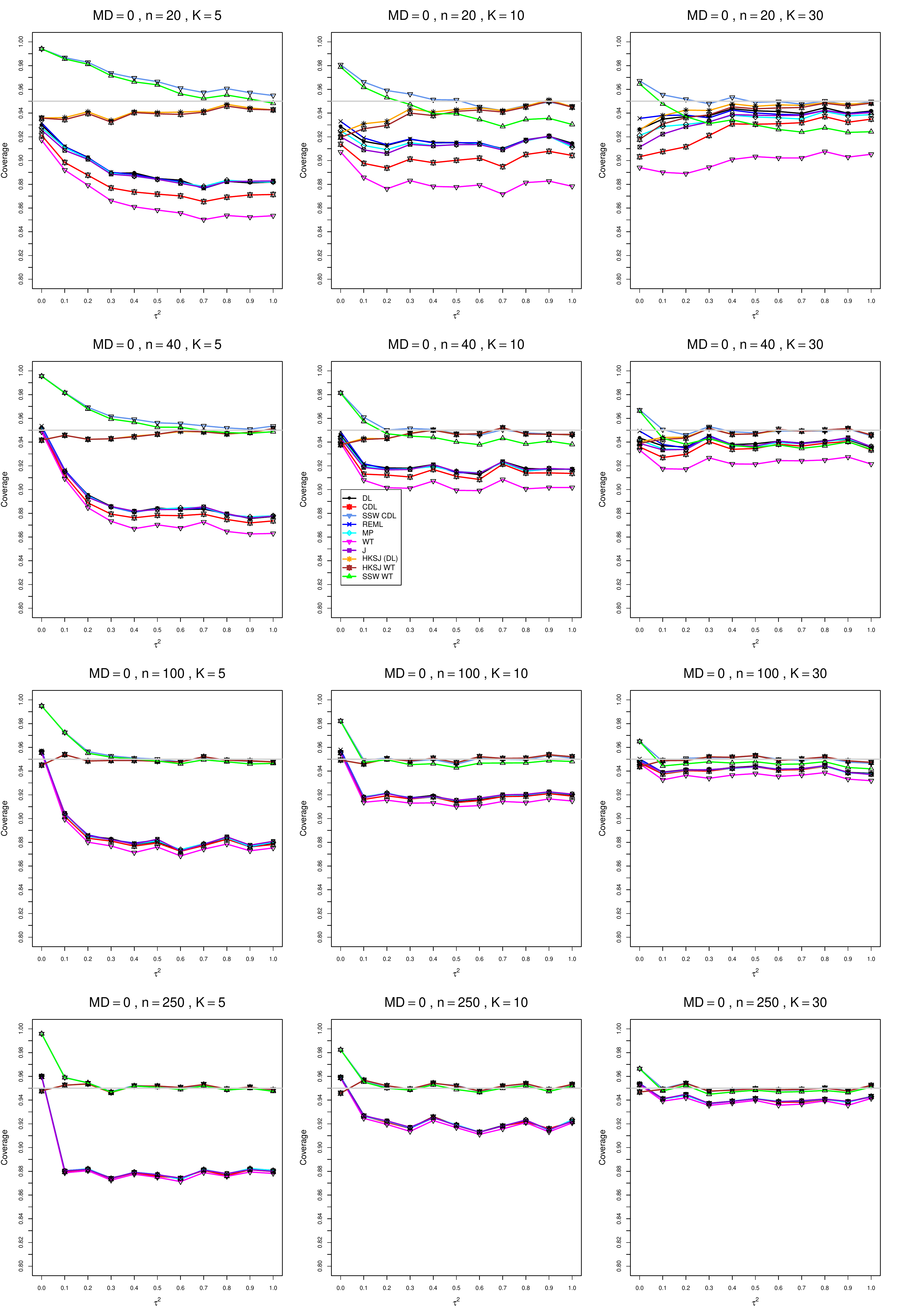}
	\caption{Coverage of 95\% confidence intervals for the $\mu = 0$ for the  between-studies variance $\tau^2 = 0.0(0.1)1.0$, $q=0.75$, $\sigma_C^2=1$, $\sigma_T^2=1$,  equal study sizes $n=20,\;40,\;100,\;250$.
		\label{CovThetaMD0_S1_1q075}}
\end{figure}

\begin{figure}[t]
	\centering
	\includegraphics[scale=0.33]{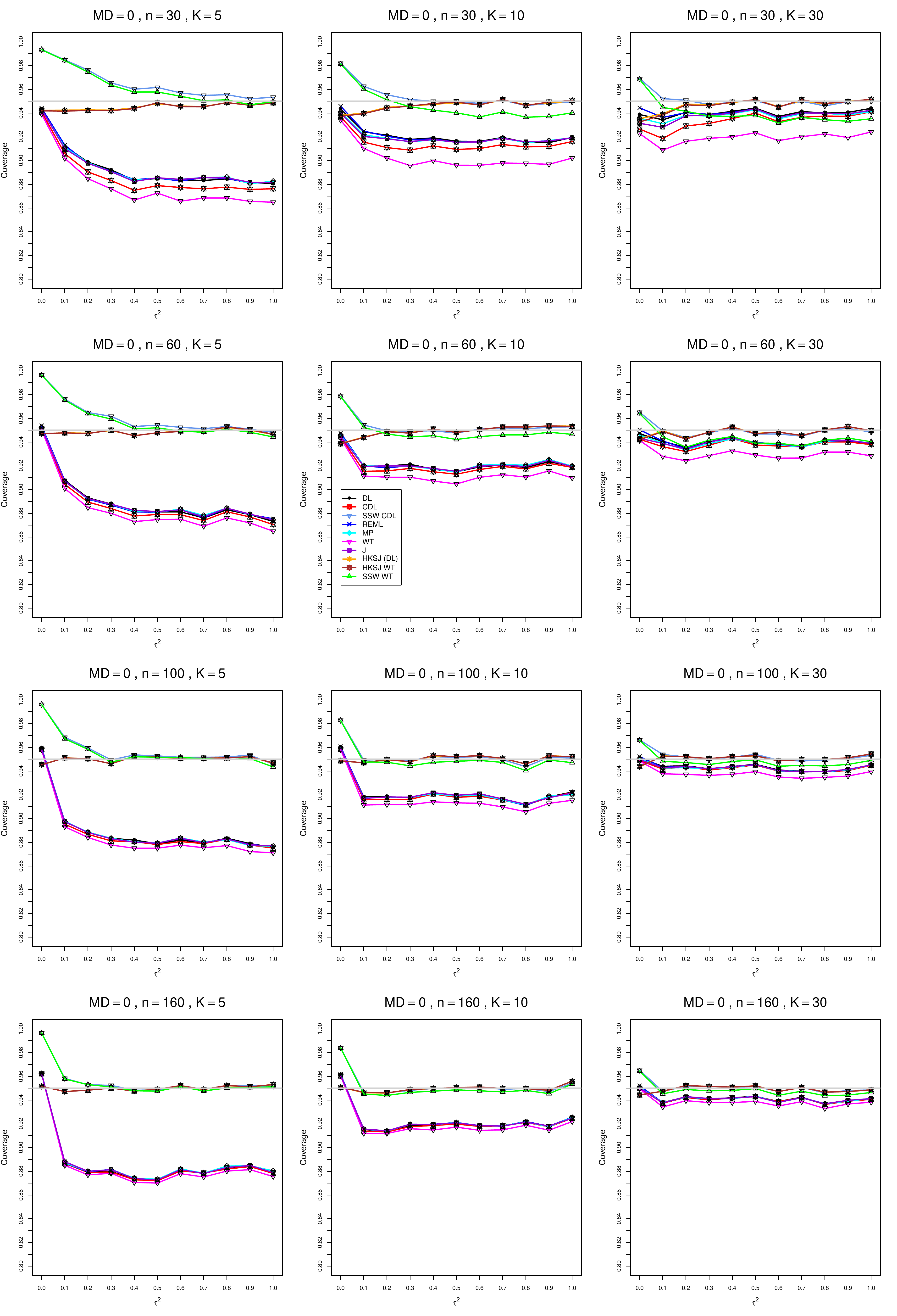}
	\caption{Coverage of 95\% confidence intervals for the $\mu = 0$ for the  between-studies variance $\tau^2 = 0.0(0.1)1.0$, $q=0.75$, $\sigma_C^2=1$, $\sigma_T^2=1$, unequal studies of average size $\bar{n}=30,\;60,\;100,\;160$.
		\label{CovThetaMD0_S1_1unequalq075}}
\end{figure}

\clearpage

\begin{figure}[t]
	\centering
	\includegraphics[scale=0.33]{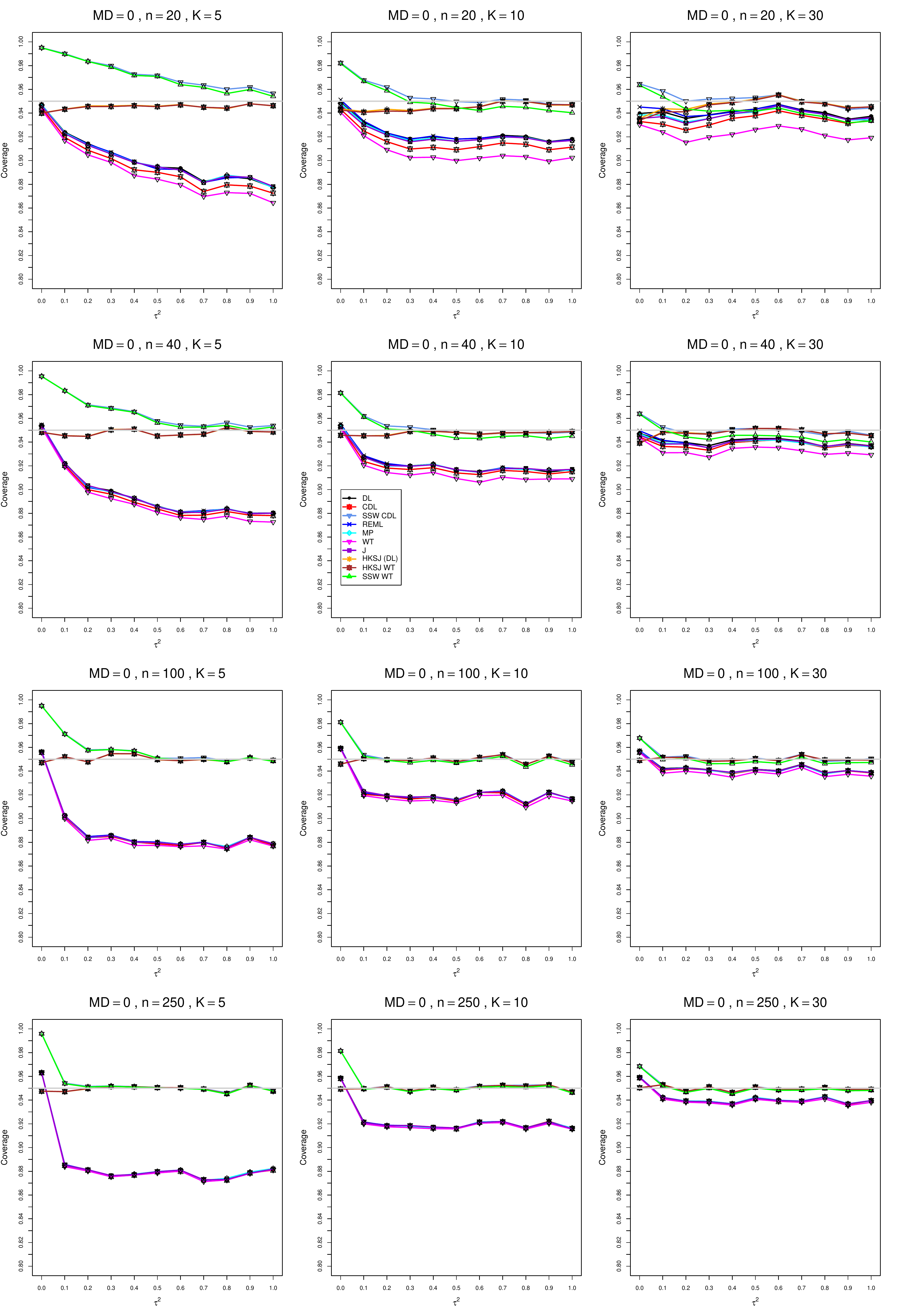}
	\caption{Coverage of 95\% confidence intervals for the $\mu = 0$ for the between-studies variance $\tau^2 = 0.0(0.1)1.0$ for, $q=0.5$, $\sigma_C^2=1$, $\sigma_T^2=2$,  equal study sizes $n=20,\;40,\;100,\;250$.
		\label{CovThetaMD0_S1_2}}
\end{figure}

\begin{figure}[t]
	\centering
	\includegraphics[scale=0.33]{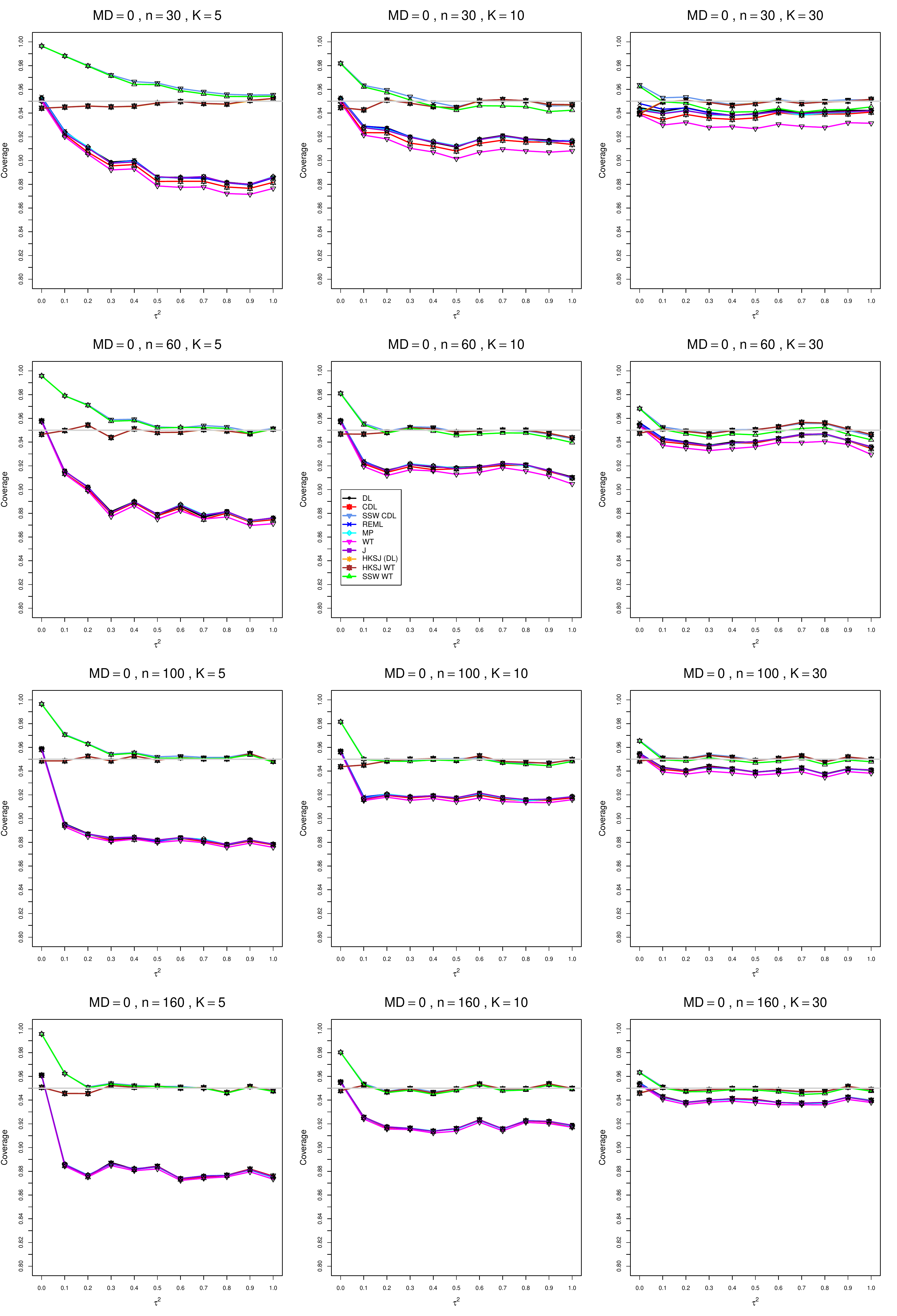}
	\caption{Coverage of 95\% confidence intervals for the $\mu = 0$ for the between-studies variance $\tau^2 = 0.0(0.1)1.0$, $q=0.5$, $\sigma_C^2=1$, $\sigma_T^2=2$, unequal studies of average size $\bar{n}=30,\;60,\;100,\;160$.
		\label{CovThetaMD0_S1_2unequal}}
\end{figure}

\clearpage

\begin{figure}[t]
	\centering
	\includegraphics[scale=0.33]{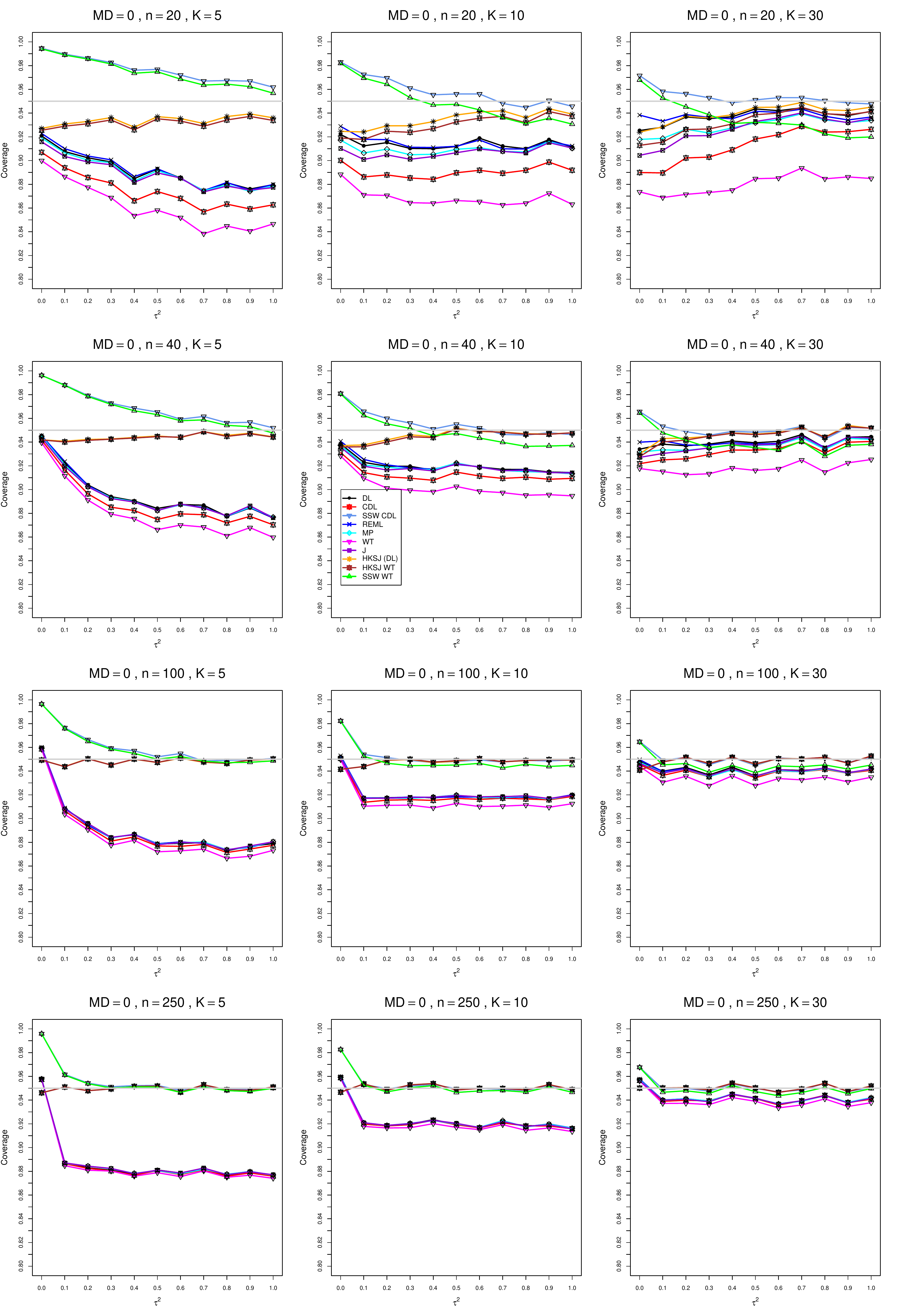}
	\caption{Coverage of 95\% confidence intervals for the $\mu = 0$ for the  between-studies variance $\tau^2 = 0.0(0.1)1.0$, $q=0.75$, $\sigma_C^2=1$, $\sigma_T^2=2$,  equal study sizes $n=20,\;40,\;100,\;250$.
		\label{CovThetaMD0_S1_2q075}}
\end{figure}

\begin{figure}[t]
	\centering
	\includegraphics[scale=0.33]{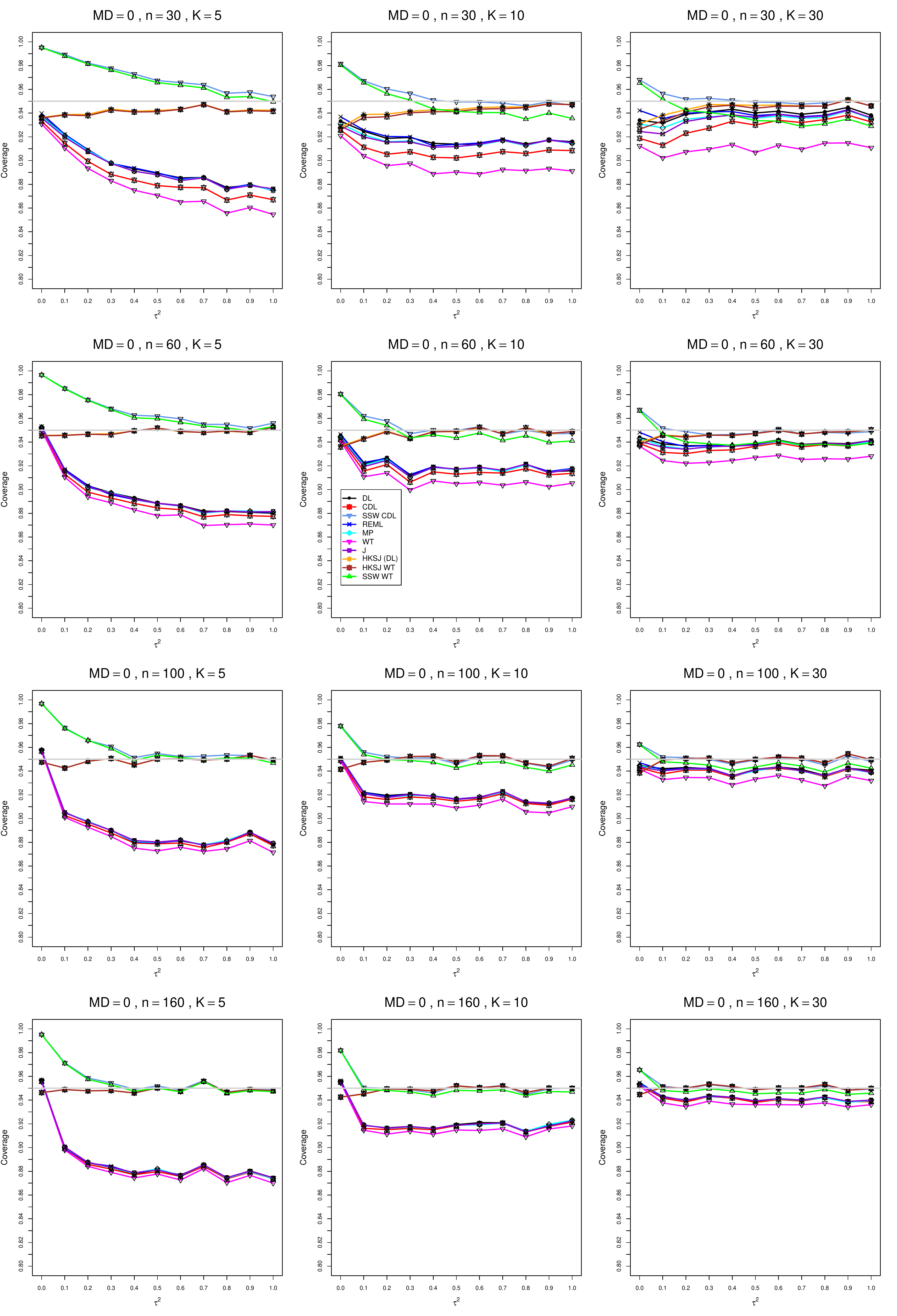}
	\caption{Coverage of 95\% confidence intervals for the $\mu = 0$ for the  between-studies variance $\tau^2 = 0.0(0.1)1.0$, $q=0.75$, $\sigma_C^2=1$, $\sigma_T^2=2$, unequal studies of average size $\bar{n}=30,\;60,\;100,\;160$.
		\label{CovThetaMD0_S1_2unequalq075}}
\end{figure}


\clearpage
\renewcommand{\thefigure}{B3.\arabic{figure}}
\setcounter{figure}{0}
\setcounter{section}{0}

\section*{B3. Bias and mean squared error of point estimators $\hat{\mu}$ for $\tau^2 = 0.0(0.01)0.1$, $\sigma_{C}^2=1$, $\sigma_{T}^2=1,\;2$.}
For bias of $\mu$, each figure corresponds to a value of $\mu (= 0, 0.2, 0.5, 1, 2)$, a value of $q (= .5, .75)$, a value of $\tau^2 = 0.0(0.01)0.1$, a value of $\sigma_{C}^2=1$, a value of $\sigma_{T}^2=1,\;2$ , and a set of values of $n$ (= 20, 40, 100, 250) or $\bar{n}$ (= 30, 60, 100, 160).\\
Figures for mean squared error (expressed as the ratio of the MSE of SSW to the MSEs of the inverse-variance-weighted estimators that use the MP or WT estimator of $\tau^2$) use the above values of $\mu$ and q but only n = 20, 40, 100, 250.\\
Each figure contains a panel (with $\tau^2$ on the horizontal axis) for each combination of n (or $\bar{n}$) and $K (=5, 10, 30)$.\\
The point estimators of $\mu$ are
\begin{itemize}
	\item DL (DerSimonian-Laird)
	\item REML (restricted maximum likelihood)
	\item MP (Mandel-Paule)
	\item WT (Corrected Mandel-Paule moment estimator based on Welch-type approximation for Q distribution)
	\item J (Jackson)
	\item CDL (Corrected DerSimonian-Laird)
	\item SSW (sample-size weighted)
\end{itemize}

\begin{figure}[t]
	\centering
	\includegraphics[scale=0.33]{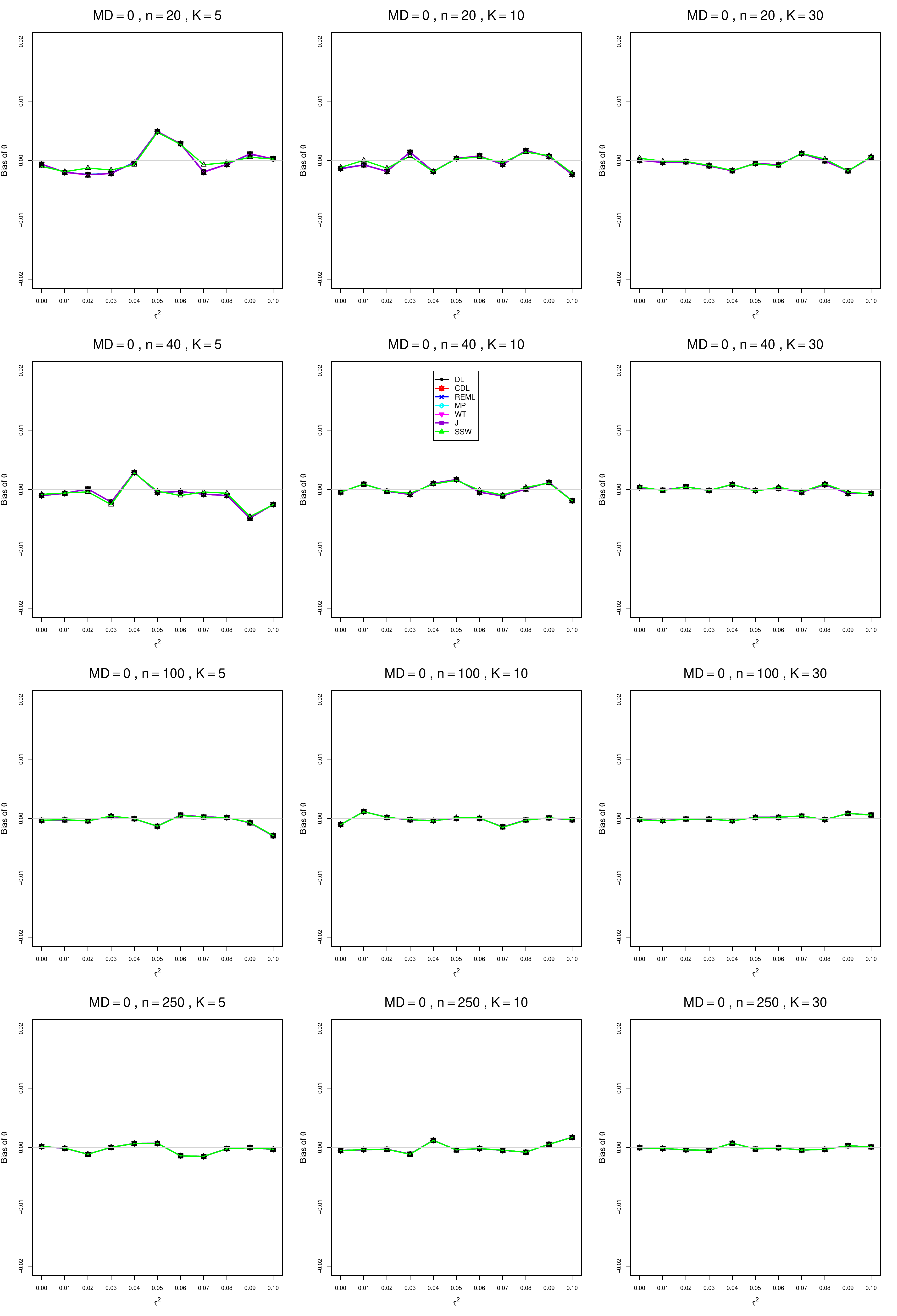}
	\caption{Bias of the estimation of  $\mu = 0$ for between-studies variance $\tau^2 = 0.0(0.01)0.1$, $q=0.5$, $\sigma_C^2=1$, $\sigma_T^2=1$,  equal study sizes $n=20,\;40,\;100,\;250$.
		\label{BiasThetaMD0_S1_1_small_tau2}}
\end{figure}

\begin{figure}[t]
	\centering
	\includegraphics[scale=0.33]{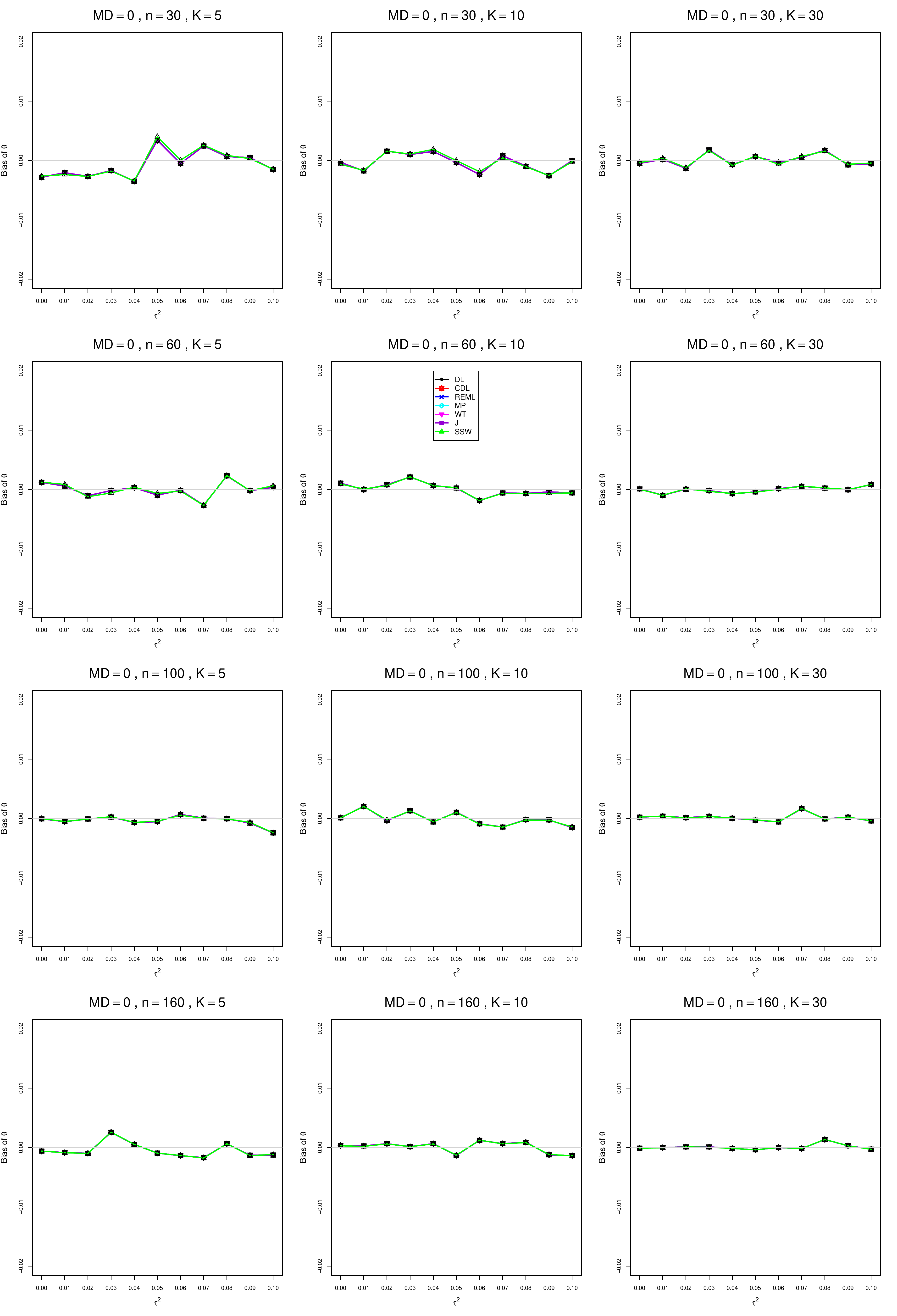}
	\caption{Bias of the estimation of  $\mu = 0$ for between-studies variance $\tau^2 = 0.0(0.01)0.1$, $q=0.5$, $\sigma_C^2=1$, $\sigma_T^2=1$, unequal studies of average size $\bar{n}=30,\;60,\;100,\;160$.
		\label{BiasThetaMD0_S1_1unequal_small_tau2}}
\end{figure}

\begin{figure}[t]\centering
	\includegraphics[scale=0.35]{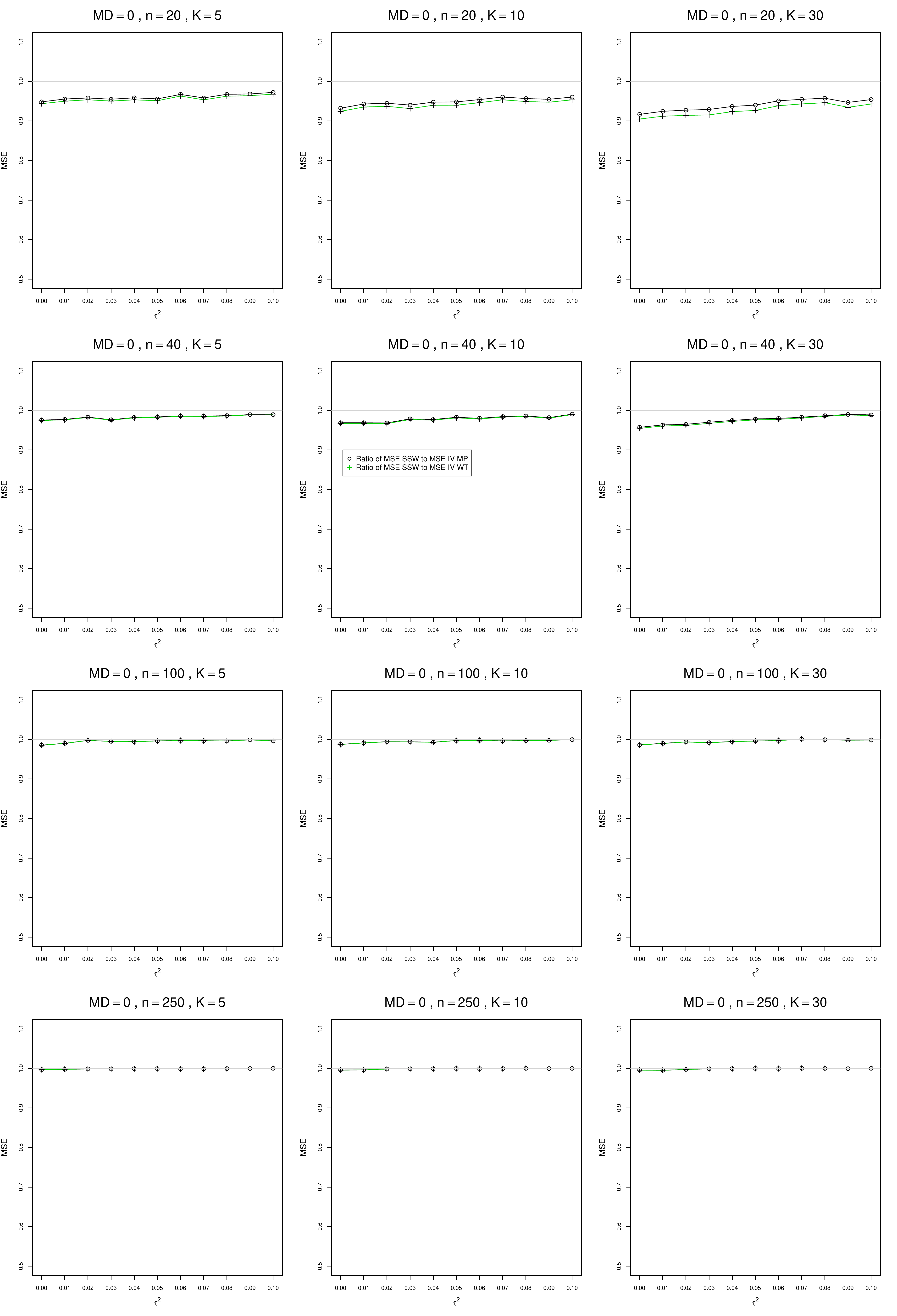}
	\caption{Ratio of mean squared errors of the fixed-weights to mean squared errors of inverse-variance estimator for $\mu=0$, $q=0.5$, $\sigma_C^2=1$, $\sigma_T^2=1$, $n=20,\;40,\;100,\;250$.
		\label{RatioOfMSEwithMD0fromMPandCMPSigma2T1andSigma2C1_small_tau2}}
\end{figure}


\begin{figure}[t]
	\centering
	\includegraphics[scale=0.33]{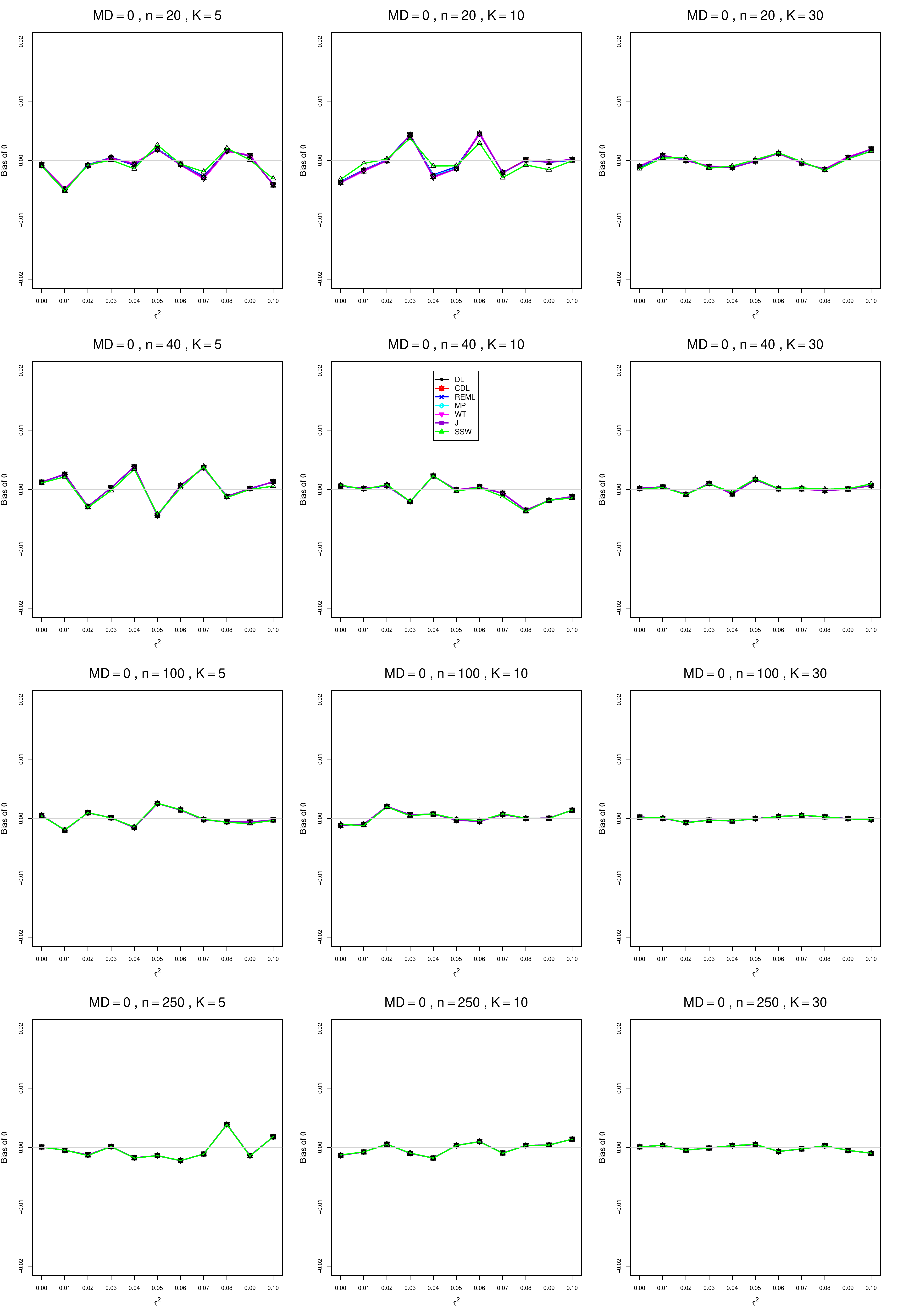}
	\caption{Bias of the estimation of  $\mu = 0$ for between-studies variance $\tau^2 = 0.0(0.01)0.1$, $q=0.75$, $\sigma_C^2=1$, $\sigma_T^2=1$,  equal study sizes $n=20,\;40,\;100,\;250$.
		\label{BiasThetaMD0_S1_1q075_small_tau2}}
\end{figure}

\begin{figure}[t]
	\centering
	\includegraphics[scale=0.33]{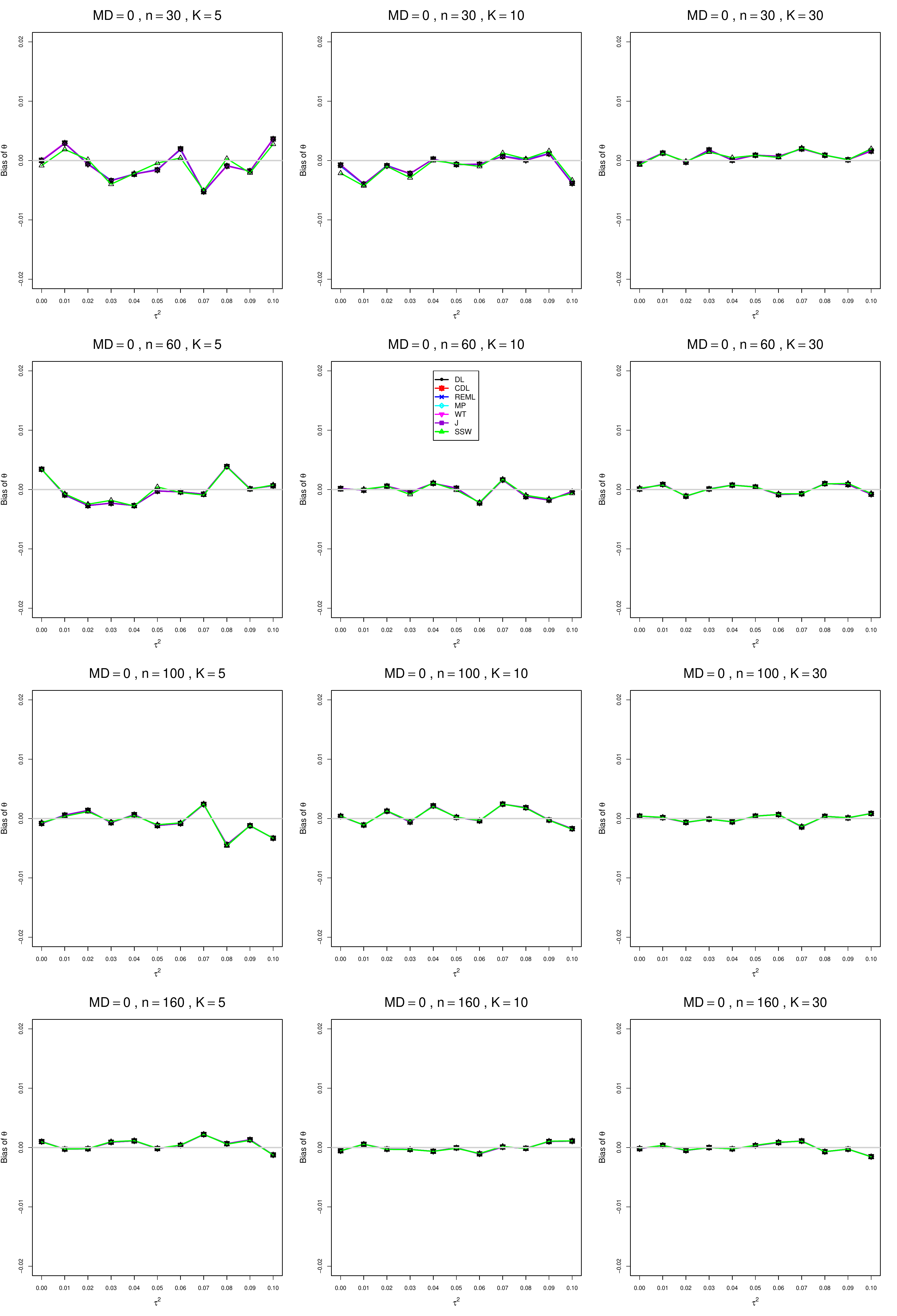}
	\caption{Bias of the estimation of  $\mu = 0$ for between-studies variance $\tau^2 = 0.0(0.01)0.1$, $q=0.75$, $\sigma_C^2=1$, $\sigma_T^2=1$, unequal studies of average size $\bar{n}=30,\;60,\;100,\;160$.
		\label{BiasThetaMD0_S1_1unequalq075_small_tau2}}
\end{figure}

\begin{figure}[t]\centering
	\includegraphics[scale=0.35]{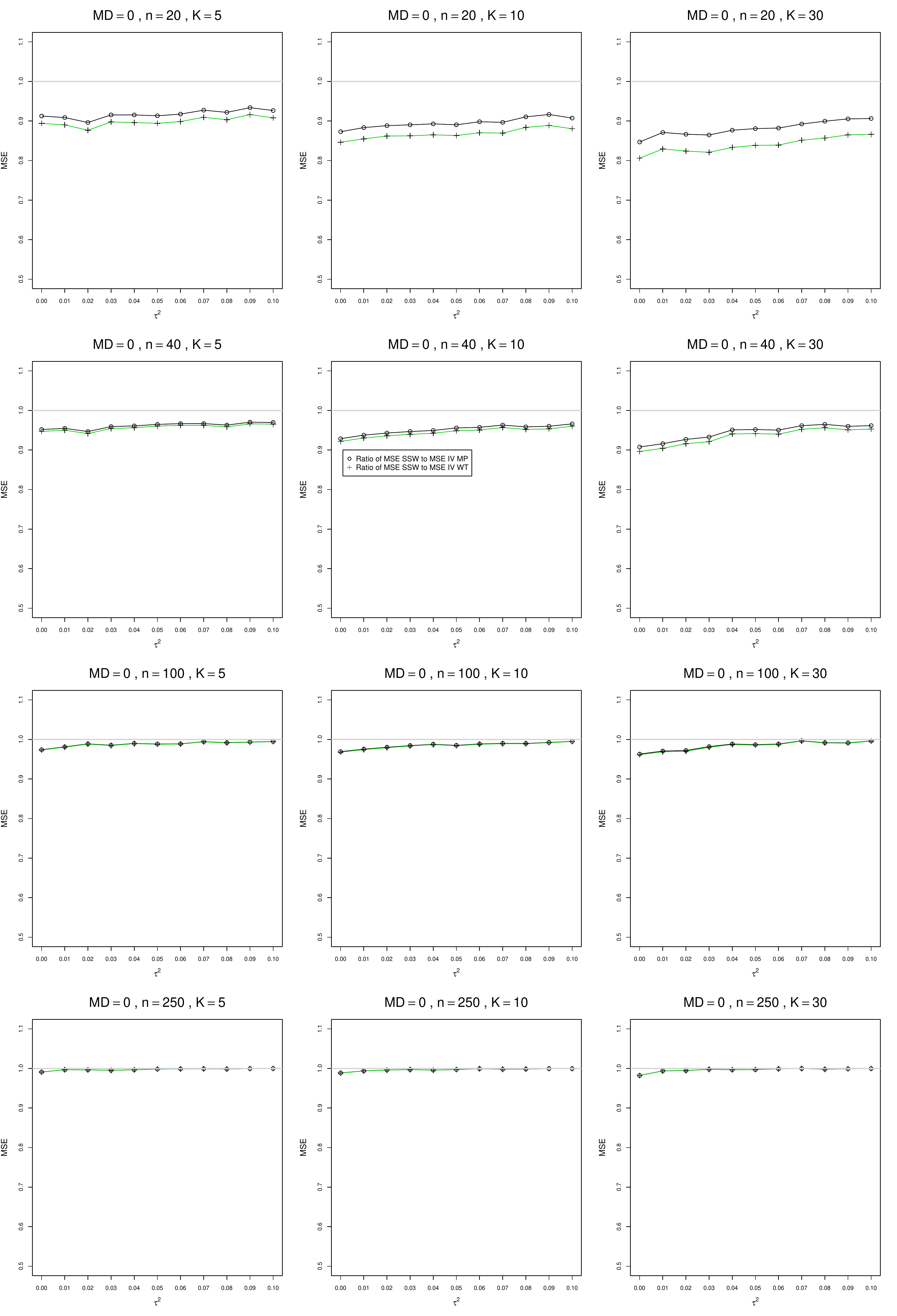}
	\caption{Ratio of mean squared errors of the fixed-weights to mean squared errors of inverse-variance estimator for $\mu=0$, for $q=0.75$, $\sigma_C^2=1$, $\sigma_T^2=1$, $n=20,\;40,\;100,\;250$.
		\label{RatioOfMSEwithMD0q075fromMPandCMPSigma2T1andSigma2C1_small_tau2}}
\end{figure}


\begin{figure}[t]
	\centering
	\includegraphics[scale=0.33]{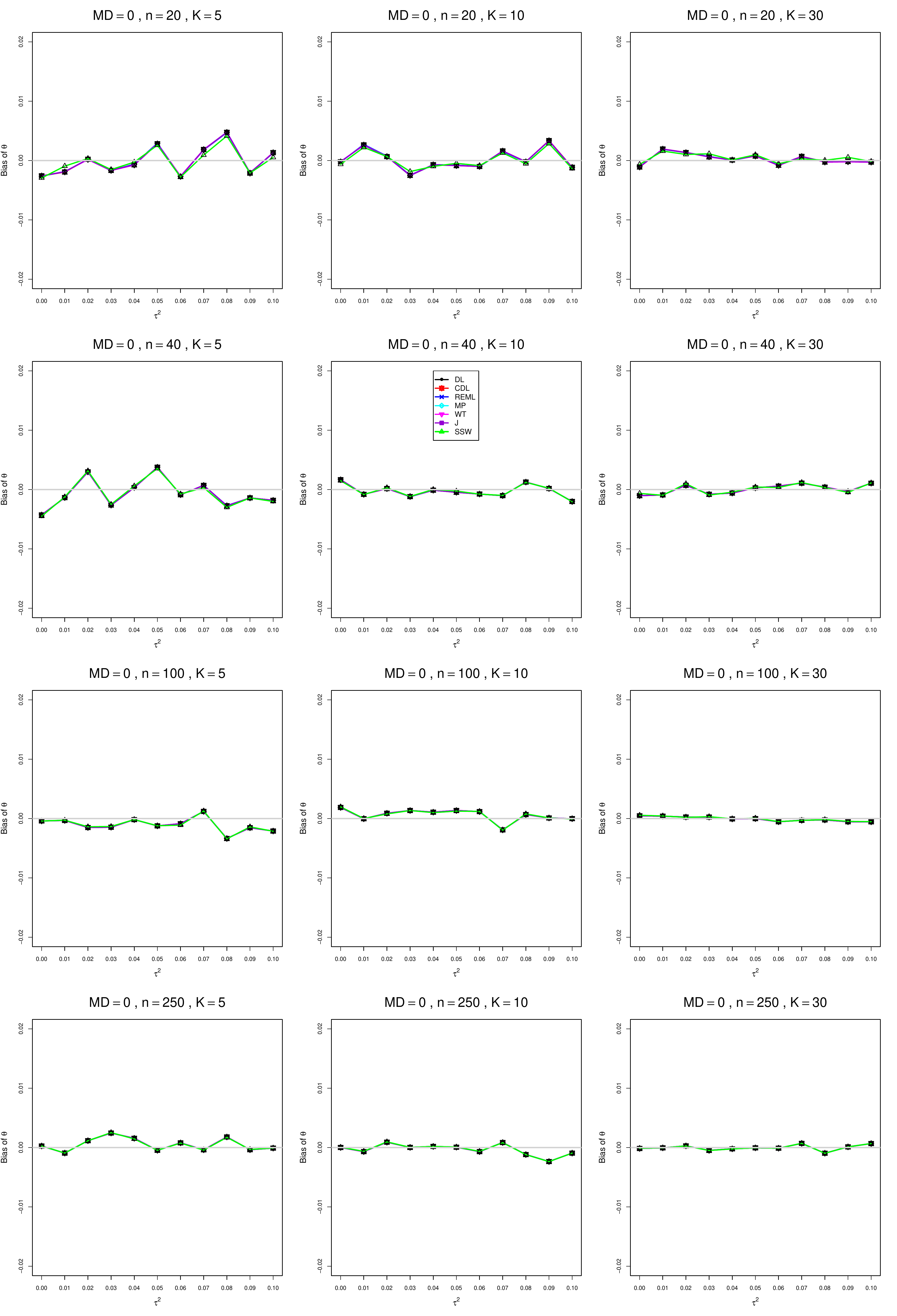}
	\caption{Bias of the estimation of  $\mu = 0$ for between-studies variance $\tau^2 = 0.0(0.01)0.1$, $q=0.5$, $\sigma_C^2=1$, $\sigma_T^2=2$,  equal study sizes $n=20,\;40,\;100,\;250$.
		\label{BiasThetaMD0_S1_2_small_tau2}}
\end{figure}

\begin{figure}[t]
	\centering
	\includegraphics[scale=0.33]{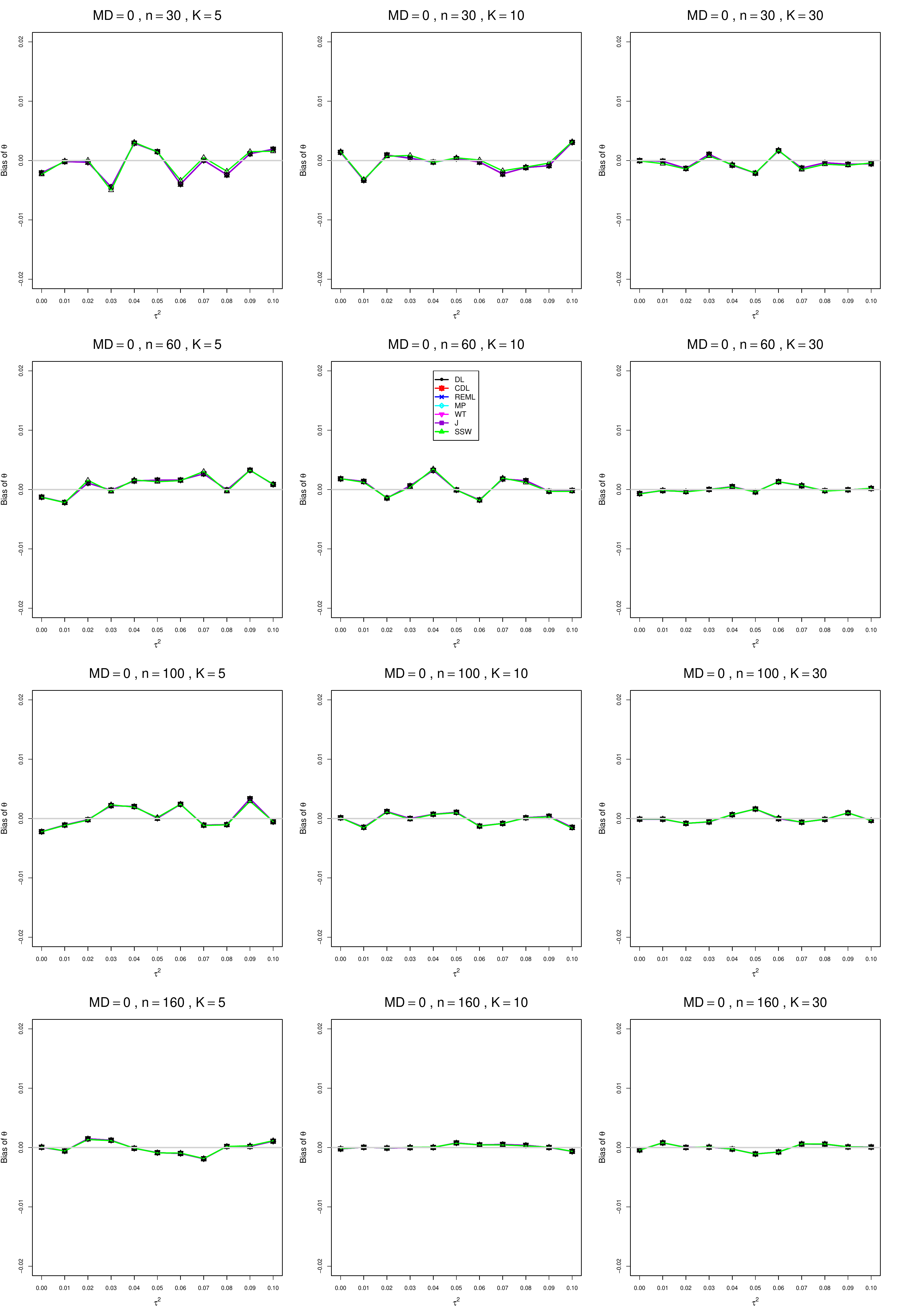}
	\caption{Bias of the estimation of  $\mu = 0$ for between-studies variance $\tau^2 = 0.0(0.01)0.1$, $q=0.5$, $\sigma_C^2=1$, $\sigma_T^2=2$, unequal studies of average size $\bar{n}=30,\;60,\;100,\;160$.
		\label{BiasThetaMD0_S1_2unequal_small_tau2}}
\end{figure}

\begin{figure}[t]\centering
	\includegraphics[scale=0.35]{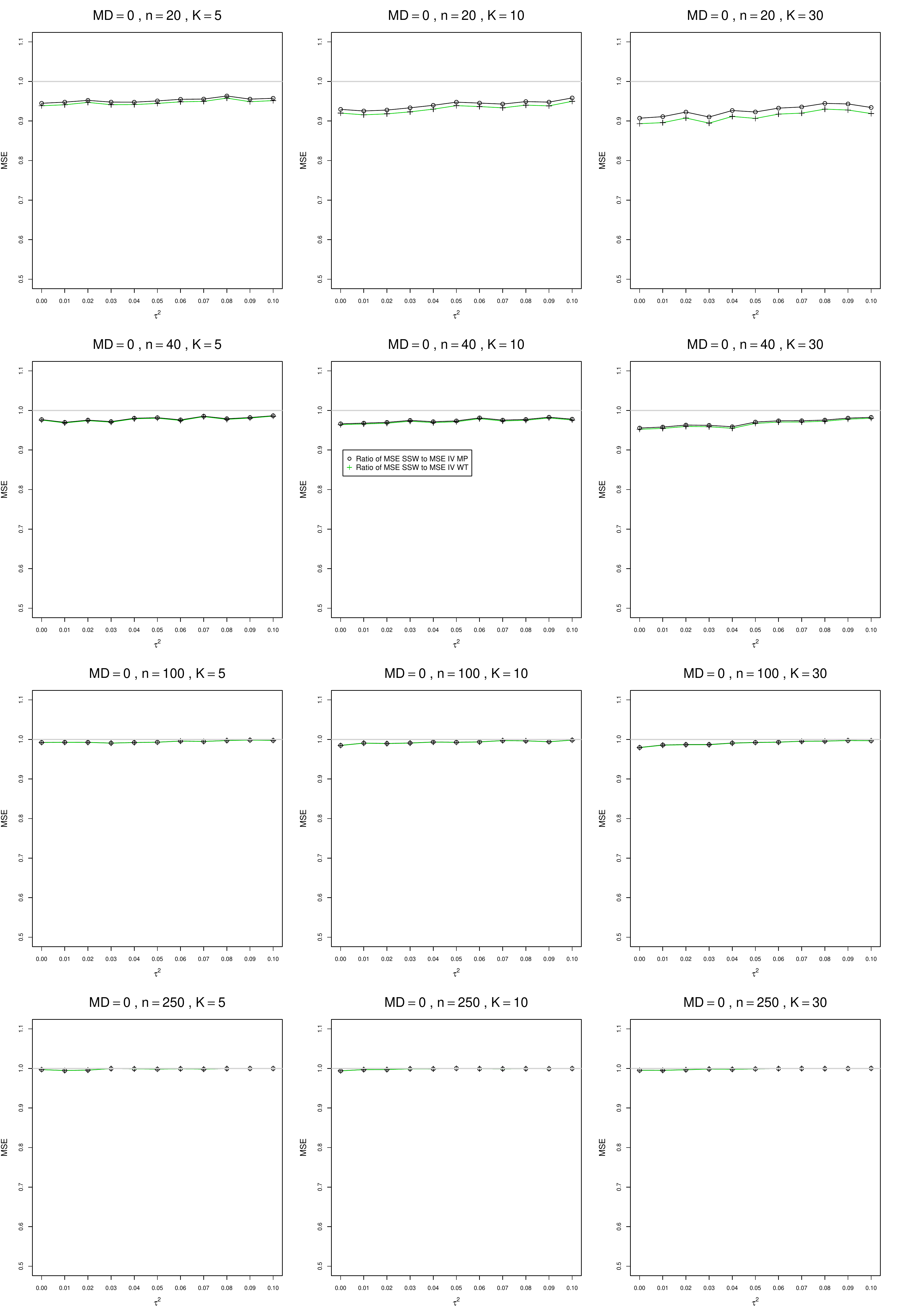}
	\caption{Ratio of mean squared errors of the fixed-weights to mean squared errors of inverse-variance estimator for $\mu=0$, $q=0.5$, $\sigma_C^2=1$, $\sigma_T^2=2$, $n=20,\;40,\;100,\;250$.
		\label{RatioOfMSEwithMD0fromMPandCMPSigma2T2andSigma2C1_small_tau2}}
\end{figure}

\clearpage

\begin{figure}[t]
	\centering
	\includegraphics[scale=0.33]{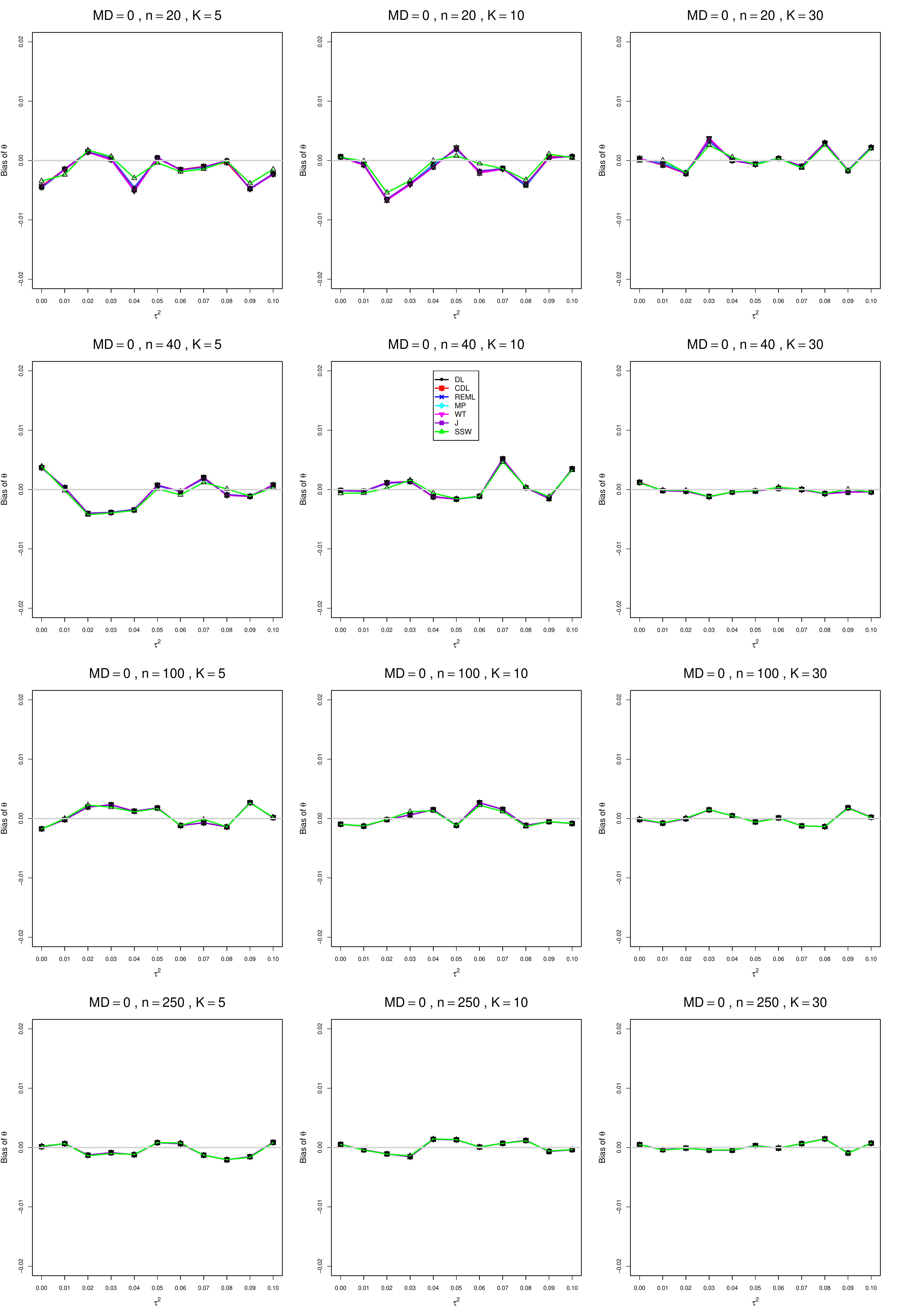}
	\caption{Bias of the estimation of  $\mu = 0$ for between-studies variance $\tau^2 = 0.0(0.01)0.1$, $q=0.75$, $\sigma_C^2=1$, $\sigma_T^2=2$,  equal study sizes $n=20,\;40,\;100,\;250$.
		\label{BiasThetaMD0_S1_2q075_small_tau2}}
\end{figure}

\begin{figure}[t]
	\centering
	\includegraphics[scale=0.33]{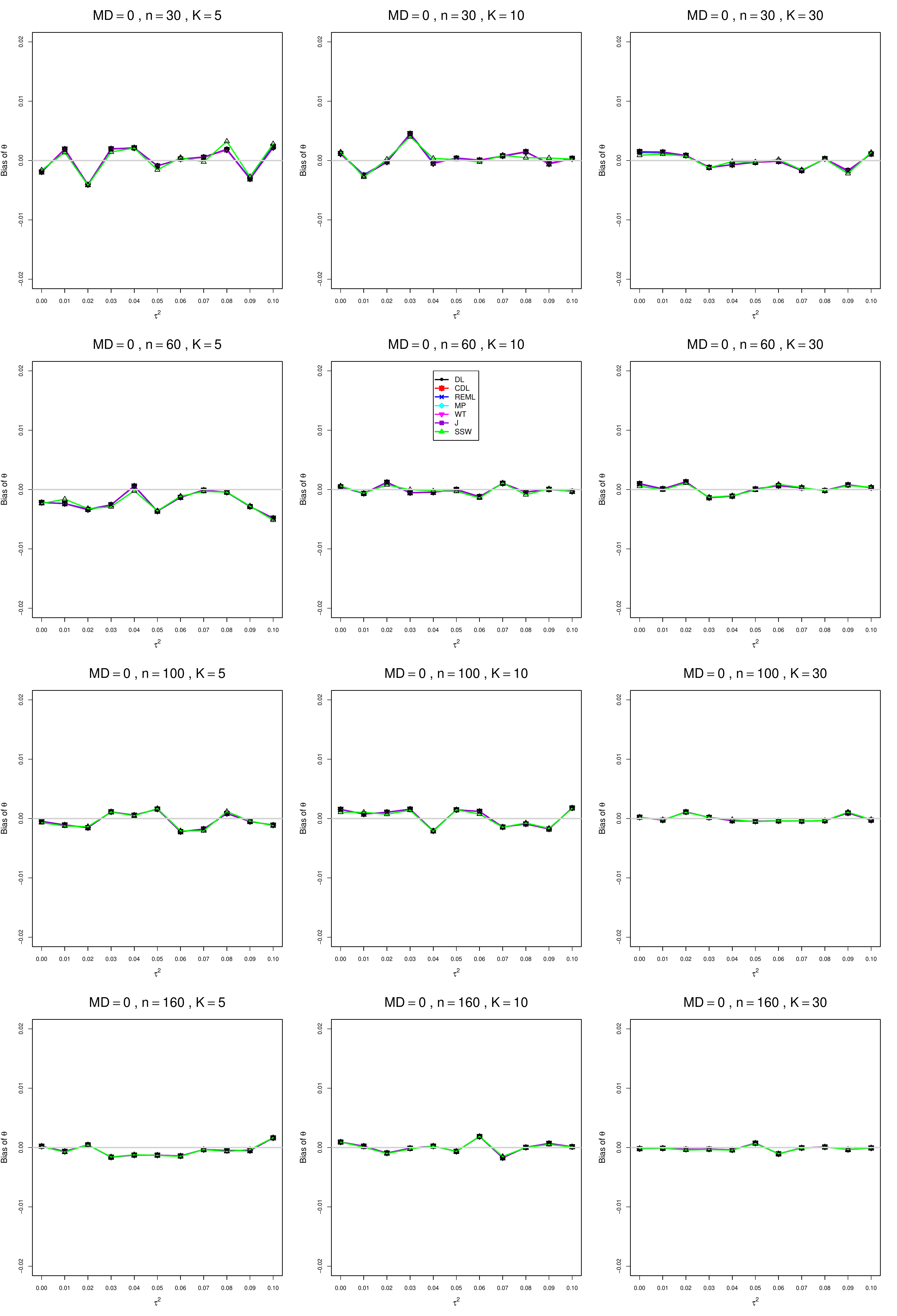}
	\caption{Bias of the estimation of  $\mu = 0$ for between-studies variance $\tau^2 = 0.0(0.01)0.1$, $q=0.75$, $\sigma_C^2=1$, $\sigma_T^2=2$, unequal studies of average size $\bar{n}=30,\;60,\;100,\;160$.
		\label{BiasThetaMD0_S1_2unequalq075_small_tau2}}
\end{figure}

\begin{figure}[t]\centering
	\includegraphics[scale=0.35]{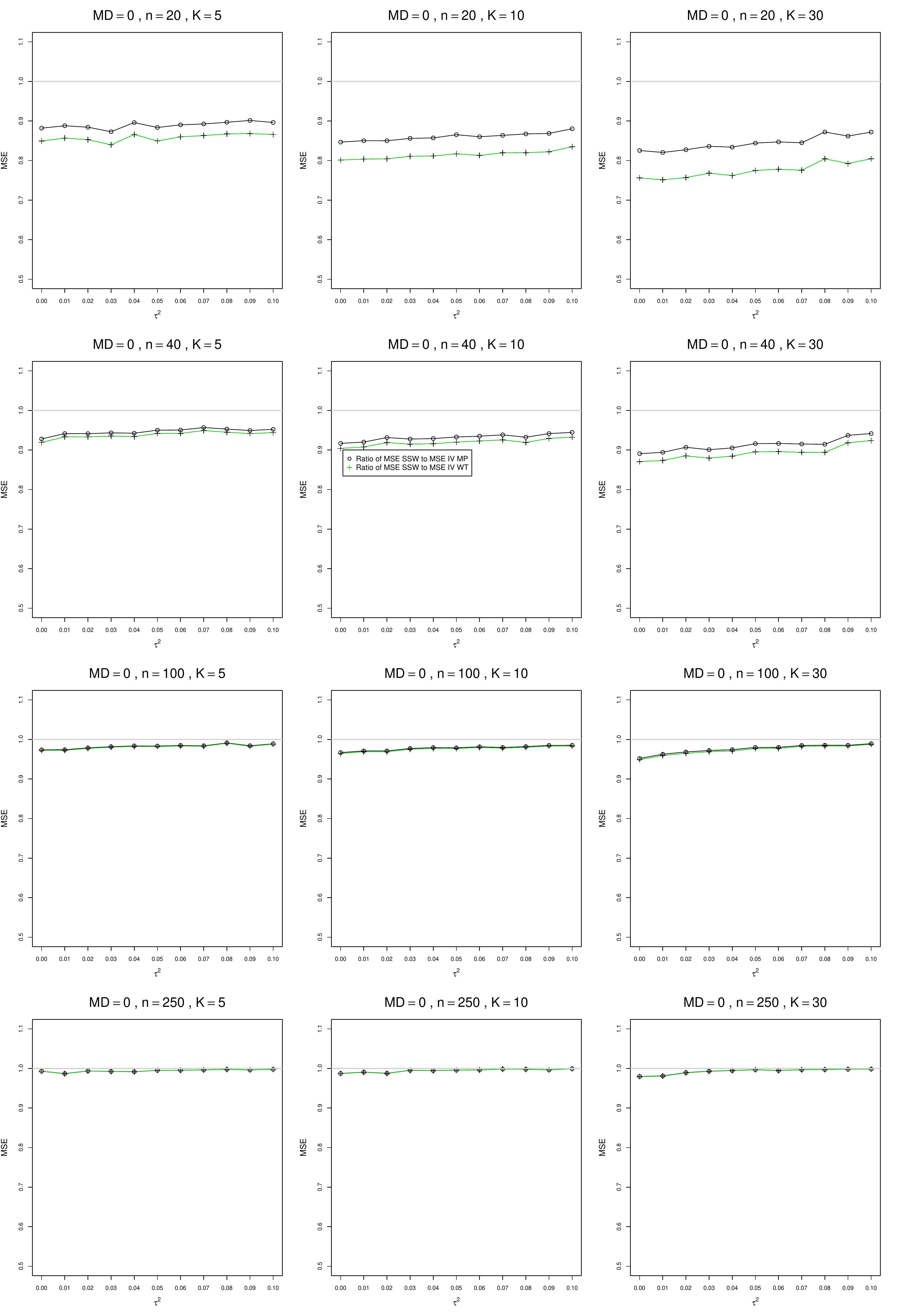}
	\caption{Ratio of mean squared errors of the fixed-weights to mean squared errors of inverse-variance estimator for $\mu=0$, for $q=0.75$,$\sigma_C^2=1$, $\sigma_T^2=2$, $n=20,\;40,\;100,\;250$. 
		\label{RatioOfMSEwithMD0q075fromMPandCMPSigma2T2andSigma2C1_small_tau2}}
\end{figure}

\clearpage
\setcounter{section}{0}
\renewcommand{\thefigure}{B4.\arabic{figure}}
\setcounter{figure}{0}
\setcounter{section}{0}
\section*{B4. Coverage of $\hat{\mu}$ for $\tau^2 = 0.0(0.01)0.1$, $\sigma_{C}^2=1$, $\sigma_{T}^2=1,\;2$.}
For coverage of $\mu$, each figure corresponds to a value of $\mu (= 0, 0.2, 0.5, 1, 2)$, a value of $q (= .5, .75)$, a value of $\tau^2 = 0.0(0.01)0.1$, a value of $\sigma_{C}^2=1$, a value of $\sigma_{T}^2=1,\;2$ , and a set of values of $n$ (= 20, 40, 100, 250) or $\bar{n}$ (= 30, 60, 100, 160).\\
Each figure contains a panel (with $\tau^2$ on the horizontal axis) for each combination of n (or $\bar{n}$) and $K (=5, 10, 30)$.\\
The interval estimators of $\mu$ are the companions to the inverse-variance-weighted point estimators
\begin{itemize}
	\item DL (DerSimonian-Laird)
	\item REML (restricted maximum likelihood)
	\item MP (Mandel-Paule)
	\item WT (Corrected Mandel-Paule moment estimator based on Welch-type approximation for Q distribution)
	\item J (Jackson)
	\item CDL (Corrected DerSimonian-Laird)
\end{itemize}
and
\begin{itemize}
	\item HKSJ (Hartung-Knapp-Sidik-Jonkman)
	\item HKSJ WT (HKSJ with WT estimator of $\tau^2$)
	\item SSW (SSW as center and half-width equal to critical value from $t_{K-1}$
\end{itemize}
times estimated standard deviation of SSW with $\hat{\tau}^2$ = $\hat{\tau}^2_{WT}$

\begin{figure}[t]
	\centering
	\includegraphics[scale=0.33]{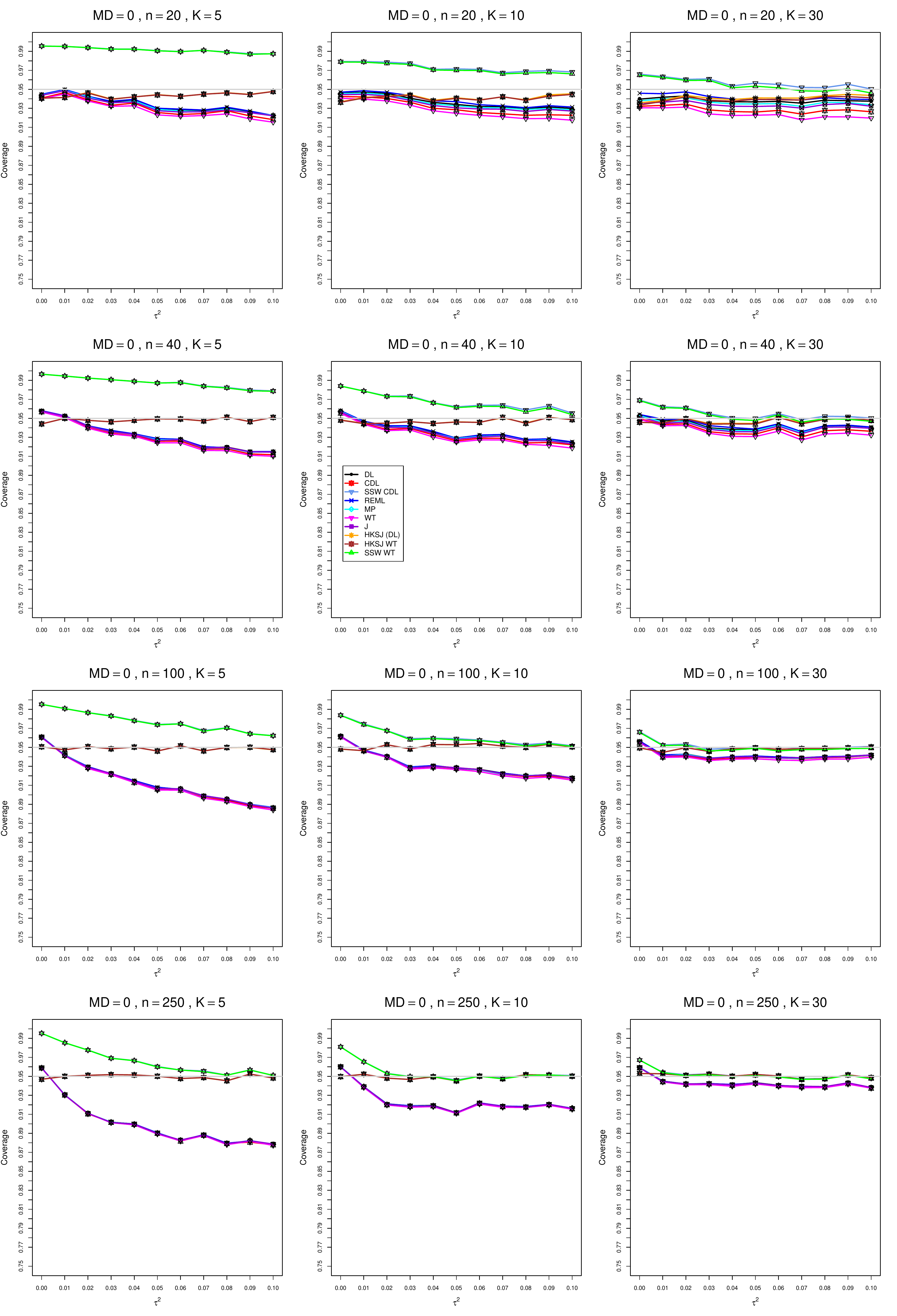}
	\caption{Coverage of 95\% confidence intervals for the $\mu = 0$ for the between-studies variance $\tau^2 = 0.0(0.01)0.1$ for, $q=0.5$, $\sigma_C^2=1$, $\sigma_T^2=1$,  equal study sizes $n=20,\;40,\;100,\;250$.
		\label{CovThetaMD0_S1_1_small_tau2}}
\end{figure}

\begin{figure}[t]
	\centering
	\includegraphics[scale=0.33]{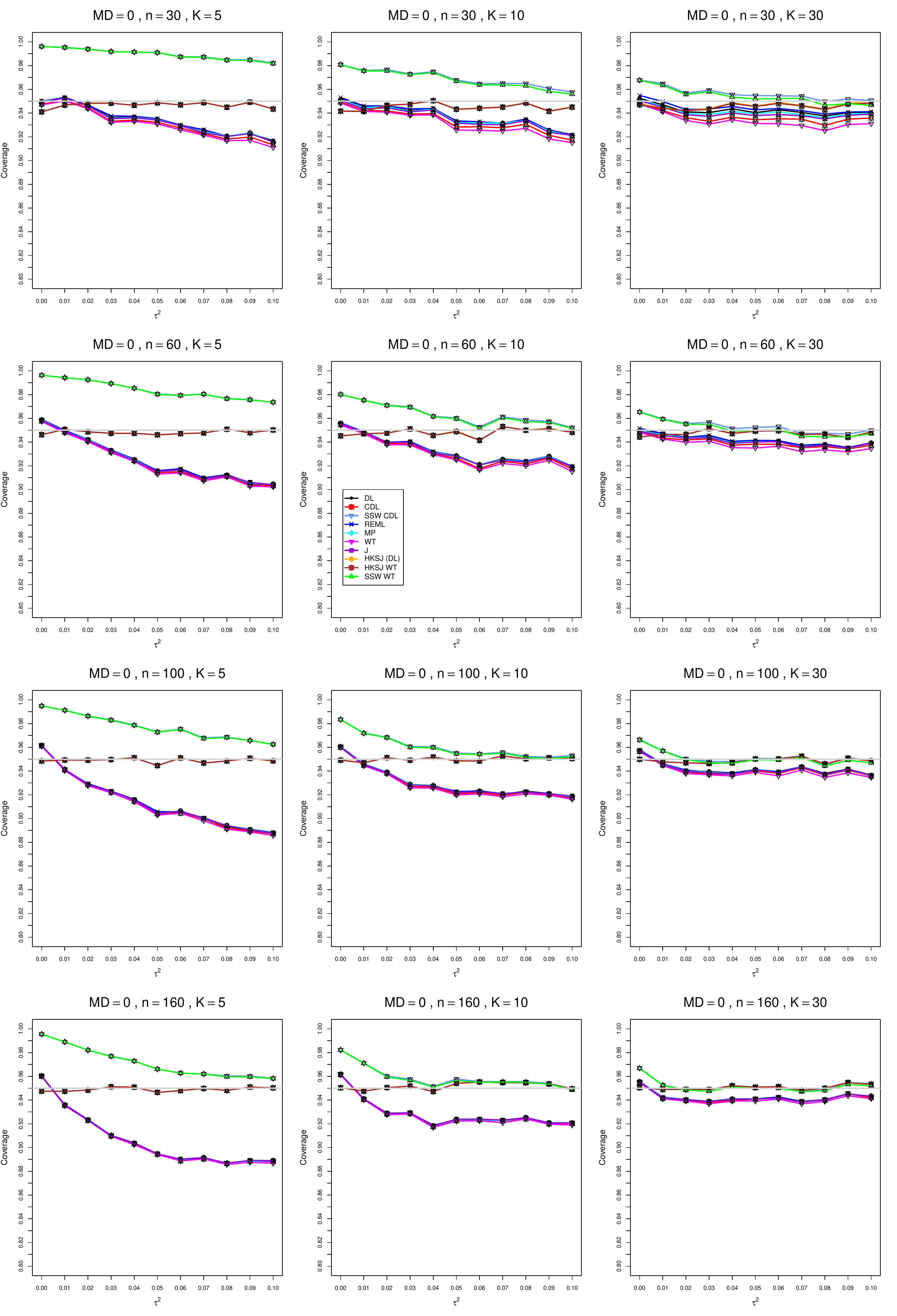}
	\caption{Coverage of 95\% confidence intervals for the $\mu = 0$ for the between-studies variance $\tau^2 = 0.0(0.01)0.1$, $q=0.5$, $\sigma_C^2=1$, $\sigma_T^2=1$, unequal studies of average size $\bar{n}=30,\;60,\;100,\;160$.
		\label{CovThetaMD0_S1_1unequal_small_tau2}}
\end{figure}


\begin{figure}[t]
	\centering
	\includegraphics[scale=0.33]{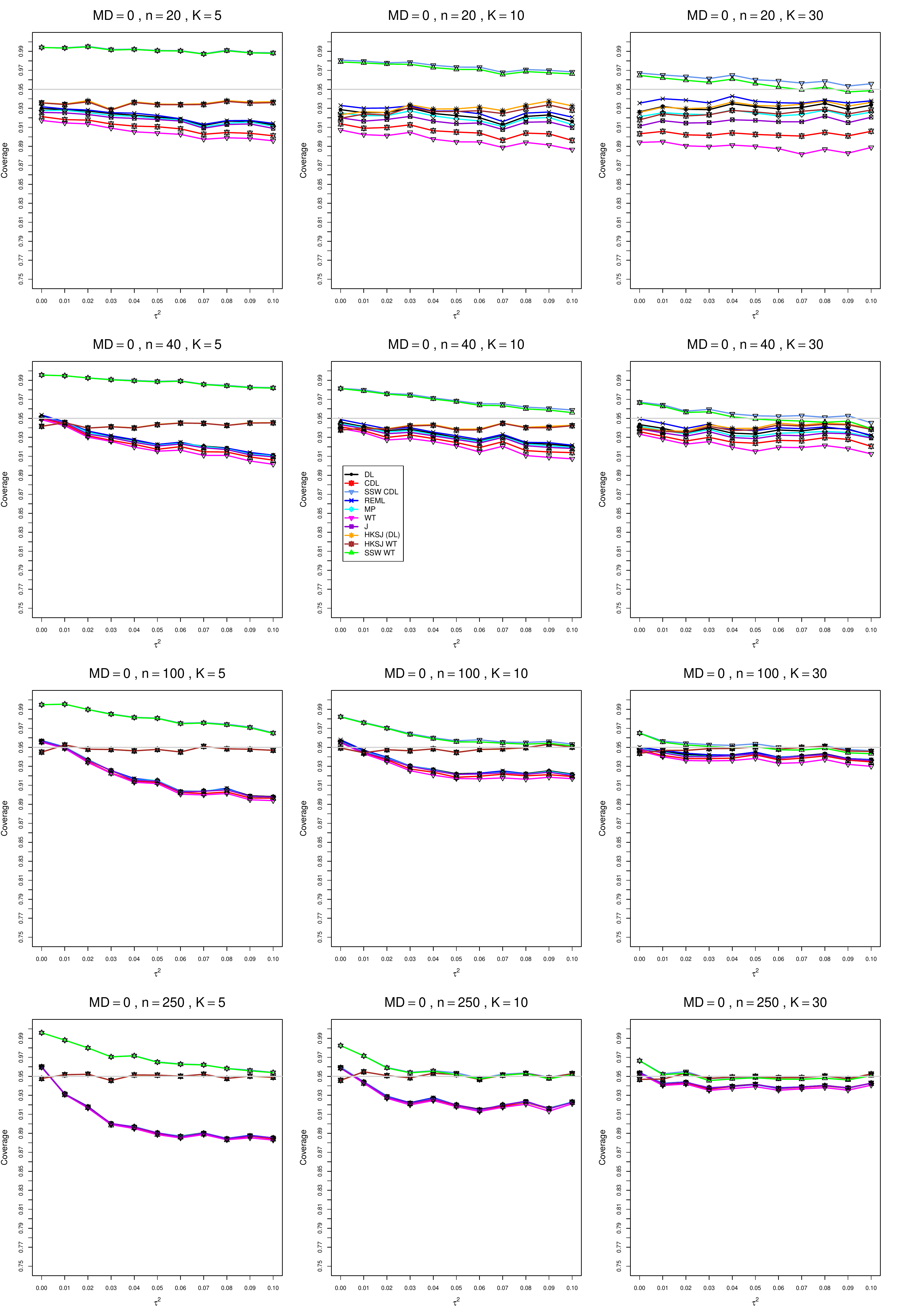}
	\caption{Coverage of 95\% confidence intervals for the $\mu = 0$ for the  between-studies variance $\tau^2 = 0.0(0.01)0.1$, $q=0.75$, $\sigma_C^2=1$, $\sigma_T^2=1$,  equal study sizes $n=20,\;40,\;100,\;250$.
		\label{CovThetaMD0_S1_1q075_small_tau2}}
\end{figure}

\begin{figure}[t]
	\centering
	\includegraphics[scale=0.33]{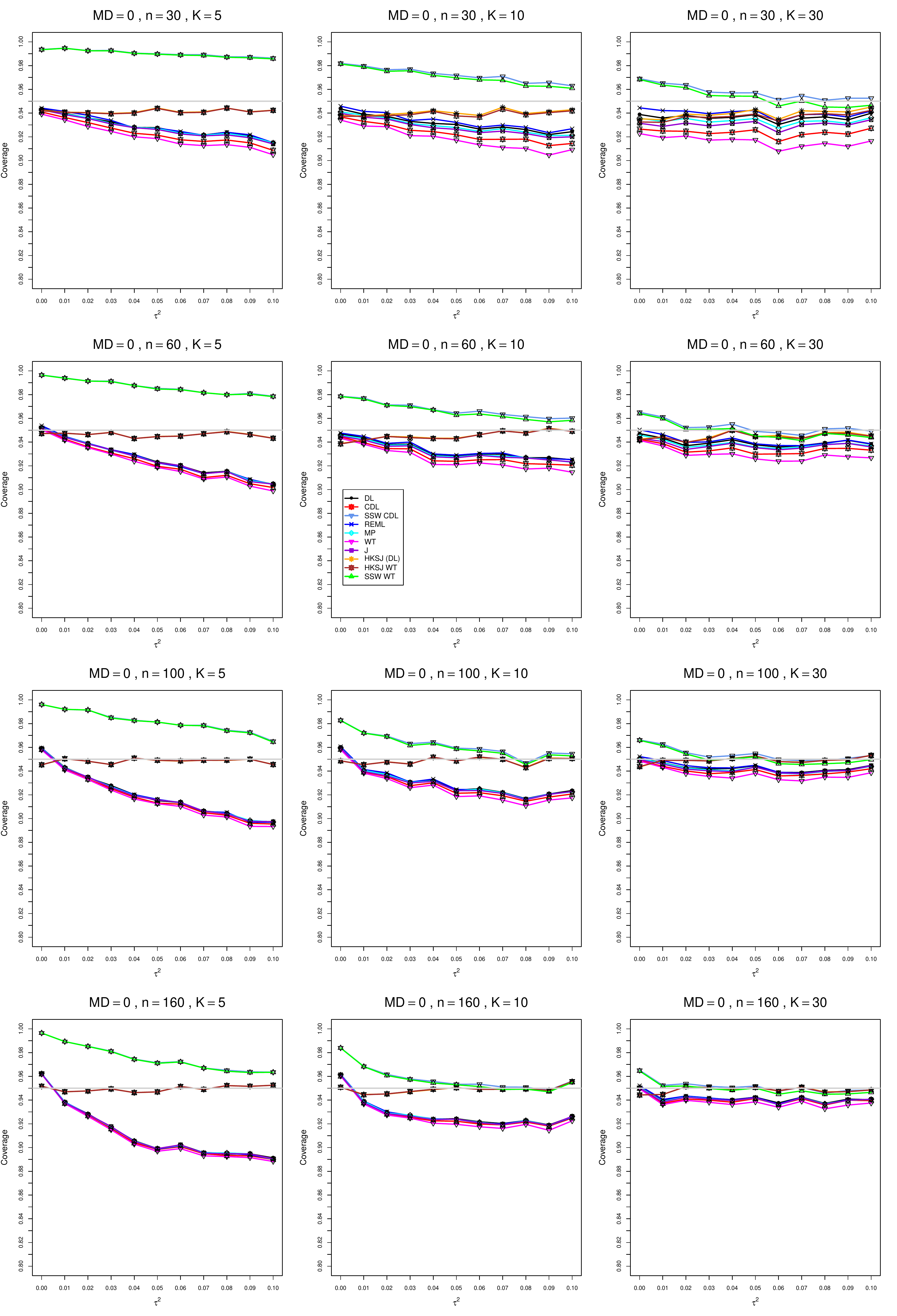}
	\caption{Coverage of 95\% confidence intervals for the $\mu = 0$ for the  between-studies variance $\tau^2 = 0.0(0.01)0.1$, $q=0.75$, $\sigma_C^2=1$, $\sigma_T^2=1$, unequal studies of average size $\bar{n}=30,\;60,\;100,\;160$.
		\label{BiasThetaMD0_S1_1unequalq075_small_tau2_unequal_sample_size}}
\end{figure}


\begin{figure}[t]
	\centering
	\includegraphics[scale=0.33]{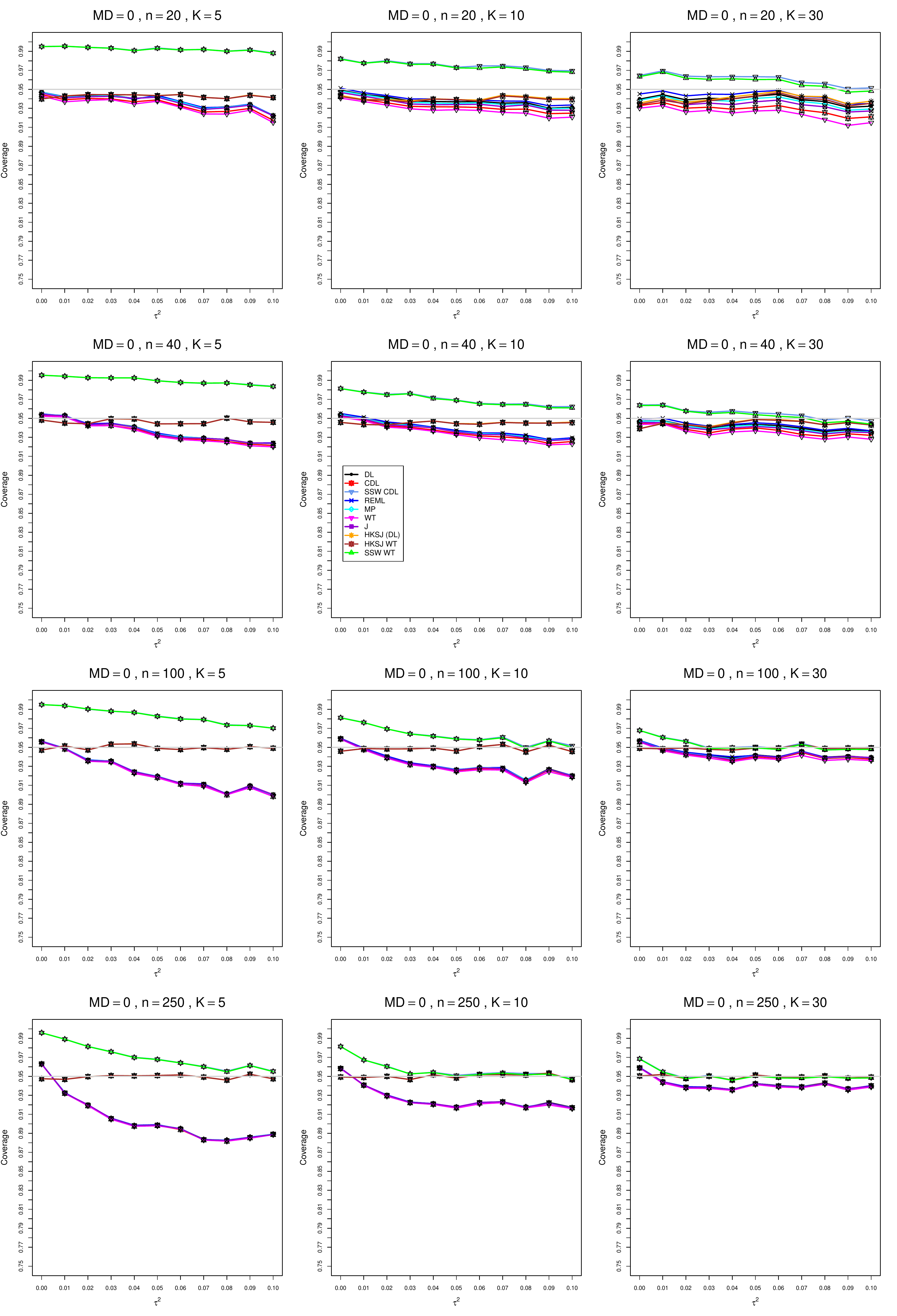}
	\caption{Coverage of 95\% confidence intervals for the $\mu = 0$ for the between-studies variance $\tau^2 = 0.0(0.01)0.1$ for, $q=0.5$, $\sigma_C^2=1$, $\sigma_T^2=2$,  equal study sizes $n=20,\;40,\;100,\;250$.
		\label{CovThetaMD0_S1_2_small_tau2}}
\end{figure}

\begin{figure}[t]
	\centering
	\includegraphics[scale=0.33]{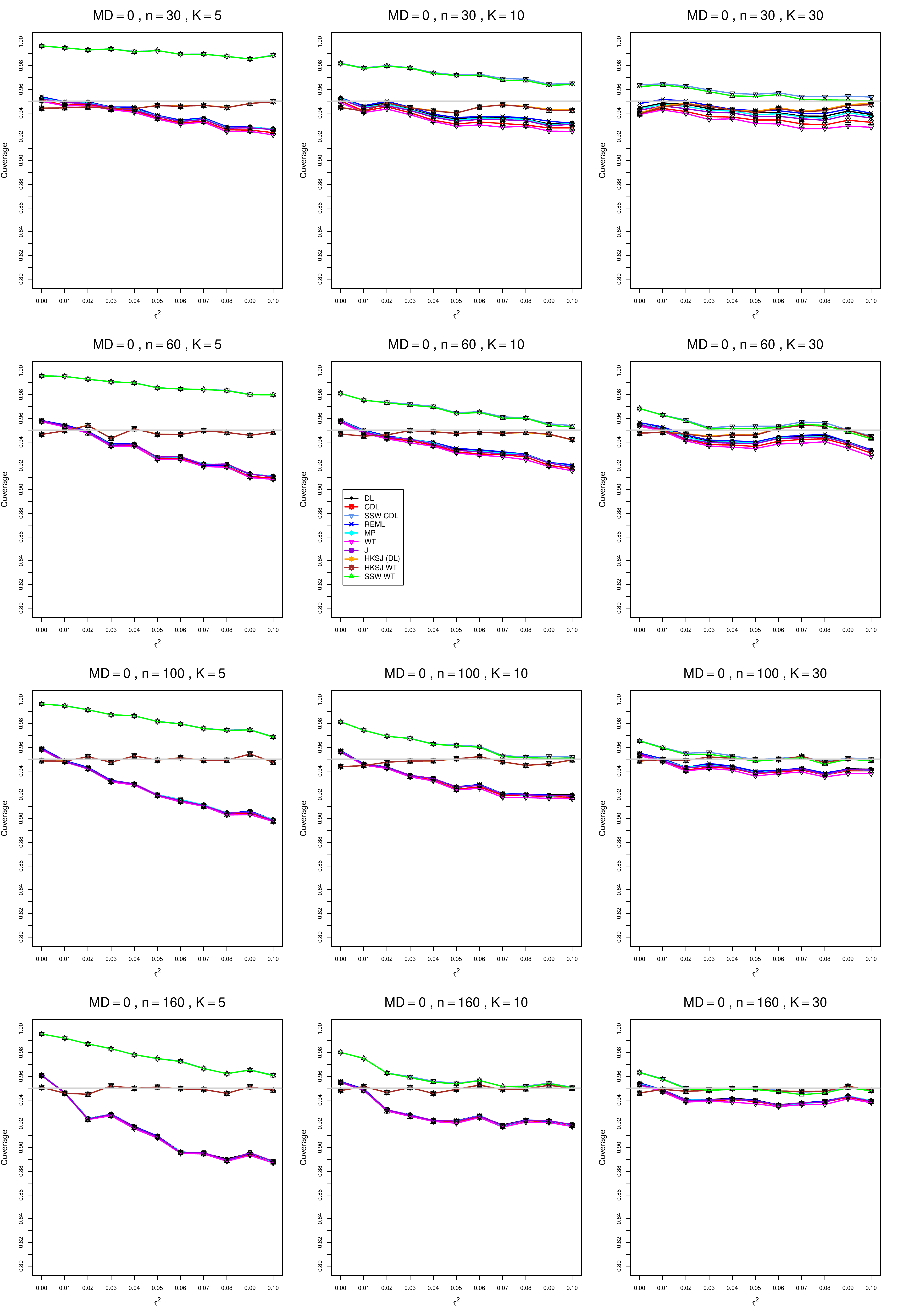}
	\caption{Coverage of 95\% confidence intervals for the $\mu = 0$ for the between-studies variance $\tau^2 = 0.0(0.01)0.1$, $q=0.5$, $\sigma_C^2=1$, $\sigma_T^2=2$, unequal studies of average size $\bar{n}=30,\;60,\;100,\;160$.
		\label{CovThetaMD0_S1_2unequal_small_tau2}}
\end{figure}


\begin{figure}[t]
	\centering
	\includegraphics[scale=0.33]{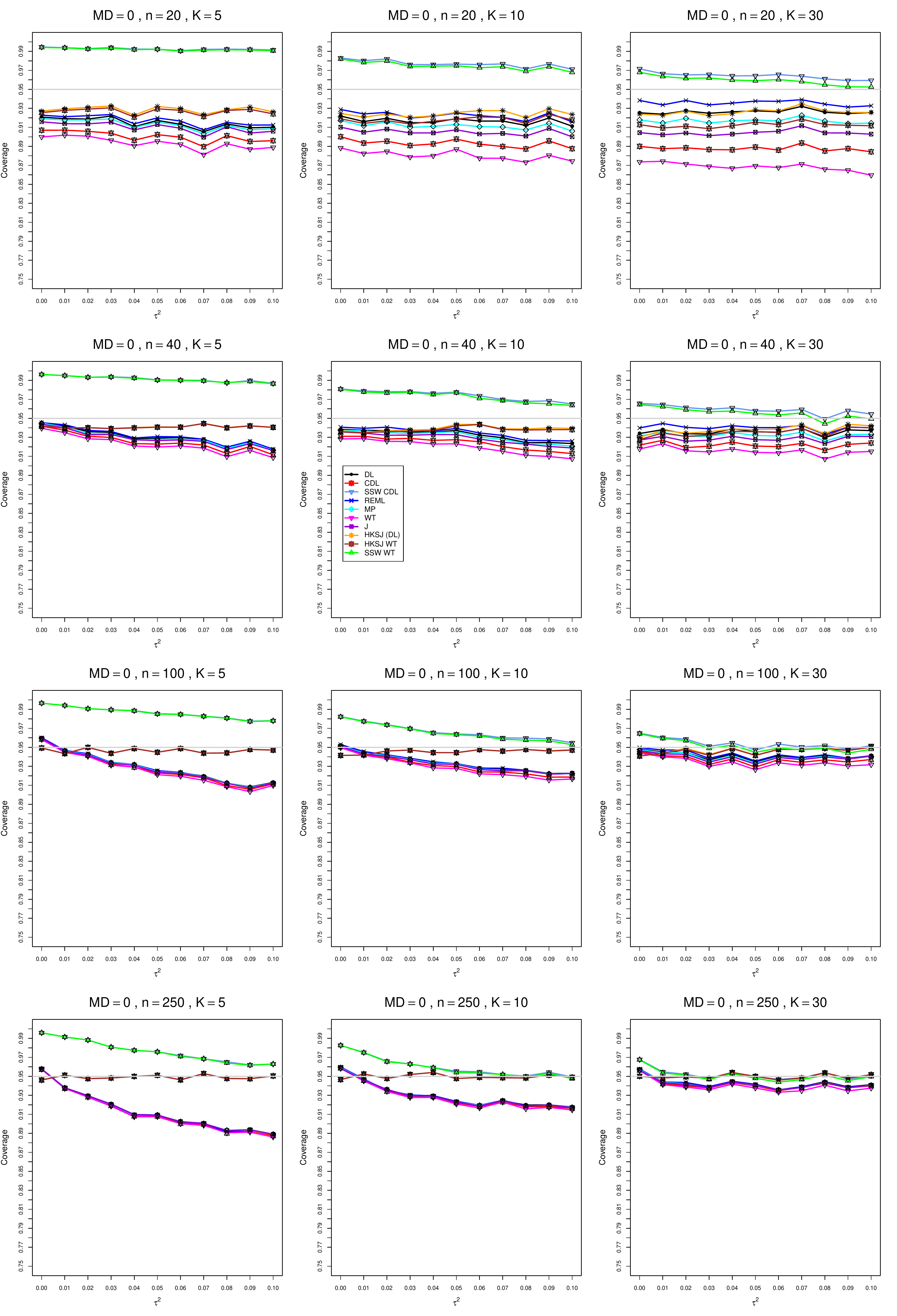}
	\caption{Coverage of 95\% confidence intervals for the $\mu = 0$ for the  between-studies variance $\tau^2 = 0.0(0.01)0.1$, $q=0.75$, $\sigma_C^2=1$, $\sigma_T^2=2$,  equal study sizes $n=20,\;40,\;100,\;250$.
		\label{CovThetaMD0_S1_2q075_small_tau2}}
\end{figure}

\begin{figure}[t]
	\centering
	\includegraphics[scale=0.33]{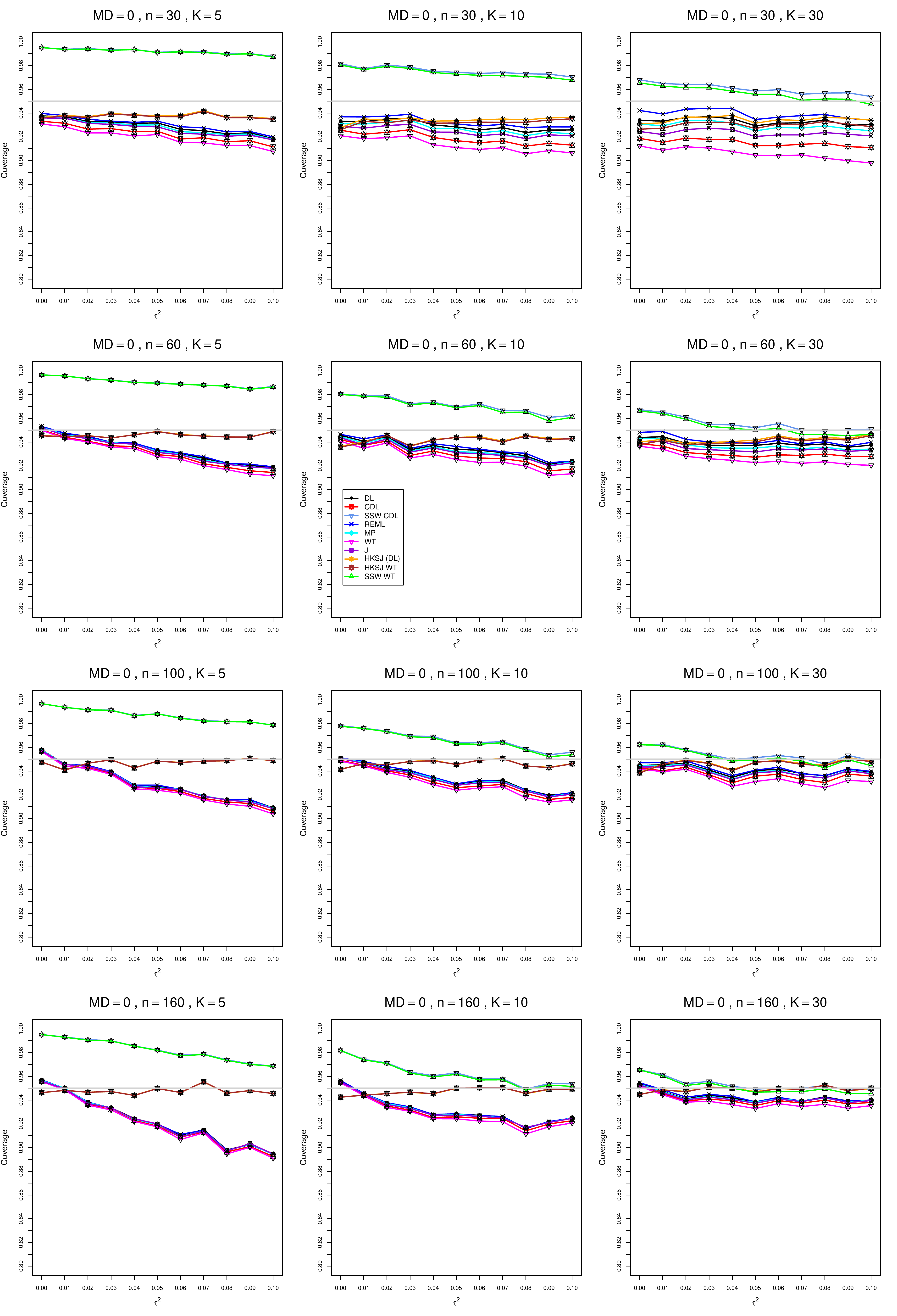}
	\caption{Coverage of 95\% confidence intervals for the $\mu = 0$ for the  between-studies variance $\tau^2 = 0.0(0.01)0.1$, $q=0.75$, $\sigma_C^2=1$, $\sigma_T^2=2$, unequal studies of average size $\bar{n}=30,\;60,\;100,\;160$.
		\label{CovThetaMD0_S1_2unequalq075_small_tau2}}
\end{figure}


\clearpage
\renewcommand{\thefigure}{B5.\arabic{figure}}
\setcounter{figure}{0}
\setcounter{section}{0}
\section*{B5. Bias and mean squared error of point estimators $\hat{\mu}$ for $\tau^2 = 0.0(0.1)1.0$, $\sigma_{C}^2=10$, $\sigma_{T}^2=10,\;20$.}
For bias of $\mu$, each figure corresponds to a value of $\mu (= 0, 0.2, 0.5, 1, 2)$, a value of $q (= .5, .75)$, a value of $\tau^2 = 0.0(0.01)0.1$, a value of $\sigma_{C}^2=10$, a value of $\sigma_{T}^2=10,\;20$ , and a set of values of $n$ (= 20, 40, 100, 250) or $\bar{n}$ (= 30, 60, 100, 160).\\
Figures for mean squared error (expressed as the ratio of the MSE of SSW to the MSEs of the inverse-variance-weighted estimators that use the MP or WT estimator of $\tau^2$) use the above values of $\mu$ and q but only n = 20, 40, 100, 250.\\
Each figure contains a panel (with $\tau^2$ on the horizontal axis) for each combination of n (or $\bar{n}$) and $K (=5, 10, 30)$.\\
The point estimators of $\mu$ are
\begin{itemize}
	\item DL (DerSimonian-Laird)
	\item REML (restricted maximum likelihood)
	\item MP (Mandel-Paule)
	\item WT (Corrected Mandel-Paule moment estimator based on Welch-type approximation for Q distribution)
	\item J (Jackson)
	\item CDL (Corrected DerSimonian-Laird)
	\item SSW (sample-size weighted)
\end{itemize}

\begin{figure}[t]
	\centering
	\includegraphics[scale=0.33]{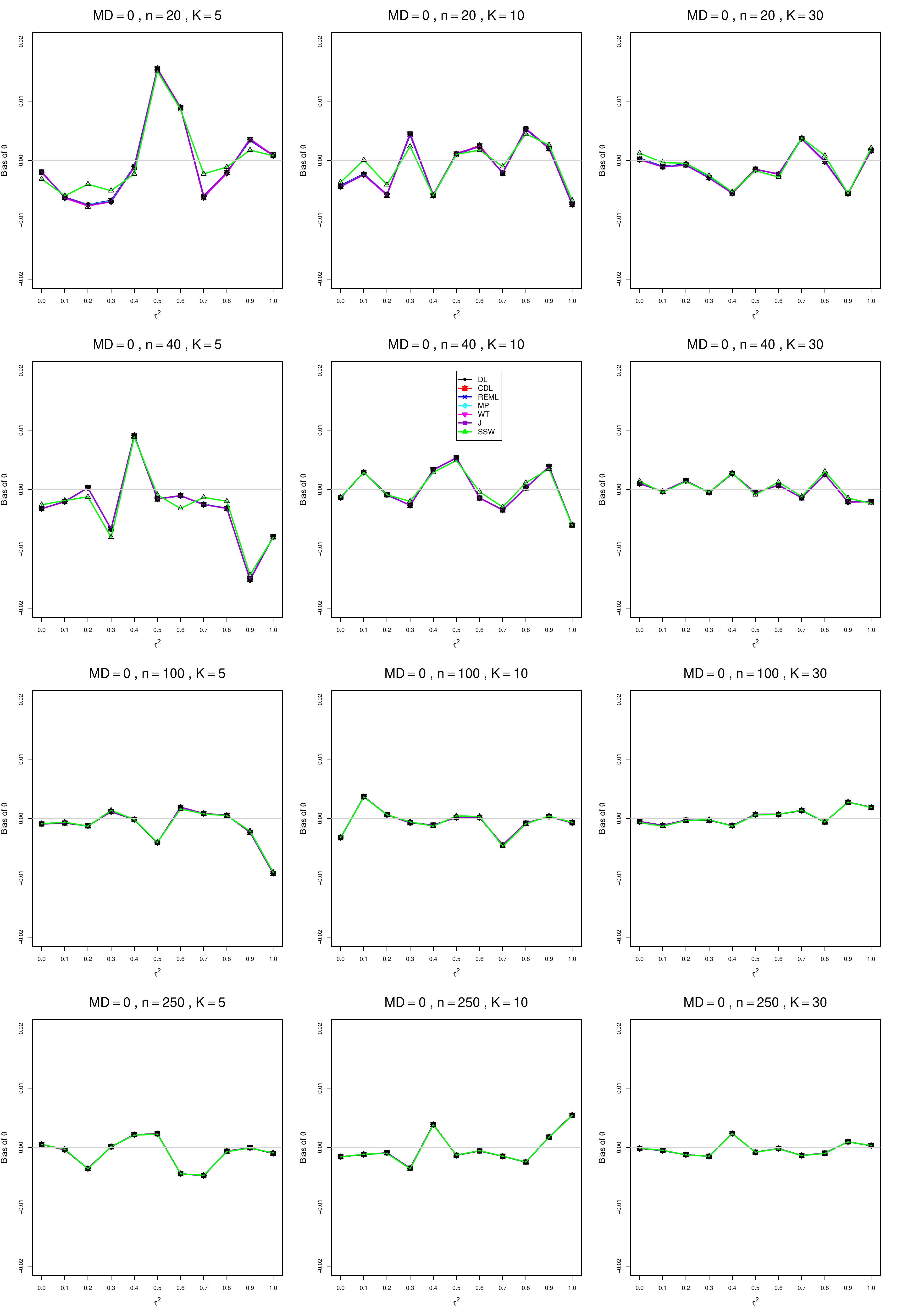}
	\caption{Bias of the estimation of  $\mu = 0$ for between-studies variance $\tau^2 = 0.0(0.1)1.0$, $q=0.5$, $\sigma_C^2=10$, $\sigma_T^2=10$,  equal study sizes $n=20,\;40,\;100,\;250$.
		\label{BiasThetaMD0_S10_10}}
\end{figure}

\begin{figure}[t]
	\centering
	\includegraphics[scale=0.33]{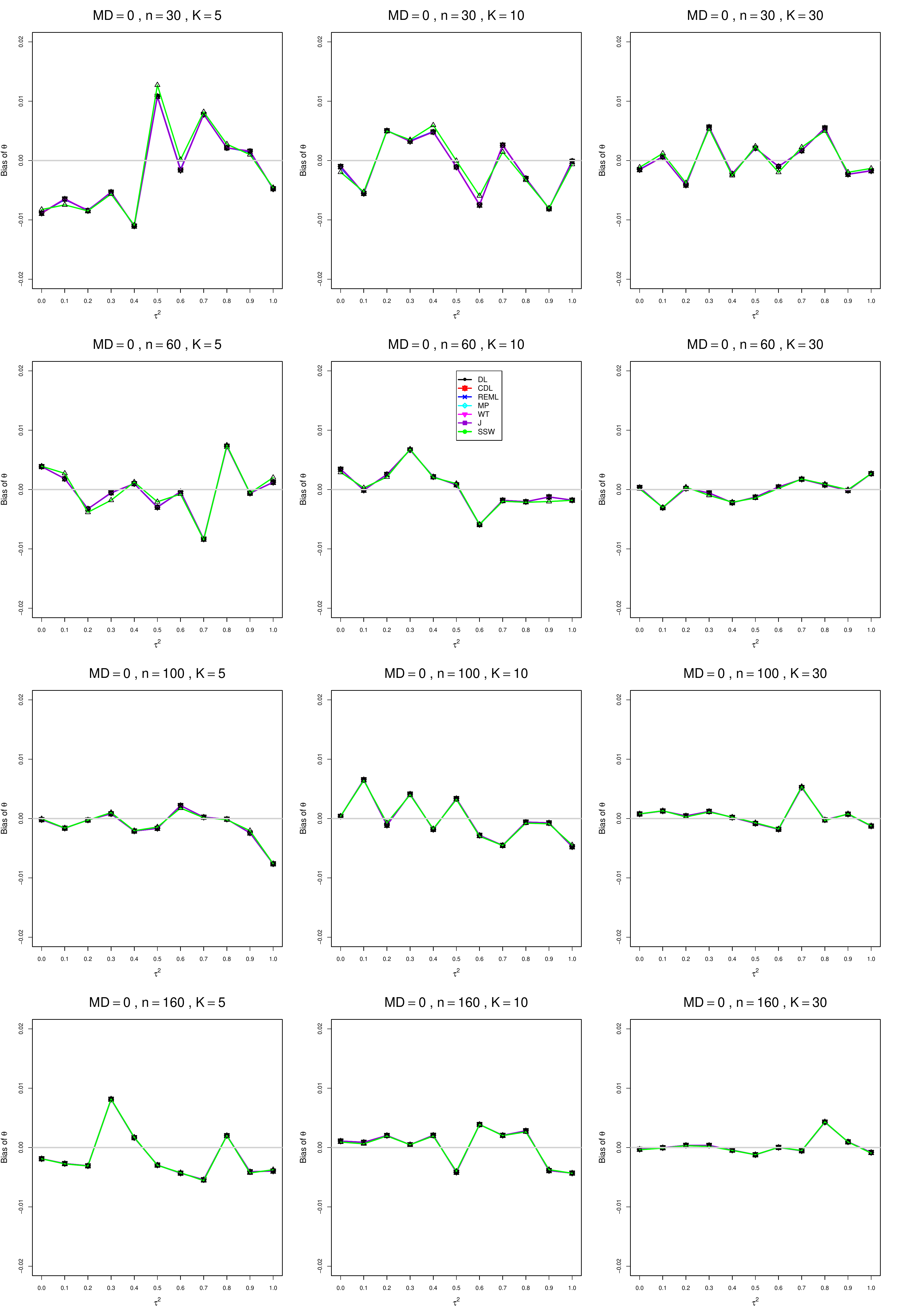}
	\caption{Bias of the estimation of  $\mu = 0$ for between-studies variance $\tau^2 = 0.0(0.1)1.0$, $q=0.5$, $\sigma_C^2=10$, $\sigma_T^2=10$, unequal studies of average size $\bar{n}=30,\;60,\;100,\;160$.
		\label{BiasThetaMD0_S10_10unequal}}
\end{figure}

\begin{figure}[t]\centering
	\includegraphics[scale=0.35]{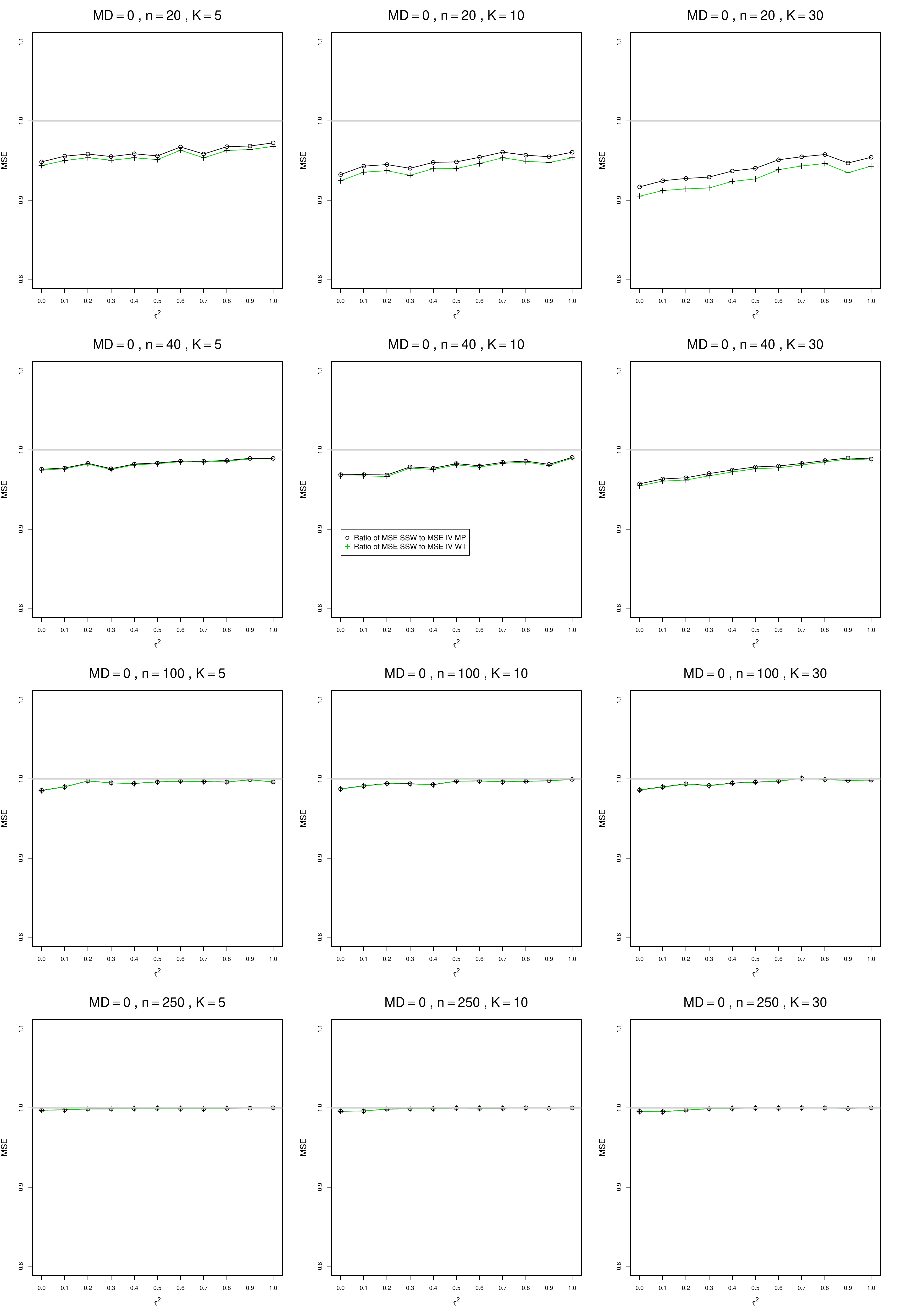}
	\caption{Ratio of mean squared errors of the fixed-weights to mean squared errors of inverse-variance estimator for $\mu=0$, $q=0.5$,  $\sigma_C^2=10$, $\sigma_T^2=10$, $n=20,\;40,\;100,\;250$. 
		\label{RatioOfMSEwithMD0fromMPandCMPSigma2T10andSigma2C10}}
\end{figure}


\begin{figure}[t]
	\centering
	\includegraphics[scale=0.33]{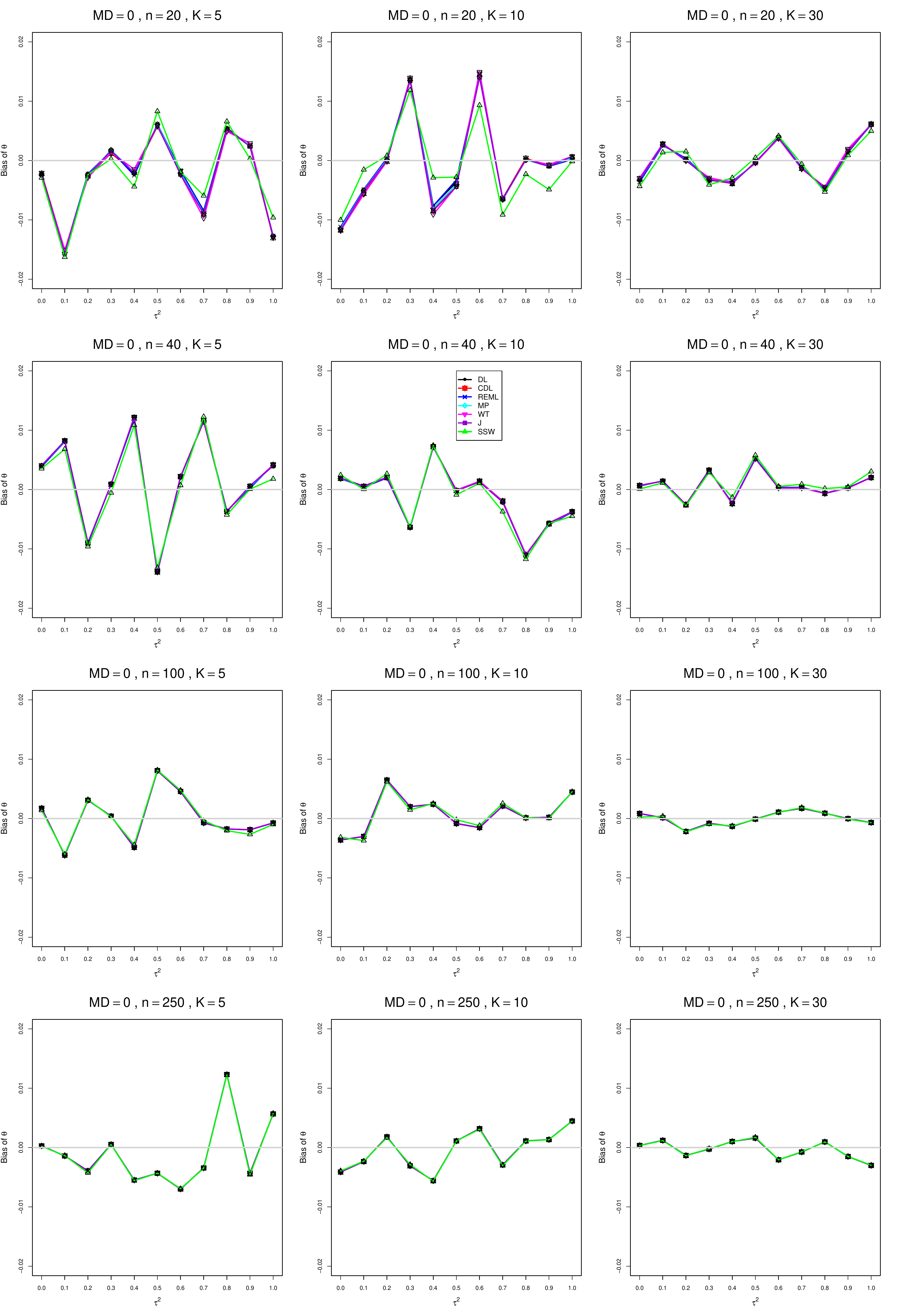}
	\caption{Bias of the estimation of  $\mu = 0$ for between-studies variance $\tau^2 = 0.0(0.1)1.0$, $q=0.75$, $\sigma_C^2=10$, $\sigma_T^2=10$,  equal study sizes $n=20,\;40,\;100,\;250$.
		\label{BiasThetaMD0_S10_10q075}}
\end{figure}

\begin{figure}[t]
	\centering
	\includegraphics[scale=0.33]{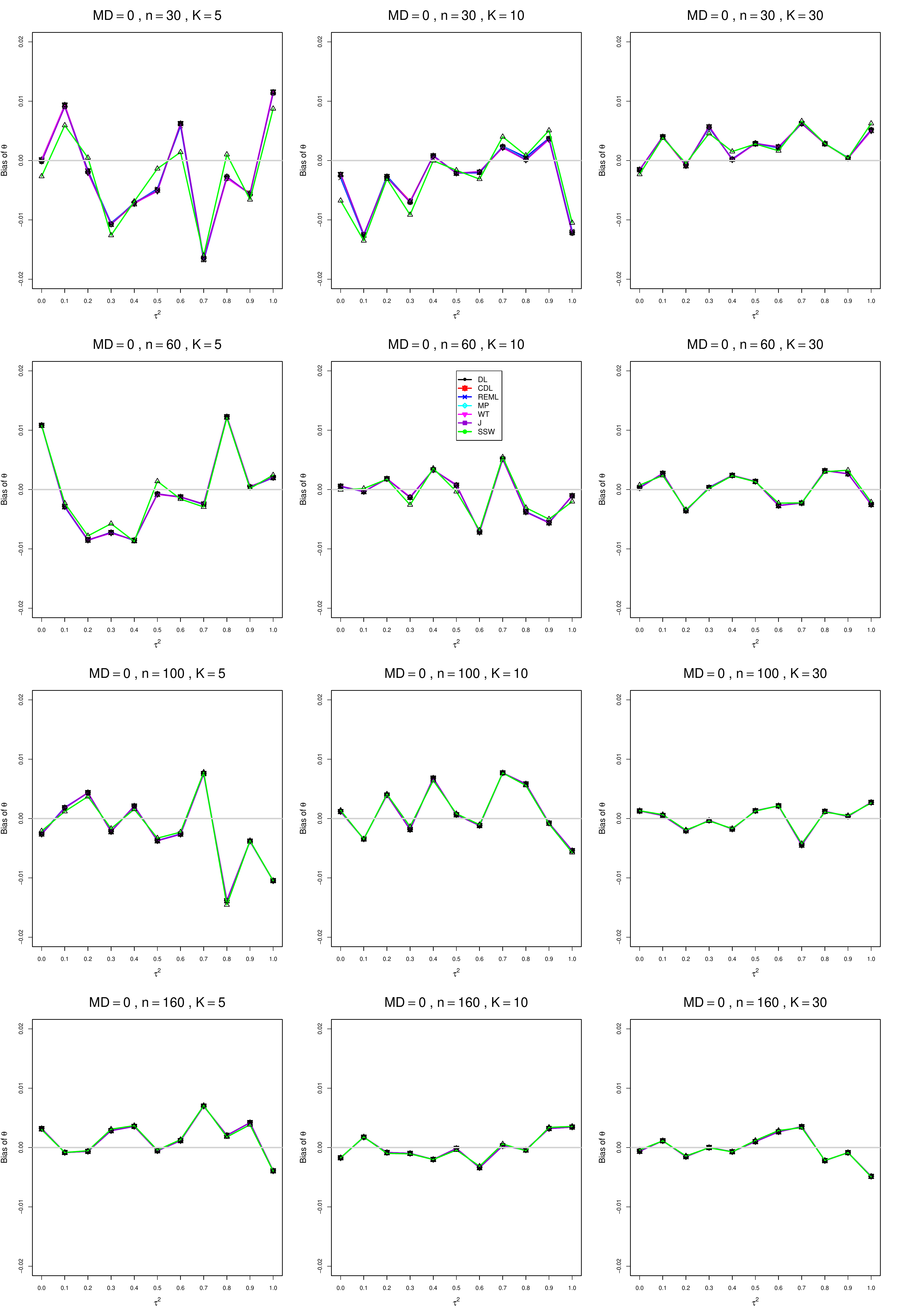}
	\caption{Bias of the estimation of  $\mu = 0$ for between-studies variance $\tau^2 = 0.0(0.1)1.0$, $q=0.75$, $\sigma_C^2=10$, $\sigma_T^2=10$, unequal studies of average size $\bar{n}=30,\;60,\;100,\;160$.
		\label{BiasThetaMD0_S10_10unequalq075}}
\end{figure}

\begin{figure}[t]\centering
	\includegraphics[scale=0.35]{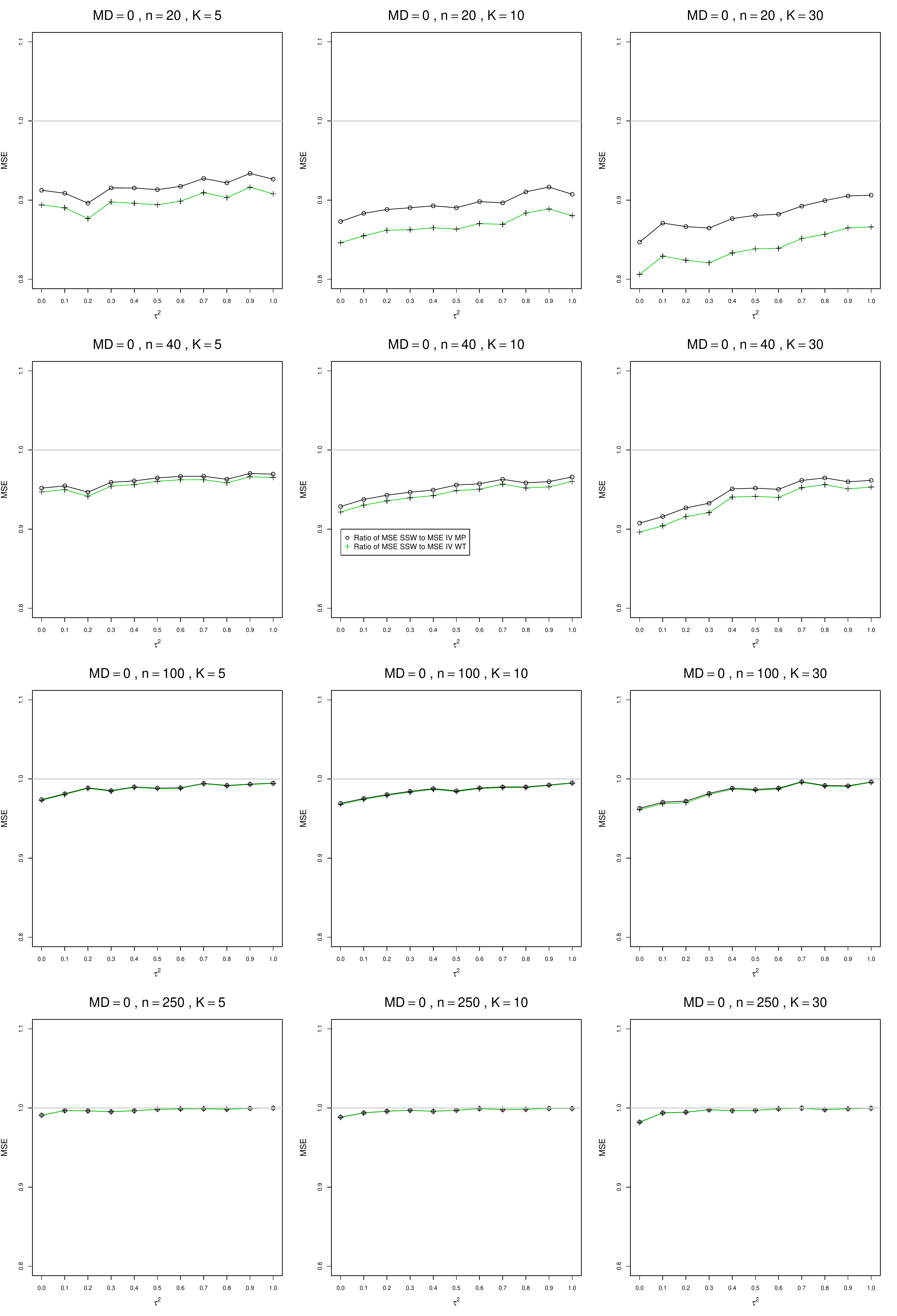}
	\caption{Ratio of mean squared errors of the fixed-weights to mean squared errors of inverse-variance estimator for $\mu=0$, for $q=0.75$,$\sigma_C^2=10$, $\sigma_T^2=10$, $n=20,\;40,\;100,\;250$.
		\label{RatioOfMSEwithMD0q075fromMPandCMPSigma2T10andSigma2C10}}
\end{figure}


\begin{figure}[t]
	\centering
	\includegraphics[scale=0.33]{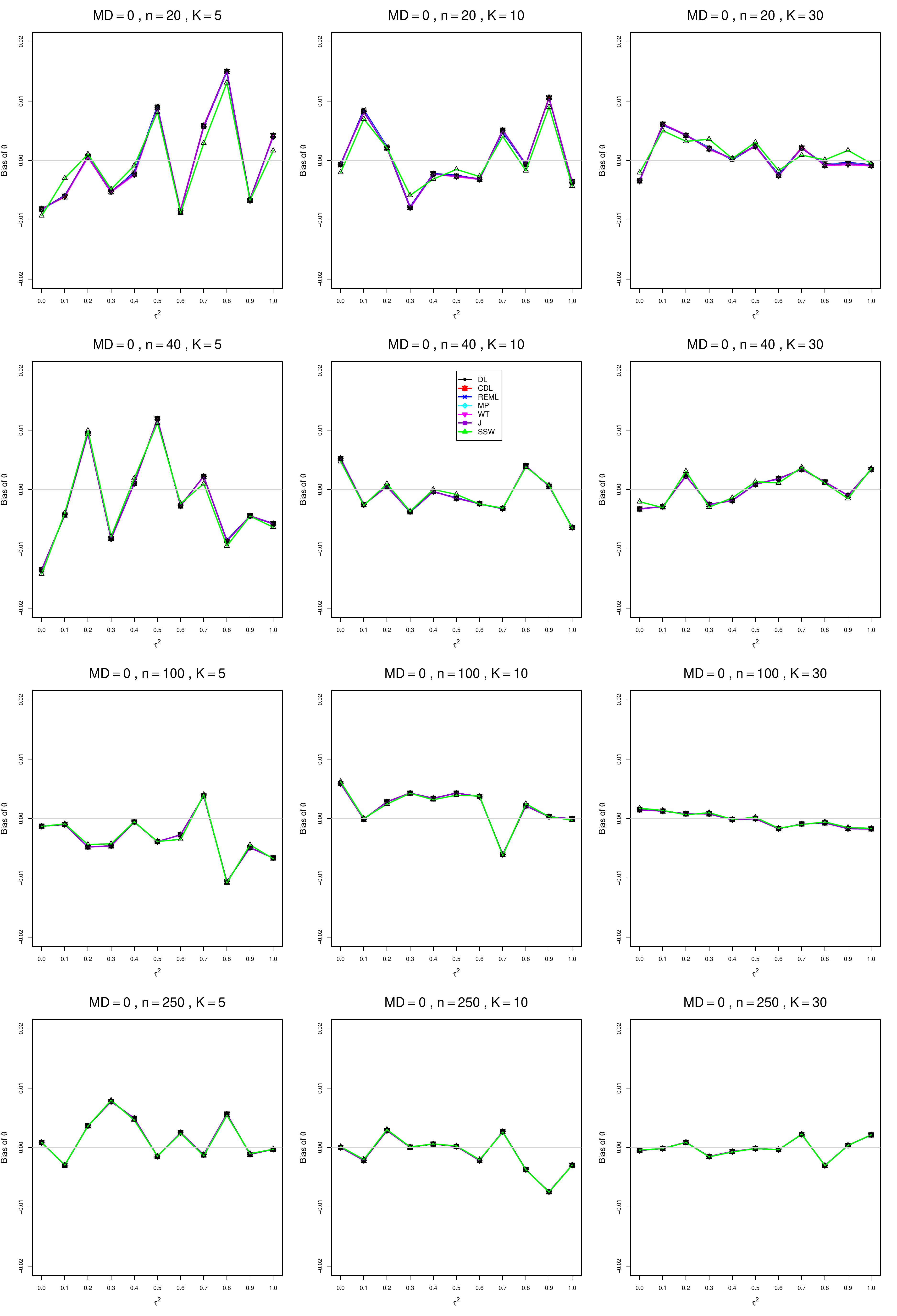}
	\caption{Bias of the estimation of  $\mu = 0$ for between-studies variance $\tau^2 = 0.0(0.1)1.0$, $q=0.5$, $\sigma_C^2=10$, $\sigma_T^2=20$,  equal study sizes $n=20,\;40,\;100,\;250$.
		\label{BiasThetaMD0_S10_20}}
\end{figure}

\begin{figure}[t]
	\centering
	\includegraphics[scale=0.33]{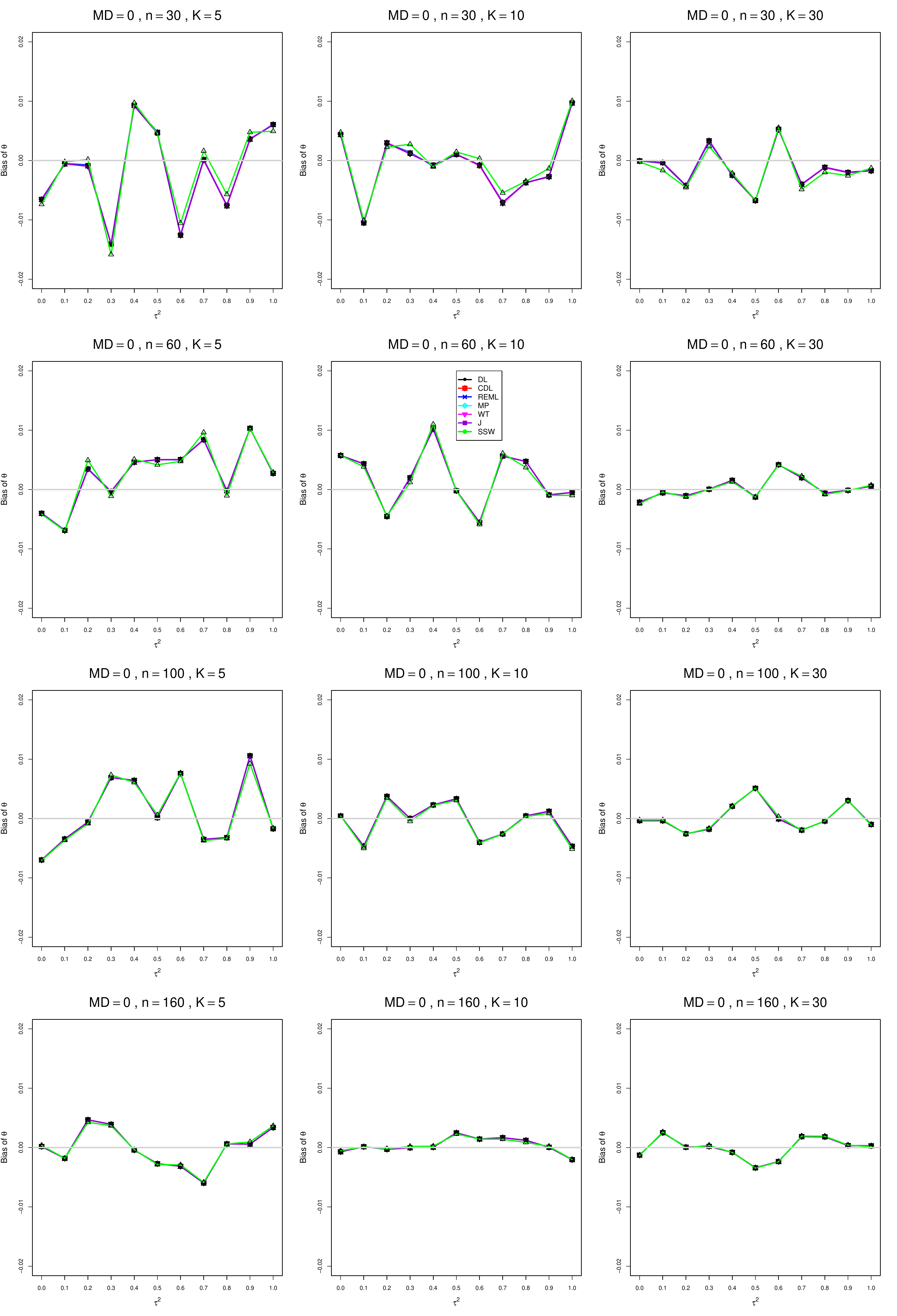}
	\caption{Bias of the estimation of  $\mu = 0$ for between-studies variance $\tau^2 = 0.0(0.1)1.0$, $q=0.5$, $\sigma_C^2=10$, $\sigma_T^2=20$, unequal studies of average size $\bar{n}=30,\;60,\;100,\;160$.
		\label{BiasThetaMD0_S10_20unequal}}
\end{figure}

\begin{figure}[t]\centering
	\includegraphics[scale=0.35]{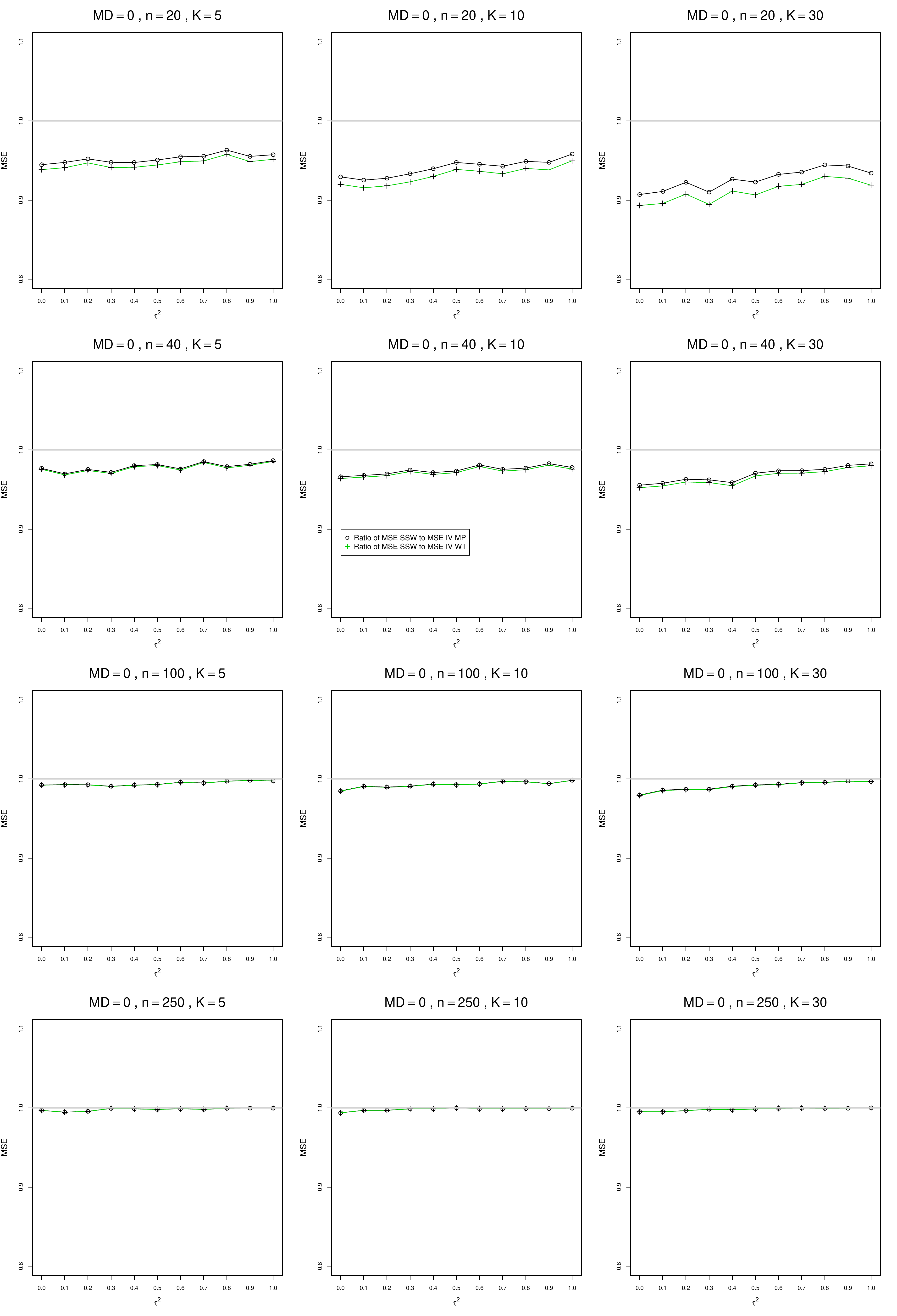}
	\caption{Ratio of mean squared errors of the fixed-weights to mean squared errors of inverse-variance estimator for $\mu=0$, $q=0.5$,  $\sigma_C^2=10$, $\sigma_T^2=20$, $n=20,\;40,\;100,\;250$. 
		\label{RatioOfMSEwithMD0fromMPandCMPSigma2T20andSigma2C10}}
\end{figure}


\begin{figure}[t]
	\centering
	\includegraphics[scale=0.33]{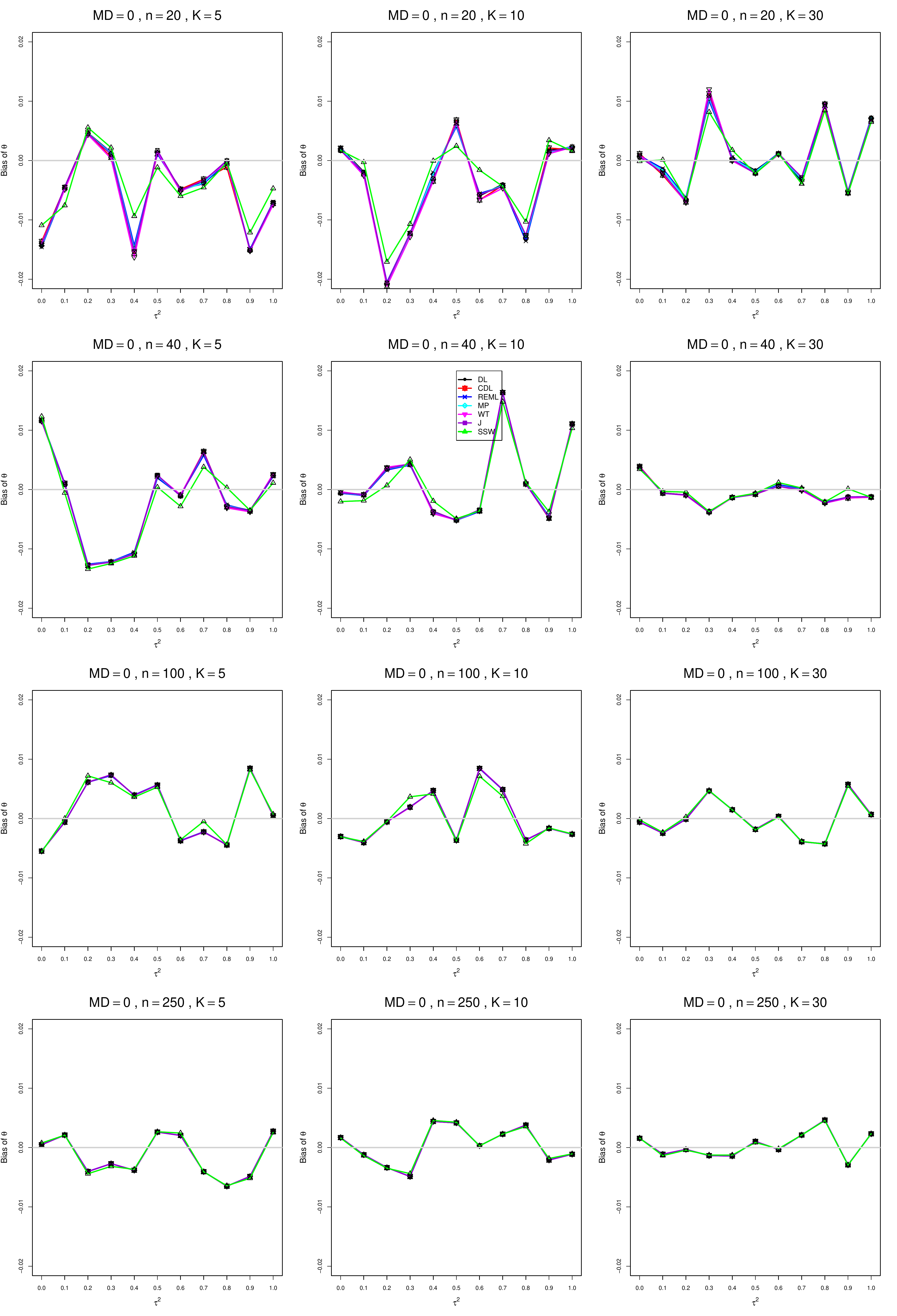}
	\caption{Bias of the estimation of  $\mu = 0$ for between-studies variance $\tau^2 = 0.0(0.1)1.0$, $q=0.75$, $\sigma_C^2=10$, $\sigma_T^2=20$,  equal study sizes $n=20,\;40,\;100,\;250$.
		\label{BiasThetaMD0_S10_20q075}}
\end{figure}

\begin{figure}[t]
	\centering
	\includegraphics[scale=0.33]{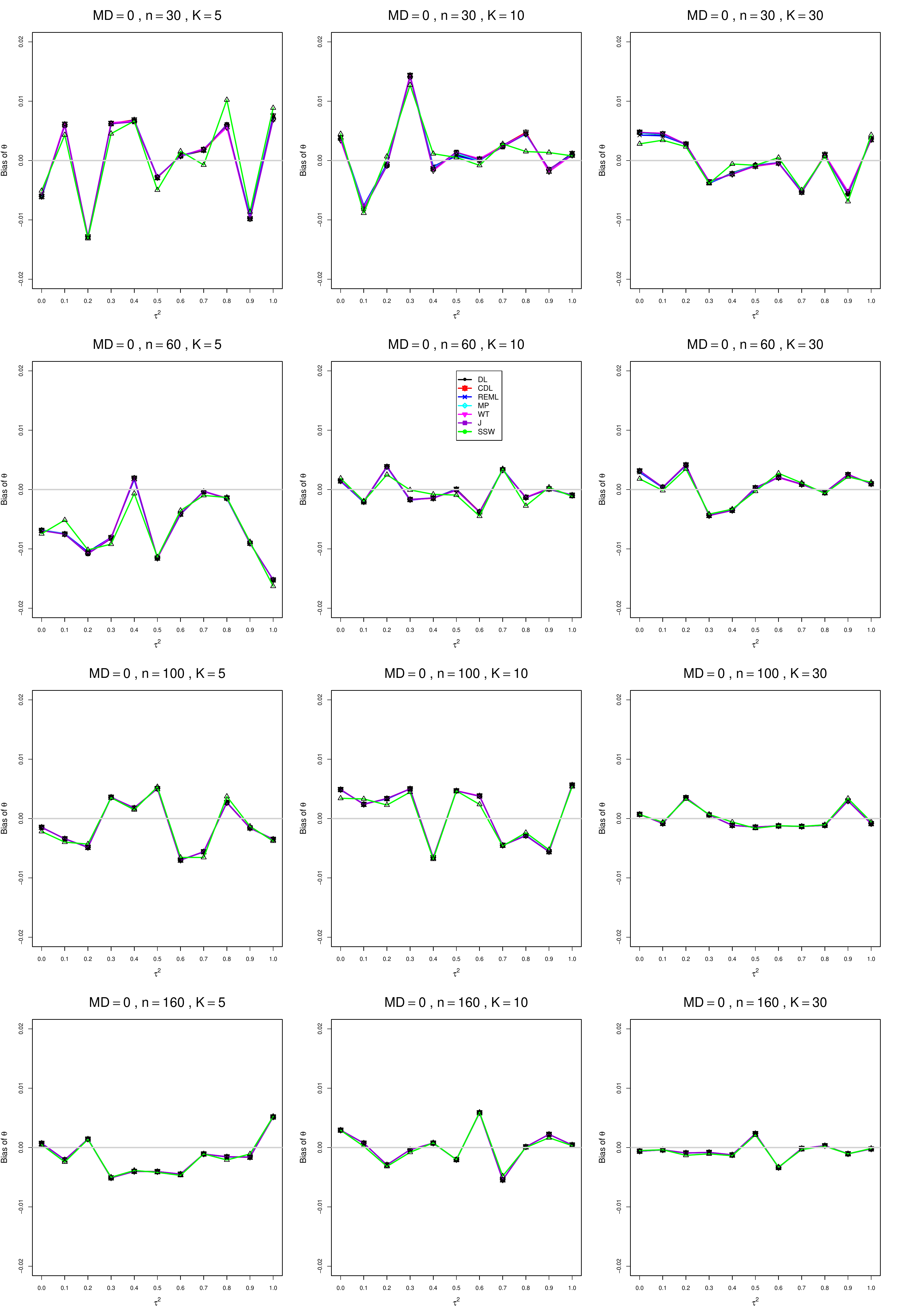}
	\caption{Bias of the estimation of  $\mu = 0$ for between-studies variance $\tau^2 = 0.0(0.1)1.0$, $q=0.75$, $\sigma_C^2=10$, $\sigma_T^2=20$, unequal studies of average size $\bar{n}=30,\;60,\;100,\;160$.
		\label{BiasThetaMD0_S10_20unequalq075}}
\end{figure}

\begin{figure}[t]\centering
	\includegraphics[scale=0.35]{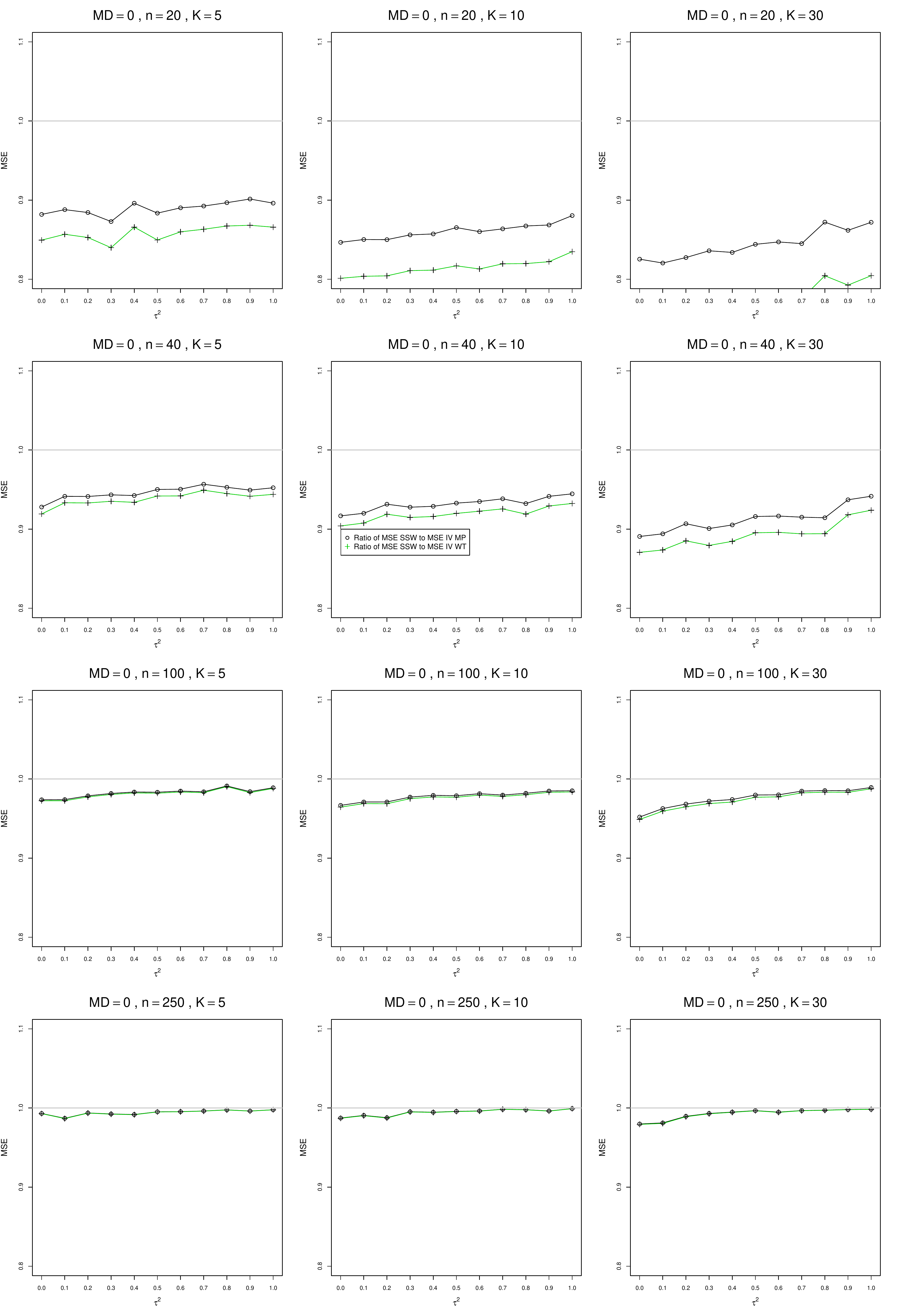}
	\caption{Ratio of mean squared errors of the fixed-weights to mean squared errors of inverse-variance estimator for $\mu=0$, for $q=0.75$,$\sigma_C^2=10$, $\sigma_T^2=20$, $n=20,\;40,\;100,\;250$. 
		\label{RatioOfMSEwithMD0q075fromMPandCMPSigma2T20andSigma2C10}}
\end{figure}

\clearpage
\setcounter{section}{0}
\renewcommand{\thefigure}{B6.\arabic{figure}}
\setcounter{figure}{0}
\setcounter{section}{0}
\section*{B6. Coverage of $\hat{\mu}$ for $\tau^2 = 0.0(0.1)1.0$, $\sigma_{C}^2=10$, $\sigma_{T}^2=10,\;20$.}
For coverage of $\mu$, each figure corresponds to a value of $\mu (= 0, 0.2, 0.5, 1, 2)$, a value of $q (= .5, .75)$, a value of $\tau^2 = 0.0(0.1)1.0$, a value of $\sigma_{C}^2=10$, a value of $\sigma_{T}^2=10,\;20$ , and a set of values of $n$ (= 20, 40, 100, 250) or $\bar{n}$ (= 30, 60, 100, 160).\\
Each figure contains a panel (with $\tau^2$ on the horizontal axis) for each combination of n (or $\bar{n}$) and $K (=5, 10, 30)$.\\
The interval estimators of $mu$ are the companions to the inverse-variance-weighted point estimators
\begin{itemize}
	\item DL (DerSimonian-Laird)
	\item REML (restricted maximum likelihood)
	\item MP (Mandel-Paule)
	\item WT (Corrected Mandel-Paule moment estimator based on Welch-type approximation for Q distribution)
	\item J (Jackson)
	\item CDL (Corrected DerSimonian-Laird)
\end{itemize}
and
\begin{itemize}
	\item HKSJ (Hartung-Knapp-Sidik-Jonkman)
	\item HKSJ WT (HKSJ with WT estimator of $\tau^2$)
	\item SSW (SSW as center and half-width equal to critical value from $t_{K-1}$
\end{itemize}
times estimated standard deviation of SSW with $\hat{\tau}^2$ = $\hat{\tau}^2_{WT}$

\begin{figure}[t]
	\centering
	\includegraphics[scale=0.33]{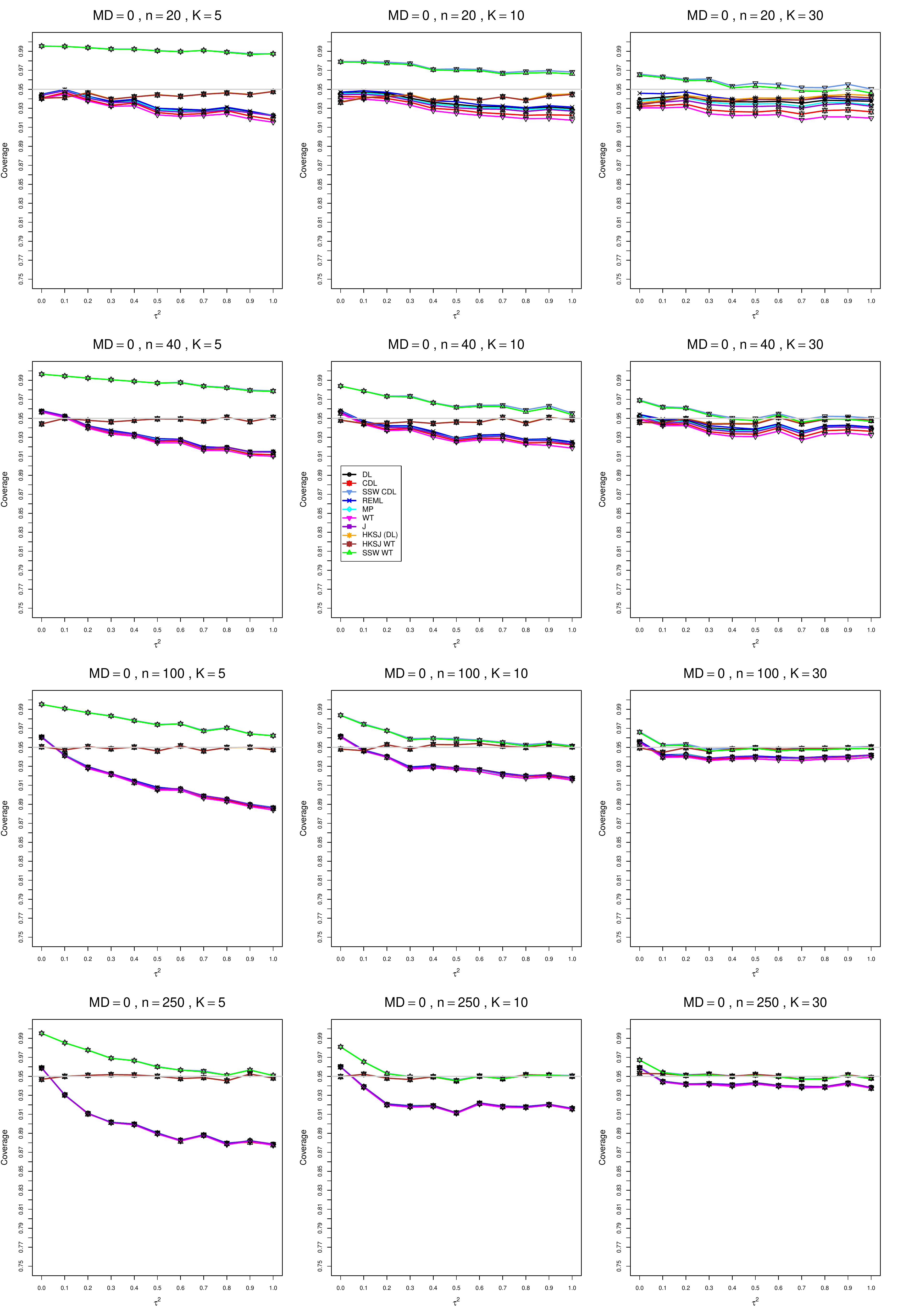}
	\caption{Coverage of 95\% confidence intervals for the $\mu = 0$ for the between-studies variance $\tau^2 = 0.0(0.1)1.0$ for, $q=0.5$, $\sigma_C^2=10$, $\sigma_T^2=10$,  equal study sizes $n=20,\;40,\;100,\;250$.
		\label{CovThetaMD0_S10_10}}
\end{figure}

\begin{figure}[t]
	\centering
	\includegraphics[scale=0.33]{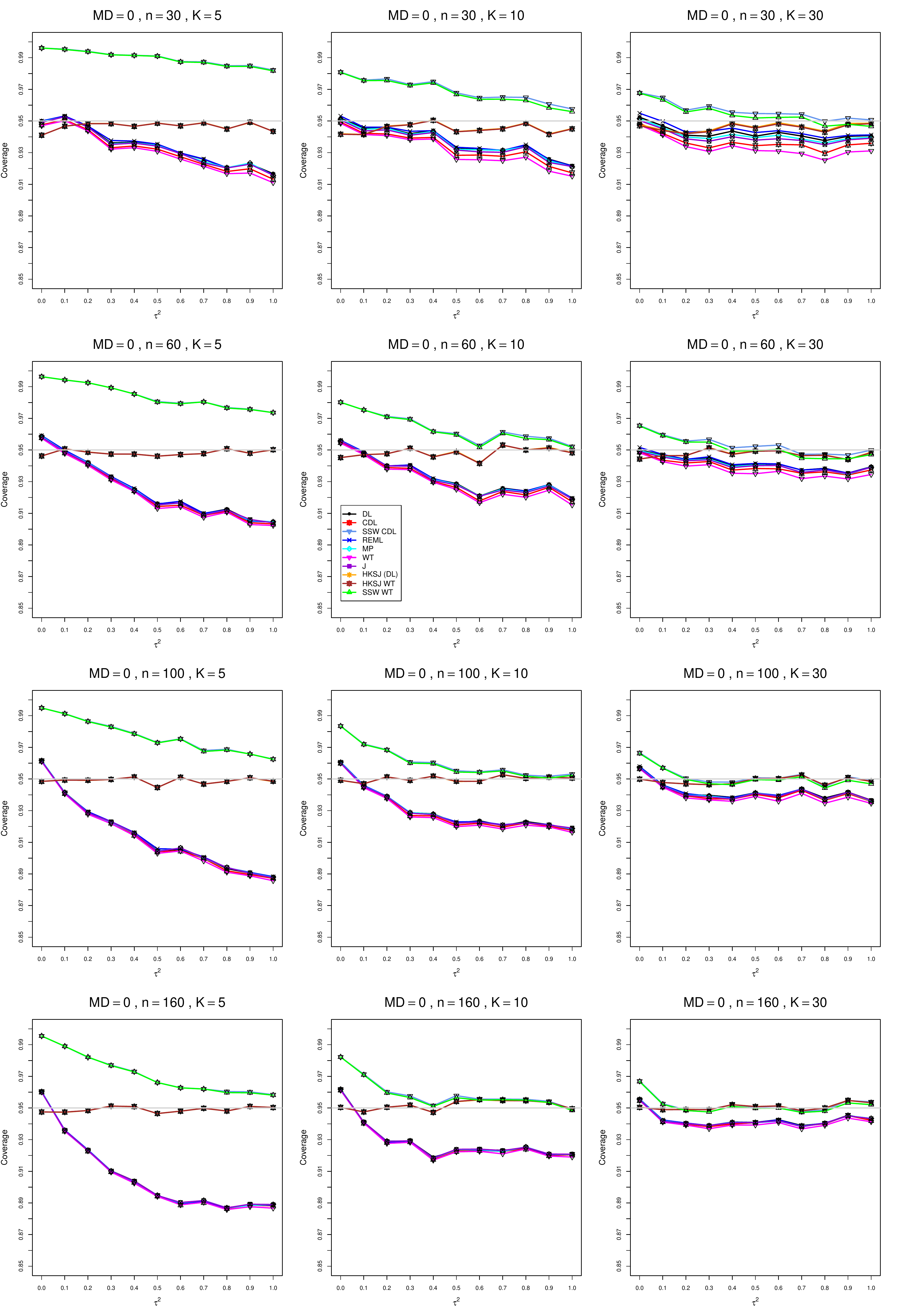}
	\caption{Coverage of 95\% confidence intervals for the $\mu = 0$ for the between-studies variance $\tau^2 = 0.0(0.1)1.0$, $q=0.5$, $\sigma_C^2=10$, $\sigma_T^2=10$, unequal studies of average size $\bar{n}=30,\;60,\;100,\;160$.
		\label{CovThetaMD0_S10_10unequal}}
\end{figure}


\begin{figure}[t]
	\centering
	\includegraphics[scale=0.33]{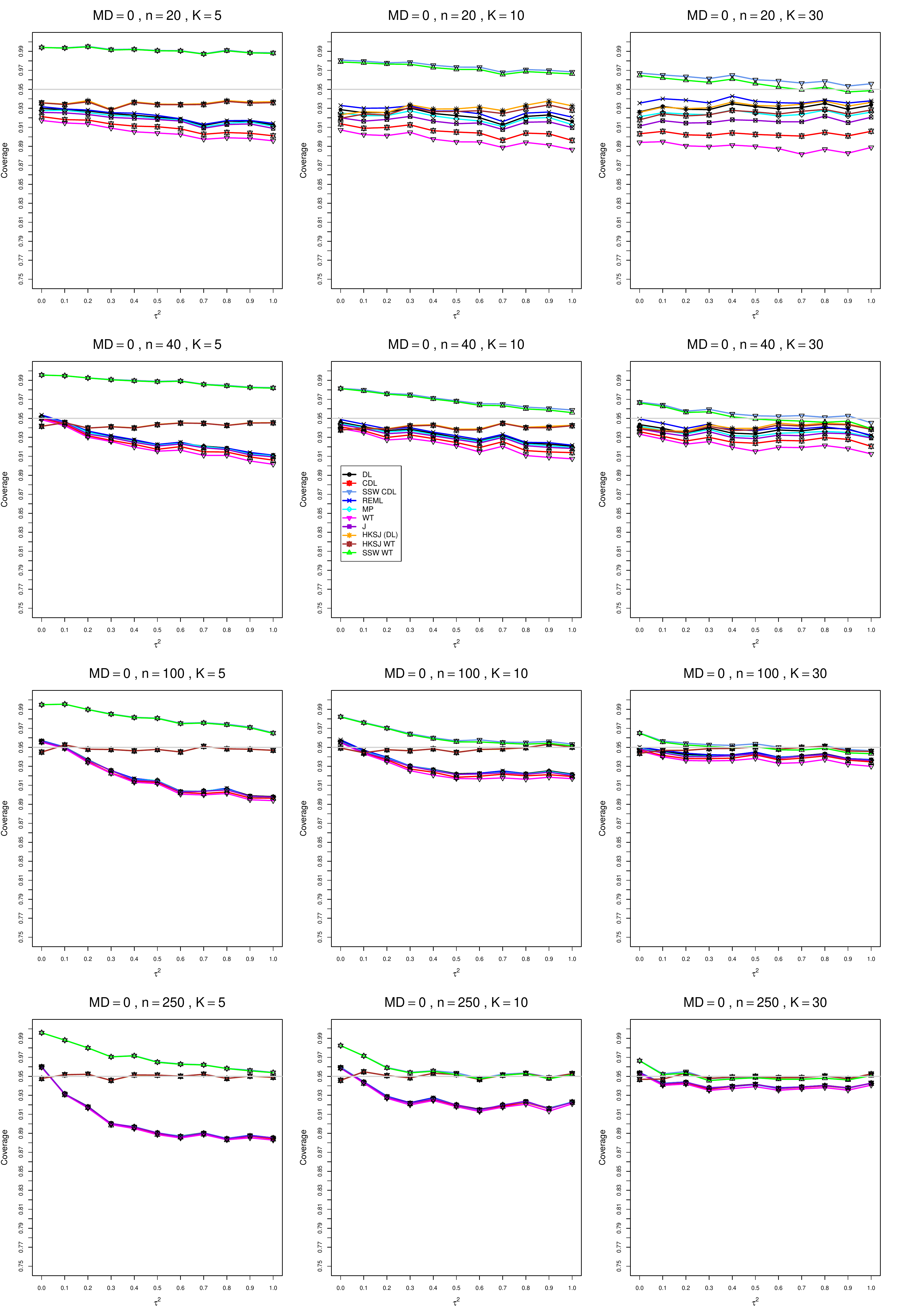}
	\caption{Coverage of 95\% confidence intervals for the $\mu = 0$ for the  between-studies variance $\tau^2 = 0.0(0.1)1.0$, $q=0.75$, $\sigma_C^2=10$, $\sigma_T^2=10$,  equal study sizes $n=20,\;40,\;100,\;250$.
		\label{CovThetaMD0_S10_10q075}}
\end{figure}

\begin{figure}[t]
	\centering
	\includegraphics[scale=0.33]{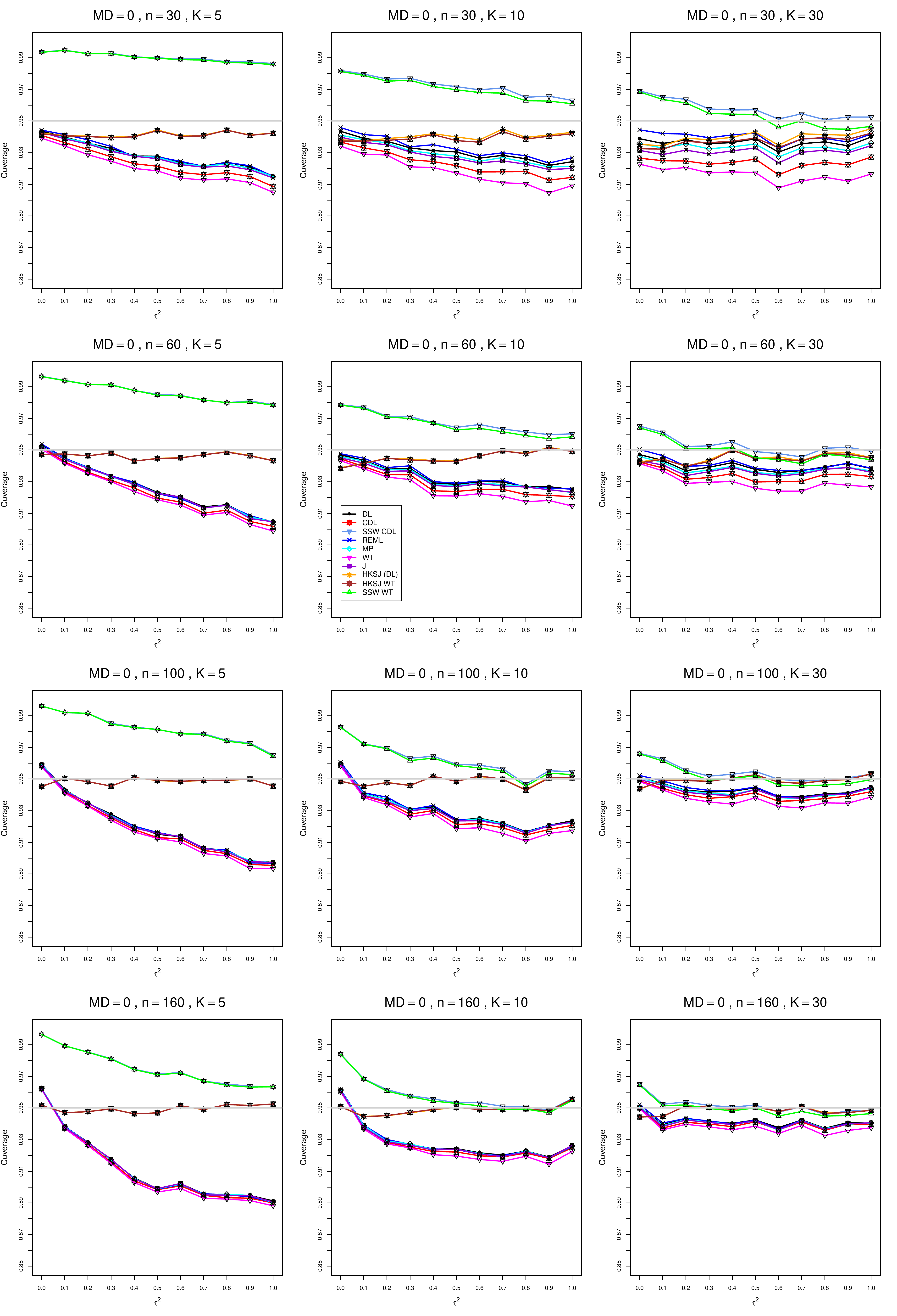}
	\caption{Coverage of 95\% confidence intervals for the $\mu = 0$ for the  between-studies variance $\tau^2 = 0.0(0.1)1.0$, $q=0.75$, $\sigma_C^2=10$, $\sigma_T^2=10$, unequal studies of average size $\bar{n}=30,\;60,\;100,\;160$.
		\label{CovThetaMD0_S10_10unequalq075}}
\end{figure}

\clearpage

\begin{figure}[t]
	\centering
	\includegraphics[scale=0.33]{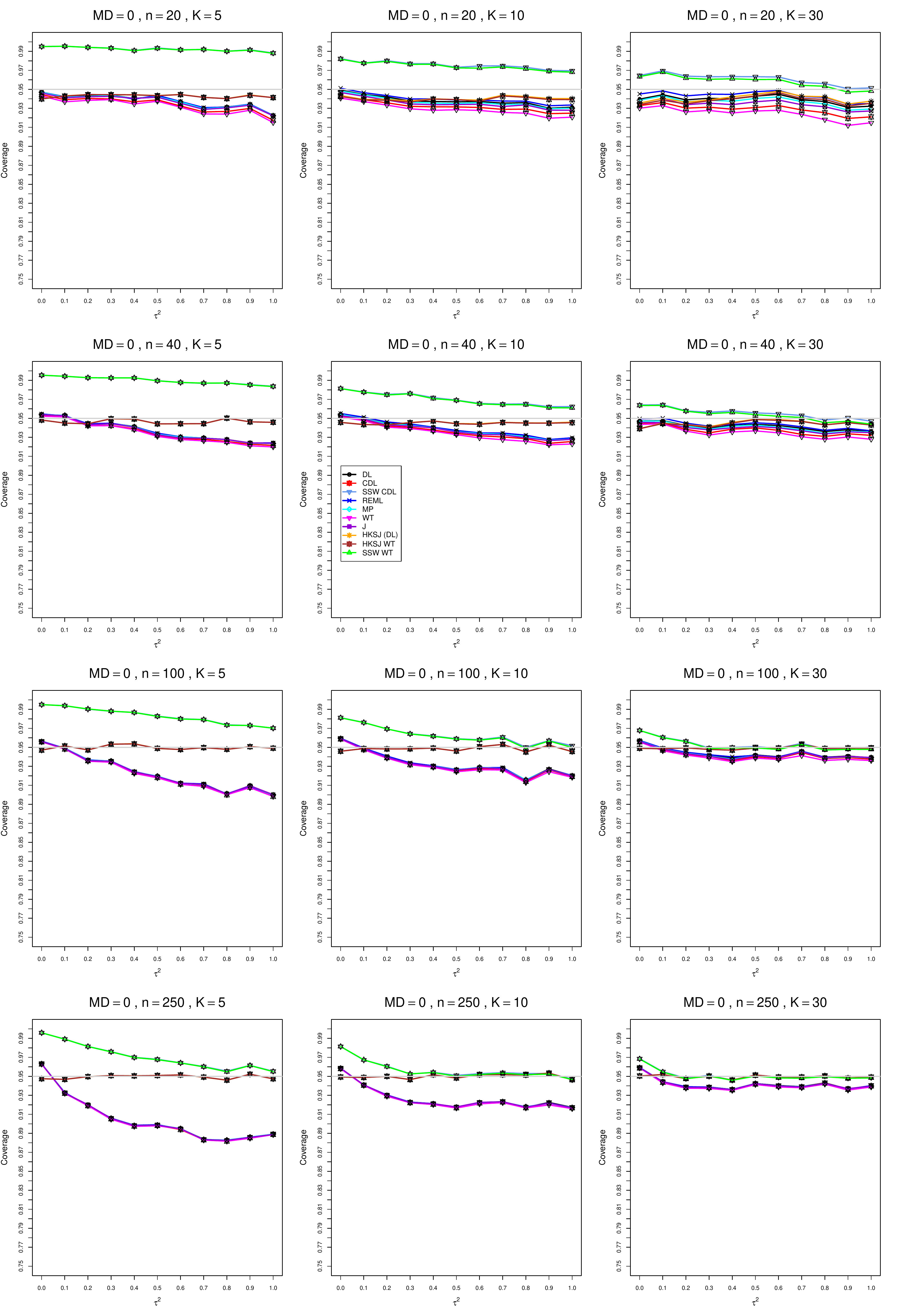}
	\caption{Coverage of 95\% confidence intervals for the $\mu = 0$ for the between-studies variance $\tau^2 = 0.0(0.1)1.0$ for, $q=0.5$, $\sigma_C^2=10$, $\sigma_T^2=20$,  equal study sizes $n=20,\;40,\;100,\;250$.
		\label{CovThetaMD0_S10_20}}
\end{figure}

\begin{figure}[t]
	\centering
	\includegraphics[scale=0.33]{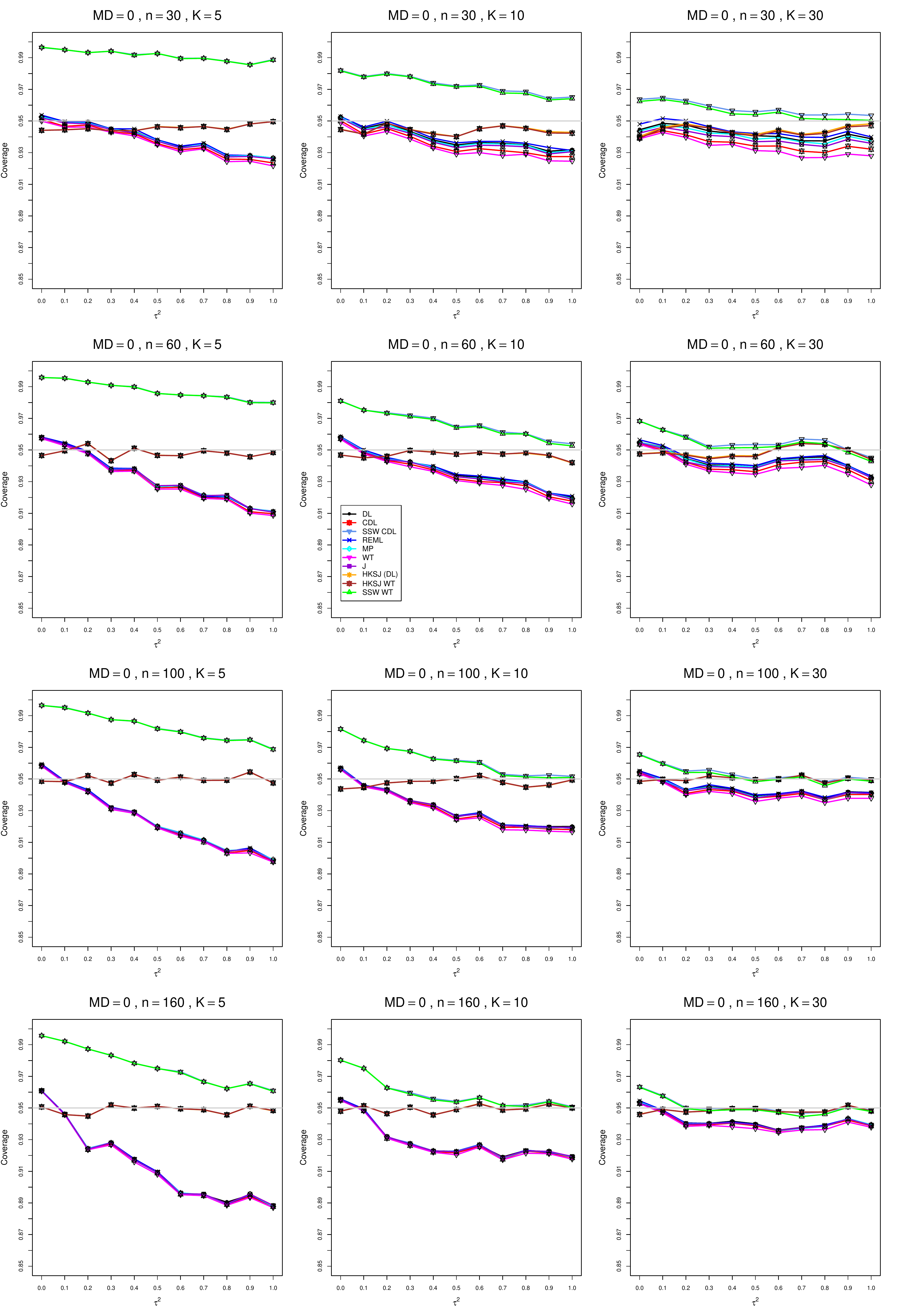}
	\caption{Coverage of 95\% confidence intervals for the $\mu = 0$ for the between-studies variance $\tau^2 = 0.0(0.1)1.0$, $q=0.5$, $\sigma_C^2=10$, $\sigma_T^2=20$, unequal studies of average size $\bar{n}=30,\;60,\;100,\;160$.
		\label{CovThetaMD0_S10_20unequal}}
\end{figure}


\begin{figure}[t]
	\centering
	\includegraphics[scale=0.33]{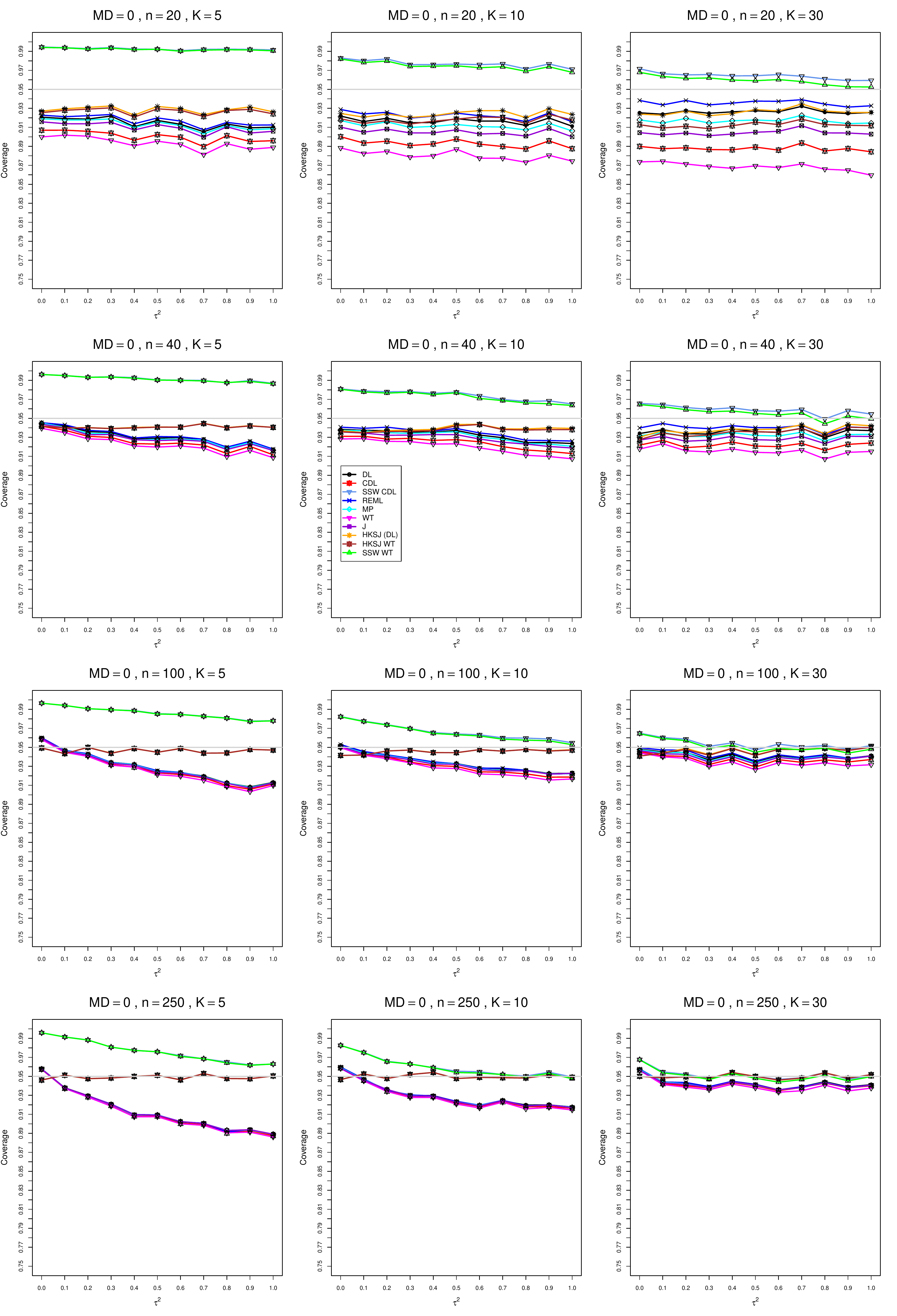}
	\caption{Coverage of 95\% confidence intervals for the $\mu = 0$ for the  between-studies variance $\tau^2 = 0.0(0.1)1.0$, $q=0.75$, $\sigma_C^2=10$, $\sigma_T^2=20$,  equal study sizes $n=20,\;40,\;100,\;250$.
		\label{CovThetaMD0_S10_20q075}}
\end{figure}

\begin{figure}[t]
	\centering
	\includegraphics[scale=0.33]{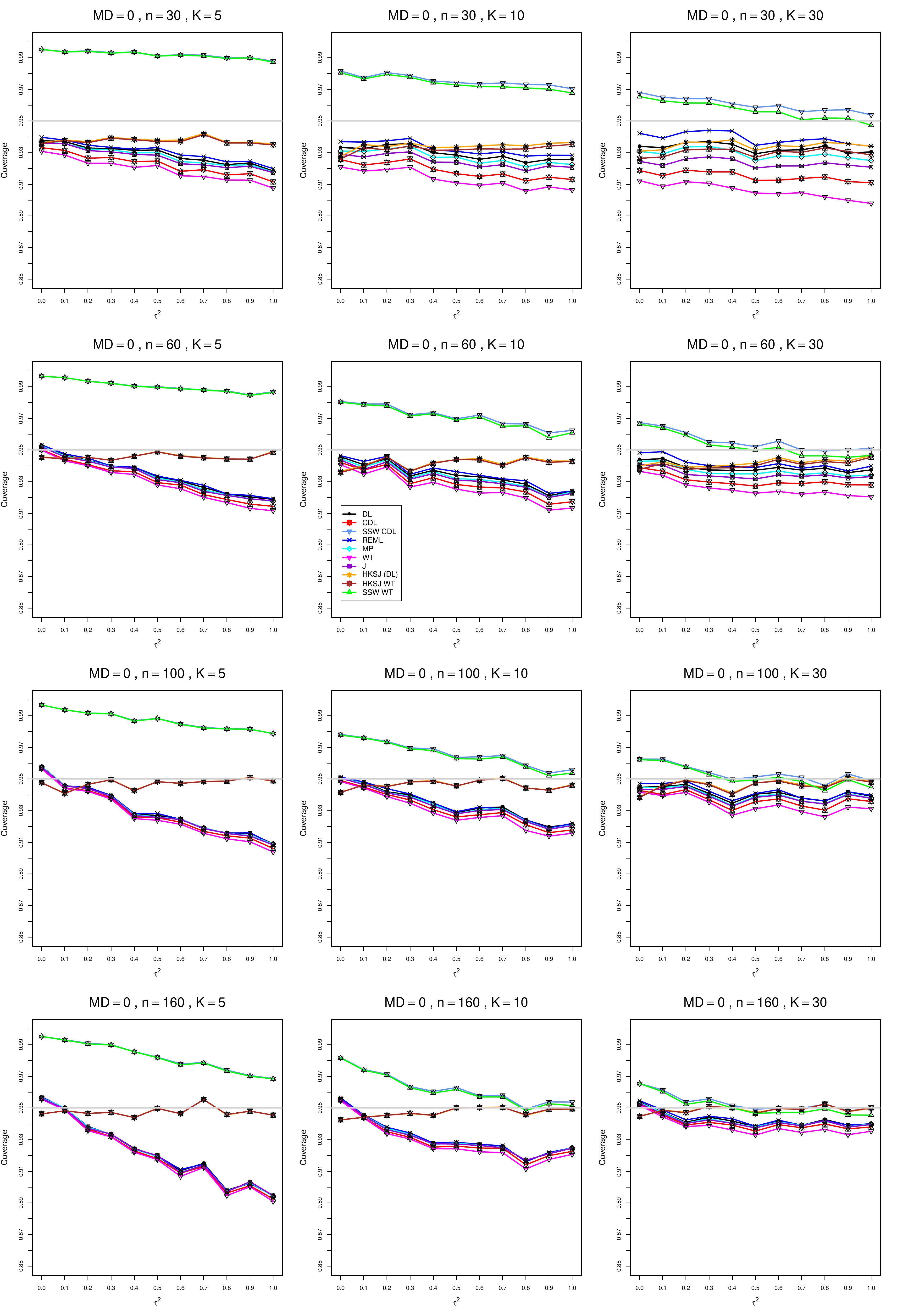}
	\caption{Coverage of 95\% confidence intervals for the $\mu = 0$ for the  between-studies variance $\tau^2 = 0.0(0.1)1.0$, $q=0.75$, $\sigma_C^2=10$, $\sigma_T^2=20$, unequal studies of average size $\bar{n}=30,\;60,\;100,\;160$.
		\label{CovThetaMD0_S10_20unequalq075}}
\end{figure}

\end{document}